\RequirePackage{fix-cm}
\documentclass[twocolumn,epjc3,twoside]{svjour3}

\makeatletter
\let\cl@chapter\undefined
\makeatletter

\smartqed  %
\RequirePackage{graphicx}
\usepackage[aboveskip=1pt,belowskip=0pt,justification=centering,hypcap=true]{subcaption}
\usepackage{tocvsec2}
\usepackage{pdflscape}
\usepackage{amsfonts,amssymb}
\usepackage{amsmath,bbm}
\usepackage{colonequals}
\usepackage[shortlabels]{enumitem}
\usepackage{makecell}
\usepackage{multirow}
\usepackage{xcolor}
\usepackage{dpfloat}
\usepackage{xspace}
\usepackage{textcomp}
\usepackage{hyperref}
\usepackage{rotating}
\usepackage{cite}
\usepackage[switch]{lineno}
\hypersetup{
    colorlinks,
    linkcolor={red!50!black},
    citecolor={blue!50!black},
    urlcolor={blue!80!black}
}
\usepackage[T1]{fontenc}
\usepackage[capitalise]{cleveref}
\usepackage{adjustbox}

\usepackage{footmisc}

\crefname{section}{Sec.}{Secs.}
\Crefname{section}{Section}{Sections}

\crefname{subsection}{Sec.}{Secs.}
\Crefname{subsection}{Section}{Sections}

\newcommand{\pink}[1]{{\textcolor[rgb]{1,0,0.5}{#1}}}

\newcommand{\tobedone}[1]{}

\newcommand{\comment}[2]{\tobedone{{\small \pink{[Comment (by {#1}):\ {#2}]}}}}

\definecolor{decompred}{rgb}{1, 0, 0}
\definecolor{decompgreen}{rgb}{0.0, 0.5, 0.0}
\definecolor{decompyellow}{rgb}{0.98, 0.63, 0.0}
\definecolor{decompblue}{rgb}{0, 0, 1}
\definecolor{decompgrey}{rgb}{0.5, 0.5, 0.5}

\crefformat{appendix}{#2#1#3}
\makeatletter
\def\@mkboth#1#2{}
\newlength\appendixwidth
\preto\appendix{\addtocontents{toc}{\protect\patchl@section}}
\newcommand{\patchl@section}{%
  \settowidth{\appendixwidth}{\textbf{Appendix }}%
  \addtolength{\appendixwidth}{1.5em}%
  \patchcmd{\l@section}{1.5em}{\appendixwidth}{}{}%
}
\makeatother

\DeclareRobustCommand{\mposd}{\ensuremath{\textsc{Mpos}{\delta}}\xspace}
\DeclareRobustCommand{\mpos}{\ensuremath{\textsc{Mpos}}\xspace}
\DeclareRobustCommand{\pos}{\ensuremath{\textsc{Pos}}\xspace}
\DeclareRobustCommand{\dpos}{\ensuremath{\textsc{Dpos}}\xspace}
\DeclareRobustCommand{\krk}{\ensuremath{\textsc{Krk}}\xspace}
\DeclareRobustCommand{\krkdy}{\ensuremath{\textsc{KrkDy}}\xspace}

\DeclareRobustCommand{\dis}{\ensuremath{\textsc{Dis}}\xspace}
\DeclareRobustCommand{\phys}{\ensuremath{\textsc{Phys}}\xspace}
\DeclareRobustCommand{\aversa}{\ensuremath{\textsc{Aversa}}\xspace}
\DeclareRobustCommand{\rFS}{\ensuremath{\mathrm{FS}}\xspace}

\DeclareRobustCommand{\ctnlo}{\texttt{CT18NLO}\xspace}
\DeclareRobustCommand{\mshtnlo}{\texttt{MSHT20nlo}\xspace}
\DeclareRobustCommand{\nnpdf}{\texttt{NNPDF40}\xspace}
\DeclareRobustCommand{\nnpdfmc}{\texttt{NNPDF40MC}\xspace}

\DeclareRobustCommand{\alphas}{\alpha_{\mathrm{s}}}

\DeclareRobustCommand{\msbar}{\ensuremath{{\overline{\mathrm{MS}}}}\xspace}
\DeclareRobustCommand{\order}[1]{\mathcal{O}\!\left(#1\right)}
\DeclareRobustCommand{\muf}{\mu_{\rF}}
\DeclareRobustCommand{\mur}{\mu_{\rR}}

\DeclareMathOperator{\finite}{fin}
\DeclareMathOperator{\cumulant}{c}

\DeclareRobustCommand{\dd}{\ensuremath{\mathrm{d}}}
\DeclareRobustCommand{\rF}{\ensuremath{\mathrm{F}}}
\DeclareRobustCommand{\rK}{\ensuremath{\mathrm{K}}}
\DeclareRobustCommand{\rP}{\ensuremath{\mathrm{P}}}

\DeclareRobustCommand{\pqq}{\ensuremath{p_{qq}(z)}}
\DeclareRobustCommand{\pqg}{\ensuremath{p_{qg}(z)}}
\DeclareRobustCommand{\pgq}{\ensuremath{p_{gq}(z)}}
\DeclareRobustCommand{\pgg}{\ensuremath{p_{gg}(z)}}

\DeclareRobustCommand{\ensuremathrm}[1]{\ensuremath{\mathrm{#1}}\xspace}

\DeclareRobustCommand{\rF}{\ensuremathrm{F}}

\DeclareRobustCommand{\rK}{\ensuremathrm{K}}

\DeclareRobustCommand{\rR}{\ensuremathrm{R}}

\DeclareRobustCommand{\rT}{\ensuremathrm{T}}

\DeclareRobustCommand{\rFS}{\mathrm{FS}}

\DeclareRobustCommand{\alphas}{\alpha_{\mathrm{s}}}

\DeclareRobustCommand{\nf}{\ensuremath{n_{f}}}
\DeclareRobustCommand{\cf}{\ensuremath{C_{F}}}
\DeclareRobustCommand{\ca}{\ensuremath{C_{A}}}

\DeclareRobustCommand{\tr}{\ensuremath{T_{R}}}
\DeclareRobustCommand{\tR}{\ensuremath{T_{R}}}

\DeclareRobustCommand{\dd}{\ensuremath{\mathrm{d}}}

\DeclareRobustCommand{\GeV}{\ensuremathrm{GeV}\xspace}

\DeclareRobustCommand{\pqq}{p_{qq}(z)}
\DeclareRobustCommand{\pqg}{p_{qg}(z)}
\DeclareRobustCommand{\pgq}{p_{gq}(z)}
\DeclareRobustCommand{\pgg}{p_{gg}(z)}

\DeclareRobustCommand{\pt}{\ensuremath{p_\rT}\xspace}

\DeclareRobustCommand{\ptl}[1]{\ensuremath{p_{\rT}^{\ell_{#1}}}\xspace}
\DeclareRobustCommand{\pttau}[1]{\ensuremath{p_{\rT}^{\tau_{#1}}}\xspace}

\newcommand{\atlas}{\textsc{Atlas}\xspace}
\newcommand{\cms}{\textsc{CMS}\xspace}

\newcommand{\lhapdf}{\textsc{Lhapdf}\xspace}

\newcommand{\herwigseven}{\textsf{Herwig~7}\xspace}

\DeclareRobustCommand{\deductor}{\textsc{Deductor}\xspace}

\mathchardef\mhyphen="2D

\DeclareRobustCommand{\lo}{\ensuremath{\text{LO}}\xspace}

\DeclareRobustCommand{\nklo}{\ensuremath{\mathrm{N^{k}LO}}\xspace}

\newcommand{\qbar}{{\bar q}}

\newcommand{\krknlo}{{\textsf{KrkNLO}}\xspace}

\usepackage[normalem]{ulem}
\journalname{Eur. Phys. J. C}

\begin{document}

\title{Factorisation schemes for proton PDFs}

\author{S. Delorme\thanksref{e1,addr1,addr2}
        \and
        A. Kusina\thanksref{e2,addr1} 
        \and
        A. Si\'odmok\thanksref{e3,addr3,addr4} 
        \and
        J. Whitehead\thanksref{e4,addr1,addr3} 
}

\thankstext{e1}{e-mail: stephane.delorme@us.edu.pl}
\thankstext{e2}{e-mail: aleksander.kusina@ifj.edu.pl}
\thankstext{e3}{e-mail: andrzej.siodmok@uj.edu.pl}
\thankstext{e4}{e-mail: james.whitehead@uj.edu.pl}

\institute{Institute of Nuclear Physics, Polish Academy of Sciences,
31-342 Kraków, Poland \label{addr1}
\and
Institute of Physics, University of Silesia, Katowice, Poland \label{addr2}
\and
Jagiellonian University, ul. prof. Stanislawa Łojasiewicza 11, 30-348 Kraków, Poland \label{addr3}
\and
Theoretical Physics Department, CERN, 1211 Geneva 23, Switzerland \label{addr4}
}

\preprintnumbers{CERN-TH-2025-037/IFJPAN-IV-2025-3/MCNET-25-02}

\date{Submitted: 30 January 2025 / Revised: 4 May 2025}

\maketitle

\begin{abstract}
Beyond leading-order, perturbative QCD requires a choice of factorisation scheme
to define the
parton distribution functions (PDFs) and hard-process cross-section.
The modified minimal-subtraction (\msbar) scheme 
has long been adopted as the default choice due to its simplicity.
Alternative schemes have been proposed with specific purposes,
including, recently, PDF positivity and NLO parton-shower matching.
In this paper we assemble these schemes in a common notation for the first time.
We perform a detailed comparison of their features,
both analytically and numerically,
and estimate the resulting factorisation-scheme uncertainty
for LHC phenomenology.


\end{abstract}

\tableofcontents

\section{Introduction}
\label{intro}

Since the first next-to-leading-order (NLO) QCD calculations \cite{Altarelli:1978id,Bardeen:1978yd,Altarelli:1979ub,Curci:1980uw,Furmanski:1981cw,Politzer:1981vc,Stevenson:1986cu,Ramalho:1983aq,Sterman:1986aj,Aurenche:1986fe,Aurenche:1986ff,Aversa:1988mm,Aversa:1988vb,Chyla:1989ik}
it has been recognised that the factorisation of hadronic cross-sections
into a perturbative (short-distance) partonic cross-section,
and universal (long-distance) parton distribution functions
is not unique beyond leading-order in QCD.
The choice of which terms at each perturbative order to treat
as universal defines a PDF factorisation scheme (FS).
PDFs in different schemes are related to each other by convolution
with a transformation kernel, which is defined
perturbatively up to the relevant order in $\alphas$.

While NLO calculations using different schemes have the same formal perturbative accuracy,
their predictions may nevertheless differ through the inclusion of different higher-order terms.
These higher-order terms, especially when logarithmic, may be numerically large,
and their effects significant.

Early calculations favoured the DIS scheme~\cite{Altarelli:1978id},
defined to absorb all perturbative corrections
to the deep inelastic scattering cross-section into the parton distribution functions.
Subsequently, the modified-minimal-subtraction (\msbar) scheme \cite{Bardeen:1978yd} came
into favour due to its simplicity.
Today, \msbar is the `default' factorisation scheme
in use for the overwhelming majority of QCD calculations and predictions.

In the intervening period,
a number of alternative factorisation schemes have been proposed with a range of %
motivations.
These include recent proposals for a new scheme 
to enforce the positivity of PDFs \cite{Candido:2020yat},
schemes for NLO calculations matched to parton showers 
\cite{Jadach:2015mza,Jadach:2016acv,Jadach:2016qti,Jadach:2011cr,Sarmah:2024hdk,Nagy:2022bph},
and a scheme aiming to separate short- from long-distance corrections
using arguments about dimensional regularisation and confinement
\cite{Oliveira:2013aug}.

In this paper, we summarise the different schemes, their motivations,
and collect their definitions in relation to \msbar PDFs in 
consistent conventions using a unified notation.
We compare the schemes both analytically and numerically,
identify common elements,
assess their numerical significance
at the level of PDFs and coefficient functions, and consider
the possible impact of the factorisation-scheme choice for
practical phenomenology.
We conclude by identifying a small number of discrete
parameters which span much of the deviation
between the schemes.

\section{Factorisation schemes}
\label{sec:facSchemes}
We work within the formalism of collinear factorisation~\cite{Bodwin:1984hc,Collins:1985ue,Collins:1998rz}
which relates a hadronic cross-section
to the convolution of a perturbatively-calculable coefficient function, $C$,
independent of hadronic physics,
with
non-perturbative PDFs $f_i$ describing the distribution of parton $i$
within the hadron (here, the proton).
For DIS-like processes, with a single PDF, this can schematically be written as:
\begin{equation}
\sigma_{lh}%
(\muf, \mur) = 
\sigma_0
\sum_{i}
f_{i} (\muf) \otimes C_i (\muf, \mur) \, ,
\label{eq:factDIS}
\end{equation}
and for Drell--Yan-like processes, with two PDFs, as:
\begin{align}
\sigma_{hh}%
(\muf, \mur) = 
{} & 
\sigma_0
\sum_{i,j}
 f_{i}(\muf) \otimes f_{j}(\muf) \otimes C_{ij}(\muf, \mur) \, ,
\label{eq:factDY}
\end{align}
where higher-twist terms of the order of $\order{\Lambda_{\mathrm{QCD}}^2/\muf^2}$ are neglected, indices $i, j$ run over partonic flavours, %
$\muf$ and $\mur$ denote factorisation and renormalisation scales respectively,
and the convolution $\otimes$ is defined as:
\begin{equation}
\label{eq:convdef}
\begin{split}
    (f \otimes g)(x) &\colonequals \int_x^1 \frac{\dd z}{z} \; f(z) \, g\left(\frac{x}{z}\right) \\
                     &\equiv \int_0^1 \dd z  \, \int_0^1 \dd y \; \delta(x-yz)\, f(y) \, g(z).
\end{split}
\end{equation}
For the result to be factorisation-scheme independent,
the factorisation-scheme-dependent PDFs and coefficient functions
must each compensate for the factorisation-scheme dependence of the other.

\subsection{Notation and definitions}
\label{subsec:fsnotation}

\subsubsection{Factorisation scheme transformations}
\label{subsubsec:fstransformations}

We define PDFs in different factorisation schemes according to their relationship
with PDFs in the \msbar scheme, via a transformation of the form
\begin{align}
	\label{eq:fstransformations_faFSvec}
	\mathbf{f}^{\rFS}
	&=
	\mathbb{K}^{\msbar \to \rFS}
	\otimes
	\mathbf{f}^{\msbar},
\end{align}
where explicitly
\begin{align}
	\label{eq:fstransformations_faFS}
	f^{\rFS}_a (x, \mu)
	&=
	\sum_b
	\int_x^1
	\frac{\dd z}{z}
	\;
	\mathbb{K}^{\msbar \to \rFS}_{ab} \left(z, \mu \right) \ 
	f_b^{\msbar} \left(\frac{x}{z}, \mu \right),
\end{align}
performed locally for each scale $\mu$.

We expand the matrix of convolution kernels $\mathbb{K}^{\msbar \to \rFS}_{ab}$ perturbatively as%
\footnote{In general we will suppress the possible explicit $\mu$-dependence of
$\rK^{\msbar \to \rFS}_{ab}(z, \mu)$.}
\begin{align}
 \notag
	\label{eq:fstransformations_pertexp}
	\mathbb{K}^{\msbar \to \rFS}_{ab} \left(z, \mu\right)
	= {} &
	\delta_{ab} \; \delta(1-z)
	+
	\frac{\alphas(\mu)}{2\pi}
	\;
	\rK^{\msbar \to \rFS}_{ab}(z, \mu)
	\\ & \quad + \order{\alphas^2}.
\end{align}
Equivalently, for convenience, 
we can express the same transformation using
\begin{align}
	(xf)(z, \muf) \equiv z \, f(z, \muf)
\end{align}
instead of $f(z,\muf)$
(as is provided for example by the \lhapdf library \cite{Buckley:2014ana}). Then,
\begin{align}
	(xf)^{\rFS}_a (x, \muf) & {} 
        \\ \notag
	=
	{} \sum_b \int_x^1 & \dd z\;
	\mathbb{K}^{\msbar \to \rFS}_{ab} \left(z, \muf\right) \ 
	(xf_b)^{\msbar} \left(\frac{x}{z}, \muf\right).
	\label{eq:fstransformations_pertexp_xf}
\end{align}

To order $\alphas$, this transformation has perturbative inversion
\begin{align}
	\mathbb{K}^{\rFS \to \msbar}_{ab} \left(z, \mu\right)
	= {} &
	\delta_{ab} \; \delta(1-z)
	-
	\frac{\alphas(\mu)}{2\pi}
	\;
	\rK^{\msbar \to \rFS}_{ab}(z)
	\\ \notag
        & \quad + \order{\alphas^2},
	\label{eq:fstransformations_pertexpinv}
\end{align}
and so 
\begin{align}
	\rK^{\rFS \to \msbar}_{ab}(z) = -\rK^{\msbar \to \rFS}_{ab}(z).
\end{align}
We therefore specify only the `forward' transformation.%
\footnote{In \cref{app:pert_inversion} we test the validity of this 
perturbative inversion numerically.}

Other perturbative expansions 
in $\alphas$
are given according to the convention
\begin{align}
F(\alphas, \mu)
&=
\notag
F^{(0)}(\mu) + \left(\frac{\alphas(\mu)}{2\pi}\right) F^{(1)}(\mu) + \dots 
\\ &= 
\sum_{k} 
\left(\frac{\alphas(\mu)}{2\pi}\right)^k 
F^{(k)}(\mu).
\label{eq:pertexpconv}
\end{align}

\subsubsection{Coefficient functions}
\label{subsubsec:coefffuncs}

We write the hadronic cross-section for
hadron--hadron boson-production process $pp \to V + X$,
differential with respect to the invariant mass (squared) of the boson,
within the framework of collinear factorisation as
\begin{align}
	\label{eq:sigmaAB_coefffun}
    \notag
    & \frac{\dd {\sigma}}{\dd M^2}
    (P_1, P_2) {} = {}
    \sigma^{(0)} \frac{M^2}{s} 
    \int_0^1 \dd \xi_1 \, \dd \xi_2 \, \dd z
    \; \delta\left(\xi_1 \xi_2 z - \frac{M^2}{s}\right)
    \\ 
    & \quad \sum_{a,b} f_a^\rFS(\xi_1, \muf) \, f_b^\rFS(\xi_2, \muf) \;
    C_{ab}^{\rFS}(z, M; \muf, \mur)
\end{align}
for partonic flavours $a,b$,
$\rFS$-scheme parton distribution functions $f^{\rFS}_a,f^{\rFS}_b$,
incoming hadronic momenta $P_1, P_2$
and collinear momentum fractions
$\xi_{1,2}$.
The partonic coefficient function $C_{ab}^{\rFS}(z, M; \muf, \mur)$
is normalised to the pointlike Born cross-section by the $\sigma^{(0)}$ factor, and
may be expanded perturbatively;
for the processes we consider, the leading-order contribution
is $\delta(1-z)$ for the active partonic flavours.

Since the left-hand-side of \cref{eq:sigmaAB_coefffun}
is explicitly independent of the (unphysical)
choice of factorisation-scheme,
so must the right-hand-side be.
Rewriting the double-convolution integral as a matrix in flavour-space
using the notation of \cref{eq:fstransformations_faFSvec},
\begin{align}
	\label{eq:cftransformations_faFSvec}
	& \left[  \left.\mathbf{f}^{\rFS}\right.^{\intercal}
    \otimes {} 
    \mathbb{C}^{\rFS}
    \otimes
    \mathbf{f}^{\rFS}
    \right] \left( \frac{M^2}{s} \right)
 \\ \notag
    & {} \equiv
	\left.
	\mathbf{f}^{\msbar}\right.^{\intercal}
    \otimes
    \left.\mathbb{K}^{\msbar \to \rFS}\right.^{\intercal}
    \otimes
    \mathbb{C}^{\rFS}
    \otimes
    \mathbb{K}^{\msbar \to \rFS}
	\otimes
	\mathbf{f}^{\msbar}
 \\ \label{eq:cftransformations_faFSvec_result}
    & {} =
	\left.
	\mathbf{f}^{\msbar}\right.^{\intercal}
    \otimes
    \left.\mathbb{K}^{\msbar \to \rFS}\right.^{\intercal}
    \otimes
    \left(\left.\mathbb{K}^{\msbar \to \rFS}\right.^{\intercal}\right)^{-1}
    \\ \notag & {} \qquad {}
    \otimes
    \mathbb{C}^{\msbar}
    \otimes
    \left(\mathbb{K}^{\msbar \to \rFS}\right)^{-1}
    \otimes
    \mathbb{K}^{\msbar \to \rFS}
	\otimes
	\mathbf{f}^{\msbar}
 \\ \notag
    & {} \equiv
    \left[  \left.\mathbf{f}^{\msbar}\right.^{\intercal}
    \otimes {} 
    \mathbb{C}^{\msbar}
    \otimes
    \mathbf{f}^{\msbar}
    \right] \left( \frac{M^2}{s} \right)
 ,
\end{align}
where factorisation-scheme independence provides the equality in \cref{eq:cftransformations_faFSvec_result}.
Therefore
\begin{align}
    \mathbb{C}^{\rFS}
    & {} =
    \left.\mathbb{K}^{\rFS \to \msbar}\right.^{\intercal}
    \otimes
    \mathbb{C}^{\msbar}
    \otimes
    \mathbb{K}^{\rFS \to \msbar},
\end{align}
or to NLO in perturbation theory,
\begin{align}
    C_{ab}^{\rFS(0)}
    & {} =
    C_{ab}^{\msbar(0)}
    \\
    C_{ab}^{\rFS(1)}
    & {} =
    C_{ab}^{\msbar(1)}
    \\ \notag {} &
    - \sum_{c} \left[ \rK^{\msbar \to \rFS}_{ca} \otimes C^{\msbar(0)}_{cb}
    + C^{\msbar(0)}_{ac} \otimes \rK^{\msbar \to \rFS}_{cb} \right].
\end{align}
For processes with flavour-diagonal leading-order coefficient functions this reduces to 
\begin{align}
    C_{ab}^{\msbar(1)}
    - \rK^{\msbar \to \rFS}_{ba} \otimes C^{\msbar(0)}_{bb}
    - C^{\msbar(0)}_{aa} \otimes \rK^{\msbar \to \rFS}_{ab}.
\end{align}
For the processes considered here, only one of the transition terms above will contribute
unless $a=b$; since the coefficient functions are consistently normalised to be $\delta(1-z)$
at leading-order, explicit dependence on $C^{\msbar(0)}$ will drop out completely.

\subsubsection{DGLAP evolution}
\label{subsubsec:fsdglap}

The DGLAP equations~\cite{Altarelli:1977zs,Gribov:1972ri,Dokshitzer:1977sg}
familiar from \msbar calculations also hold in
other factorisation schemes~\cite{White:2005wm},
albeit with modified splitting functions $\mathbb{P}^{\rFS}$:
\begin{align}
	\label{eq:dglap}
	\mu^2 \frac{\partial}{\partial \mu^2} \mathbf{f}^{\rFS}(\mu)
	  = {} &
    \mathbb{P}^{\rFS} (\mu) 
	\otimes
	\mathbf{f}^{\rFS} (\mu) .
\end{align}
The relationship between the splitting functions in factorisation scheme FS and \msbar{} 
can be obtained upon taking the logarithmic derivative $\partial / \partial (\log \mu^2)$
of \cref{eq:fstransformations_faFSvec}. 
\begin{align}
	\label{eq:dglapkernels}
	\mathbb{P}^{\rFS} (\mu)
	= {} &
    \mu^2 \left( \frac{\partial}{\partial \mu^2} 
    \mathbb{K}^{\msbar \to \rFS} (\mu) \right)
    \otimes
    \mathbb{K}^{\rFS \to \msbar} (\mu)
 \\ & {} \notag
    + 
    \mathbb{K}^{\msbar \to \rFS} (\mu)
    \otimes
    \mathbb{P}^{\msbar} (\mu) 
    \otimes
    \mathbb{K}^{\rFS \to \msbar} (\mu)
\end{align}
or, using the perturbative expansion of \cref{eq:fstransformations_pertexp}
and expanding according to the convention of \cref{eq:pertexpconv},

\begin{align}
    \rP^{\rFS(0)}_{ab}
    & {} =
    \rP^{\msbar(0)}_{ab} \equiv 0
    \\
    \rP^{\rFS(1)}_{ab}
    & {} =
    \rP^{\msbar(1)}_{ab}
    + \mu^2 \frac{\partial}{\partial \mu^2} 
    \rK_{ab}^{\msbar \to \rFS} (\mu) 
    \\
    \rP^{\rFS(2)}_{ab}
    & {} =
    \rP^{\msbar(2)}_{ab}
    + \mu^2 \frac{\partial}{\partial \mu^2} 
    \rK_{ab}^{\msbar \to \rFS(2)} (\mu)  
    \\ \notag
    & {}
    - \sum_c \left( \mu^2 \frac{\partial}{\partial \mu^2} 
    \rK_{ac}^{\msbar \to \rFS} (\mu)\right) \otimes \rK_{cb}^{\msbar \to \rFS} (\mu)
    \\ \notag
    & {}
    + \frac{\beta^{(2)}}{2\pi} \rK_{ab}^{\msbar \to \rFS} (\mu) 
    \\ \notag
    & {}
    + \sum_c \Bigl(
    \rK_{ac}^{\msbar \to \rFS} (\mu) \otimes \rP^{\msbar(1)}_{cb}
    \\ \notag
    & {}
    \hphantom{+ \sum_c \Bigl(
    \rK}
    -
    \rP^{\msbar(1)}_{ac} \otimes \rK_{cb}^{\msbar \to \rFS} (\mu)
    \Bigr).
\end{align}
The modifications to the DGLAP kernels therefore arise
in turn from
any explicit scale-dependence of the transformation
kernels $\rK^{\msbar\to\rFS}(z;\mu)$,
the QCD $\beta$-function,
and the \msbar DGLAP kernels of the input partons.

For the NLO factorisation-scheme transformations considered in our paper,
in which there is no explicit dependence within $\rK$ on $\mu$,\footnote{Note that in \cref{eq:fstransformations_pertexp}, and elsewhere,
we suppress the possible explicit $\mu$-dependence of
$\rK^{\msbar \to \rFS}_{ab}(z, \mu)$; the factorisation
schemes we consider here have no such dependence.
In general, there is no obstacle to it; such a dependence
features in the \deductor shower-oriented schemes \cite{Nagy:2022bph}.
}
the PDFs in alternative schemes therefore obey a
DGLAP evolution equation that is modified only at NLO and above.

Therefore, the DGLAP evolution and factorisation scheme transformation
commute, up to remainder that is higher-orders in perturbation theory.
Specifically, 
a PDF evolved 
from its input parametrisation
using the \msbar \nklo
evolution equations and then transformed into another scheme
will not satisfy precisely the same evolution equations as a
PDF transformed at its input scale and then evolved
using the modified DGLAP equations.%
\footnote{
A recent proposal to resolve the factorisation-scheme-dependence of DGLAP evolution
was advanced in \cite{Lappi:2023lmi,Lappi:2024dvv}, in which the perturbative relationship between
six linearly-independent DIS structure functions and the proton PDFs was
inverted to derive a `physical', scheme-independent DGLAP-type evolution equation
for structure functions as a function of the physical (momentum-transfer) scale.
}
In the former case, the effective  terms
$\rP^{\rFS(k+2)}_{ab}$ and higher are non-zero,
while in the latter case, the series expansion of the kernels is truncated
at the chosen order.

\subsubsection{Momentum sum rule}
\label{subsubsec:momsumrule}

To compare factorisation schemes on a like-for-like basis,
we consistently impose the momentum sum-rule \cite{Collins:1981uw}
\begin{align}
	\label{eq:momsumrule_PDFs}
	\sum_a \int_0^1 \xi \, f_{a}^{\rFS} (\xi, \mu) \; \dd \xi = 1 ,
\end{align}
or equivalently at the level of the transformation kernels,
\begin{align}
	\label{eq:momsumrule_kernels}
    \sum_a \int_0^1 \; z \, \rK_{ab}^{\msbar\to\rFS}(z) \; \dd z = 0
\end{align}
for all flavours $b$.

Except where otherwise specified by the authors of the scheme,
this is imposed through a virtual-type modification of the $\delta(1-z)$
component of the flavour-diagonal transformation kernels,
\begin{align}
	\label{eq:momsumrule_deltas}
    \rK_{bb}^{\prime\msbar\to\rFS}(z)
    & = \rK_{bb}^{\msbar\to\rFS}(z) 
    \\ \notag {} & \quad - 
    \delta(1-z) \sum_a \int_0^1 \; z' \, \rK_{ab}^{\msbar\to\rFS}(z') \; \dd z'.
\end{align}

\subsubsection{Flavour thresholds}
\label{subsubsec:flavthresh}

Within \cref{eq:fstransformations_faFS} the
$\rK_{qg}^{\msbar\to\rFS}$ kernels act upon an input $\msbar$
gluon PDF to produce a contribution to an output (FS scheme)
quark PDF.
Heavy quark and antiquark flavours%
\footnote{For the purpose of this section,
`heavy' means charm and bottom,
depending on the conventions adopted by the PDF-fitting group.}
$q_f, \qbar_f$ are typically chosen to have no `intrinsic'
contribution to the proton PDF; that is, they are chosen to be
zero below some threshold scale $\mu_f$ of the order of the quark mass
$m_{f}$,
and generated purely by DGLAP evolution above this scale.

For such PDFs we
enforce explicitly that the output, transformed, PDF 
is also zero below the threshold, i.e.\
\begin{align}
	\mathbb{K}^{\msbar \to \rFS}_{q_f b} \left(z, \mu\right)
	= {} &
	\Theta\left[ \mu > \mu_f \right]
	\, 
    \mathbb{K}^{\msbar \to \rFS}_{q_f b} \left(z, \mu\right);
\end{align}
this leads in general to a discontinuity at the threshold,%
\footnote{Since we expect in general to introduce a discontinuity
in the transformed PDFs at the quark-thresholds,
in practice, we split the \lhapdf grids at the thresholds and compute
the two limits
$\lim_{\mu \to \mu_f\pm} \left[ f^{\rFS}_{a} (x, \mu) \right]$
separately to store in the interpolation grids.
The PDF values within each grid are therefore computed with a single value of
$\nf$ and a single set of active kernels, so represent a continuous function
which may be approximated with an interpolation by standard methods.
Further technical details are given in \cref{sec:app_validation}.}
in both $f_g$ and $f_{q_f}$.
The mixing from the gluon 
via the transformation convolution
is turned on suddenly, even if the DGLAP evolution of the input 
\msbar PDF has a continuous solution (possible at NLO
through the choice $\mu_f = m_{f}$).
Similar discontinuities at the flavour thresholds occur naturally at NNLO
and beyond,
as a consequence of constant terms in matching conditions~\cite{Buza:1996wv,Stavreva:2012bs,Bertone:2017djs}.

The `turning on' of the transformation from the gluon at 
the threshold $\mu = \mu_f$
also requires $\nf$ to be incremented by 1 to maintain momentum conservation \cref{eq:momsumrule_kernels}. 
This happens because the expression on the right-hand-side of 
\cref{eq:momsumrule_deltas} contains additional non-zero
kernels above the threshold; i.e.\
\begin{align}
	\nf(\mu) = \sum_{f} \Theta[ \mu > \mu_f ].
\end{align}
This is equivalent to the usual definition of the
variable-flavour-number scheme (VFNS), e.g.~\cite{Aivazis:1993kh,Aivazis:1993pi,Thorne:1997ga,Thorne:2006qt,Cacciari:1998it}, used for the solution
of the DGLAP evolution equations.

\subsection{Summary of scheme motivations}
\label{subsec:schememotivations}

\subsubsection{\msbar scheme}
\label{subsubsec:MSbar}

The \msbar scheme \cite{Bardeen:1978yd} is used for its
theoretical simplicity and practical convenience.
Parton distributions in the \msbar scheme
are defined as renormalised expectation values
of the partonic number operator in the 
hadronic state \cite{Curci:1980uw,Collins:1981uw,CTEQ:1993hwr}.
For partonic calculations in perturbation theory
this choice is `minimal' by virtue of 
absorbing only the 
$\varepsilon$-pole and universal numerical factors
into the PDF renormalisation coefficents
defining the PDF,
and no finite terms.

\subsubsection{DIS scheme}
\label{subsubsec:DIS}
The \textsc{Dis} scheme \cite{Altarelli:1978id} uses the freedom to choose a factorisation scheme to absorb the higher-order DIS coefficient functions into the PDFs,
   including the gluon-initiated partonic contribution,
   extending the leading-order relation for the structure function
   \begin{align}
       F_2(x, \mu^2) \equiv x \, \sum_{f} Q_f^2 \, \left( f_{q_f}^{\dis} (x, \mu) + f_{\qbar_f}^{\dis}(x, \mu) \right)
   \end{align}
   to all orders of perturbation theory.
   Calculations for other processes must therefore be adjusted by the DIS matrix-elements to restore perturbative accuracy. 
   
   We follow \cite{Martin:1998np} and adopt the convention in which the \dis-scheme gluon
   PDF is determined by
   the extension to all Mellin moments of the momentum conservation constraint on the second moment,
   which implies 
   \begin{alignat}{3}
     \label{eq:DIS_def_qg}
     \rK_{qg}^{\msbar\to\dis} {} = {} && -       & \rK_{qq}^{\msbar\to\dis}
     \\
     \label{eq:DIS_def_gg}
     \rK_{gg}^{\msbar\to\dis} {} = {} && - 2 \nf & \rK_{gq}^{\msbar\to\dis}.
   \end{alignat}
   This is equivalent to strengthening \cref{eq:momsumrule_PDFs}
   from a sum-rule that holds only upon integration to a constraint
   local in $\xi$,%
   \footnote{
   The effect of this may be seen graphically in \cref{fig:momentum_view_a}.
   }
    \begin{align}
    \sum_a \xi \, f_{a}^{\dis} (\xi, \mu) & {} = \sum_a \xi \, f_{a}^{\msbar} (\xi, \mu)
    &
    \text{for all }\xi\in[0,1].
    \label{eq:DIS_localsumrule}
    \end{align}
   
\subsubsection{\krk scheme}
\label{subsubsec:Krk}
The \krkdy scheme \cite{Jadach:2015mza,Jadach:2016acv} uses for its factorisation scheme transformation the collinear convolution terms arising from the integral of Catani--Seymour subtraction dipoles \cite{Catani:1996vz}
over the unresolved phase-space
for colour-singlet final states. By absorbing these collinear counterterms into the PDFs rather than including them in the hard-process, matched NLO-plus-parton-shower calculations can be implemented using only positive multiplicative weights (the \krknlo method \cite{Jadach:2015mza,Jadach:2016qti,Sarmah:2024hdk}).

The full \krk scheme extends the \krkdy scheme to all partonic flavours~\cite{Jadach:2016acv}, defining a transformation for the gluon PDF using the gluon-gluon fusion Higgs-production process in the infinite-top-mass limit~\cite{Jadach:2016qti}.

\subsubsection{\dpos/\pos/\mpos/\mposd schemes}
\label{subsubsec:POS}
In \cite{Candido:2020yat} the authors identify the origin of negativity
   in \msbar-scheme coefficient functions to be an oversubtraction within the \msbar scheme.
   This arises from a mismatch of scales for collinear emission between the hard-process and the \msbar subtraction
   term and leads to terms proportional to $\log (1-z)$ for off-diagonal splittings, which become arbitrarily negative for $z \to 1$.
   
   In the \dpos and \pos schemes the subtraction is performed at a modified scale (defined correspondingly based on DIS and DY processes) to ensure the coefficient function remains positive.
   This is argued to be inadequate in \cite{Collins:2021vke} and the argument is refined in \cite{Candido:2023ujx}.
   
   The \mpos scheme is a modification of the \pos scheme which enforces momentum conservation by adding a choice of `soft' function to the diagonal elements. In this work, we will also refer to the
   \mposd
   scheme, which we define to be identical to \mpos save for the imposition of 
   momentum-conservation through a virtual-like delta-function contribution
   instead of a soft function (to be comparable with all other schemes).%
   \footnote{The significance of this change is explored in \cref{subsubsec:CompPDFsImpMomCons}.}

\subsubsection{\aversa scheme}
\label{subsubsec:Aversa}
The \aversa scheme \cite{Aversa:1988vb} aims to remove large universal logarithmic terms of kinematic origin from coefficient functions
   and move them instead into the PDFs (and fragmentation functions),
   in order to improve the perturbative convergence and scale-dependence
   of NLO predictions.

\subsubsection{\phys scheme}
\label{subsubsec:Phys}
The \phys scheme \cite{Oliveira:2013aug} aims to remove finite ${\varepsilon}/{\varepsilon}$ contributions
   of IR origin which emerge in the \msbar scheme from long-distance interactions between massless QCD partons.
   Because of confinement it is argued that any terms related to long-distance
   QCD interactions between partons must be unphysical,
   and should be removed (hence `\phys').

\subsubsection{Others}
\label{subsubsec:otherFS}

Finally, we do not include in our comparisons, but wish to mention, the 
\textsc{Deductor} family of `shower-oriented' schemes \cite{Nagy:2016pwq,Nagy:2017ggp,Nagy:2022bph}
and the Ramalho--Sterman \cite{Ramalho:1983aq} and Sterman \cite{Sterman:1986aj}
schemes.

The \textsc{Deductor} family of `shower-oriented' schemes 
\cite{Nagy:2016pwq,Nagy:2017ggp,Nagy:2022bph}
for use with
parton-shower evolution in the \textsc{Deductor} framework \cite{Nagy:2014mqa,Nagy:2015hwa,Nagy:2016pwq,Nagy:2017ggp,Nagy:2019pjp,Nagy:2022xae,Nagy:2022bph}
is defined by a transformation from the \msbar scheme that is dependent on the
ordering variable used within the shower, so that the DGLAP evolution of the PDFs
matches that of the initial-state evolution within the shower.

The Ramalho--Sterman scheme outlined in \cite{Ramalho:1983aq} was introduced
as an alternative to the DIS scheme to ensure that the Mellin
moments of the Drell--Yan process were bounded in $N$, by absorbing
terms with moments proportional to $\log^2 N$
(arising from plus-distribution factors $\mathcal{D}_1$,%
\footnote{
See \cref{eq:Dk_kernels}.}
associated with soft-gluons)
into the PDF.

Sterman \cite{Sterman:1986aj} defines
two factorisation schemes, for the Drell--Yan and DIS processes separately,
in which
the real and virtual contributions
are separately finite, and all `threshold logarithm' distribution terms $\mathcal{D}_k$
are absorbed into the PDFs, curing the coefficient functions of their
large-$N$ logarithms in Mellin space.
Factorisation-scheme independence then allows the identification and
all-order resummation of the threshold logarithms.

\subsection{Scheme definitions}
\label{subsec:definitions}
Throughout we adopt the convention of decomposing convolution
kernels as arise up to NLO in QCD as:
\begin{align}
\label{eq:tableDecomp}
	K(z) =  {} & 
	\sum_{k=0}^1 a_k \, \mathcal{D}_k (z) 
	+ b(z) \log (1-z)
	+ c(z) \log z
	\\ \notag & {} \quad
	+ P(z)
	- \Delta \, \delta(1-z)
\end{align}
where $b(z), c(z), P(z)$ are rational
functions and, concretely, for all the kernels we consider, 
can be expressed as Laurent series
with at most simple poles at $z=0,1$.
The distributions $\mathcal{D}_k(z)$
are defined as
\begin{equation}
    \label{eq:Dk_kernels}
	\mathcal{D}_k(z) = \left[ \frac{\log^k (1-z)}{1-z} \right]_+
\end{equation}
where the `plus-distribution' regularisation,
for functions otherwise singular at $z=1$,
is defined by its action upon integration against
a smooth function $f$ as
\begin{align}
    \int_x^1 \dd z \, f(z) \left[ \frac{g(z)}{1-z} \right]_+
    & {} =
    \int_x^1 \dd z \, \frac{f(z) - f(1)}{1 - z} \, g(z)
    \\ \notag 
    & {} \qquad - f(1) \int_0^x \dd z \, \frac{g(z)}{1-z}.
\end{align}

Transformation kernels $\rK^{\rFS}_{ab}(x)$ for the
considered schemes are introduced in \cref{sec:schemeComparisons}
and listed in this convention in \cref{tab:Kqq,tab:Kqg,tab:Kgq,tab:Kgg};
each column is labelled according to the term it accompanies.
For ease of reference, the LO DGLAP splitting functions
used within the tables
are given in \cref{sec:app_notation}.

\renewcommand{\arraystretch}{1.266} %
\begin{table*}[p]
	\begin{center}
		\begin{tabular}{ |c|c|c|c|c|c|c| } 
			\hline
			$ \cf^{-1} \rK_{qq}^{\msbar\to\rFS} $ 
			& $\mathcal{D}_1$ 
			& $\mathcal{D}_0$
			& $\log (1-z)$
			& $\log z$
			& $P(z)$
			& $- \delta(1-z)$
			\\ 
			\hline
			\textsc{Aversa}
			& %
			$ 2 $
			& %
			$ -\frac{3}{2} $
			& %
			$ -(1+z) $
			& %
			$ - \pqq $
			& %
			$ 3 + 2 z $
			& %
			$ \frac{\pi^2}{3} + \frac{9}{2} $
			\\ 
			\textsc{Dis}
			& %
			$ 2 $
			& %
			$ -\frac{3}{2} $
			& %
			$ -(1+z) $
			& %
			$ - \pqq $
			& %
			$ 3 + 2 z $
			& %
			$ \frac{\pi^2}{3} + \frac{9}{2} $
			\\
			\textsc{Krk}
			& %
			$ 4 $
			& %
			& %
			$ -2(1+z) $
			& %
			$ - \pqq $
			& %
			$ 1 - z $
			& %
			$ \frac{\pi^2}{3} + \frac{17}{4} $
			\\
			\textsc{KrkDY}
			& %
			$ 4 $
			& %
			& %
			$ -2 (1+z) $
			& %
			$ - \pqq $
			& %
			$ 1 - z $
			& %
			$ \frac{\pi^2}{3} +\frac{11}{4} $
			\\
			\dpos
			& %
			& %
			& %
			& %
			& %
			& %
			\\
			\pos
			& %
			& %
			& %
			& %
			& %
			& %
			\\
			\mpos
			& %
			& %
			& %
			& %
			& %
			$ \frac{350}{3}z^2(1-z)^2 $
			& %
			\\
			\mposd
			& %
			& %
			& %
			& %
			& %
			& %
			$ -\frac{35}{18} $
			\\
			\textsc{Phys}
			& %
			$ 2 $
			& %
			& %
			$ -(1+z) $
			& %
			& %
			$ 1 - z $
			& %
			$ \frac{11}{4} $
			\\
			\hline
		\end{tabular}
	\end{center}
	\caption{Transformation kernels $\rK^{\rFS}_{qq}(x)$, in the notation of \cref{eq:tableDecomp}.
    The terms in each column are the coefficients of the function or distribution in the column heading.
    Note that $\rK^{\pos}_{qq} = \rK^{\dpos}_{qq} \equiv 0$.}
	\label{tab:Kqq}
\end{table*}

\begin{table*}[p]
	\begin{center}
		\begin{tabular}{ |c|c|c|c|c|c|c| } 
			\hline
			$ \tr^{-1} \rK_{qg}^{\msbar\to\rFS} $ 
			& $\mathcal{D}_1$ 
			& $\mathcal{D}_0$
			& $\log (1-z)$
			& $\log z$
			& $P(z)$
			& $- \delta(1-z)$
			\\ 
			\hline
			\textsc{Aversa}
			& %
			& %
			& %
			$ \pqg $
			& %
			$ - \pqg $
			& %
			& %
			\\
			\textsc{Dis}
			& %
			& %
			& %
			$ \pqg $
			& %
			$ -\pqg $
			& %
			$ -4 \pqg + 3 $
			& %
			\\
			\textsc{Krk}
			& %
			& %
			& %
			$ 2 \pqg $
			& %
			$ -\pqg $
			& %
			$ -\pqg + 1 $
			& %
			\\
			\textsc{KrkDY}
			& %
			& %
			& %
			$ 2 \pqg $
			& %
			$ -\pqg $
			& %
			$ -\pqg + 1 $
			& %
			\\
			\dpos
			& %
			& %
			& %
			$ \pqg $
			& %
			$ - \pqg $
			& %
			$ - \pqg $
			& %
			\\
			\pos
			& %
			& %
			& %
			$ 2 \pqg $
			& %
			$ - \pqg $
			& %
			$ - \pqg $
			& %
			\\
			\mpos
			& %
			& %
			& %
			$ 2 \pqg $
			& %
			$ - \pqg $
			& %
			$ - \pqg $
			& %
			\\
			\mposd
			& %
			& %
			& %
			$ 2 \pqg $
			& %
			$ - \pqg $
			& %
			$ - \pqg $
			& %
			\\
			\textsc{Phys}
			& %
			& %
			& %
			$ \pqg $
			& %
			& %
			$ - \pqg + 1 $
			& %
			\\
			\hline
		\end{tabular}
	\end{center}
	\caption{Transformation kernels $\rK^{\rFS}_{qg}(x)$, in the notation of \cref{eq:tableDecomp}.}
	\label{tab:Kqg}
\end{table*}

\begin{table*}[p]
	\begin{center}
		\begin{tabular}{ |c|c|c|c|c|c|c| } 
			\hline
			$ \cf^{-1} \rK_{gq}^{\msbar\to\rFS} $ 
			& $\mathcal{D}_1$ 
			& $\mathcal{D}_0$
			& $\log (1-z)$
			& $\log z$
			& $P(z)$
			& $- \delta(1-z)$
			\\ 
			\hline
			\textsc{Aversa}
			& %
			& %
			& %
			$ \pgq $
			& %
			$ - \pgq $
			& %
			$ - \frac{4}{3} $
			& %
			\\
			\textsc{Dis}
			& %
			$ -2 $
			& %
			$ \frac{3}{2} $
			& %
			$ 1 + z $
			& %
			$ \pqq $
			& %
			$ -3 - 2 z $
			& %
			$ -\frac{\pi^2}{3} - \frac{9}{2} $
			\\
			\textsc{Krk}
			& %
			& %
			& %
			$ 2 \pgq $
			& %
			$ -\pgq $
			& %
			$ z $
			& %
			\\
			\pos
			& %
			& %
			& %
			$ 2 \pgq $
			& %
			$ - \pgq $
			& %
			$ - \pgq $
			& %
			\\
			\mpos
			& %
			& %
			& %
			$ 2 \pgq $
			& %
			$ - \pgq $
			& %
			$ - \pgq $
			& %
			\\
			\mposd
			& %
			& %
			& %
			$ 2 \pgq $
			& %
			$ - \pgq $
			& %
			$ - \pgq $
			& %
			\\
			\textsc{Phys}
			& %
			& %
			& %
			$ \pgq $
			& %
			& %
			$ z $
			& %
			\\
			\hline
		\end{tabular}
	\end{center}
	\caption{Transformation kernels $\rK^{\rFS}_{gq}(x)$, in the notation of \cref{eq:tableDecomp}.
		Note that $\rK^{\krkdy}_{gq} = \rK^{\dpos}_{gq} \equiv 0$.
	}
	\label{tab:Kgq}
\end{table*}

\renewcommand{\arraystretch}{1.4} %
\begin{table*}[p]
	\begin{center}
		\begin{tabular}{ |c|c|c|c|c|c|c| } 
			\hline
			$ \ca^{-1} \rK_{gg}^{\msbar\to\rFS} $ 
			& $\mathcal{D}_1$ 
			& $\mathcal{D}_0$
			& $\log (1-z)$
			& $\log z$
			& $P(z)$
			& $- \delta(1-z)$
			\\ 
			\hline
			\textsc{Aversa}
			& %
			$ 2 $
			& %
			& %
			$ -2 $
			& %
			$ -2 \frac{z}{1-z} $
			& %
			& %
			$ \frac{\pi^2}{3} 
			+ 1
			- \frac{5}{6} \frac{\tr \nf}{\ca} $
			\\
			$ (- 2\nf \tr)^{-1} \, \rK_{gg}^{\msbar\to\text{DIS}} $
			& %
			& %
			& %
			$ \pqg $
			& %
			$ -\pqg $
			& %
			$ -4 \pqg + 3 $
			& %
			\\
			\textsc{Krk}
			& %
			$ 4 $
			& %
			& %
			$ 4 \left(\frac{1}{z} - 2 + z(1-z)\right) $
			& %
			$ -2\pgg $
			& %
			& %
			$ \frac{\pi^2}{3} 
			+ \frac{341}{72}
			- \frac{59}{36} \frac{\tr \nf}{\ca} $
			\\
			\textsc{MPos}
			& %
			& %
			& %
			& %
			& %
			$ \frac{475}{3}\frac{\tR \nf}{\ca} z^2(1-z)^2 $
			& %
			\\
			\mposd
			& %
			& %
			& %
			& %
			& %
			& %
			$ -\frac{95}{36}\frac{\tR \nf}{\ca} $
			\\
			\textsc{Phys}
			& %
			$ 2 $
			& %
			& %
			$ 2 \left(\frac{1}{z} - 2 + z(1-z)\right) $
			& %
			& %
			& %
			$ \frac{203}{72} 
			- \frac{29}{36} \frac{\tr \nf}{\ca} $
			\\
			\hline
		\end{tabular}
	\end{center}
	\caption{Transformation kernels $\rK^{\rFS}_{gg}(x)$, in the notation of \cref{eq:tableDecomp}.
	Note that $\rK^{\krkdy}_{gg} = \rK^{\pos}_{gg} = \rK^{\dpos}_{gg} \equiv 0$.}
	\label{tab:Kgg}
\end{table*}

\begin{table*}[p]
	\begin{center}
		\begin{tabular}{ |c|c|c|c|c|c|c| } 
			\hline
			$\cf^{-1} C_{q}^{(1)\rFS}$ 
			& $\mathcal{D}_1$ 
			& $\mathcal{D}_0$
			& $\log (1-z)$
			& $\log z$
			& $P(z)$
			& $- \delta(1-z)$
			\\ 
			\hline
			\textsc{$\msbar$}
			& %
			$ 2 $
			& %
			$ -\frac{3}{2} $
			& %
			$ -(1+z) $
			& %
			$ - \pqq $
			& %
			$ 3 + 2z $
			& %
			$ \frac{\pi^2}{3} + \frac{9}{2} $
			\\
			\dis, \aversa
			& %
			& %
			& %
			& %
			& %
			& %
			\\
			\krk
			& %
			$ -2 $
			& %
			$ -\frac{3}{2} $
			& %
			$ 1 + z $
			& %
			& %
			$ 2 + 3 z $
			& %
			$ \frac{1}{4} $
			\\
			\krkdy
			& %
			$ -2 $
			& %
			$ -\frac{3}{2} $
			& %
			$ 1 + z $
			& %
			& %
			$ 2 + 3 z $
			& %
			$ \frac{7}{4} $
			\\
			\pos, \dpos
			& %
			$ 2 $
			& %
			$ -\frac{3}{2} $
			& %
			$ -(1+z) $
			& %
			$ -\pqq $
			& %
			$ 3 + 2 z $
			& %
			$ \frac{\pi^2}{3} + \frac{9}{2} $
			\\
			\mpos
			& %
			$ 2 $
			& %
			$ -\frac{3}{2} $
			& %
			$ -(1+z) $
			& %
			$ -\pqq $
			& %
			$ 3 + 2 z - \frac{350}{3} z^2 (1-z)^2 $
			& %
			$ \frac{\pi^2}{3}+\frac{9}{2} $
			\\
			\mposd
			& %
			$ 2 $
			& %
			$ -\frac{3}{2} $
			& %
			$ -(1+z) $
			& %
			$ -\pqq $
			& %
			$ 3 + 2 z $
			& %
			$ \frac{\pi^2}{3}+\frac{9}{2} + \frac{35}{18} $
			\\
			\phys
			& %
			& %
			$ -\frac{3}{2} $
			& %
			& %
			$ -\pqq $
			& %
			$ 2 + 3 z $
			& %
			$ \frac{\pi^{2}}{3} + \frac{7}{4} $
			\\
			\hline
			$\tr^{-1} C_{g}^{(1)\rFS}$ 
			& $\mathcal{D}_1$ 
			& $\mathcal{D}_0$
			& $\log (1-z)$
			& $\log z$
			& $P(z)$
			& $- \delta(1-z)$
			\\ 
			\hline
			\textsc{$\msbar$}
			& %
			& %
			& %
			$ \pqg $
			& %
			$ - \pqg $
			& %
			$ -4 \pqg + 3 $
			& %
			\\
			\dis
			& %
			& %
			& %
			& %
			& %
			& %
			\\
			\aversa
			& %
			& %
			& %
			& %
			& %
			$ -4 \pqg + 3 $
			& %
			\\
			\krk, \krkdy
			& %
			& %
			& %
			$ -\pqg $
			& %
			& %
			$ -3 \pqg + 2 $
			& %
			\\
			\pos
			& %
			& %
			& %
			$ - \pqg $
			& %
			& %
			$ -3 \pqg + 3 $
			& %
			\\
			\dpos
			& %
			& %
			& %
			& %
			& %
			$ -3 \pqg + 3 $
			& %
			\\
			\mpos
			& %
			& %
			& %
			$ -\pqg $
			& %
			& %
			$ -3\pqg + 3 $
			& %
			\\
		      \phys
			& %
			& %
			& %
			& %
			$ -\pqg $
			& %
			$ -3\pqg + 2 $
			& %
			\\
			\hline
		\end{tabular}
	\end{center}
	\caption{DIS coefficient functions for each factorisation scheme, in the notation of \cref{eq:tableDecomp}.
    The terms in each column are the coefficients of the function or distribution in the column heading.}
 \label{tab:CoDIS}
\end{table*}

\begin{table*}[p]
	\begin{center}
		\begin{tabular}{ |c|c|c|c|c|c|c| } 
			\hline
			$\cf^{-1} D_{q\bar{q}}^{(1)\rFS}$ 
			& $\mathcal{D}_1$ 
			& $\mathcal{D}_0$
			& $\log (1-z)$
			& $\log z$
			& $P(z)$
			& $- \delta(1-z)$
			\\ 
			\hline
			\msbar
			& %
			$ 8 $
			& %
			& %
			$ -4(1+z) $
			& %
			$ - 2\pqq $
			& %
			& %
			$ -\frac{2\pi^2}{3} + 8 $
			\\
			\aversa
			& %
			$ 4 $
			& %
			$ 3 $
			& %
			$ -2 (1+z) $
			& %
			& %
			$ -6 - 4 z $
			& %
			$ -\frac{4\pi^2}{3} - 1 $
			\\
			\dis
			& %
			$ 4 $
			& %
			$ 3 $
			& %
			$ -2(1+z) $
			& %
			& %
			$ -6 - 4z $
			& %
			$ -\frac{4\pi^2}{3} - 1 $
			\\
			\krk
			& %
			& %
			& %
			& %
			& %
			$ -2 + 2z$
			& %
			$ -\frac{4\pi^2}{3} - \frac{1}{2} $
			\\
			\krkdy
			& %
			& %
			& %
			& %
			& %
			$ -2 + 2z$
			& %
			$ -\frac{4\pi^2}{3} + \frac{5}{2} $
			\\
			\pos, \dpos
			& %
			$ 8 $
			& %
			& %
			$ -4(1+z) $
			& %
			$ -2\pqq $
			& %
			& %
			$ -\frac{2\pi^2}{3} + 8 $
			\\
			\mpos
			& %
			$ 8 $
			& %
			& %
			$ -4(1+z) $
			& %
			$ -2 \pqq $
			& %
			$ - \frac{700}{3} z^2 (1-z)^2 $
			& %
			$ -\frac{2\pi^2}{3} + 8 $
			\\
			\phys
			& %
			$ 4 $
			& %
			& %
			$ -2(1 + z) $
			& %
			$ -2 \pqq $
			& %
			$ -2 + 2 z $
			& %
			$ -\frac{2\pi^2}{3} + \frac{5}{2} $
			\\
			\hline
			$\tr^{-1} D_{qg}^{(1)\rFS}$ 
			& $\mathcal{D}_1$ 
			& $\mathcal{D}_0$
			& $\log (1-z)$
			& $\log z$
			& $P(z)$
			& $- \delta(1-z)$
			\\ 
			\hline
			\msbar
			& %
			& %
			& %
			$ 2 \pqg $
			& %
			$ - \pqg $
			& %
			$ \frac{1}{2} + 3z - \frac{7}{2}z^{2} $
			& %
			\\
			\aversa
			& %
			& %
			& %
			$ \pqg $
			& %
			& %
			$ \frac{1}{2} + 3z - \frac{7}{2}z^{2} $
			& %
			\\
			\dis
			& %
			& %
			& %
			$ \pqg $
			& %
			& %
			$ \frac{3}{2} - 5z + \frac{9}{2}z^{2} $
			& %
			\\
			\krk, \krkdy
			& %
			& %
			& %
			& %
			& %
			$ \frac{1}{2} + z - \frac{3}{2}z^{2} $
			& %
			\\
			\dpos
			& %
			& %
			& %
			$ \pqg $
			& %
			& %
			$ \frac{3}{2} + z - \frac{3}{2}z^{2} $
			& %
			\\
			\pos
			& %
			& %
			& %
			& %
			& %
			$ \frac{3}{2} + z - \frac{3}{2}z^{2} $
			& %
			\\
			\mpos
			& %
			& %
			& %
			& %
			& %
			$ \frac{3}{2} + z - \frac{3}{2}z^{2} $
			& %
			\\
			\phys
			& %
			& %
			& %
			$ \pqg $
			& %
			$ - \pqg $
			& %
			$ \frac{1}{2} + z - \frac{3}{2}z^{2} $
			& %
			\\
			\hline
		\end{tabular}
	\end{center}
	\caption{Drell--Yan coefficient functions for each factorisation scheme, in the notation of \cref{eq:tableDecomp}.}
    \label{tab:CoDY}
\end{table*}

\begin{table*}[p]
	\begin{center}      
 \begin{adjustbox}{angle=90}
		\begin{tabular}{ |c|c|c|c|c|c|c| } 
			\hline
			$\cf^{-1} C_{gq}^{(1)\rFS}$ 
			& $\mathcal{D}_1$ 
			& $\mathcal{D}_0$
			& $\log (1-z)$
			& $\log z$
			& $P(z)$
			& $- \delta(1-z)$
			\\ 
			\hline
			\msbar
			& %
			& %
			& %
			$2 \pgq$
			& %
			$- \pgq$
			& %
			$
			-\frac{3}{2} \frac{1}{z}
			+ 3 - \frac{1}{2} z$
			& %
			\\
			\aversa
			& %
			& %
			& %
			$ \pgq $
			& %
			& %
			$ 
			-\frac{3}{2} \frac{1}{z}
			+ \frac{13}{3} - \frac{1}{2} z$
			& %
			\\
			\dis
			& %
			$ 2 $
			& %
			$ -\frac{3}{2} $
			& %
			$ 2\pgq - 1 - z $
			& %
			$ - \pgq - \pqq $
			& %
			$
			- \frac{3}{2} \frac{1}{z}
			+ 6 + \frac{3}{2} z $
			& %
			$ \frac{\pi^2}{3} + \frac{9}{2} $
			\\
			\krk
			& %
			& %
			& %
			& %
			& %
			$
			- \frac{3}{2} \frac{1}{z}
			+ 3 - \frac{3}{2} z
			$
			& %
			\\
			\pos
			& %
			& %
			& %
			& %
			& %
			$
			\frac{(z+1)^2}{2z}
			=
			\frac{1}{2}\frac{1}{z}
			+ 1
			+ \frac{1}{2}z
			$
			& %
			\\
			\mpos
			& %
			& %
			& %
			& %
			& %
			$
			\frac{1}{2}\frac{1}{z}
			+ 1
			+ \frac{1}{2}z
			$
			& %
			\\
			\phys
			& %
			& %
			& %
			$\pgq$
			& %
			$- \pgq$
			& %
			$
			- \frac{3}{2} \frac{1}{z}
			+ 3 - \frac{3}{2} z
			$
			& %
			\\
			\hline
			$\ca^{-1} C_{gg}^{(1)\rFS}$ 
			& $\mathcal{D}_1$ 
			& $\mathcal{D}_0$
			& $\log (1-z)$
			& $\log z$
			& $P(z)$
			& $- \delta(1-z)$
			\\ 
			\hline
			\textsc{$\msbar$}
			& %
			$ 8 $
			& %
			& %
			$ 8 \left( \frac{1}{z} - 2 + z(1-z) \right) $
			& %
			$ - 4 \pgg $
			& %
			$ - \frac{11}{3} \frac{(1-z)^3}{z} $
			& %
			$ - \frac{2\pi^2}{3} - \frac{11}{3} $
			\\
			\aversa
			& %
			$ 4 $
			& %
			& %
			$ 8 \left( \frac{1}{z} - \frac{3}{2} + z(1-z) \right) $
			& %
			$ -4 \left( \frac{1}{z} - 1 + z(1-z) \right)$
			& %
			$ - \frac{11}{3} \frac{(1-z)^3}{z} $
			& %
			$ - \frac{4\pi^2}{3} - \frac{17}{3} + \frac{5}{3} \frac{\tR \nf}{\ca} $
			\\
			\dis
			& %
			$ 8 $
			& %
			& %
            $ 8 \left( \frac{1}{z} - 2 + z(1-z) \right) + 4 \frac{\tR \nf} {\ca} \pqg $
			& %
			$ - 4 \pgg - 4 \frac{\tR \nf}{\ca} \pqg $
			& %
            $ - \frac{11}{3} \frac{(1-z)^3}{z} + 32 \frac{\tR \nf}{\ca} z(1-z)
			- 4 \frac{\tR \nf}{\ca} $
			& %
			$ - \frac{2\pi^2}{3} - \frac{11}{3} $
			\\
			\krk
			& %
			& %
			& %
			& %
			& %
			$ - \frac{11}{3} \frac{(1-z)^3}{z} $
			& %
			$ - \frac{4\pi^2}{3} - \frac{473}{36} + \frac{59}{18} \frac{\tR \nf}{\ca}$
			\\
			\dpos
			& %
			$ 8 $
			& %
			& %
			$ 8 \left( \frac{1}{z} - 2 + z(1-z) \right) $
			& %
			$ - 4 \pgg $
			& %
			$ - \frac{11}{3} \frac{(1-z)^3}{z} $
			& %
			$ - \frac{2\pi^2}{3} - \frac{11}{3} $
			\\
			\pos
			& %
			$ 8 $
			& %
			& %
			$ 8 \left( \frac{1}{z} - 2 + z(1-z) \right) $
			& %
			$ - 4 \pgg $
			& %
			$ - \frac{11}{3} \frac{(1-z)^3}{z} $
			& %
			$ - \frac{2\pi^2}{3} - \frac{11}{3} $
			\\
			\mpos
			& %
			$ 8 $
			& %
			& %
			$ 8 \left( \frac{1}{z} - 2 + z(1-z) \right) $
			& %
			$ - 4 \pgg $
			& %
			$ - \frac{11}{3} \frac{(1-z)^3}{z} - \frac{950}{3} \frac{\tR \nf}{\ca} z^2 (1-z)^2$
			& %
			$ - \frac{2\pi^2}{3} - \frac{11}{3} $
			\\
			\phys
			& %
			$ 4 $
			& %
			& %
			$ 4 \left( \frac{1}{z} - 2 + z(1-z) \right) $
			& %
			$ - 4 \pgg $
			& %
			$ - \frac{11}{3} \frac{(1-z)^3}{z} $
			& %
			$ - \frac{2\pi^2}{3} - \frac{335}{36} + \frac{29}{18} \frac{\tR \nf}{\ca} $
			\\
			\hline
		\end{tabular}
  \end{adjustbox}
	\end{center}
	\caption{Higgs coefficient functions for each factorisation scheme, in the notation of \cref{eq:tableDecomp}.}
 \label{tab:CoHiggs}
\end{table*}

\section{Scheme comparisons}
\label{sec:schemeComparisons}

In this section we examine the effects of the transformations
given in \cref{tab:Kqq,tab:Kqg,tab:Kgq,tab:Kgg}.
In \cref{subsec:CompPDFs} we simply show the transformed PDFs themselves,
at `low' (2 GeV) and `characteristic LHC' (100 GeV) scales.
In \cref{subsec:CompPDFsDecomp}
we break down each transformation into the constituent contributions from the decomposition of \cref{eq:tableDecomp},
and further according to the flavour of the input-PDF.
In \cref{subsec:CompPDFsMomCons} we examine the effect
of the transformations on the momentum sum-rule, and
in \cref{subsec:NumSumRules} on the number sum-rules.
\Cref{subsec:CompPos} discusses the positivity of PDFs in different schemes.
Finally, in \cref{subsec:CompCFs} we examine the consequence of each
of the considered factorisation schemes on the coefficient functions of the
DIS, Drell--Yan and (gluon-fusion) Higgs production processes.

Throughout this section, we emphasise that there are three
inequivalent methods of obtaining a PDF set in a given factorisation
scheme:
\begin{enumerate}[(i)]
    \item by direct fitting in the desired scheme;
    \item by transforming a PDF fitted in another scheme at the input-scale, and performing DGLAP evolution in the new scheme, as discussed in \cref{subsubsec:fsdglap};
    \item by transforming a PDF fitted and DGLAP-evolved in another scheme locally at each scale, using \cref{eq:fstransformations_faFS}.
\end{enumerate}
We restrict ourselves to PDFs obtained using (iii), applied to modern PDF sets
in common use, fitted and DGLAP-evolved in the \msbar scheme.%
\footnote{Historically, the differences between \dis-scheme PDFs obtained via
the former and the latter methods motivated
simultaneous independent fits in the \msbar and \dis schemes
using the same data and methodology
\cite{Lai:1994bb,Lai:1996mg,Lai:1999wy}.}

\subsection{PDFs}
\label{subsec:CompPDFs}

\begin{figure*}[p]
\centering
\begin{subfigure}[t]{\textwidth}
\makebox[\textwidth][c]{
    \includegraphics[width=0.47\textwidth]{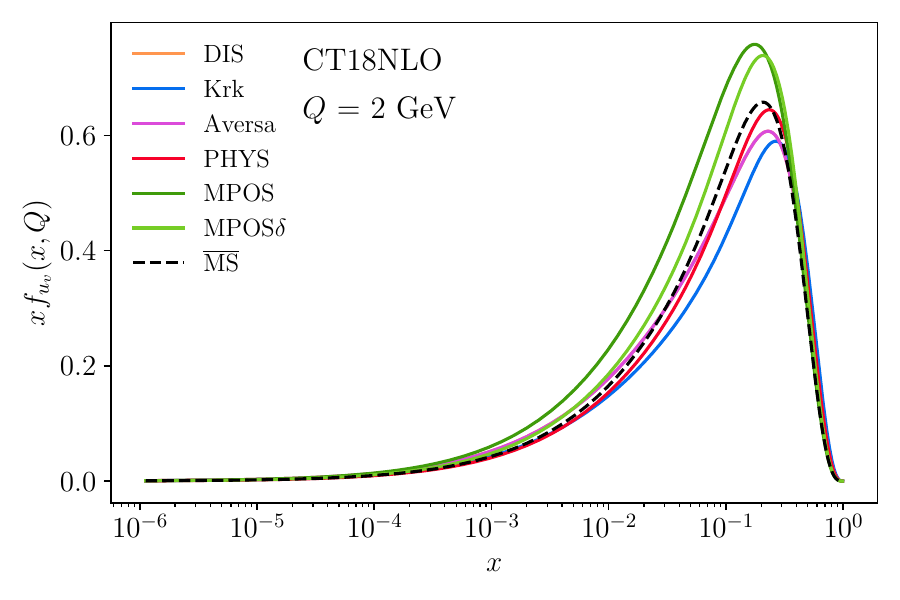} \hfill
    \includegraphics[width=0.47\textwidth]{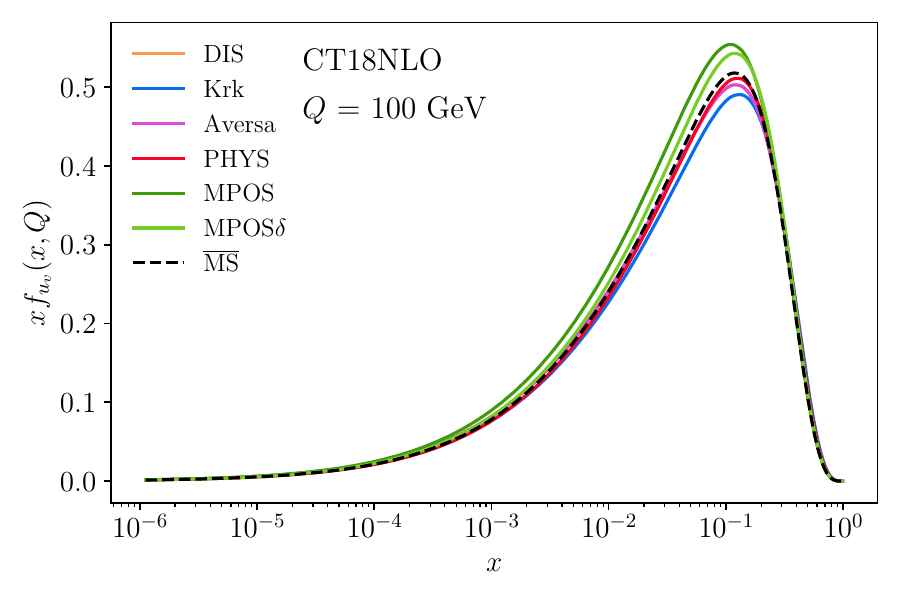}
}
\caption{up valence-quark distribution, $u_v$\label{fig:PDF_compare_uval}}
\end{subfigure}
\\
\begin{subfigure}[t]{\textwidth}
\makebox[\textwidth][c]{
\includegraphics[width=0.47\textwidth]{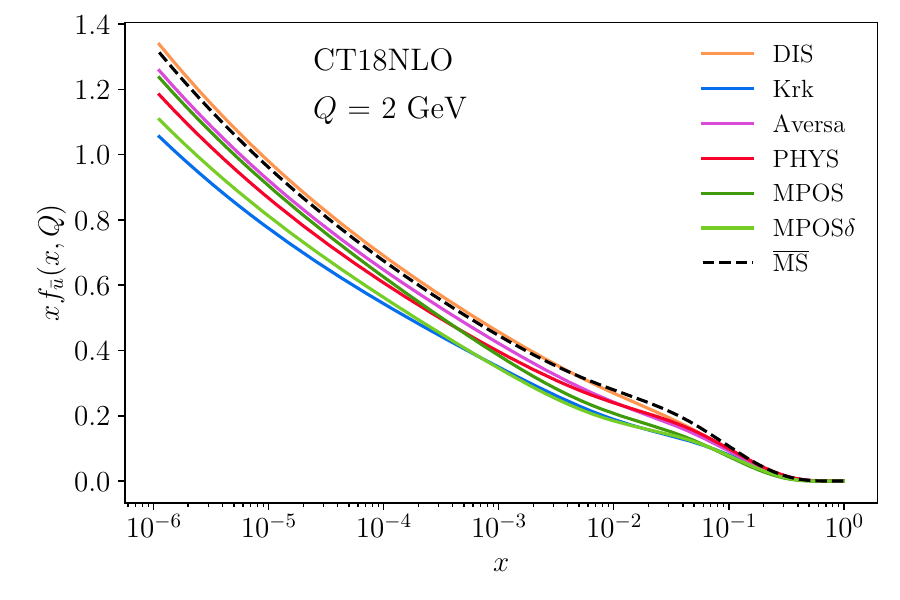} \hfill
\includegraphics[width=0.47\textwidth]{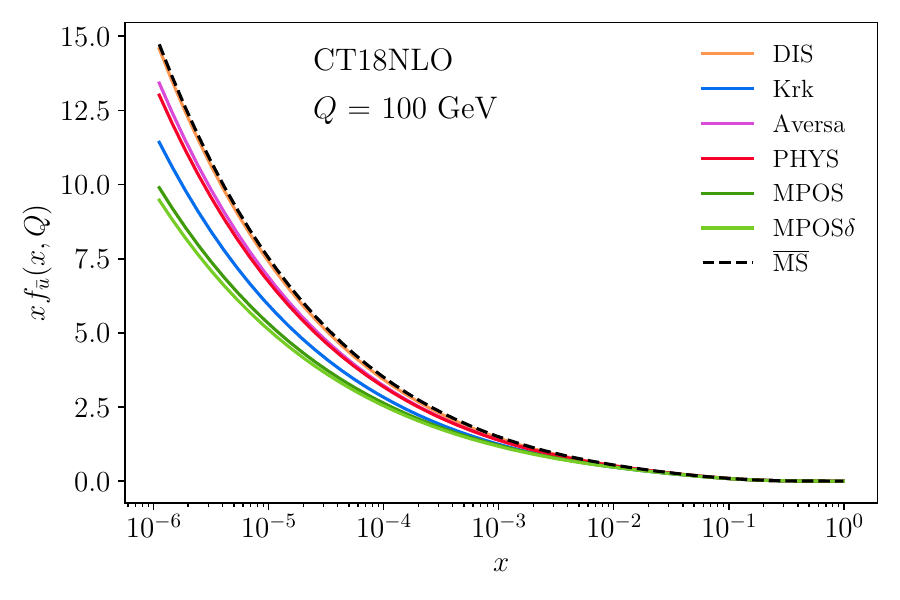}
}
\caption{up-antiquark distribution, $\bar{u}$\label{fig:PDF_compare_ub}}
\end{subfigure}
\\
\begin{subfigure}[t]{\textwidth}
\makebox[\textwidth][c]{
\includegraphics[width=0.47\textwidth]{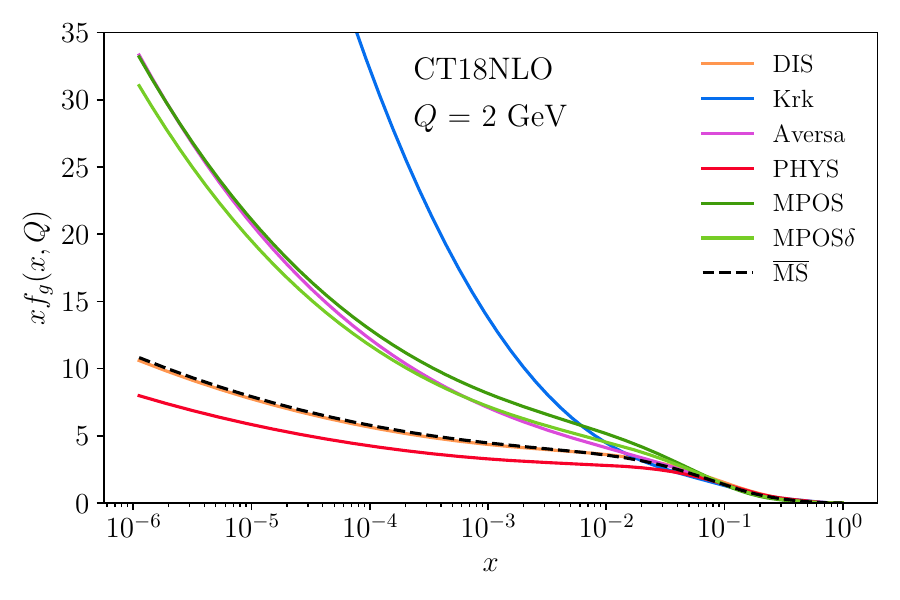} \hfill
\includegraphics[width=0.47\textwidth]{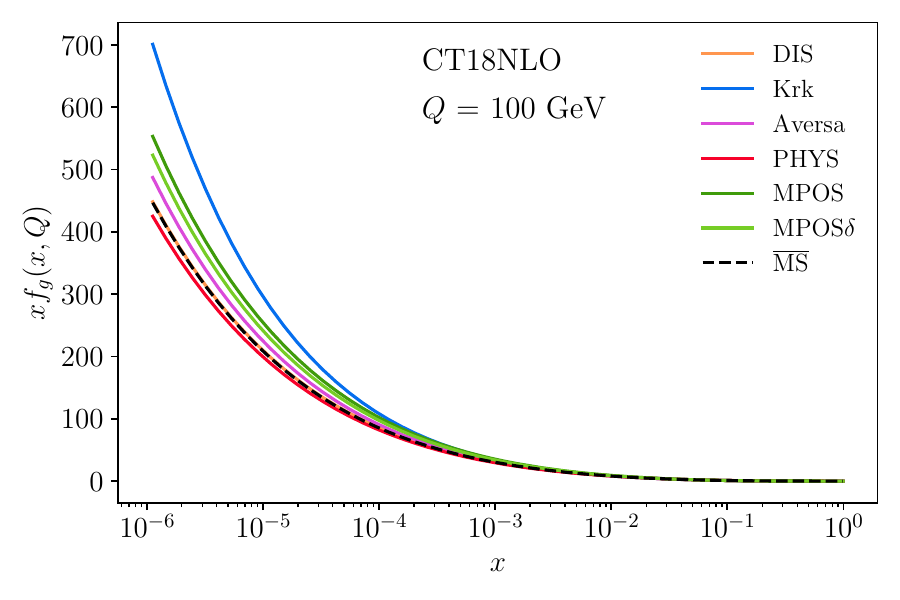}
}
\caption{gluon distribution, $g$\label{fig:PDF_compare_g}}
\end{subfigure}
\\
\begin{subfigure}[t]{\textwidth}
\makebox[\textwidth][c]{
\includegraphics[width=0.47\textwidth]{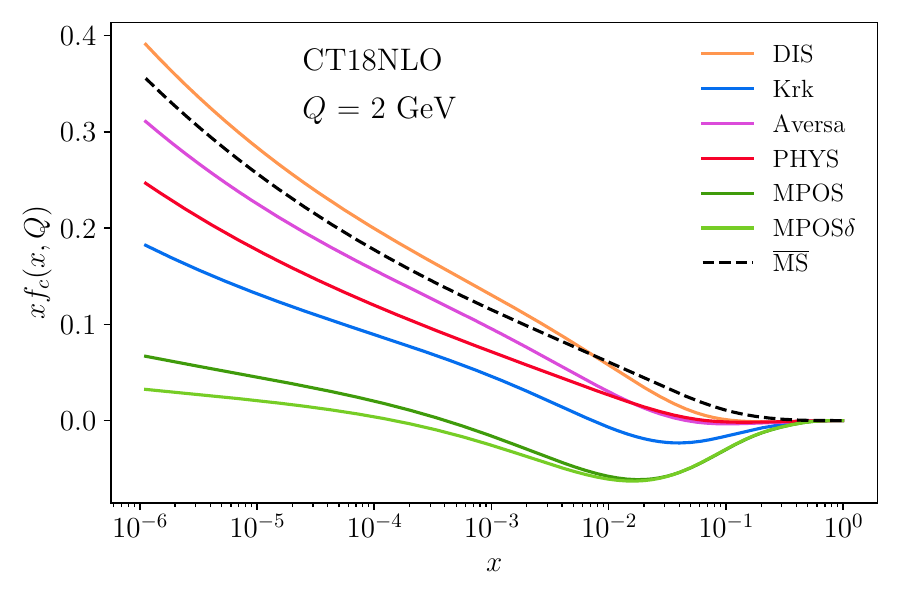} \hfill
\includegraphics[width=0.47\textwidth]{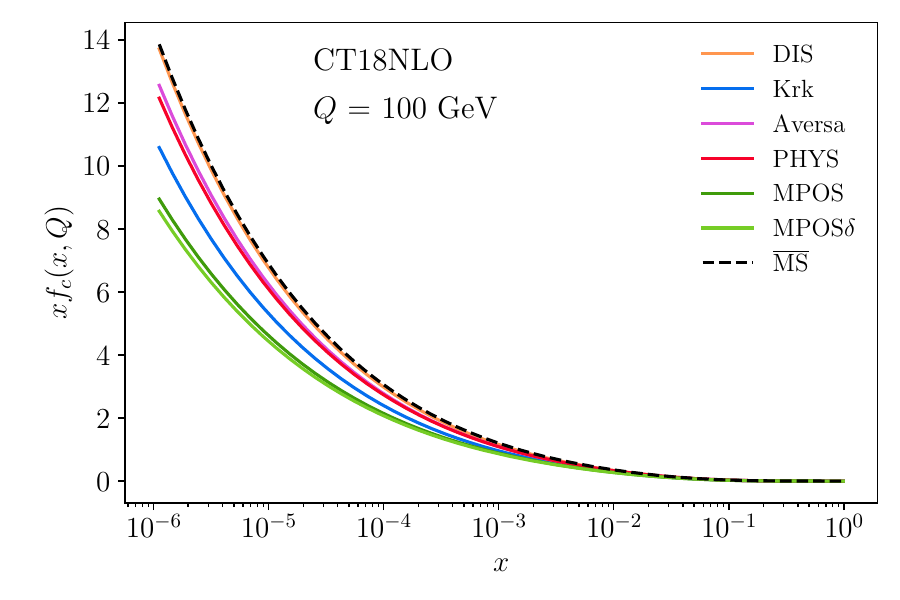}
}
\caption{charm-quark distribution, $c$\label{fig:PDF_compare_c}}
\end{subfigure}
\caption{
Comparison of transformed \ctnlo PDFs in different schemes for $u$-valence, %
$\bar{u}$, gluon and charm, at $Q=2$ GeV (left) and $100$ GeV (right).
The remaining flavours are presented in \cref{fig:PDF_compare_CT18_2} in \cref{app:PDF_plots}.
}
\label{fig:PDF_compare}
\end{figure*}

In \cref{fig:PDF_compare} we show the results of numerically
transforming \msbar PDFs fitted by the CTEQ collaboration
(\ctnlo \cite{Hou:2019efy})
into the discussed schemes using the kernels given in
\cref{tab:Kqq,tab:Kgq,tab:Kqg,tab:Kgg}.
Similar plots produced using PDF sets fitted
by other groups are given in 
\cref{fig:PDF_compare_NNPDF40MC} (\nnpdfmc \cite{Cruz-Martinez:2024cbz})
and
\cref{fig:PDF_compare_MSHT20nlo} (\mshtnlo \cite{Bailey:2020ooq}) in \cref{app:PDF_plots}.
Details of the numerical codes used to perform the convolutions
are given in \cref{sec:app_validation}.

The most dramatic difference can be seen in the low-$x$ gluon distribution, in \cref{fig:PDF_compare_g}.
This is especially pronounced for the \krk\ and \mpos\ (\mposd) schemes at low scales, for which the low-$x$ gluon is up to an order of magnitude
larger than in the \msbar and other schemes.
As we will show in~\cref{subsec:CompPDFsDecomp}, this effect is primarily due to
the form of $\rK^{\msbar \to \rFS}_{gg}$. 
This behaviour persists at higher scales but is substantially reduced in 
magnitude.
An example illustrating the effect of this on phenomenology is shown in 
\cref{fig:pheno_H}.

Another apparent feature is the difference in the $u$-valence (and $d$-valence) quark peak at large $x\sim0.2$ which can be observed at low 
scales (see \cref{fig:PDF_compare_uval,fig:xfdvxQ_CT18_2}).
This is partly washed out at higher scales by the DGLAP evolution.
The light quarks and anti-quarks 
(\cref{fig:PDF_compare_ub,fig:xfdxQ_CT18_2,fig:xfdbxQ_CT18_2,fig:xfsxQ_CT18_2})
exhibit differences at low $x$-values which persist to large factorisation scales.

The charm quark PDF (\cref{fig:PDF_compare_c}) shows a relatively large spread between the schemes 
and interestingly, for some schemes is negative.
This will be discussed further in \cref{subsec:CompPos}.
These features are however washed out by the evolution and at large scales, the charm PDF resembles the PDFs of the light sea quarks, cf.~\cref{fig:PDF_compare_ub}.

For completeness, we provide the remaining flavours for the transformed \ctnlo PDFs
in \cref{fig:PDF_compare_CT18_2}  in \cref{app:PDF_plots}.
As a general comment, for all flavours,
we observe that the differences between the schemes
decline as the factorisation scale $Q$ increases.
This follows naturally from the fact that
the PDF factorisation-scheme transformations of 
\cref{eq:fstransformations_faFS}
modify the PDFs at $\order{\alphas}$, and $\alphas(Q)$
decreases by a factor of approximately 2.5 between
$Q = 2 \; \GeV$ and $Q = 100 \; \GeV$.
This effect continues at higher scales,
where the schemes converge.

In general, the observations summarised above for \ctnlo also hold for
the
\nnpdfmc %
and \mshtnlo %
PDF sets,
shown in 
\cref{fig:PDF_compare_NNPDF40MC}
and
\cref{fig:PDF_compare_MSHT20nlo}
respectively.
At high-scales, the differences between the discussed PDF-sets are modest.
This is to be expected, as the DGLAP evolution `forgets'
initial conditions with the rising scale.
However, at low-$Q$ we observe several differences, especially prominent between
\mshtnlo and \ctnlo at $x \lesssim 10^{-4}$.
These are particularly visible in the charm, gluon, and light sea distributions.
In particular, for the 
\mpos, \mposd and \krk schemes, 
the \mshtnlo charm PDF is decreasing rather than increasing as $x \to 0$,
and becomes significantly negative. 
These features are driven by the much larger \mshtnlo \msbar
gluon distribution at scales close to the 
input scale $Q_0 \sim 1 \; \GeV$
(shown in \cref{fig:PDF_compare_g,fig:xfgxQ_MSHT20nlo}).
The dependence of the transformed PDFs on the features of the input PDFs
is discussed in further detail in \cref{subsec:CompPDFsDecomp}.

\subsection{Anatomy of transformed PDFs}
\label{subsec:CompPDFsDecomp}

We can look in more detail at how the PDFs in different schemes 
are assembled
from the input PDFs by
the transformation of Eq.~\eqref{eq:fstransformations_faFS}.
This is non-trivial, due to the nature of the convolution operation
and its interaction with the different input \msbar PDFs, and allows us to assess
the effects of different terms within the scheme-transformations.

To illuminate the effect of the different factorisation-scheme transformations
on proton PDFs, we consider a simultaneous decomposition by flavour and by contribution type.
Separating \cref{eq:fstransformations_faFS} into the contributions
to each (output) flavour according to the
input flavours we get, for quarks of flavour $f$,
\begin{align}
\label{eq:fstransformations_faFS_quark}
	f^{\rFS}_{q_f} (x, \muf)
	&= f^{\msbar}_{q_f} (x, \muf)
	 \\ &+ \frac{\alphas}{2\pi} \left[  \nonumber
  \rK^{\msbar \to \rFS}_{qg} \otimes f^{\msbar}_g
  + \rK^{\msbar \to \rFS}_{qq} \otimes f^{\msbar}_{q_f}
 \right],
\end{align}
and for the gluon,
\begin{alignat}{3}
 \label{eq:fstransformations_faFS_gluon}
 &	f^{\rFS}_g (x, \muf)
	= f^{\msbar}_g (x, \muf)
	\\ \notag
 & {} + \frac{\alphas}{2\pi} \left[
  \rK^{\msbar \to \rFS}_{gg} \otimes f^{\msbar}_g
  + \rK^{\msbar \to \rFS}_{gq} 
  \otimes \sum_{f} (f^{\msbar}_{q_f} + f^{\msbar}_{\qbar_f})
 \right].
\end{alignat}

In \cref{fig:PDF_decomposition_uv,fig:PDF_decomposition_ub,fig:PDF_decomposition_g,fig:PDF_decomposition_c}
we show a decomposition of the transformed PDFs based on the terms from 
\cref{tab:Kqq,tab:Kqg,tab:Kgq,tab:Kgg}.
To simplify the plots, the terms of the decomposition of \cref{eq:tableDecomp} used in
\cref{tab:Kqq,tab:Kqg,tab:Kgq,tab:Kgg} are recombined according to their asymptotic
behaviour as $z\to 0, 1$:
\begin{align}
 \label{eq:tableDecomp_plot}
	K(z) = {} & \sum_{k=0}^1 a_k \, \mathcal{D}_k (z) + b(z) \log (1-z) && {\color{red} \boldsymbol{\cdots \cdots}}
 \\ \notag
              & {} + c(z) \log z + P(z)                                 && {\color{decompyellow} \boldsymbol{- - -}}
 \\ \notag
              & - \Delta \, \delta(1-z)                                 && {\color{decompgreen} \boldsymbol{\cdot \, \mhyphen\mhyphen\mhyphen \cdot \mhyphen\mhyphen\mhyphen}}
\end{align}
The leading-order contribution to the transformed PDF is simply the input 
\msbar PDF itself (from \cref{eq:fstransformations_pertexp}), and is
plotted separately, either as a solid grey line,
or combined with the NLO $\delta$-function
contribution as
\begin{align}
    \hphantom{K(z) = {}}
    \left( 1 - \frac{\alphas}{2\pi} \Delta \right) \delta(1-z).
    &&
    {\color{decompblue} \boldsymbol{\cdot \, \mhyphen\mhyphen\mhyphen \cdot \mhyphen\mhyphen\mhyphen}}
    \label{eq:tableDecomp_combdelta}
\end{align}
This combination is shown separately to demonstrate that in many cases the
LO term in \cref{eq:fstransformations_faFS_gluon,eq:fstransformations_faFS_quark}
is mostly cancelled by the $\delta(1-z)$ terms responsible for momentum conservation.

To see more clearly how the new PDFs are built,
in each plot, the contributions to the transformed PDF are
plotted in aggregate (left-hand panel), as well as 
decomposed according to those
from input quark-type PDFs (central panel), 
and the input gluon PDF (right-hand panel).

For valence quark PDFs, the only quark-type contribution is from the corresponding \msbar PDF of the same flavour,
while for gluon PDFs the quark-type contribution is that of the transformed quark singlet PDF.
In each panel, the relevant contributions are again decomposed according to \cref{eq:tableDecomp_plot,eq:tableDecomp_combdelta}.

The sum of contributions from each kernel is plotted in black, both to give the total
NLO contribution arising from each input flavour (central/right-hand panels),
and to give the total overall PDF (left-hand panel).%
\footnote{Note that in order to obtain the full transformed `output' PDF (black line in left-hand panel),
we can either sum all the curves from the left-hand panel apart from the blue one (which is a combination of green and grey),
or sum the combined NLO contributions from input quarks (black line in central panel)
and gluons (black line in right-hand panel)
together with the LO contribution from the input \msbar PDF (grey line in left-hand panel).}
\subsubsection{Valence quark PDFs}
\label{subsubsec:CompqPDFs}
\begin{figure*}[p]
\begin{leftfullpage}
\centering
\includegraphics[width=\textwidth]{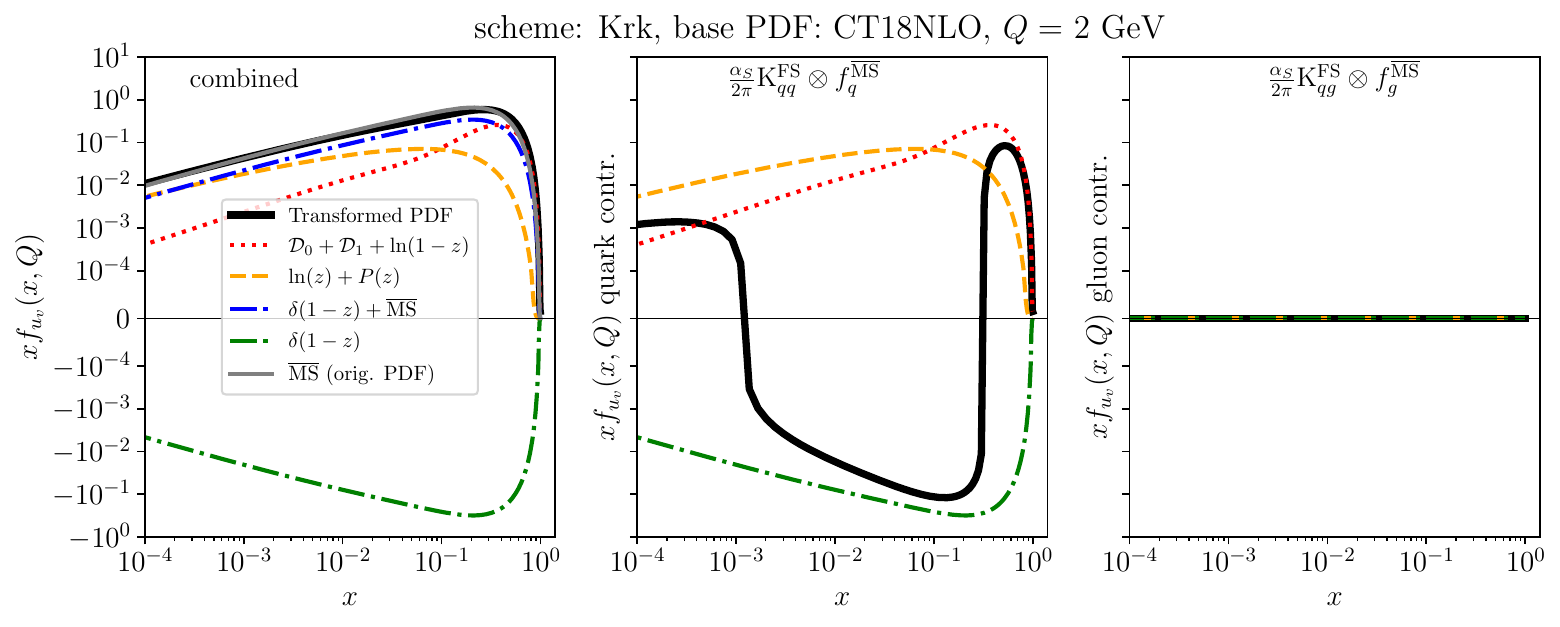}
\\
\includegraphics[width=\textwidth]{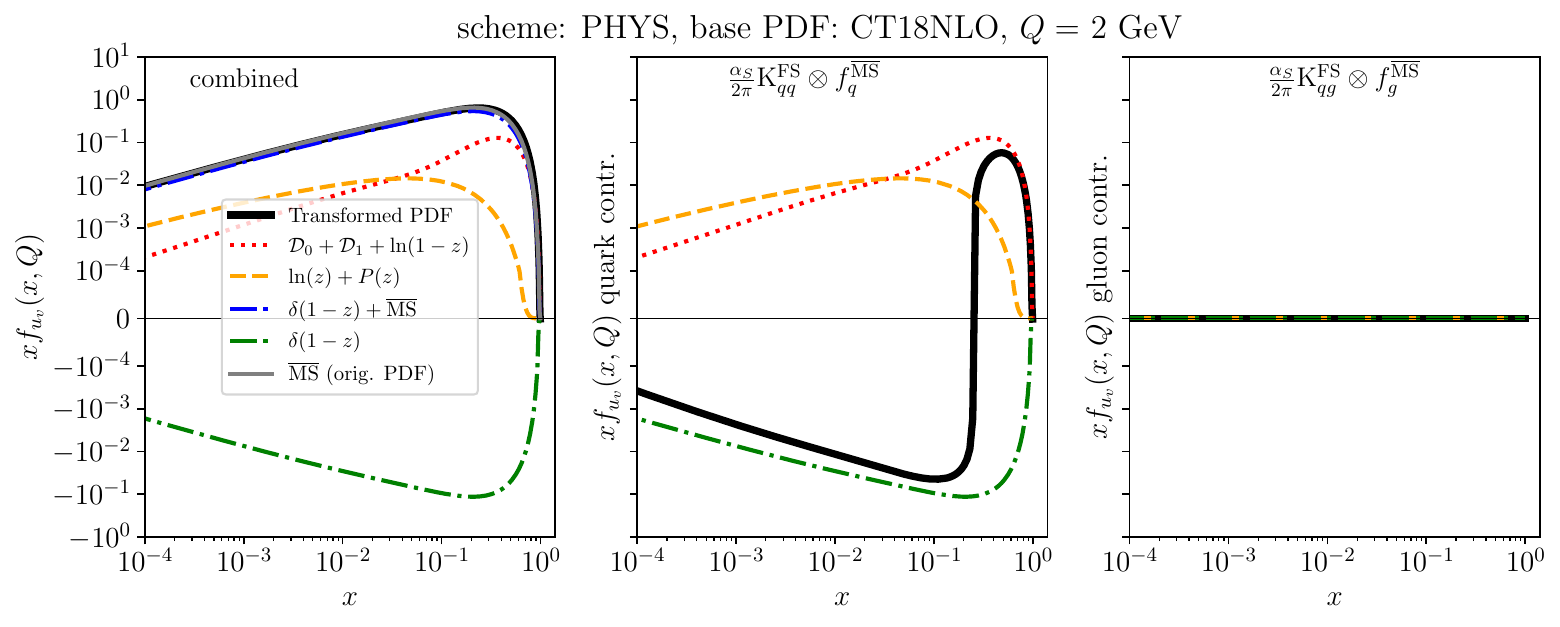}
\\
\includegraphics[width=\textwidth]{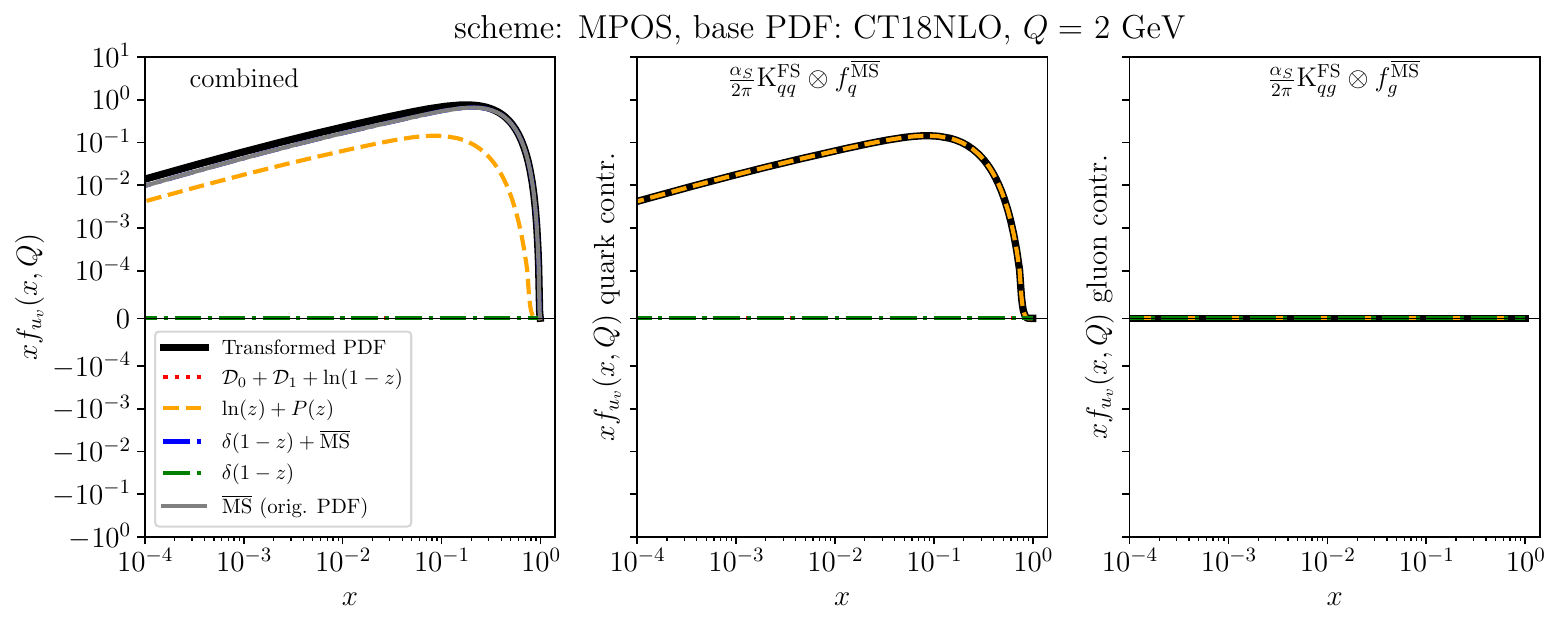}
\caption{Decomposition of transformed $u_v$-quark PDF in the \krk, \mpos, and \phys{} schemes at factorisation scale $Q=2$ GeV,
as described in \cref{subsec:CompPDFsDecomp}.
}
\label{fig:PDF_decomposition_uv}
\end{leftfullpage}
\end{figure*}

In \cref{fig:PDF_decomposition_uv} we show the decomposition of the transformed up-quark valence PDF
 ${u_v}$, 
\begin{align}
 \label{eq:fstransformations_faFS_valence}
    f_{u_v} (x, Q)
    {} & =
    f_u(x,Q) - f_{\bar{u}} (x,Q),
\end{align}
in
the \krk (1st row), 
\mpos (2nd row),
and \phys (3rd row) schemes at 
factorisation scale $\mu=2$ GeV.
The remaining schemes are presented in \cref{fig:PDF_decomposition_uv_2} in \cref{app:extra_decomp_plots}. 
The same transformed $u_v$ PDFs were previously shown
in \cref{fig:PDF_compare_uval}, using a linear scale for the $y$-axis.

It follows from \cref{eq:fstransformations_faFS_quark} that for quark valence
PDFs
the contribution from the gluon cancels,
\begin{align}
f^{\rFS}_{u_v} (x, \muf)
	&= f^{\msbar}_{u_v} (x, \muf) + \frac{\alphas}{2\pi} \left[ 
    \rK^{\msbar \to \rFS}_{qq} \otimes f^{\msbar}_{u_v}
 \right].
\end{align}
The gluon contribution to the valence quark distribution can indeed be seen to be zero
in the right-hand panels of \cref{fig:PDF_decomposition_uv,fig:PDF_decomposition_uv_2}.

As is expected from a perturbative transformation,
in all schemes the dominant contribution to the total PDF
(black line in left-hand panel)
is given by the leading order (input) \msbar contribution (grey line).
This remains true at higher scales and for the schemes not shown here
(see \cref{app:extra_decomp_plots}).
The net NLO contribution to the transformed PDF is visible as the
black line in the central panels.

For the \krk, \dis and \aversa schemes there is a large cancellation between
the input \msbar distribution
(the LO contribution to the transformed PDF)
and the NLO $\delta(1-z)$ contribution 
(compare blue and grey lines in the decomposition figures).
This is responsible in part for the vertical shifts in the valence peak
visible in \cref{fig:PDF_compare_uval,fig:xfdvxQ_CT18_2}, and can be attributed to the large 
$\delta(1-z)$ coefficients in \cref{tab:Kqq}.%
\footnote{Note the minus-sign in the heading of the $\delta(1-z)$ column (equivalently
in front of $\Delta$ in \cref{eq:tableDecomp}).}
The $\delta(1-z)$ terms also dominate the vertical shift in other schemes where they are partially mitigated by the positive contribution
from the $\mathcal{D}_k$ distribution and $\log(1-z)$-terms.
\mposd is the only scheme with opposite (positive) sign of $\delta(1-z)$ coefficient which causes the upwards vertical shift of the valence peak visible in \cref{fig:PDF_compare_uval}.
For the \mpos scheme, the $P(z)$ contribution fulfils the same role
of restoring momentum-conservation 
and shifts the peak in the same upward vertical
direction, but the shape-effect of the soft-function shifts
the valence peak towards lower $x$
(this will be further discussed in \cref{subsubsec:CompPDFsImpMomCons}).

For the \aversa, \dis, \phys, and \krk schemes the convolution
with the $\mathcal{D}_k$ distribution and $\log(1-z)$ terms
shifts the valence peak horizontally to higher $x$ values,
and is the dominant NLO contribution at high-$x$ (red line in central panel).
This shift is largest for the \krk scheme due to the distribution contribution
(coefficients $a_1$ and $b$) being exactly twice as large as for the \phys and \aversa/\dis schemes.

The difference in $c(z)\log z + P(z)$ between \phys and \krk schemes is driven by 
the absence of the $ - \pqq \log z$ contribution cf. \cref{tab:Kqq};
this has a noticeable shape effect at large-$x$.
The effect of the specific choice of $P(z)$ is significant and can be seen comparing
the \krk scheme shown in \cref{fig:PDF_decomposition_uv}  to \aversa/\dis, shown in \cref{fig:PDF_decomposition_uv_2} (yellow lines).

At $Q = 100 \; \GeV$ the contributions and conclusions are very similar
(see \cref{fig:PDF_decomposition_100_uv}).

In this section we focused on $u$-valence PDFs, but similar
observations hold for $d$-valence PDFs (\cref{fig:xfdvxQ_CT18_2}).

\subsubsection{Sea quark PDFs}
\label{subsubsec:CompqbPDFs}

\begin{figure*}[p]
\centering
\includegraphics[width=\textwidth]{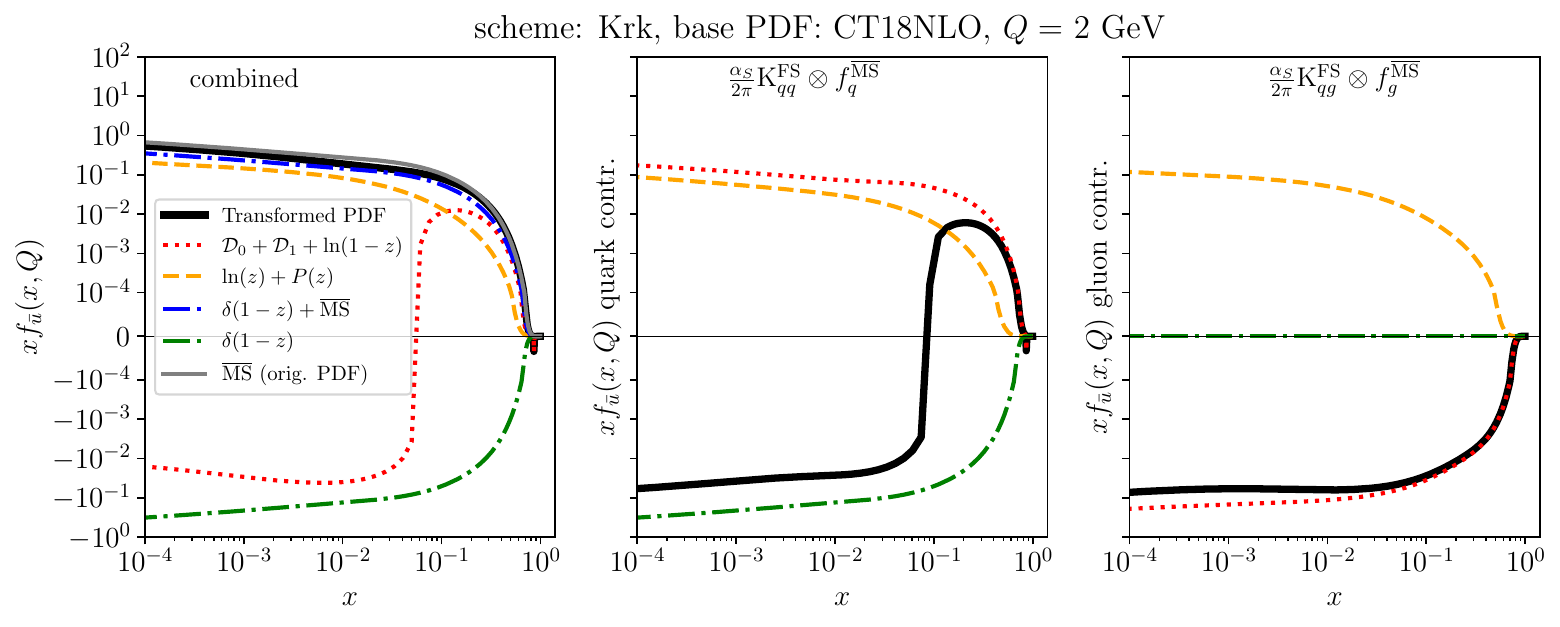}
\\
\includegraphics[width=\textwidth]{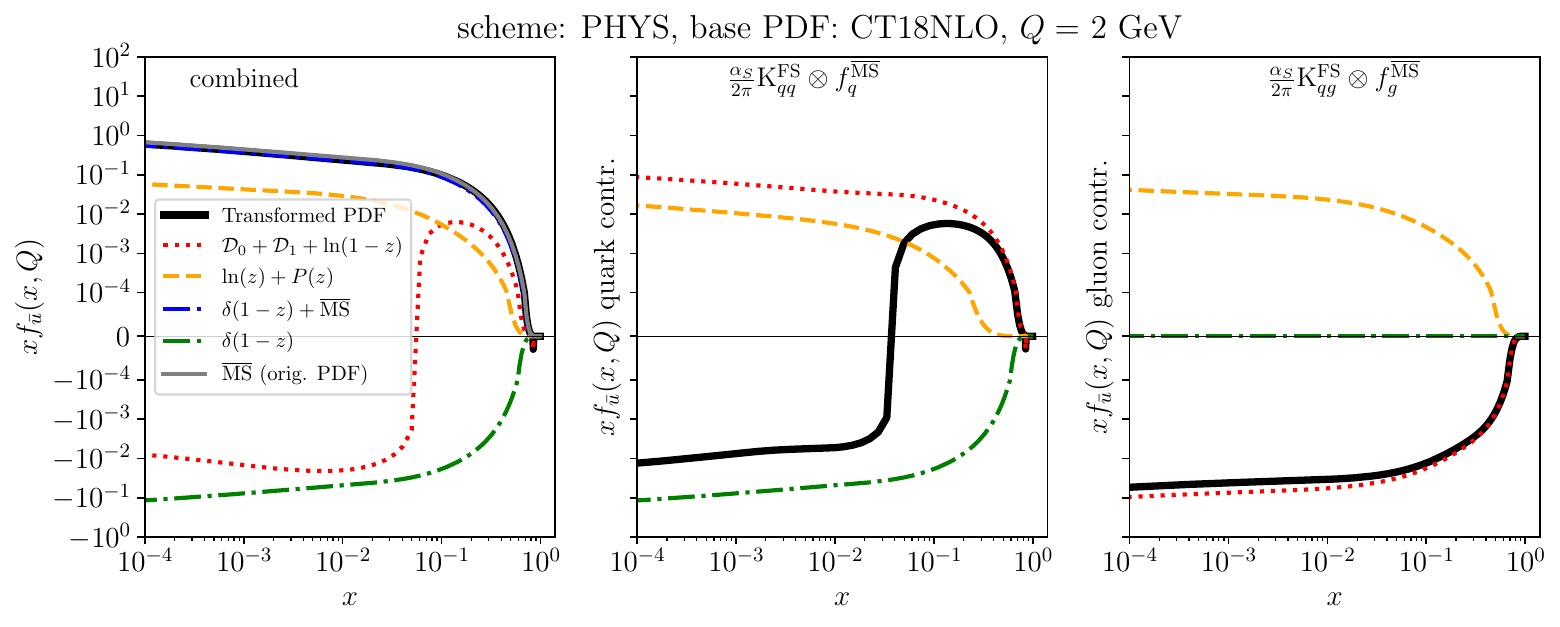}
\\
\includegraphics[width=\textwidth]{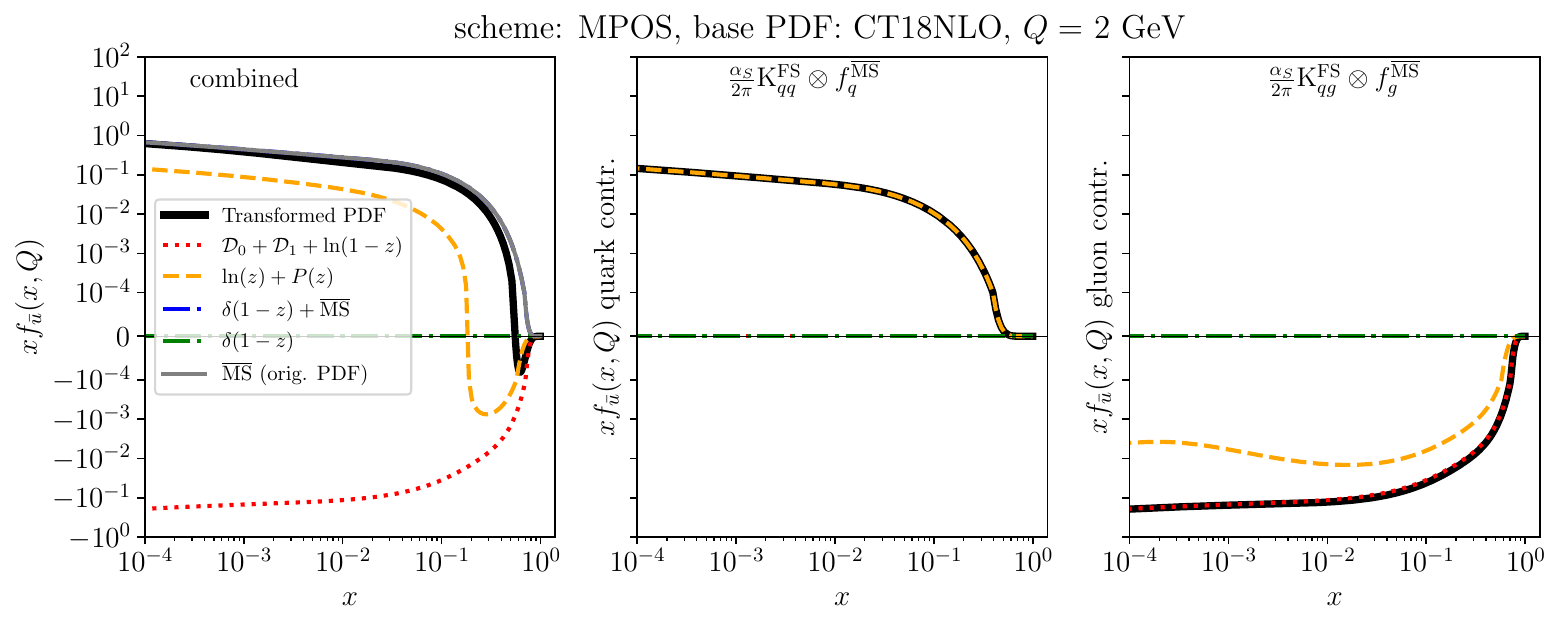}
\caption{Decomposition of transformed $u$-antiquark PDF in the \krk, \mpos, and \phys{} schemes at factorisation scale $Q=2$ GeV,
as described in \cref{subsec:CompPDFsDecomp}.
}
\label{fig:PDF_decomposition_ub}
\end{figure*}

In \cref{fig:PDF_decomposition_ub} we show the decomposition of the 
transformed $\bar{u}$ PDFs;
the remaining schemes are shown in \cref{fig:PDF_decomposition_ub_2}.
Unlike the valence-quark PDFs, this receives contributions from the \msbar gluon PDF through
the $qg$ kernel,
in addition to those from the \msbar $\bar{u}$ PDF through the $qq$ kernel.

At high-$x$, the contributions from the gluon are dominated by the $\log(1-z)$
contribution (red line in the right panel), which is negative, and is either equal to 
$\pqg \log(1-z)$ (\aversa, \dis, \phys)
or $2 \pqg \log(1-z)$ (\krk, \mpos, \mposd)
as shown in \cref{tab:Kqg};
the gluon contribution to the transformed PDFs is therefore qualitatively very similar.
At low-$x$, the $-\pqg \log z$ contribution
common to all schemes
is positive and can be seen
directly in the \aversa scheme plot in \cref{fig:PDF_decomposition_ub_2} (yellow line in right panel).
With the exception of the \aversa scheme (for which $P(z) \equiv 0$)
there is a common $-\pqg$ contribution to $P(z)$
in all schemes (scaled by a factor of 4 for \dis).
This is negative and largely cancels with the $- \pqg \log z$ contribution
to give a residual (smaller) negative
contribution (visible in the \mpos
and \mposd schemes).
This reduction in magnitude leads the overall contribution from the gluon PDF (black line in right panel)
to closely match the $\log(1-z)$ contribution for \mpos/\mposd.
The addition of 1 in $P(z)=-\pqg+1$ in the \krk and \phys schemes
is sufficient to turn the $P(z)$ and 
$P(z) + c(z) \log z$ contributions positive
(visible in the \phys and \krk schemes respectively; 
yellow line in right panel).

The contributions from the input sea-quark PDFs are qualitatively similar
for the \krk, \phys, \aversa and \dis schemes (see \cref{tab:Kqq} and black line in the middle panel of \cref{fig:PDF_decomposition_ub,fig:PDF_decomposition_ub_2}).
The \mpos-type schemes are off-diagonal transformations ($\rK_{qq}$ vanishes), 
apart from the terms responsible for momentum-conservation 
via a soft function or $\delta(1-z)$.
At high-$x$, the quark contributions are governed by the $\mathcal{D}_k$ distribution terms as well as by
the large $\delta(1-z)$ term in the case of \aversa/\dis/\krk;
$\mathcal{D}_0$ can be seen to contribute a characteristic bump in
\aversa/\dis.
At low-$x$, there is a large cancellation between the sum of contributions from the 
$\mathcal{D}_k+\log(1-z)$ and polynomial terms (red and yellow lines) which are positive
and the negative delta-function contributions (green line).
The net effect of the $qq$-kernels at low-$x$ is an overall-negative
contribution,
which for the  \aversa, \krk, \phys schemes is 
similar in magnitude and direction to that of the $qg$-kernels,
leading to the overall suppression visible in \cref{fig:PDF_compare_ub}.

At $Q = 100 \; \GeV$ (see \cref{fig:PDF_decomposition_100_ub}) the individual contributions are qualitatively similar,
especially from the input sea-quark,
but the negative off-diagonal contribution from the gluon is much larger
due to the growth of the gluon PDF.
This contribution grows faster than the reduction in $\alphas$ leading to the 
larger relative spread of transformed PDFs than at $Q = 2 \; \GeV$,
visible in \cref{fig:PDF_compare_ub,fig:PDF_decomposition_100_ub}.

In this section we focused on the $\bar{u}$ PDF; similar observations
hold for the remaining sea-quark flavours
(see also \cref{fig:xfdbxQ_CT18_2,fig:xfsxQ_CT18_2}).

\subsubsection{Gluon PDFs}
\label{subsubsec:CompgPDFs}

\begin{figure*}[p]
\centering
\includegraphics[width=\textwidth]{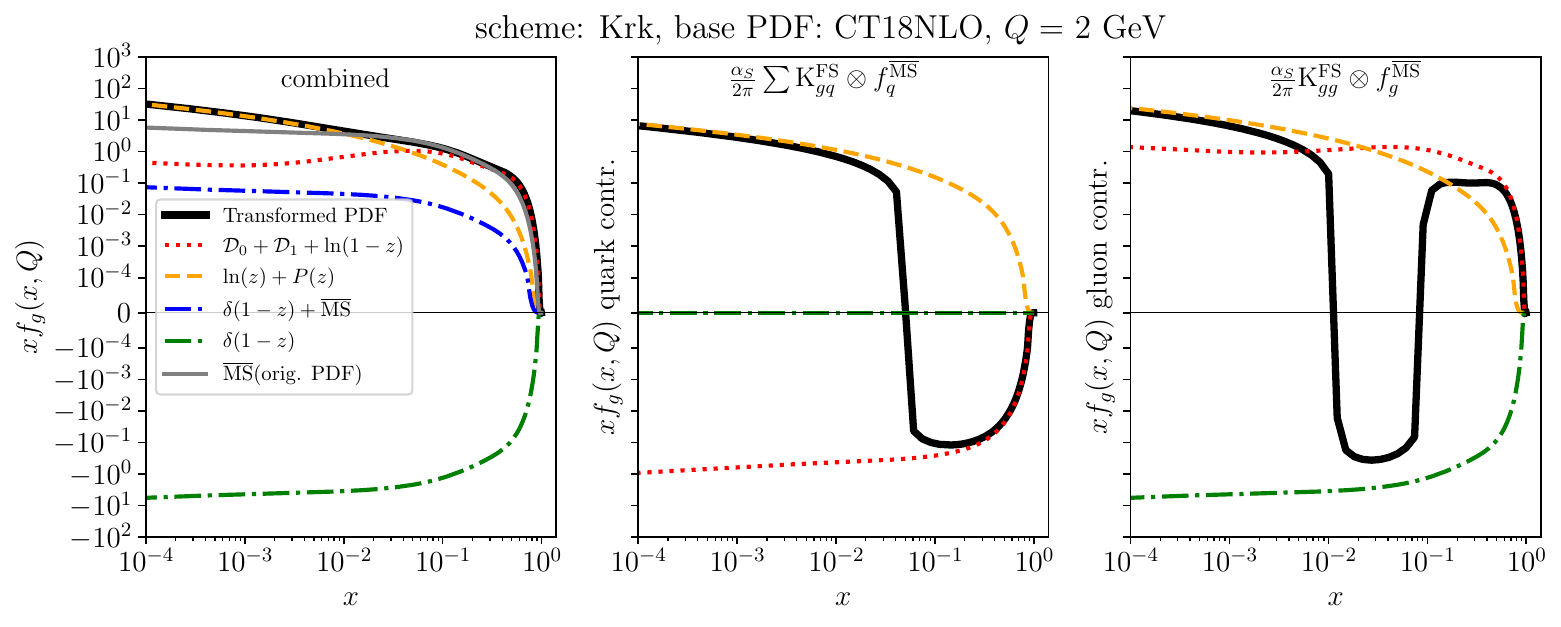}
\\
\includegraphics[width=\textwidth]{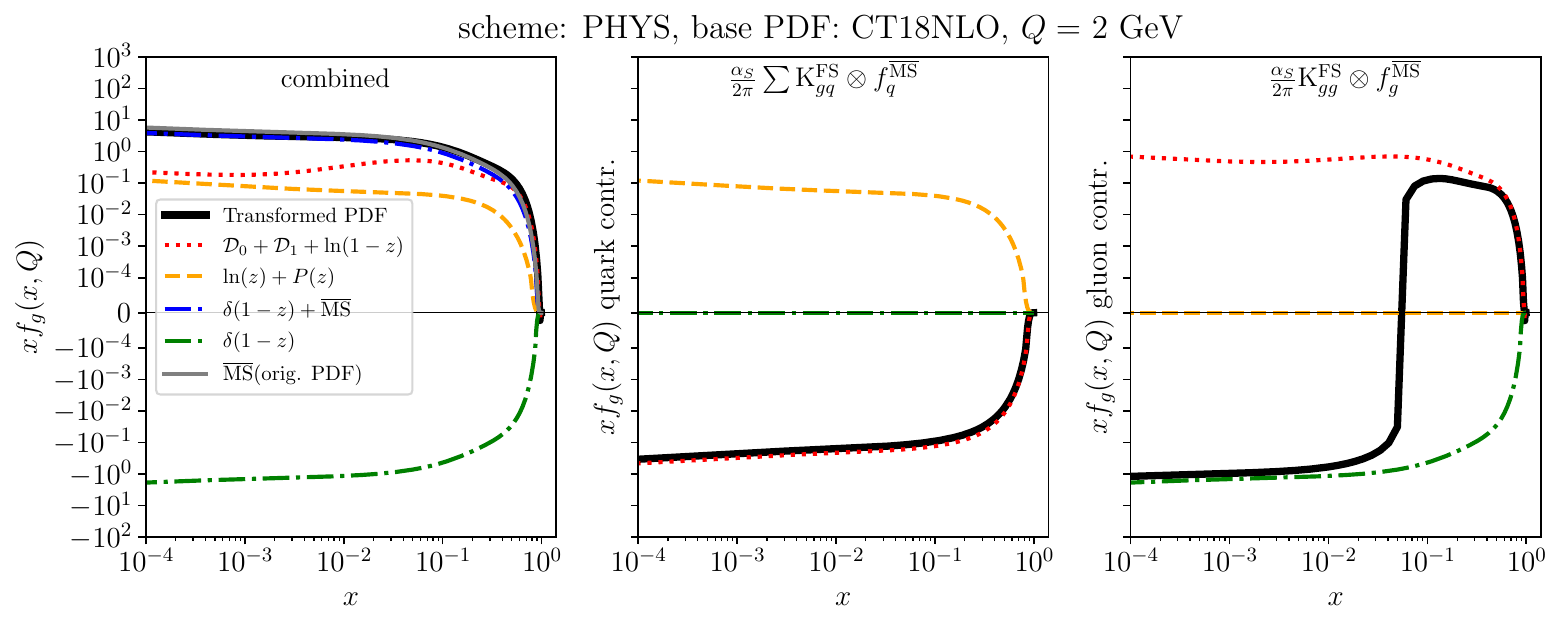}
\\
\includegraphics[width=\textwidth]{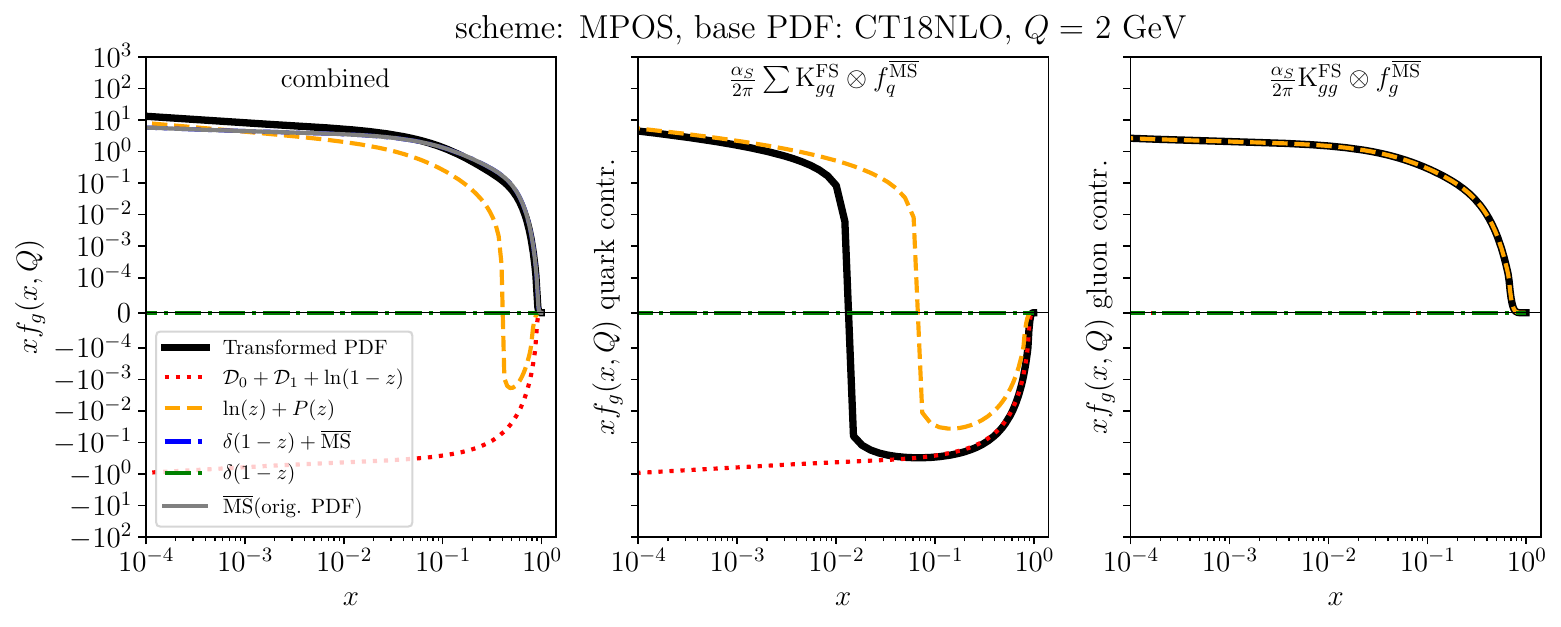}
\caption{Decomposition of transformed gluon PDF in the \krk, \mpos, and \phys schemes at factorisation scale $Q=2$ GeV,
as described in \cref{subsec:CompPDFsDecomp}.
}
\label{fig:PDF_decomposition_g}
\end{figure*}

We show the decomposition of the transformed gluon PDFs in in \cref{fig:PDF_decomposition_g};
the remaining schemes are shown in \cref{fig:PDF_decomposition_g_2}.

At low-$x$, the NLO contribution becomes comparable in magnitude to LO for
the \krk, \mpos, \aversa and \mposd schemes (compare grey and black lines in the left panel),
indicating a breakdown of perturbation theory.
This is not the case for the \dis or \phys schemes.
The behaviour at low-$x$ may be seen to be dominated by the $\log z$ contribution
of \cref{tab:Kgq,tab:Kgg}
(i.e.\ the choice of $c(z)$ in \cref{eq:tableDecomp}).
The \dis scheme behaves markedly different in this respect due to the appearance of the
$\pqq$ splitting function in the $\rK_{gq}^{\dis}$ kernel in \cref{tab:Kgq};%
\footnote{\label{footnote:DIS}
This is due to the way momentum conservation is enforced in the DIS scheme, \cref{eq:DIS_localsumrule}.
}
other schemes have either $c(z) \equiv 0$ (\phys), or
$ c(z) = \pgq $ (all other schemes).
In the flavour-diagonal contribution from the input-gluon PDF, the large
difference between the \krk scheme and the others, evident in \cref{fig:PDF_compare_g},
can be traced back to the $c(z) = -2 \pgg$ coefficient of $\log z$ in $\rK_{gg}^{\krk}$
(\cref{tab:Kgg}), which contributes a divergent low-$z$ 
term proportional to $\log (z) / z$.

At large-$x$ the shape is governed by the distribution terms $\mathcal{D}_k$,
where present.
For the $\rK_{gg}$ kernels, this contribution (red line) is exactly twice as large
for the \krk scheme as for \phys, and approximately twice as large
as for \aversa.
For the $\rK_{gq}$ kernels, the only non-zero contribution relevant at high-$x$ is
proportional to $\pgq \log(1-z)$ with coefficient
$\cf$ (\aversa, \phys)
or $2\cf$ (\krk, \pos-type),
with the exception of the \dis scheme which uniquely contains
$\mathcal{D}_{0,1}$ terms originating from the $C_q$ DIS coefficient function.%
\footref{footnote:DIS}
This leads to a region of negativity for $x \gtrsim 0.55$.

At $Q=100 \; \GeV$
(shown in \cref{fig:PDF_decomposition_g,fig:PDF_decomposition_100_g})
the effect of the contributions is qualitatively similar,
but reduced in magnitude; 
only the \krk scheme PDF is shifted significantly from the
leading-order contribution given by the input \msbar PDF,
due again to the $\log(z)/z$ contribution.

\subsubsection{Heavy quark PDFs}
\label{subsubsec:CompPDFsHeavyQ}

\begin{figure*}[p]
\centering
\includegraphics[width=\textwidth]{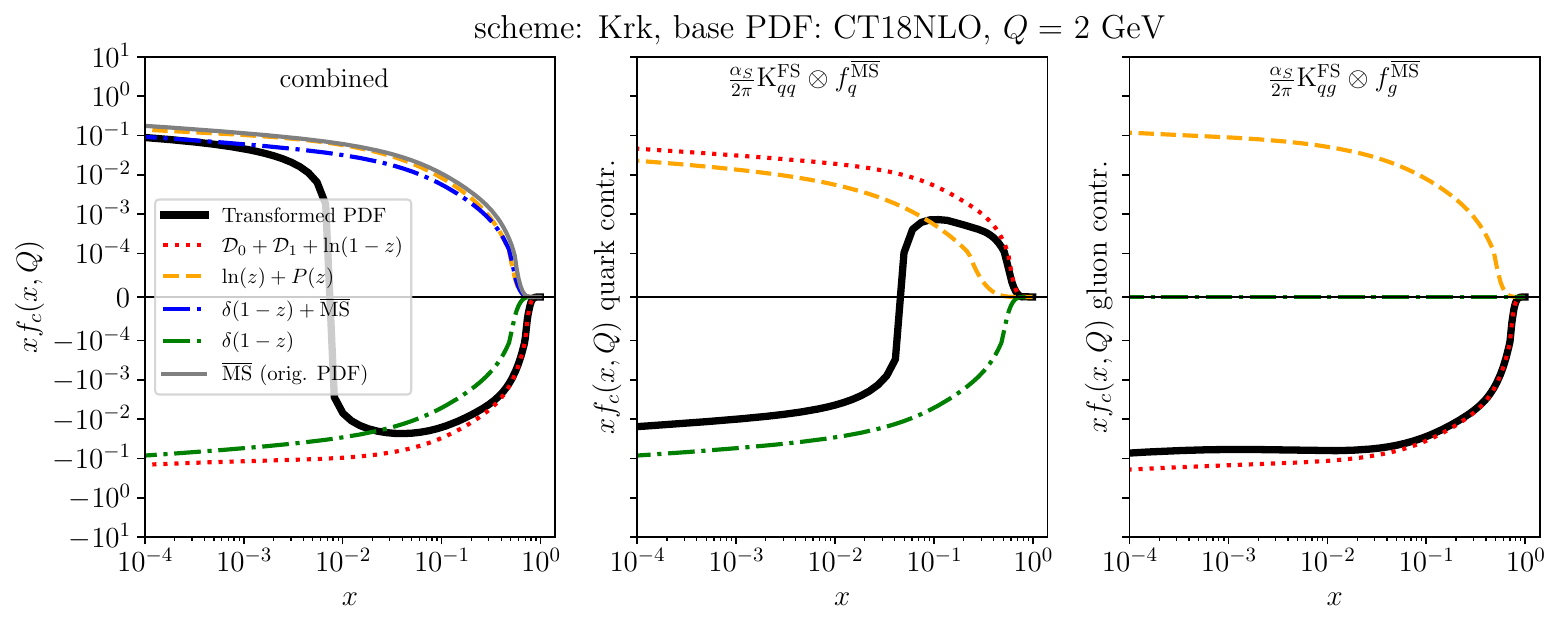}
\\
\includegraphics[width=\textwidth]{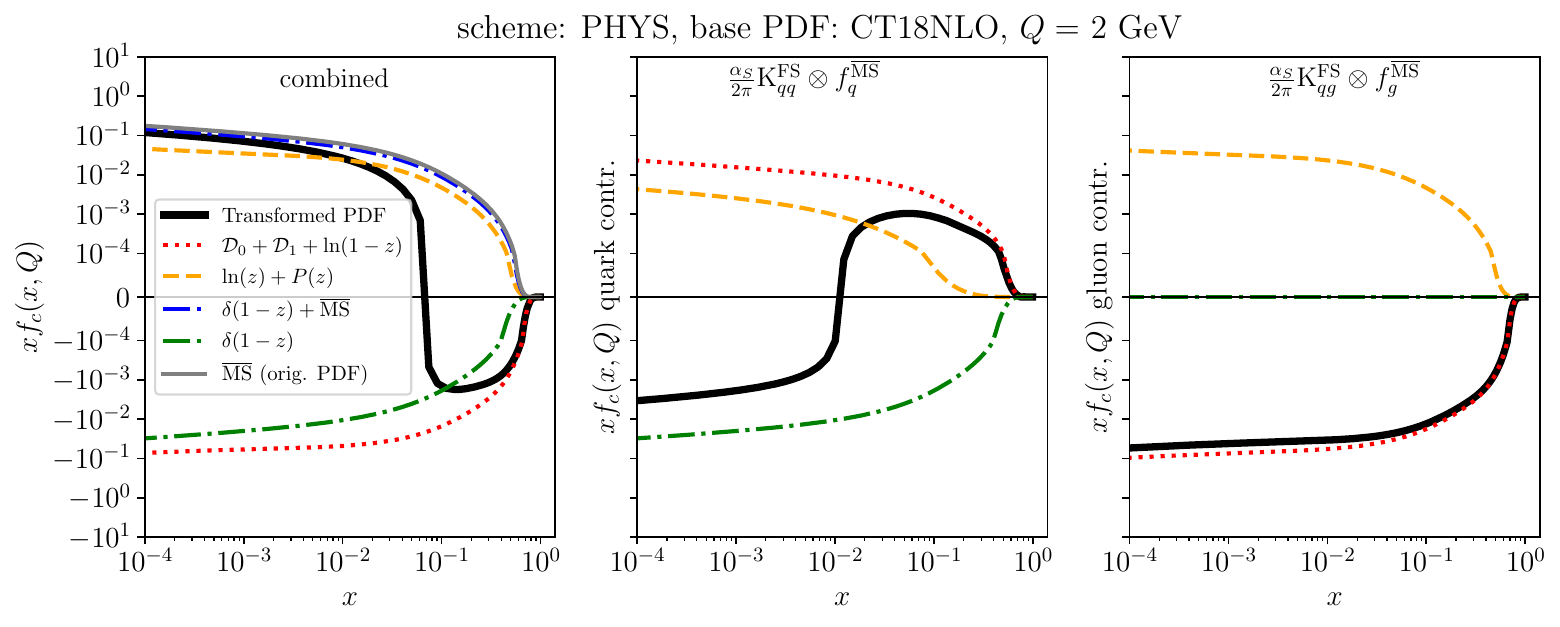}
\\
\includegraphics[width=\textwidth]{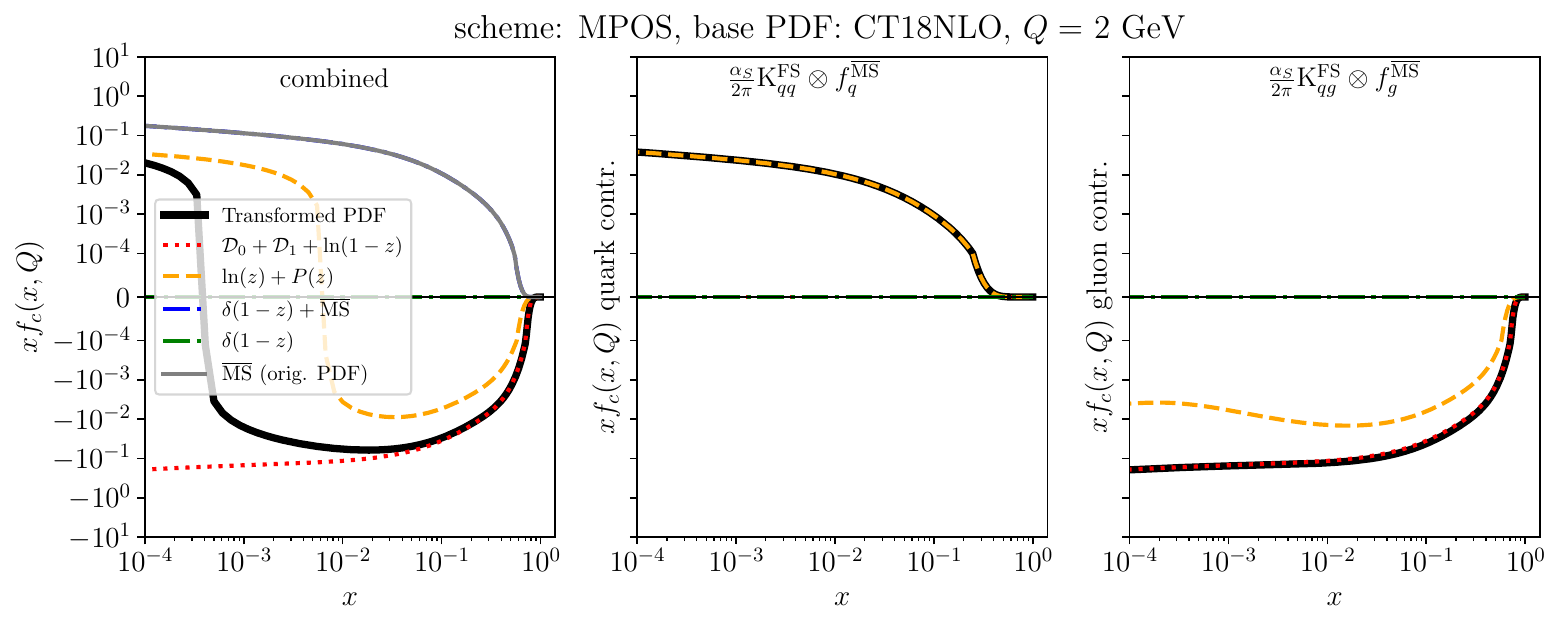}
\caption{Decomposition of transformed charm-quark PDF in the \krk, \mpos, and \phys schemes at factorisation scale $Q=2$ GeV,
as described in \cref{subsec:CompPDFsDecomp}.
}
\label{fig:PDF_decomposition_c}
\end{figure*}

\begin{figure*}[p]
\centering
\includegraphics[width=\textwidth]{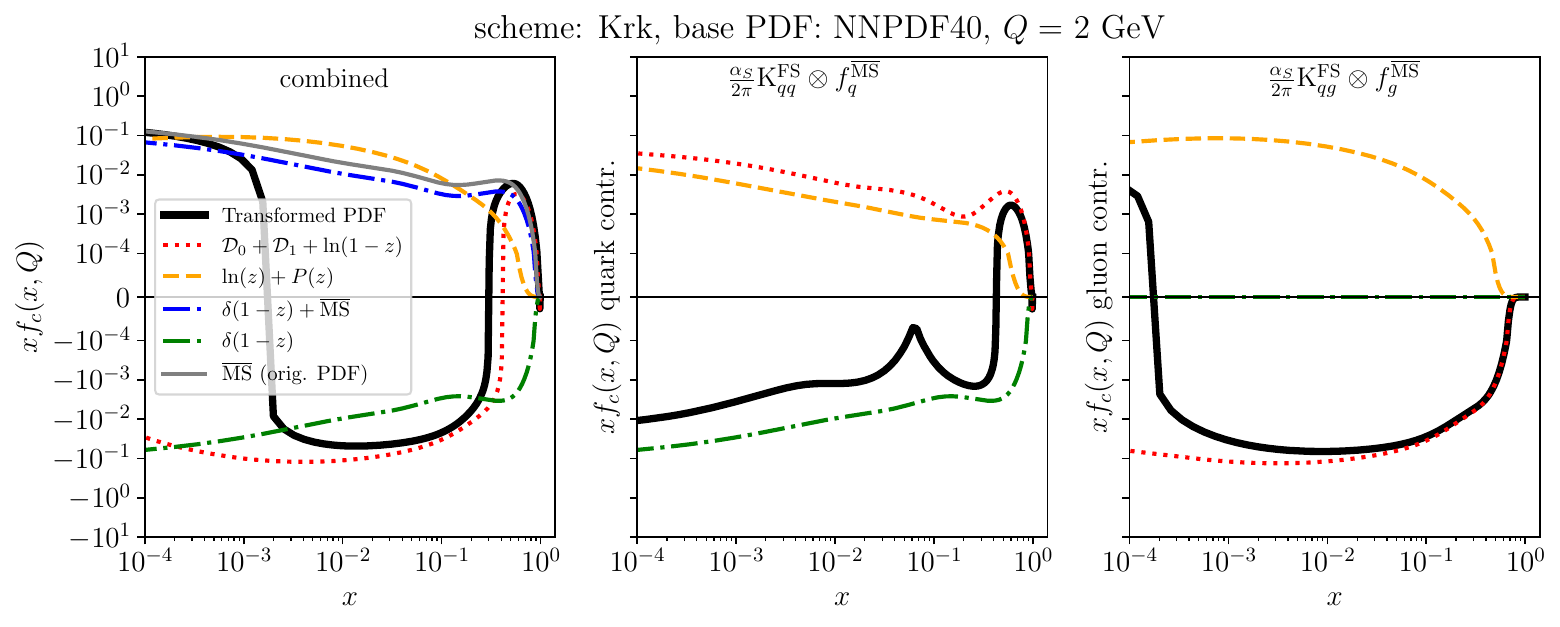}
\\
\includegraphics[width=\textwidth]{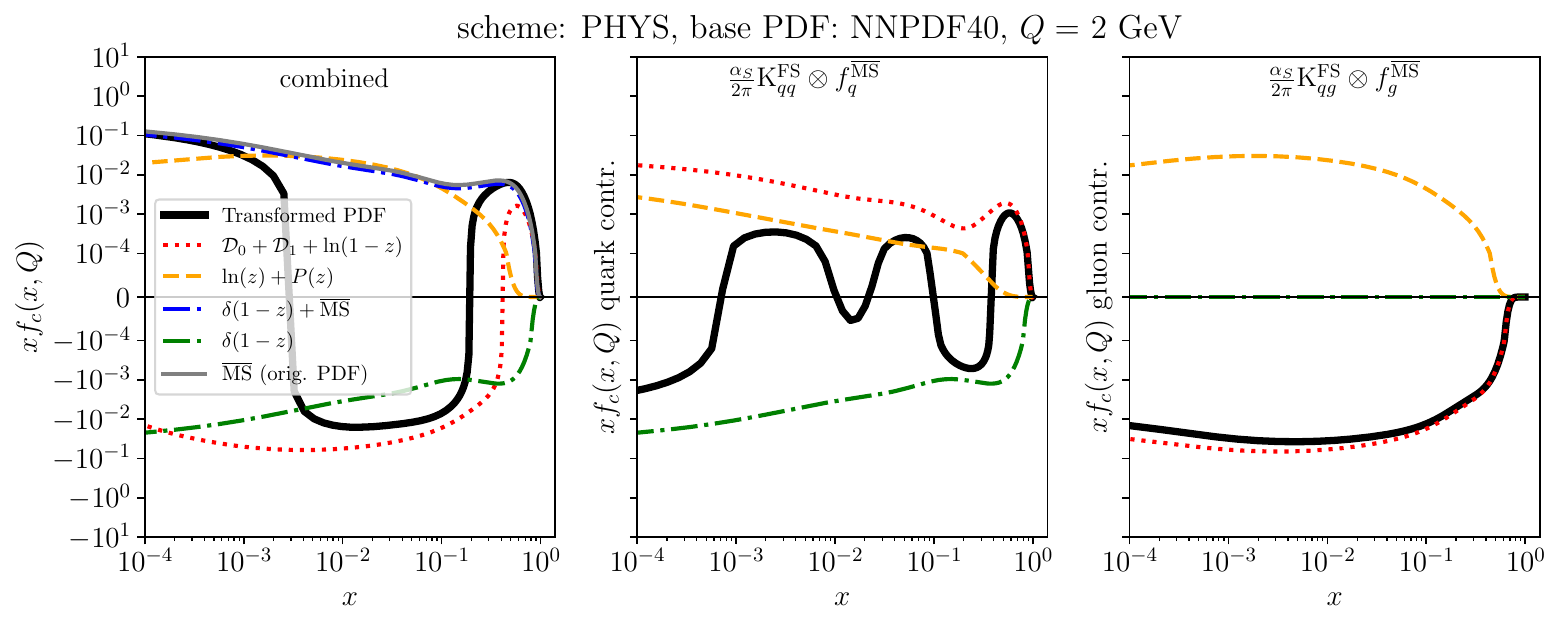}
\\
\includegraphics[width=\textwidth]{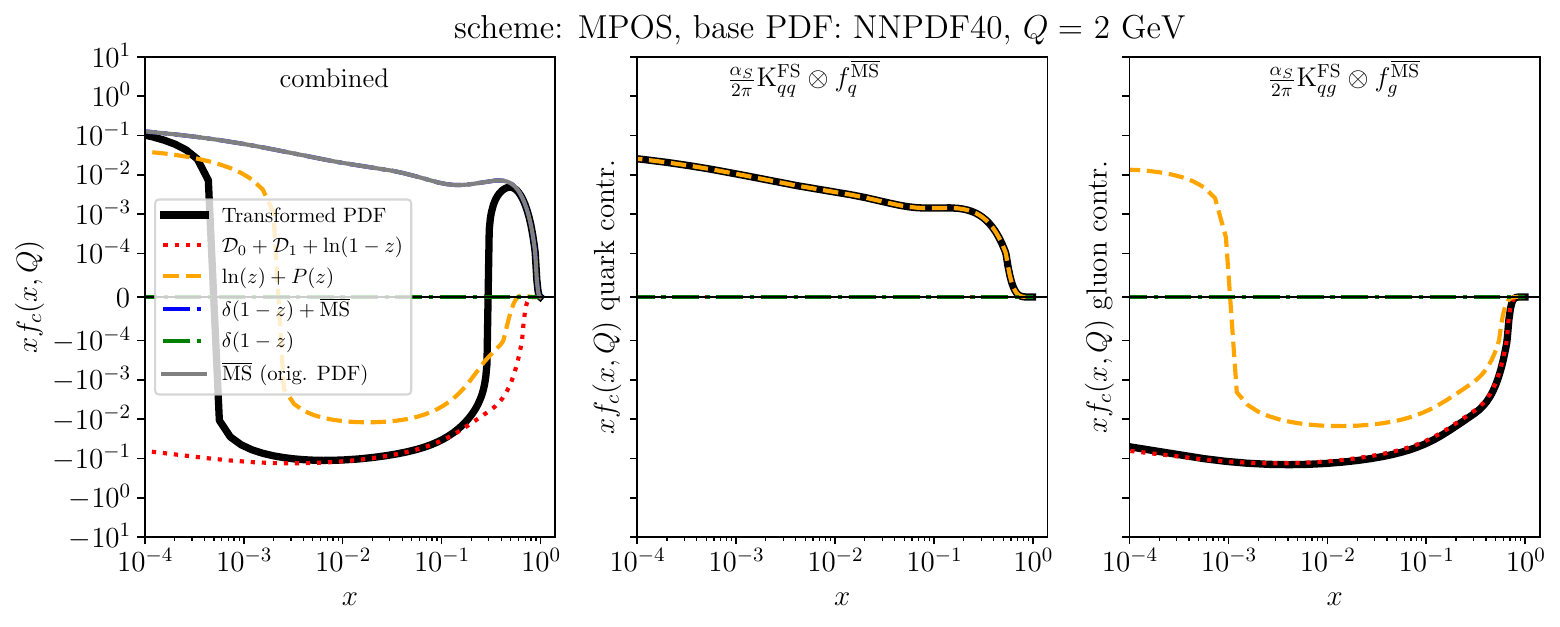}
\caption{Decomposition of transformed charm-quark PDF in the \krk, \mpos, and \phys schemes at factorisation scale $Q=2$ GeV,
as described in \cref{subsec:CompPDFsDecomp}, based on \nnpdf \msbar PDFs.
Note that in contrast with \cref{fig:PDF_decomposition_c},
the \nnpdf PDF set includes an intrinsic charm contribution.
}
\label{fig:PDF_decomposition_NNPDF40_c}
\end{figure*}

In this section we use the example of charm-quark PDF to examine how heavy-quark PDFs are constructed in the different schemes.
As discussed in \cref{subsubsec:flavthresh},
since modern PDFs use variable-flavour-number schemes (VFNS),
heavy-quark PDFs are non-zero only above the flavour thresholds,
at scales $\mu_f \gtrsim m_f$.
The behaviour at these thresholds will be discussed here and also later in \cref{subsec:CompPos}.

As before, we show the decomposition of the transformed charm-quark PDF at 
$\mu=2$ GeV for the \krk, \mpos, and \phys schemes in \cref{fig:PDF_decomposition_c},
and for the remaining schemes in \cref{fig:PDF_decomposition_c_2} in \cref{app:extra_decomp_plots}.
These can be also compared with \cref{fig:PDF_compare_c} showing the
transformed charm PDFs (using a linear scale for the $y$-axis).

Relative to the sea-quark PDFs previously discussed in \cref{subsubsec:CompqbPDFs},
the contributions from the input gluon PDF are identical,
while the contributions from the input heavy quark PDF are suppressed
by approximately an order of magnitude,
due to the proximity to the flavour threshold.

At high-$x$, all of the schemes exhibit negativity (see black line in the left panels).
This is driven by the $\log(1-z)$ terms within the $\rK_{qg}$
kernels of \cref{tab:Kqg} (red line in right panels). 
The coefficient of these terms in the
\krk and \pos-type schemes is double that of the others.
In the \krk and \phys schemes, this negative contribution is partially
mitigated with respect to the \pos-type schemes
by the inclusion of $+1$ in the $P(z)$ contribution
(see yellow line in right-hand panels).
This generates a positive contribution from the gluon input PDF
which is of a similar order of magnitude to (though smaller than)
the negative contribution from the $\log (1-z)$ terms.
For the \mpos scheme the $P(z)$ contribution is itself negative
and suppressed by an order of magnitude,
so provides no such mitigation.
As a result the region of negativity for the \mpos scheme stretches an order of magnitude further than for the \krk scheme, to
$x \gtrsim 4 \cdot 10^{-4}$.

At low-$x$, the leading-order contribution dominates except for the
\mpos-type schemes and, to a lesser extent, the \krk scheme,
due to the large negative contributions summarised above.

At $Q=100 \; \GeV$, the transformed $c$-quark PDF resembles that
of the sea-quarks (see \cref{fig:PDF_compare,fig:PDF_decomposition_100_c}).

In \cref{fig:PDF_decomposition_NNPDF40_c} we show the same decomposition plots as in
\cref{fig:PDF_decomposition_c} using the \nnpdf PDF set, which includes
an intrinsic charm-quark contribution.
The intrinsic contribution restores positivity at high-$x$ through the 
distribution contributions, while the negativity at intermediate-$x$
persists.

The negativity of the heavy-quark PDFs 
is discussed further in \cref{subsec:CompPos}.

\subsection{Momentum sum-rule}
\label{subsec:CompPDFsMomCons}

As discussed in \cref{subsubsec:momsumrule}, PDFs in the \msbar scheme
are constrained to satisfy the momentum sum-rule of \cref{eq:momsumrule_PDFs},
for all factorisation-scales $\mu$;
that is, the average momentum fractions carried by each
parton must sum to 1 (the total momentum fraction of the hadron; here, proton).

To compare factorisation schemes
on a like-for-like basis, this has been imposed by the
modification of the $\Delta$ coefficient in \cref{tab:Kqq,tab:Kqg,tab:Kgq,tab:Kgg}
according to \cref{eq:momsumrule_deltas},
to satisfy momentum conservation at the level of the transformation kernels
\cref{eq:momsumrule_kernels,}. The situation is different for 
\mpos where it is imposed through a `soft-function'
rather than a $\delta$-function, and for 
\dpos and \pos where the sum rule is not fulfilled.

In this section, we directly test the sum rule of the transformed PDFs in \cref{subsubsec:CompPDFsMomCons},
and compare the different methods for imposing it,
focusing on the \mpos and \mposd schemes, in \cref{subsubsec:CompPDFsImpMomCons}.

\subsubsection{Numerical momentum conservation}
\label{subsubsec:CompPDFsMomCons}
\begin{figure*}[tp]
\centering
\begin{subfigure}[t]{0.49\textwidth}
    \centering
    \includegraphics[width=\textwidth]{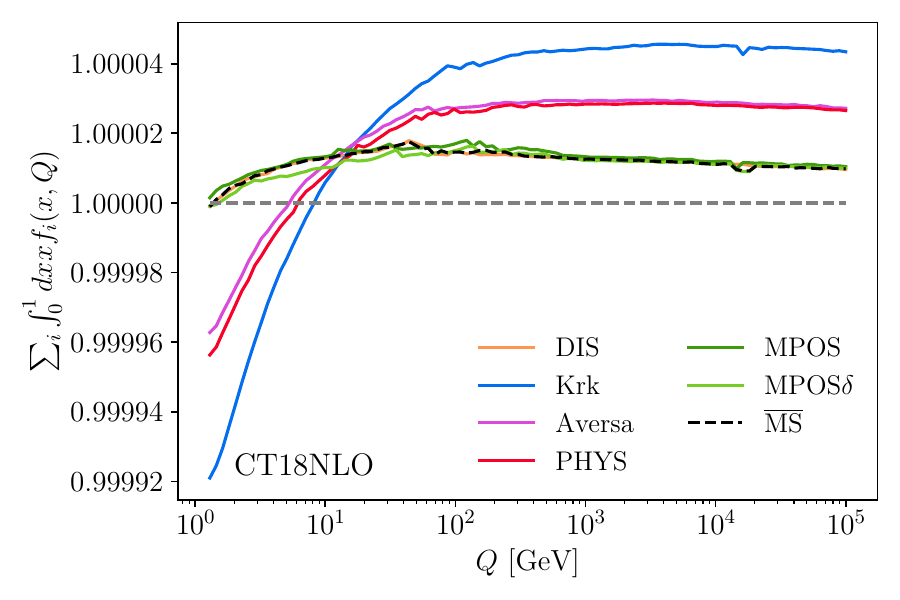}
    \caption{Factorisation schemes for which momentum-conservation is imposed.}
    \label{fig:momentum_sum_rules_a}
    \end{subfigure}
\hfill
   \begin{subfigure}[t]{0.49\textwidth}
    \centering
    \includegraphics[width=\textwidth]{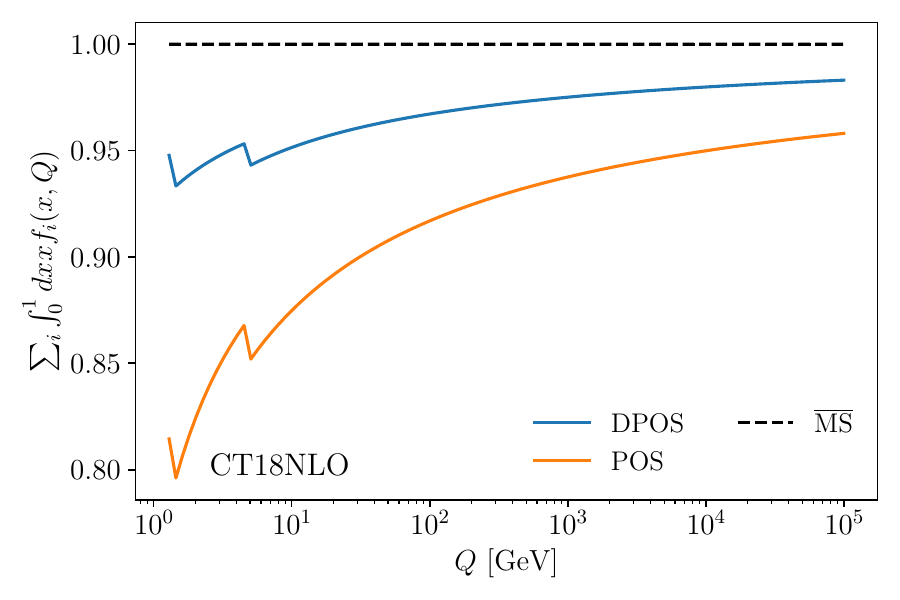}
    \caption{\pos-type schemes without momentum-conservation imposed.}
    \label{fig:momentum_sum_rules_b}
\end{subfigure}
\caption{Numerical calculation of the momentum sum-rule,
        \cref{eq:momsumrule_PDFs}, as a function of the factorisation scale $Q$,
        for \ctnlo PDFs transformed into the considered factorisation schemes.
\label{fig:momentum_sum_rules}}
\end{figure*}

\begin{figure*}[tp]
\centering
\begin{subfigure}[t]{0.49\textwidth}
    \centering
    \includegraphics[width=\textwidth]{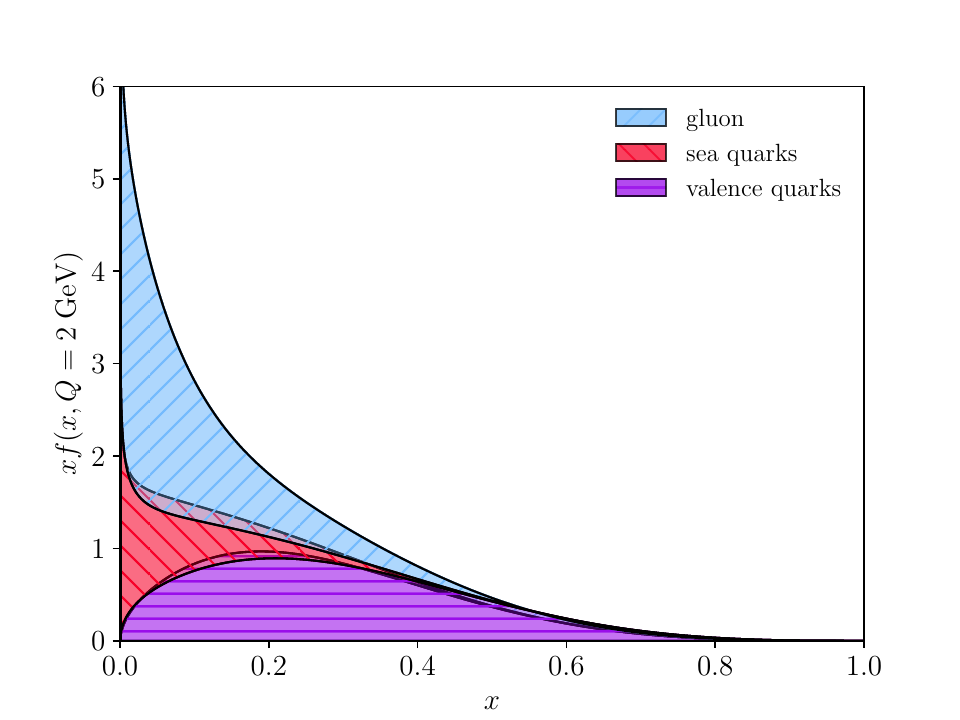}
    \caption{\msbar (background) and \dis (foreground)}
    \label{fig:momentum_view_a}
\end{subfigure}
\hfill
\begin{subfigure}[t]{0.49\textwidth}
    \centering
    \includegraphics[width=\textwidth]{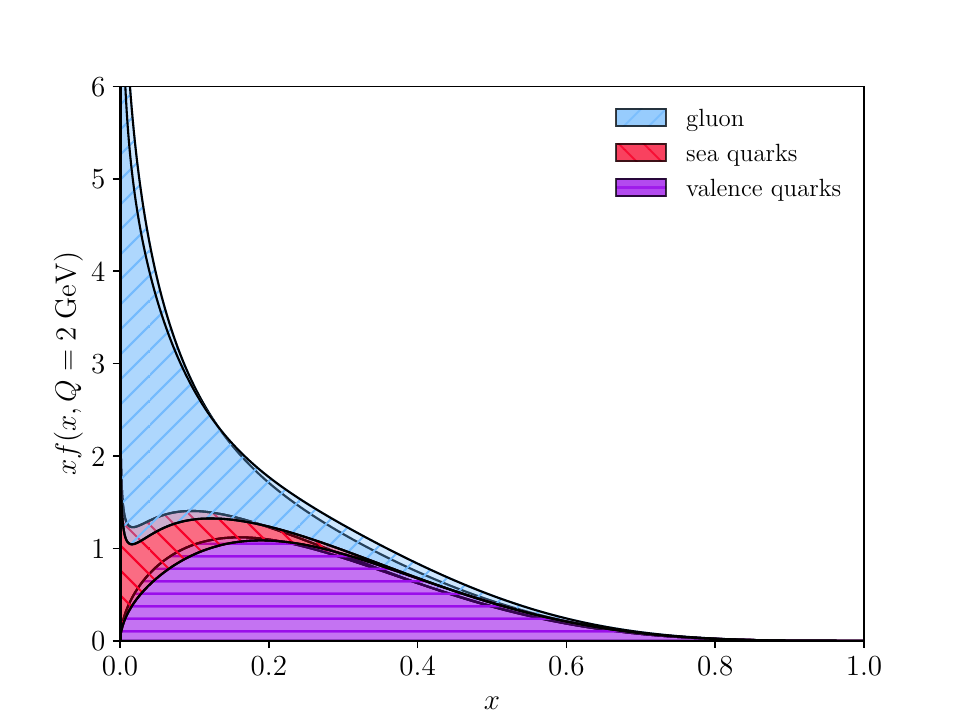}
    \caption{\mpos (background) and \mposd (foreground)}
    \label{fig:momentum_view_b}
\end{subfigure}
\caption{Graphical illustration of the momentum sum-rule;
carried momentum here corresponds to shaded area.
The momentum sum-rule \cref{eq:momsumrule_PDFs} implies that the total shaded area is 1.
\label{fig:momentum_view}}
\end{figure*}

In \cref{fig:momentum_sum_rules} we show the total momentum carried by all active flavours as a function of factorisation scale,
i.e.\ we evaluate numerically the left-hand-side of \cref{eq:momsumrule_PDFs} and compare it to 1.
In addition to being of theoretical interest, this is
a test of the numerical accuracy of the implemented PDF 
transformations and convolutions.
We also compare it to the results for the input \msbar \ctnlo PDF set~\cite{Hou:2019efy}.

In \cref{fig:momentum_sum_rules_a} we show the schemes
for which momentum conservation is imposed on the kernels.
We find that the modifications restore the sum-rule
to well within the target \lhapdf interpolation accuracy ($10^{-3}$);
for the \mpos and \dis schemes the deviation is approximately $10^{-6}$.

In \cref{fig:momentum_sum_rules_b} we show the effect of the \pos
and \dpos transformations on the momentum sum-rule; that is, for
the schemes in which momentum conservation is not explicitly
imposed.
We can deduce the magnitude of the required momentum-conservation-restoring
terms from the deviations shown here.
We see that for the \pos scheme, the deviation from
the sum-rule is approximately 20\% at the input-scale $Q_0 \sim 1 \; \GeV$,
reducing to approximately 5\% at high scales.
For the \dpos scheme, the sensitivity of the sum-rule to the scale is 
smaller but qualitatively has the same relationship with scale,
reducing from approximately 6\% to approximately 2\%.

\subsubsection{Alternative methods for imposing momentum conservation}
\label{subsubsec:CompPDFsImpMomCons}

In \cref{fig:momentum_view} we illustrate the momentum sum-rule graphically:
the left-hand-side of \cref{eq:momsumrule_PDFs} corresponds to the shaded area,
divided into valence quarks (purple), sea quarks (red), and gluons (blue).

\cref{fig:momentum_view_a} shows the effect of the strengthened
momentum sum-rule of \cref{eq:DIS_localsumrule} used to define
the \dis-scheme gluon PDF.
As a direct consequence, the overall distribution of constituent momentum fractions
(top line/total shaded area) is unchanged despite the relative
momentum fractions changing under the scheme-transformation.

The \mpos and \mposd schemes differ only in their method for restoring
the momentum sum-rule; this is an arbitrary choice,
by convention imposed by modifying $\Delta$ as in \cref{eq:momsumrule_deltas}.
Imposing momentum-conservation then
induces a modification of only the virtual contribution to the coefficient function.
Any other function normalised with respect to the second Mellin moment,
\begin{align}
    \mathcal{M} \left[f^{\mathrm{MOM}}\right] (2)
    \equiv
    \int_0^1 \dd z \; z^1 \, f^{\mathrm{MOM}} (z)
    =
    1,
\end{align}
may
be used with the same coefficient.

In \cite{Candido:2020yat} a `soft-function' was chosen,
a quartic polynomial vanishing in the $z \to 0, 1$ limits,
and normalised to satisfy $\mathcal{M}[f](2) = 1,$
\begin{align}
    f^{\mathrm{MOM}} (z) = 60 \, z^2 (1-z)^2,
\end{align}
for the \mpos scheme.\footnote{
This is a special case of a two-parameter family of
possible such choices, 
\begin{align}
f^{(m,n)\mathrm{MOM}}(z) = \frac{(m+n)!}{m! (n-1)!} \, z^{m-1} (1-z)^{n-1},
\end{align}
with $m=n=3$.
}

Here, for \mpos and \mposd, we compare the effect of the
respective choices.
As can be seen from \cref{tab:Kqq,tab:Kqg,tab:Kgq,tab:Kgg},
the off-diagonal kernels are identical in both schemes,
while the diagonal kernels are
non-zero only due to the inclusion of the momentum-conservation-restoring
terms.
As a result, the schemes differ only in the 
quark (input) contribution to the quark (output) PDF, and
the gluon (input) contribution to the gluon (output) PDF.
These are illustrated in 
\cref{fig:PDF_decomposition_uv,fig:PDF_decomposition_ub,fig:PDF_decomposition_g,fig:PDF_decomposition_c}
for the \mpos scheme, and in
\cref{fig:PDF_decomposition_uv_2,fig:PDF_decomposition_ub_2,fig:PDF_decomposition_g_2,fig:PDF_decomposition_c_2}
for the \mposd scheme.

The \mpos and \mposd schemes are compared in \cref{fig:momentum_view_b}, illustrating
the effect on the partonic momentum-fraction distributions of the choice
of delta- or soft-function. We see that the choice has an effect both on the overall distribution
of constituent momentum fractions within the proton, 
with \mpos shifting the overall distribution lower in $x$,
and on the relative momentum fractions of each flavour.

The change in the valence-quark distributions is entirely due to the
flavour-diagonal kernels, which in the \mpos and \mposd schemes
are entirely determined by the chosen momentum-conservation method.
For \mpos and \mposd this has a positive coefficient, so
increases the fractional composition of valence quarks relative
to sea quarks.

The sea-quark distributions are also affected by the
off-diagonal $qg$ kernels;
as a consequence, within the overall distribution, the \mpos and \mposd gluon PDFs take momentum directly from
the sea quarks at low-$x$, significantly changing the qualitative shape of the
overall quark distribution within the proton
(cf.\ \cref{fig:momentum_view_a}).
As a consequence the expected fraction of the proton's momentum carried by the valence quarks relative
to sea quarks is seen to be 
sensitive to the choice of factorisation scheme.

\subsection{Number sum-rules}
\label{subsec:NumSumRules}

\begin{table*}[htp]
\begin{center}
\begin{tabular}{|c|c|c|c|c|c|c|c|c|c|} 
\hline
$ $
& \aversa
& \dis
& \krk
& \krkdy
& \dpos
& \pos
& \mpos
& \mposd
& \phys
\\ \hline
$ \displaystyle \cf^{-1} \int_0^1 \rK_{qq}^{\rFS}(z) \; \dd z $
& $0$
& $0$
& $ \displaystyle -\frac{3}{2}$
& $0$
& $0$
& $0$
& $ \displaystyle \frac{35}{9}$
& $ \displaystyle \frac{35}{18}$
& $ \displaystyle - \frac{1}{2}$
\\ \hline
\end{tabular}
\end{center}
\caption{Integral of the $\rK_{qq}^{\rFS}(z)$ kernel entering the number sum rule for $N_q^{\rFS}$.}
\label{tab:intKqq}
\end{table*}

\begin{figure*}[htp]
\centering
\begin{subfigure}[t]{0.49\textwidth}
    \centering
    \includegraphics[width=\textwidth]{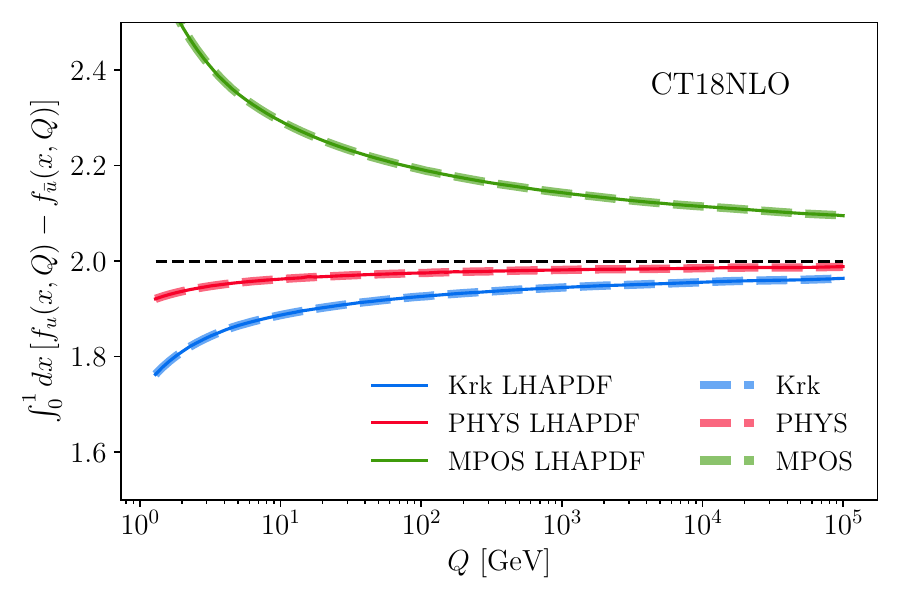}
    \caption{up-quark}
    \label{fig:number_sum_rules_a}
    \end{subfigure}
\hfill
   \begin{subfigure}[t]{0.49\textwidth}
    \centering
    \includegraphics[width=\textwidth]{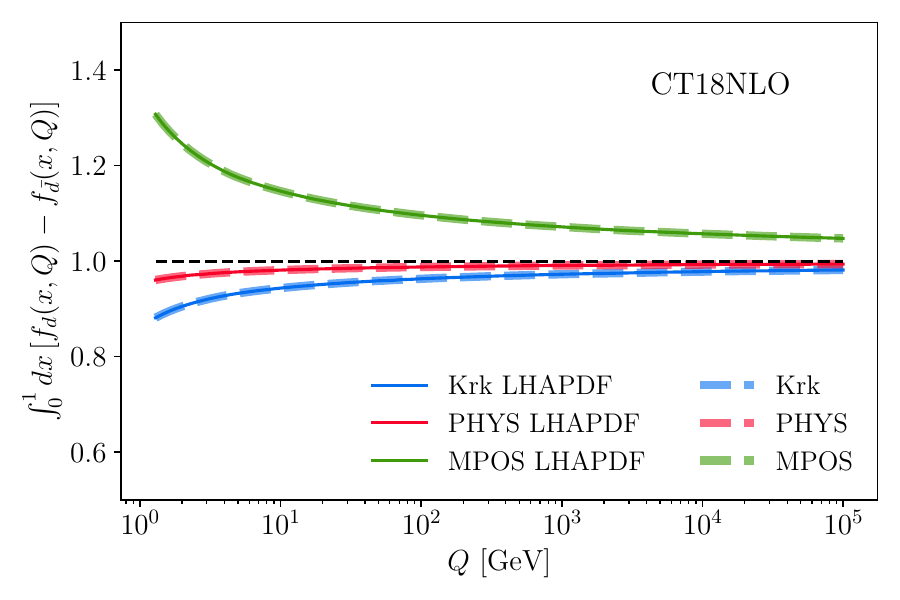}
    \caption{down-quark}
    \label{fig:number_sum_rules_b}
\end{subfigure}
\caption{Number sum rule as a function of the factorisation/renormalisation scale for (a) $u$-quark, and (b) $d$-quark. Compared are integrals obtained from \lhapdf files of the transformed PDFs and analytic integrals coming from Eq.~\eqref{eq:numbersumrule_PDFs_FS}.}
\label{fig:number_sum_rule}
\end{figure*}

In the \msbar scheme we know that the valence-quark number sum rules are fulfilled 
in the following way:
\begin{align}
	\label{eq:numbersumrule_PDFs_MSbar}
	\int_0^1 \dd x \, \left[ f_{q}^{\msbar}(x;\muf) - f_{\bar{q}}^{\msbar}(x;\muf) \right] = N_q^{\msbar},
\end{align}
where $N_q^{\msbar}=\{2,1,0\}$ for $u$, $d$, and any other quark flavour (assuming there is no asymmetric intrinsic contributions, e.g.\ for charm).
There is however, no guarantee that the same sum rule holds in a different factorisation scheme.
In an alternative factorisation scheme, the same integral can be computed to give
\begin{align}
	\label{eq:numbersumrule_PDFs1}
	N_q^{\rFS} &= \int_0^1 \dd x \, \left[ f_{q}^{\rFS}(x;\muf) - f_{\bar{q}}^{\rFS}(x;\muf) \right]
 \\ \notag {} & 
 = \int_0^1 \dd x \, \bigg[ f_{q}^{\msbar}(x;\muf) - f_{\bar{q}}^{\msbar}(x;\muf)
  \\ \notag {} & 
   + \frac{\alphas(\mu)}{2\pi} \; \sum_b \int_x^1 \frac{\dd z}{z} \, \rK^{\msbar \to \rFS}_{qb}(z) f_{b}^{\msbar}\left(\frac{x}{z};\muf\right)
  \\ \notag {} & 
  - \frac{\alphas(\mu)}{2\pi} \; \sum_b \int_x^1 \frac{\dd z}{z} \, \rK^{\msbar \to \rFS}_{\bar{q}b}(z) f_{b}^{\msbar}\left(\frac{x}{z};\muf\right)
  \bigg].
\end{align}
Using the definitions and properties of the kernels $\rK^{\msbar \to \rFS}_{ab}$ and performing some algebra we get:
\begin{align}
	\label{eq:numbersumrule_PDFs_FS}
 	N_q^{\rFS} &= 
   N_q^{\msbar} \left\{ 1 + \frac{\alphas(\mu)}{2\pi} \,
  \int_0^1 \dd z \; \rK^{\msbar \to \rFS}_{qq}(z) \right\}.
\end{align}
This means that the number sum rule is modified when changing the scheme
by terms of order $\order{\alphas}$.
Note that it will not be altered relative to the \msbar scheme
(at $\order{\alphas}$) if $\rK_{qq}$ integrates to 0,
including when $\rK_{qq}$ is given by a plus distribution.
This is the case, e.g.\ for the \dis{} scheme.

Since the modification is proportional to $\alphas(\mu)$, the sum-rule
integral is in general scale-dependent (unless the $\order{\alphas}$
correction vanishes).

We can compute this change analytically which is given by the integral of $\rK_{qq}$ kernel, see \cref{tab:intKqq}.
At NLO the modification will only affect $u$ and $d$ quarks for which, in \cref{fig:number_sum_rule}, we show the numerical value of the number sum rule for selected schemes.
We can clearly see the deviation from the \msbar result, as well as the scale-dependence.
In \cref{fig:number_sum_rule} we also compare the values of ~\cref{eq:numbersumrule_PDFs_FS} with those obtained directly from
\ctnlo PDFs transformed into the different schemes, and see perfect agreement.

We should note that the modified form of the number sum rules for different factorisation schemes does not make them `less physical' than the \msbar scheme or the \dis scheme for which the usual sum rules are fulfilled.
The values of the sum rules are perturbative quantities which receive corrections at each perturbative order.
A special property of the \msbar scheme is that these corrections are zero \cite{Collins:2011zzd,Collins:email}.

This can have practical consequences for the determination of PDFs in phenomenological fits to data. 
In the \msbar or \dis scheme,
in which the value of the sum-rule integral is constant,
the usual practice is to enforce the sum rules during the fit.
Alternatively, one can allow the fitting process to extract them from the data,
and use the consistency of the sum-rules with their expected values as a test
of the quality of the fit.%
    \footnote{This kind of check has been performed for the momentum sum rule, e.g.~\cite{Ball:2011uy}, but not in case of the number sum rules. The obtained value of the momentum sum rule was consistent with 1 both at NLO and NNLO.}
Although it would be possible to impose the modified sum-rules as constraints
on phenomenological fits of PDFs
to data in alternative factorisation schemes with $N_q^{\rFS}\neq N_q^{\msbar}$
in which the value is scale-dependent,
a more natural approach might be to
leave the PDFs unconstrained and use the sum-rules as a test of the consistency
of the extracted PDFs with their expected properties.%

\subsection{Positivity}
\label{subsec:CompPos}

Phenomenological determinations of PDFs have been found to 
give negative distributions,
typically for gluon or sea quarks at low scales and small $x$
\cite{Diehl:2019fsz,Martin:2009iq,Altarelli:1998gn,H1:2015ubc}.
Recently this possibility attracted theoretical attention,
with efforts to establish whether PDFs in the \msbar scheme must in fact be positive
\cite{Candido:2020yat,Candido:2023ujx,Collins:2021vke}.%
\footnote{The \mpos scheme discussed here was introduced in \cite{Candido:2020yat}
for this purpose.}
Imposing positivity as a constraint on PDF fits leads to reduced
uncertainties\cite{Martin:2009iq,Ball:2010de,Ball:2008by,Pumplin:2002vw}, and so if theoretically justified,
in the \msbar or any other scheme,
could be a valuable tool in the path towards the precision determination of PDFs.

We approach the question of positivity from two complementary perspectives.
First, we adapt the theoretical argument of \cite{Candido:2023ujx} and apply it to the 
factorisation schemes considered here.
Secondly, we describe the positivity of the PDFs obtained as a result of our 
transformations of \msbar PDFs into alternative factorisation schemes.

\subsubsection{Theoretical positivity}
\label{subsubsec:CompPos_theoretical}

In \cite{Candido:2023ujx} an argument is advanced, building on \cite{Altarelli:1998gn,Candido:2020yat},
that the positivity of \msbar PDFs can be guaranteed by theoretical arguments
based on properties of the implicit transformation from a basis of physical observables
to a set of PDFs.

For a given convolution kernel $K$ in the notation of \cref{eq:tableDecomp},
we adopt the terminology of \cite{Candido:2023ujx} and define its `finite piece' as
\begin{align}
    \finite [K](z) = b(z) \log(1-z) + c(z) \log z + P(z),
\end{align}
that is, the functional part which remains with
the distribution contributions ($\mathcal{D}_k$ and Dirac-delta) subtracted.%
\footnote{This is exactly analogous to the finite-piece defined in \cite{Candido:2023ujx}
and there denoted with subscript `$F$', i.e.\ $K_F$.
For the transformations and coefficient-functions considered here it can
immediately be extracted from the central three columns of \cref{tab:Kqq,tab:Kqg,tab:Kgq,tab:Kgg,tab:CoDIS,tab:CoDY,tab:CoHiggs}.
}
From the finite piece we define the `cumulant' of the kernel as%
\footnote{Note that in contrast to \cite{Candido:2023ujx}, we 
apply the absolute-value to the integrand rather than the integral.
This leads to a strengthened condition below.}
\begin{align}
  \label{eq:FScumulant_definition}
    \cumulant [K](x) = \int_x^1 \left \lvert \finite [K] (z) \right \rvert \ \frac{\dd z}{z}.
\end{align}
The argument of \cite{Candido:2023ujx} concludes that the transformation
given by the finite-piece
is perturbative, i.e.\ the magnitude of the NLO modification is smaller than
that of the input, provided that
\begin{align}
  \label{eq:FScumulants_criterion}
    \frac{\alphas(\mu)}{2\pi} \sum_b \cumulant\left[ 
    \rK_{ab}^{\rFS_1\to\rFS_2} \right] 
    \leqslant 1,
\end{align}
and the input-distribution is a positive, decreasing function.
Clearly if the transformation is perturbative in this sense,
and the input distribution is positive, the transformation gives
a positive result.

The cumulants applied in \cite{Candido:2023ujx} are calculated from the DIS and Higgs-production coefficient
functions (in the \msbar scheme), i.e.\ precisely the content of \cref{tab:CoDIS,tab:CoHiggs}:%
\footnote{We reiterate that these are presented for massless quarks only,
and the effect of mass-corrections may be significant.}
\begin{align}
  \label{eq:FScumulants_definition}
  \tilde{\mathcal{C}}_{qb}^{\rFS} (x) &= \cumulant [ C_{b}^{(1)\rFS} ], &
  \tilde{\mathcal{C}}_{gb}^{\rFS} (x) &= \cumulant [ C_{gb}^{(1)\rFS} ].
\end{align}
These are considered as transformation kernels applied to a `physical' factorisation scheme
argued to comprise measurements which are assumed to be positive and decreasing.
We adapt the analysis of \cite{Candido:2023ujx} and apply it to the factorisation schemes considered here,
focusing explicitly on the role of the finite-part cumulant relied upon in the quantitative conclusions of
\cite{Candido:2023ujx}.%
\footnote{We defer consideration of the $\delta$- and distributional components,
which differ between schemes, and of the application
of the argument to transformations between factorisation schemes, to future work.}

\begin{figure*}[htp]
\centering
\begin{subfigure}[t]{0.49\textwidth}
    \centering
    \includegraphics[width=\textwidth]{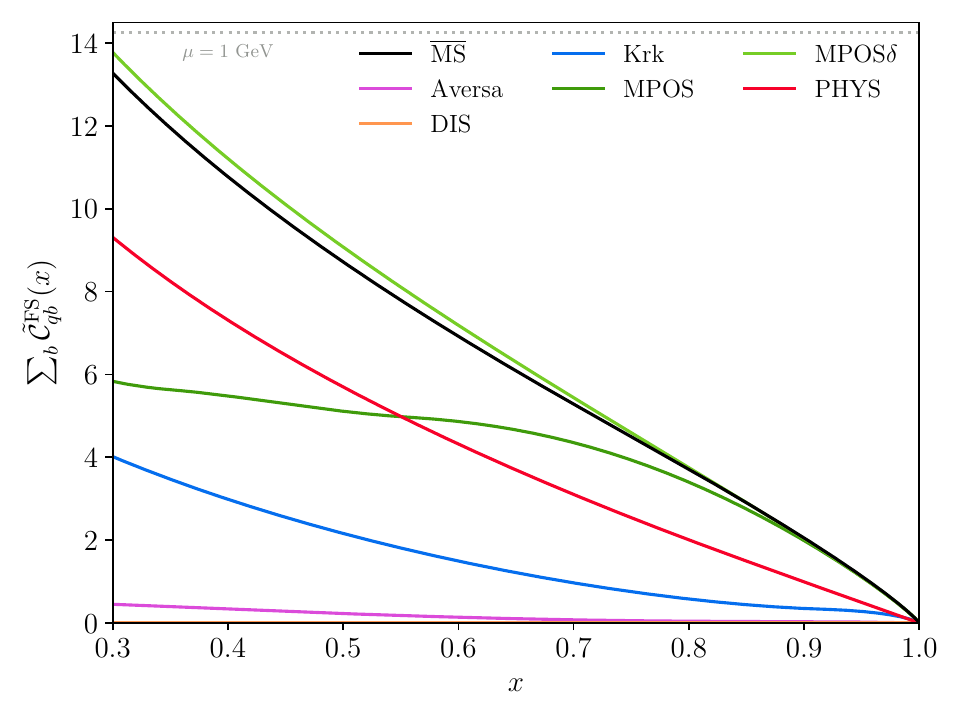}
    \caption{$\tilde{\mathcal{C}}_{qq} + \tilde{\mathcal{C}}_{qg}$}
    \label{fig:cumulant_sumq}
\end{subfigure}
\begin{subfigure}[t]{0.49\textwidth}
    \centering
    \includegraphics[width=\textwidth]{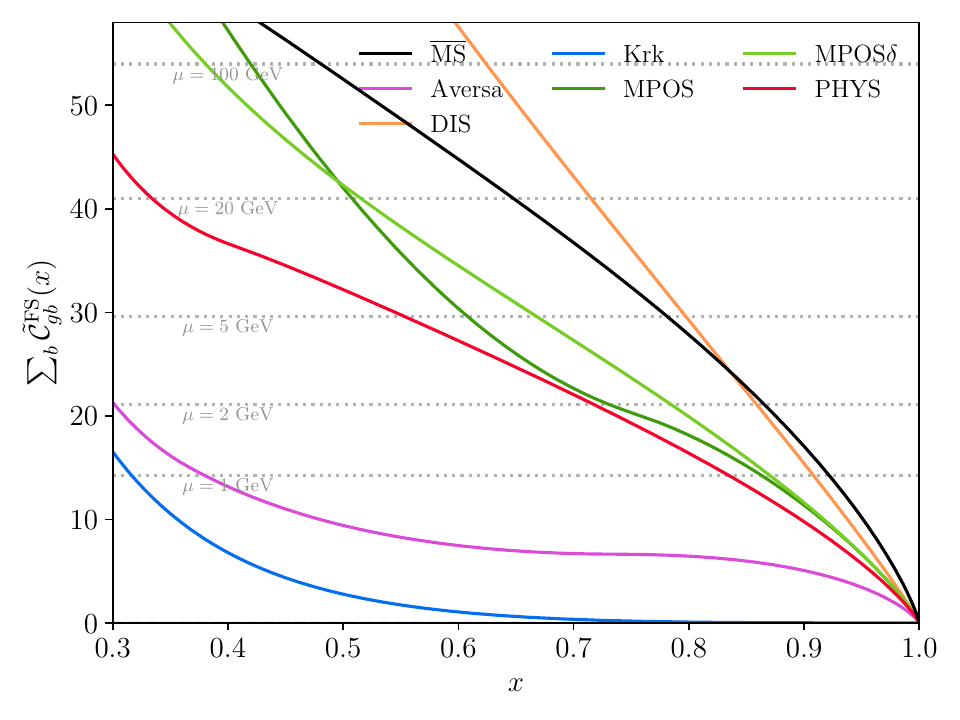}
    \caption{$\tilde{\mathcal{C}}_{gg} + 2 \nf \tilde{\mathcal{C}}_{gq}$ (here, $\nf = 5$)}
    \label{fig:cumulant_sumg}
\end{subfigure}
\caption{Cumulants, as defined in \cref{eq:FScumulants_definition}, adapted from the positivity argument of \cite{Candido:2023ujx}.
  The dotted grey lines show $2\pi / \alphas(\mu)$ at the labelled value of $\mu$.
}
\label{fig:FScumulants}
\end{figure*}

These cumulants, summed over flavours, are shown in \cref{fig:FScumulants}, together with
illustrative values of 
${2\pi}/{\alphas(\mu)}$
at relevant choices of $\mu$ to indicate the scale and $x$-values above which the criterion of
\cite{Candido:2023ujx}, summarised in \cref{eq:FScumulants_criterion}, indicates positivity.

Although this is a preliminary analysis, the modified criterion
applied (as in \cite{Candido:2023ujx}) at $x = 0.8$ leads to a similar conclusion as reached there,
of 5 GeV as a scale above which positivity is indicated in this region for the \msbar scheme.
The \dis scheme is similar, while the \mpos, \mposd and \phys schemes lead%
\footnote{Note that the omission of the $\delta$ contributions from the analysis here is
especially relevant to any comparison between the \mpos and \mposd schemes.}
to a scale of 2 GeV, and the \krk and \aversa scheme to 1 GeV.%
\footnote{Note that at scales close to the flavour-threshold of the heavy quarks,
the argument may need further refinement to include quark-mass effects
in the coefficient functions used for the definition of the cumulants.}

\subsubsection{Empirical positivity}
\label{subsubsec:CompPos_empirical}

In contrast to the approach referred to in the previous section,
which attempts to derive universal
bounds on PDFs from first principles, here we 
summarise our observations about the positivity of proton PDFs
in different factorisation schemes according to the transformations we have performed.
This is naturally sensitive to the choice of PDF set used, and potentially to
any positivity constraints imposed by the fitting group on the input \msbar PDFs.
We consider PDF sets fitted using different methods
and both with, and without positivity explicitly imposed upon the fit,
concretely
the \ctnlo PDF set in which goodness-of-fit function $\chi^2$
is directly minimised over a chosen parameter space;
the \nnpdf PDF set fitted using neural-network replicas with 
positivity imposed on non-heavy flavours above $\sqrt{5}\; \GeV$,
and the 
\nnpdfmc PDF set which resembles \nnpdf except for the absence of intrinsic
charm and the global imposition of positivity.
\comment{all}{details of above paragraph needs double-checking}

In this section we use `positive' to mean `not definitely negative',
i.e.
\begin{align}
    x f(x) > - \varepsilon,
\end{align}
for tolerance parameter $\varepsilon > 0$ representing the target numerical precision
for the numerical interpolation of the input PDF grids.
We choose $\varepsilon \approx 10^{-4}$ (for comparison
the \lhapdf 6 target (relative) interpolation uncertainty 
is $10^{-3}$ \cite{Buckley:2014ana}).
Negative PDFs values which are within $\varepsilon$ of 0 could therefore 
be set to zero
without perturbing the resulting PDF by more than the
target interpolation uncertainty, and we disregard them.

In general, we find that PDFs for all flavours in all schemes become positive at
sufficiently large $Q$.
More precisely, for all schemes the 
valence-quark-flavour PDFs are positive at all scales, while 
in some schemes the remaining light sea PDFs show slight
negativity at very low scales and high-$x$.
The gluon PDF is positive at all scales in all schemes
but the \dis scheme,
where for scales as large as 50 GeV the gluon PDF is negative for $x > 0.5$.

The heavy quarks, however, are negative close to their mass-thresholds,
in all the schemes.
This is due to the transformed quark PDFs receiving a negative contribution
from the gluon input-PDF (shown in the decomposition plots of \cref{fig:PDF_decomposition_c,fig:PDF_decomposition_c_2}).
Where the input quark distribution is small
(as is the case for perturbatively-generated heavy quarks close to their mass-thresholds),
this is sufficient to turn the resulting PDF negative.

\begin{figure*}[thp]
\centering
\begin{subfigure}[t]{0.49\textwidth}
    \centering
    \includegraphics[width=\textwidth]{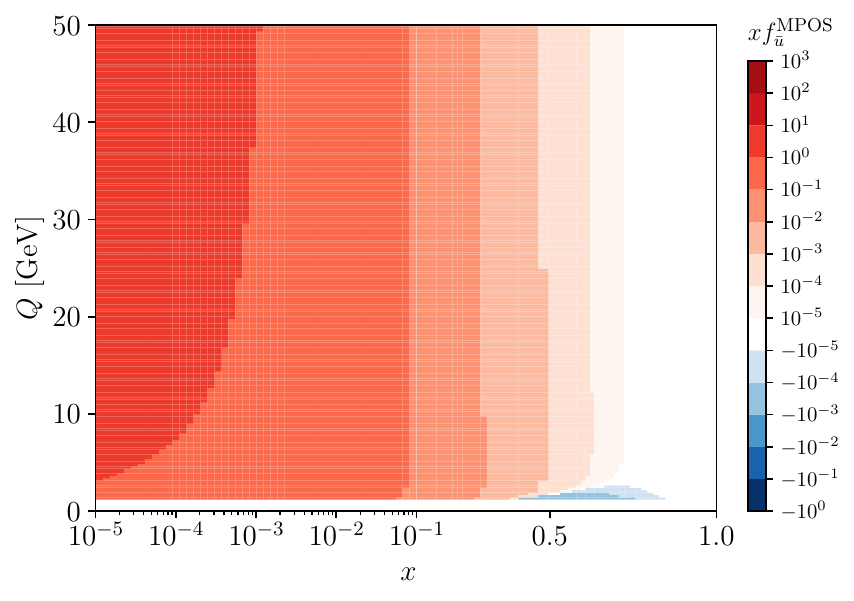}
    \caption{$\bar{u}$}
    \label{fig:2D_MPOS_ub}
    \end{subfigure}
\hfill
   \begin{subfigure}[t]{0.49\textwidth}
    \centering
    \includegraphics[width=\textwidth]{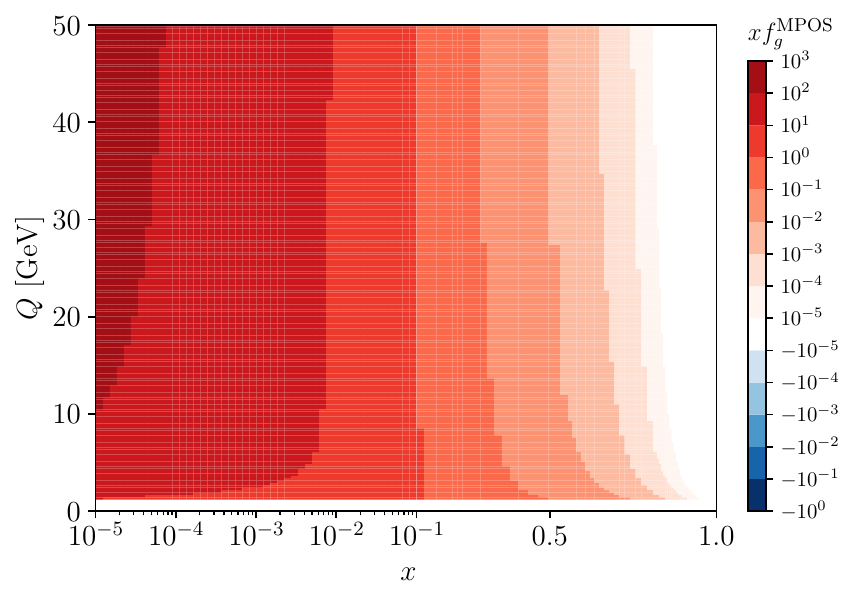}
    \caption{gluon}
    \label{fig:2D_MPOS_g}
    \end{subfigure}
\\
   \begin{subfigure}[t]{0.49\textwidth}
    \centering
    \includegraphics[width=\textwidth]{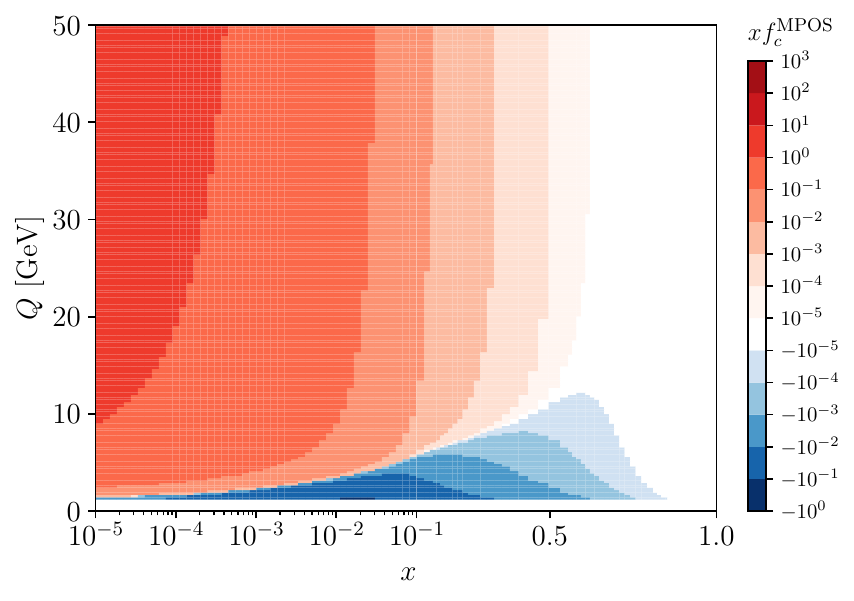}
    \caption{charm-quark}
    \label{fig:2D_MPOS_charm}
    \end{subfigure}
\hfill
   \begin{subfigure}[t]{0.49\textwidth}
    \centering
    \includegraphics[width=\textwidth]{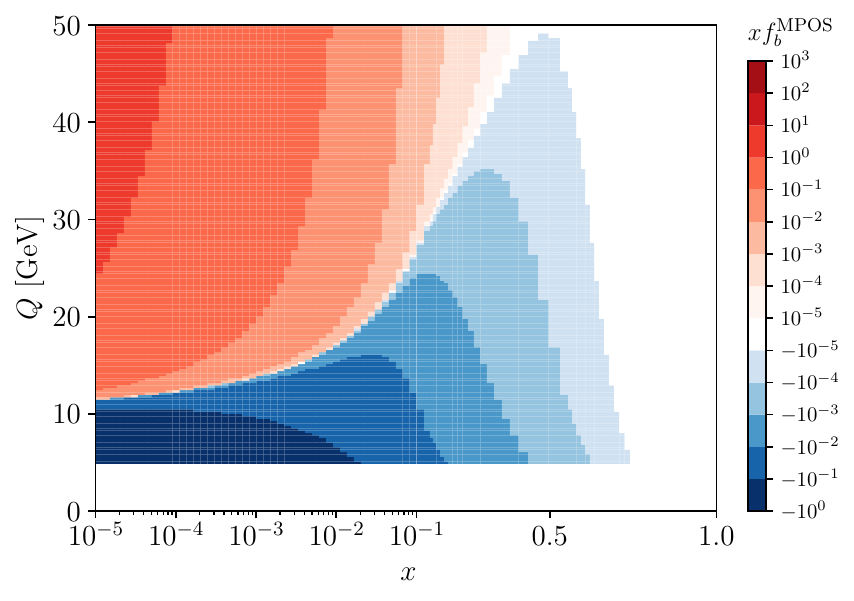}
    \caption{bottom-quark}
    \label{fig:2D_MPOS_bottom}
    \end{subfigure}
\caption{Heatmap showing the sign and order-of-magnitude
of heavy-quark \ctnlo PDFs $f_{b,c}^{\mpos}(x,Q)$ in the \mpos scheme
as a function of momentum-fraction $x$ and scale $Q$.
Note that for the purposes of the discussion of \cref{subsubsec:CompPos_empirical}
the lightest blue band shown here is treated as negligible.}
\label{fig:2D_MPOS}
\end{figure*}

For the $c$-quark, this is effectively overcome by the DGLAP evolution,
to restore a positive output PDF, at around 10 GeV.
In the case of the $b$-quark, scale required to 
restore positivity can be as high as 30 GeV.
Plots illustrating this through the signed magnitude of the 
$c$- and $b$-quark PDFs in $(x,Q)$-space,
in the \mpos scheme, are shown in \cref{fig:2D_MPOS}.

\begin{figure*}[thp]
\centering
   \begin{subfigure}[t]{0.49\textwidth}
    \centering
    \includegraphics[width=\textwidth]{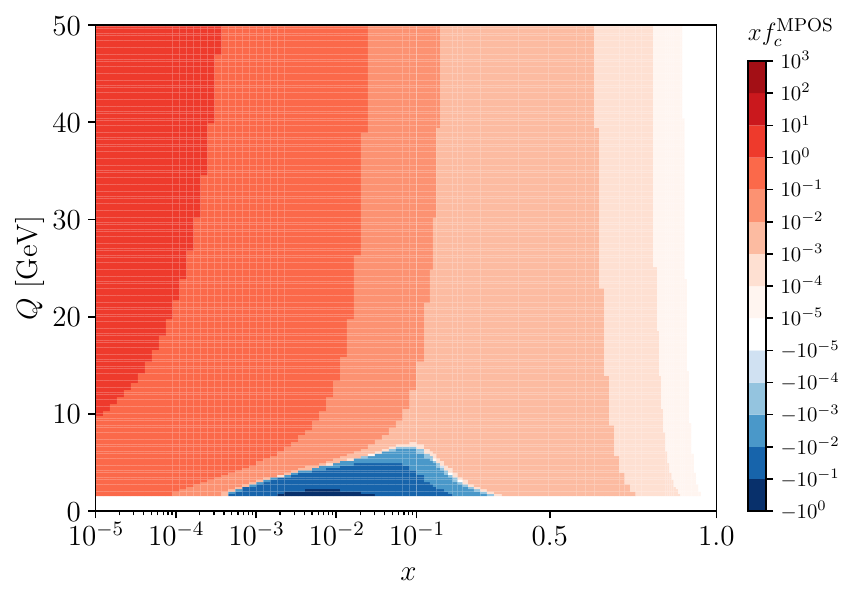}
    \caption{\nnpdf (intrinsic charm) in \mpos scheme}
    \label{fig:2D_MPOS_charm_NNPDF4}
    \end{subfigure}
\hfill
   \begin{subfigure}[t]{0.49\textwidth}
    \centering
    \includegraphics[width=\textwidth]{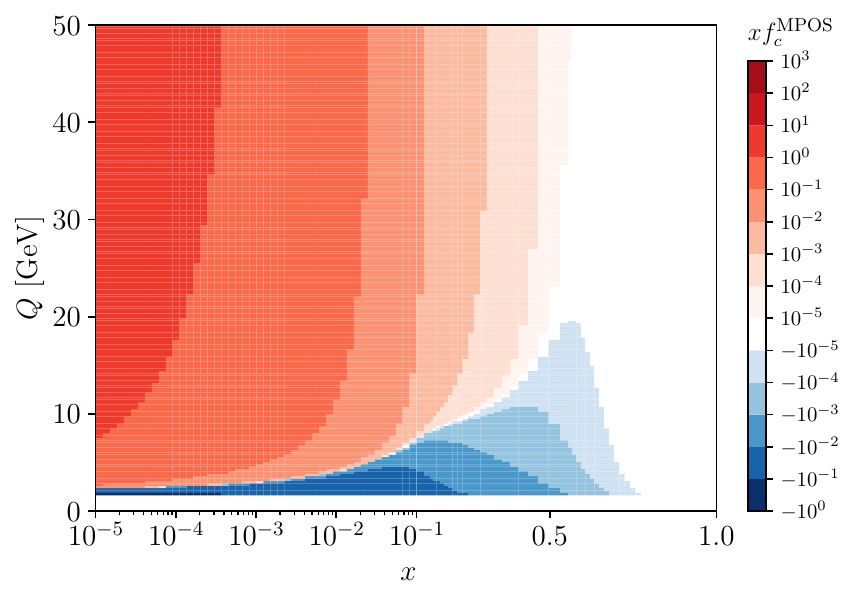}
    \caption{\nnpdfmc (perturbative charm) in \mpos scheme}
    \label{fig:2D_MPOS_charm_NNPDF4MC}
    \end{subfigure}
    \\
\begin{subfigure}[t]{0.49\textwidth}
    \centering
    \includegraphics[width=\textwidth]{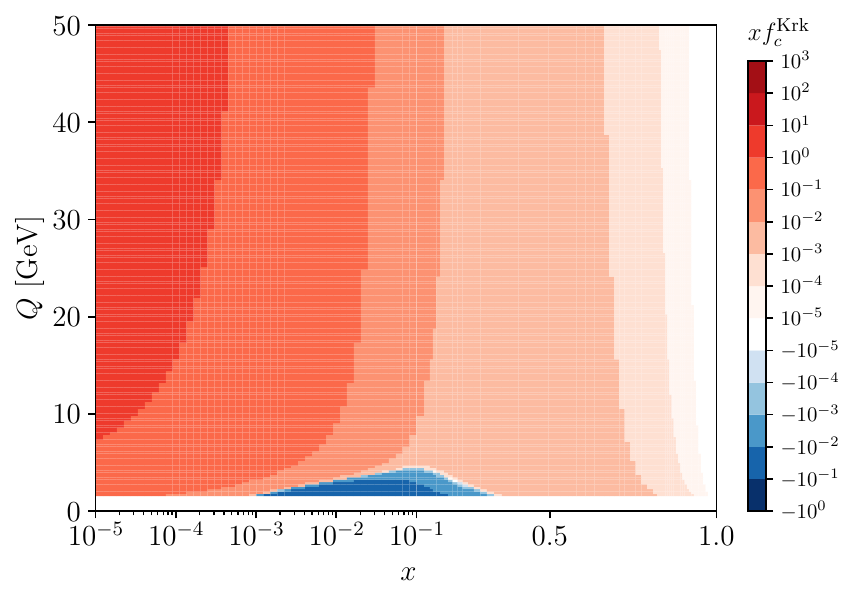}
    \caption{\nnpdf (intrinsic charm) in \krk scheme}
    \label{fig:2D_Krk_charm_NNPDF4}
    \end{subfigure}
\hfill
\begin{subfigure}[t]{0.49\textwidth}
    \centering
    \includegraphics[width=\textwidth]{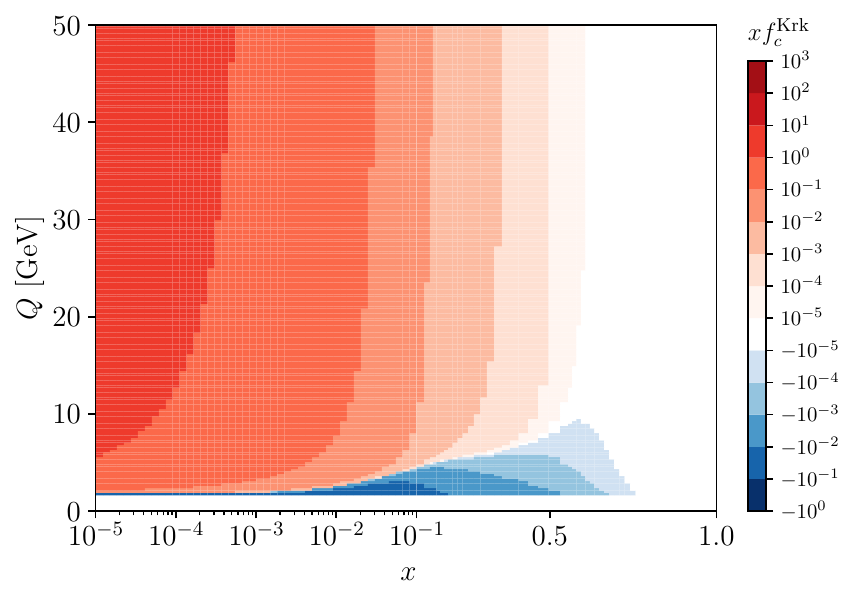}
    \caption{\nnpdfmc (perturbative charm) in \krk scheme}
    \label{fig:2D_Krk_charm_NNPDF4MC}
    \end{subfigure}
\caption{Heatmap showing the sign and order-of-magnitude
of charm-quark PDFs $f_{c}^{\krk,\mpos}(x,Q)$
in the \krk and \mpos schemes
as a function of momentum-fraction $x$ and scale $Q$,
from the \nnpdf \cite{NNPDF:2021njg}
and \nnpdfmc \cite{Cruz-Martinez:2024cbz} PDF sets.
The key relevant difference is the inclusion of an intrinsic charm
component in the \nnpdf PDF set.
Note that for the purposes of the discussion of \cref{subsubsec:CompPos_empirical}
the lightest blue band shown here is treated as negligible.}
\label{fig:2D_c}
\end{figure*}

Since this behaviour is due to the perturbatively-generated charm
being small close to its flavour-threshold, we examine the
effect of the inclusion of intrinsic charm in \cref{fig:2D_c},
comparing the \nnpdf PDF set \cite{NNPDF:2021njg},
which contains intrinsic charm,
to the closely-related \nnpdfmc set \cite{Cruz-Martinez:2024cbz}
which contains only perturbatively-generated charm.
This is sufficient to restrict the region in which negativity emerges,
and its magnitude, to intermediate-$x$ and low-$Q$
(as shown in \cref{fig:PDF_decomposition_NNPDF40_c}).

The intrinsic charm included in the \nnpdf set thus
`solves' the problem of negativity at low-$Q$ and
high-$x$, which is common to all of the PDF sets
in which the charm PDF is purely perturbative.

This cannot, however, resolve the problem of the $b$-quark PDF,
which (as shown in \cref{fig:2D_MPOS_bottom}) is very negative,
with negativity extending up to scales significantly above the mass threshold.
This invites the possibility to define
the factorisation-scheme transition kernels differently
for heavy flavours, to include quark-mass effects.

The different low-$x$ behaviour of the gluon PDFs at low scales in the different PDF sets (see \cref{fig:PDF_compare,fig:PDF_compare_NNPDF40MC,fig:PDF_compare_MSHT20nlo}),
due to the lack of constraints from data,
drives the differences observed between the heavy-flavour PDFs
at low-$x$ and $Q$.

\subsubsection{Conclusions}
\label{subsubsec:CompPos_conc}

We observe a tension between the conclusions of the argument
set out in \cite{Candido:2023ujx} and our numerical calculations:
both light sea-quarks, and heavy-quarks, exhibit negativity above
the predicted scales.

Amongst other things, this could be due to the distributional
$\mathcal{D}_k$ or $\delta$ contributions,
which differ between the schemes and which we neglect here.

Additionally, 
as indicated in \cite{Candido:2023ujx},
at the low scales which emerge from the argument,
it may not be justified to neglect the effect of the heavy quark masses
in the coefficient functions used for the definition of the cumulants.

Since we observe significant negativity above the scales predicted 
by the argument of
\cref{subsubsec:CompPos_theoretical},
imposing positivity on PDFs in these schemes across all flavours
on the basis of this argument
would be expected to substantially change the resulting PDFs.

Excluding the possibility of intrinsic $c$/$b$ content,
making the heavy-quark PDFs positive close to their
mass-thresholds in all schemes would seem to require a
significant modification to the gluon PDF.

Alternatively, another approach could be to use
a `hybrid' variable flavour number scheme,
where the number of flavours changes not at the mass threshold
but at a higher scale, e.g., at twice the heavy-quark mass~\cite{Kusina:2013slm,xFitterDevelopersTeam:2017fzy},
which should mitigate the problem.

In principle, as hinted at in \cite{Candido:2023ujx}, 
the argument given there could be adapted
to isolate constraints on individual flavours;
one approach would be to use flavour-sensitive observables,
such as $F_2^c$ \cite{H1:2012xnw,Cooper-Sarkar:2010tpx}
or tagged jet measurements \cite{Gauld:2022lem,Caletti:2022hnc,Czakon:2022wam,Caola:2023wpj},
for the positive input `physical-scheme' distributions.

\subsection{Coefficient functions}
\label{subsec:CompCFs}

As discussed in \cref{subsubsec:coefffuncs}, a factorisation-scheme transformation induces a change to the coefficient functions compensating the corresponding modification of PDFs. This mechanism restores
the factorisation-scheme independence for observable quantities
order-by-order in the coupling constant (up to the operative order in perturbation theory).
This allows us to choose a scheme at our convenience, according to any desirable properties
it exhibits, including those arising due to the effect of this compensating term on the partonic cross-sections (here represented by the coefficient functions).

In this section we consider the effect of this transformation on
the coefficient functions of three illustrative processes:
deep inelastic scattering (DIS),
the Drell--Yan process (DY)
and Higgs boson production via gluon fusion (in the large top quark mass limit).

\begin{figure*}[p]
\centering
\begin{subfigure}[t]{\textwidth}
\makebox[\textwidth][c]{
    \includegraphics[width=0.49\textwidth]{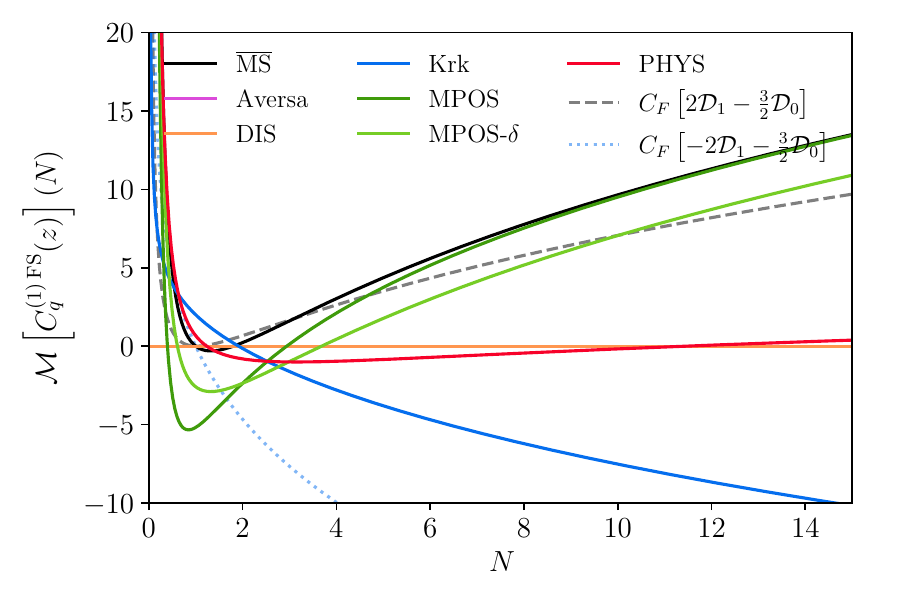} \hfill
    \includegraphics[width=0.49\textwidth]{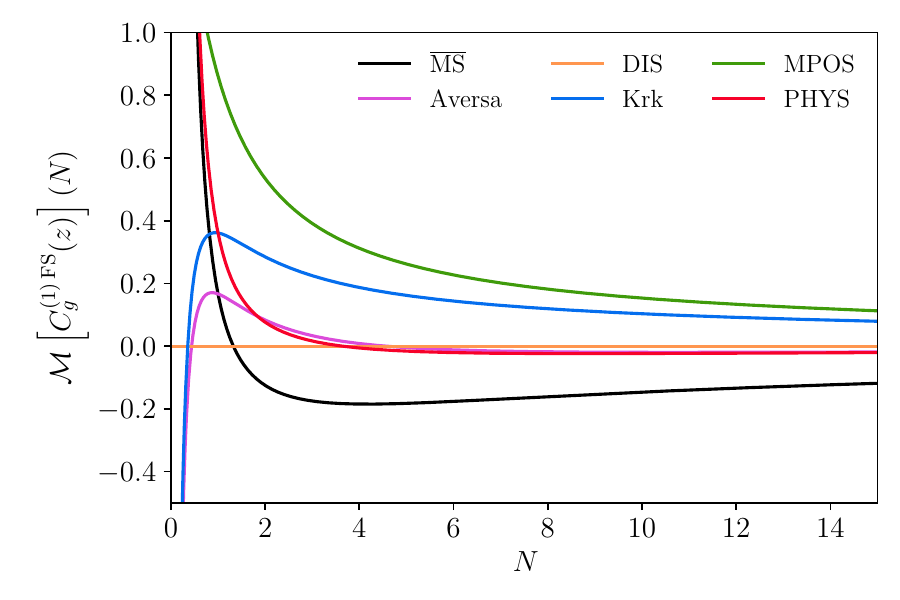}
}
\caption{DIS coefficient functions $C_q^{\rFS(1)}$, $C_g^{\rFS(1)}$\label{fig:NspaceCDIS}}
\end{subfigure}
\\
\begin{subfigure}[t]{\textwidth}
\makebox[\textwidth][c]{
    \includegraphics[width=0.49\textwidth]{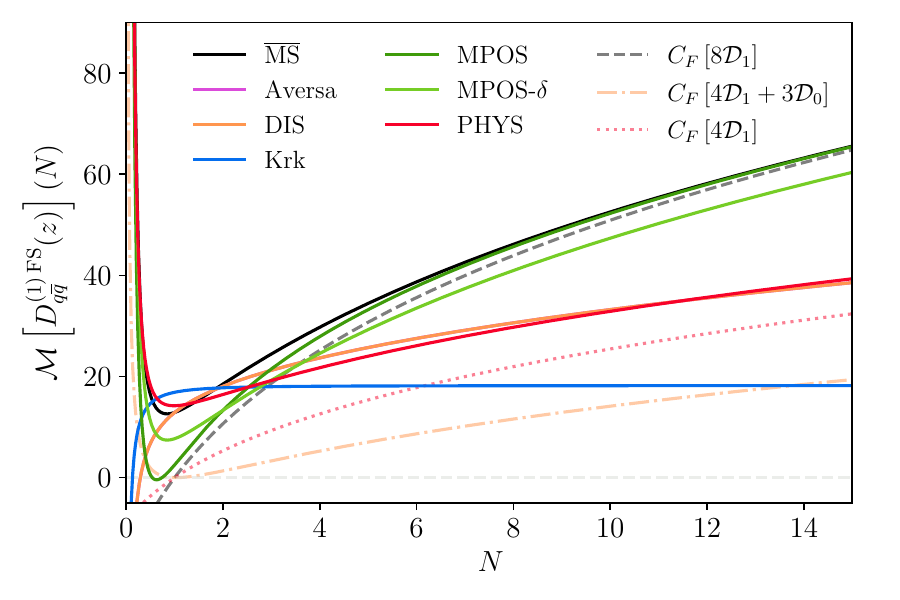}
    \hfill
    \includegraphics[width=0.49\textwidth]{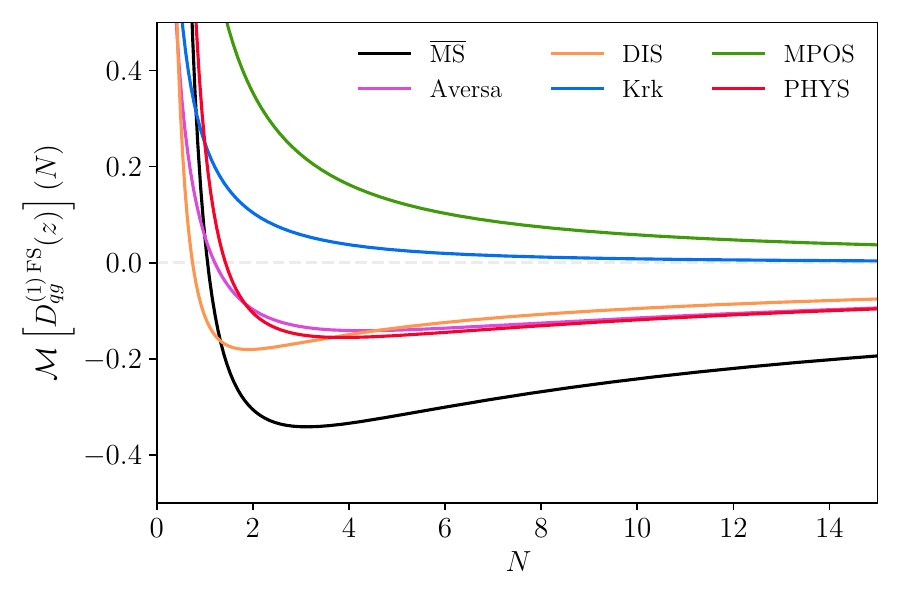}
}
\caption{Drell--Yan coefficient functions $D_{q\bar{q}}^{\rFS(1)}$, $D_{qg}^{\rFS(1)}$\label{fig:NspaceCDY}}
\end{subfigure}
\\
\begin{subfigure}[t]{\textwidth}
\makebox[\textwidth][c]{
    \includegraphics[width=0.49\textwidth]{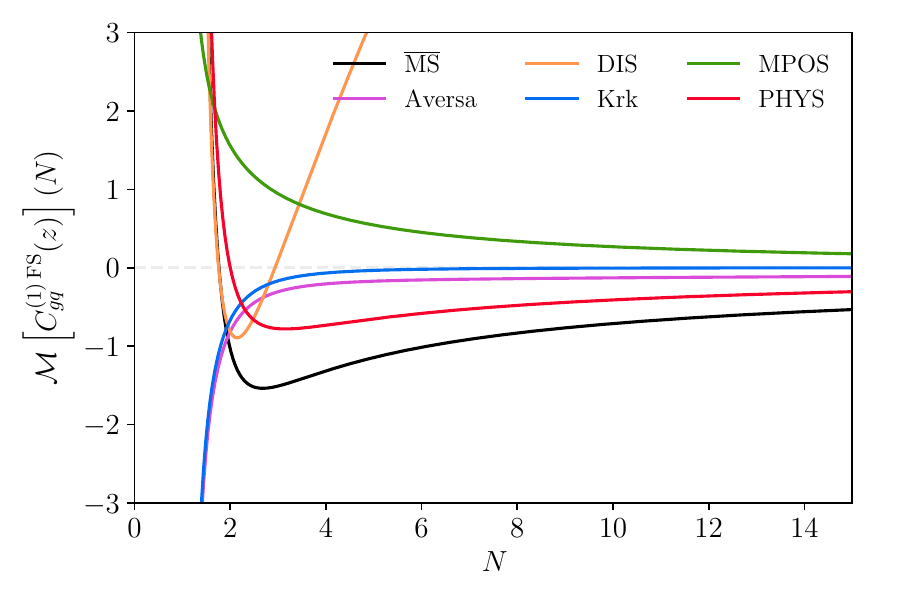} \hfill
    \includegraphics[width=0.49\textwidth]{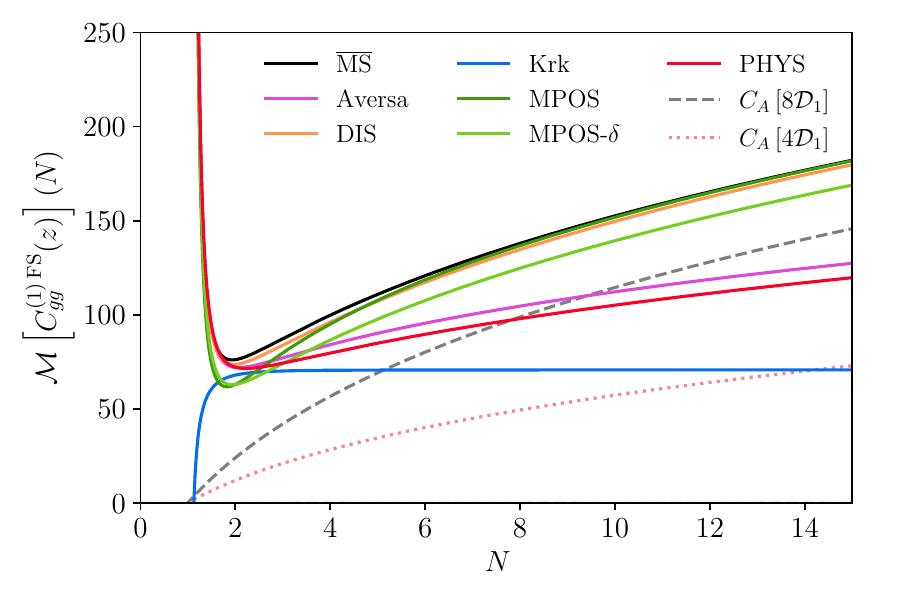}
}
\caption{Higgs coefficient functions $C_{gq}^{\rFS(1)}$, $C_{gg}^{\rFS(1)}$\label{fig:NspaceCH}}
\end{subfigure}
\caption{The NLO contributions to coefficient functions for the DIS, Drell--Yan and Higgs production processes, in Mellin space.
Where relevant, the Mellin moments of the
distributions governing the asymptotic large-$N$
behaviour have also been plotted.
Note that, the \lo term $\mathcal{M}[C_q^{\rFS(0)}](N)$, which is not plotted here,
would on these plots correspond to a constant value of $2\pi/\alphas(\mu) \approx 53$.}
\label{fig:NspaceC}
\end{figure*}

In addition to the NLO contributions to the coefficient functions themselves,
tabulated in \cref{tab:CoDY,tab:CoDIS,tab:CoHiggs},
we present plots of their Mellin transforms in Mellin-moment space,
defined as
\begin{equation}
    \mathcal{M}[f](N) \colonequals \int_0^1 z^{N-1} \, f(z) \,  \dd z
\end{equation}
for $N > 0$.
These are collected in \cref{fig:NspaceC}.

Under the Mellin transform, convolutions of functions as defined in \cref{eq:convdef},
factorise into products of their Mellin transforms. The degree of singularity
of functions and distributions in the $z \to 1$ limit is then made manifest
in the $N\to \infty$ limit in Mellin-space.

Note that for a bounded function on $[0,1]$ the $N^\text{th}$ Mellin-moment is also bounded
and decays as $N\to \infty$,
\begin{align}
    \left \vert \mathcal{M}[f](N) \right\vert
    \leqslant
    \frac{1}{N}
    \max_{[0,1]} \left\vert f \right\vert
\end{align}
whilst a distribution divergent at $z=1$ may give a constant Mellin-transform,
\begin{align}
    \mathcal{M}[\delta(1-z)](N)
    =
    1,
\end{align}
single logarithms of $N$,%
\begin{align}
\mathcal{M}[\mathcal{D}_0(z)](N) & \sim \log N
\end{align}
or double-logarithms of $N$,
\begin{align}
    \mathcal{M}[\mathcal{D}_1(z)](N) & \sim \frac{1}{2} \log^2 N.\footnotemark
\end{align}
\footnotetext{Explicit expressions for the Mellin transforms of the remaining terms used
within the decomposition of \cref{eq:tableDecomp} are given in \cref{sec:app_notation}.}%
The coefficients of \cref{eq:tableDecomp} governing the asymptotic large-$N$ behaviour
of the Mellin moments of coefficient functions,
and thus the degree of divergence as $z \to 1$,
are therefore $a_0, a_1$ and $\Delta$.%
\footnote{
As a consequence, the choice of method for imposing momentum
conservation discussed in \cref{subsec:CompPDFsMomCons} can be significant
for the $N \to \infty$ behaviour of the coefficient functions,
and may implicitly be compared
between the \mpos and \mposd schemes, see \cref{fig:NspaceC}.
}

The $\mathcal{D}_k$ contributions (`threshold logarithms') arise
from a miscancellation of soft-gluon radiation between real- and virtual-diagrams, 
since the latter, despite regularising the former, contribute as $\delta(1-z)$, i.e.\ only at $z=1$.
Since the logarithms can be large, they can overcome the $\alphas$ suppression
and lead to a breakdown of perturbation theory in the $z \approx 1$ region.
Conventionally, this is addressed via their resummation
\cite{Sterman:1986aj,Catani:1989ne,Catani:1996yz,Contopanagos:1996nh,Forte:2002ni,Manohar:2003vb,Idilbi:2005ky,Becher:2006nr,Becher:2007ty,Bonvini:2010tp,Bonvini:2012az,Forte:2021wxe}.
The differing coefficients arise from the different kinematic upper limits
allowed by phase-space constraints
on the transverse momentum of the radiated gluon at fixed boson mass
between the DIS and Drell--Yan/Higgs processes,
due to their differing kinematics
\cite{Curci:1979am,Ramalho:1983aq,Forte:2002ni}.

\subsubsection{DIS}
\label{subsubsec:CFsDIS}

Following \cite{Candido:2020yat} we consider the coefficient functions of the DIS
structure-function $F_2$,
\begin{align}
    F_2(x, Q^2)
    {} =
    x \sum_a
    Q_a^2
    \int_0^1 & \dd \xi \; \dd z
    \;
    \delta (\xi z - x) 
    \\ \notag &
    f_{a}^{\rFS}(\xi, \muf)
    \;
    C_a^{\rFS}(z, Q^2; \muf, \mur),
\end{align}
where $Q_g^2 = \sum_{q,\qbar} Q_q^2 $.

Following the reasoning of \cref{subsubsec:coefffuncs} but for a single incoming hadron,
\begin{align}
    C_q^{\rFS (0)}
    & {} = 
    \delta(1-z)
    &
    C_g^{\rFS(0)}
    & {} = 
    0
    \\
    C_q^{\rFS(1)}
    & {} = 
    C_q^{\msbar(1)} - \rK_{qq}^{\rFS}
    &
    C_g^{\rFS(1)}
    & {} = 
    C_g^{\msbar(1)} - \rK_{qg}^{\rFS} .
\end{align}

The NLO contributions to the transformed coefficient functions are given in
\cref{tab:CoDIS}, and plotted in Mellin-space in \cref{fig:NspaceCDIS}.

Notably, the \dis and \aversa schemes exactly remove the distribution terms from
the $C_q$ coefficient function, as they were designed with explicit reference to the \dis process
to absorb $C_q$ into the quark PDF.%
\footnote{Hence, the \dis and \aversa $C_q$ coefficient functions in \cref{fig:NspaceCDIS} are identical.}
The same is true for $C_g$ coefficient function in the case of the \dis scheme.
As is evident from \cref{tab:CoDIS}, in the \pos-family schemes these contributions remain unaltered compared to \msbar, while the $\mathcal{D}_1$ terms
are removed in the \phys scheme (but not $\mathcal{D}_0$).
In the \krk schemes, the coefficients of these terms are all negative.
In \cref{fig:NspaceCDIS} we see the consequence of this: asymptotically the 
\msbar and \pos-type schemes are positive at large-$N$ ($\sim \log^2 N$),
as are the \phys scheme ($\sim \log N$); the \krk scheme is negative ($\sim - \log^2 N$).
Note that, on the scale used for this plot, the Mellin-transformed \lo coefficient function 
$\mathcal{M}[C_q^{\rFS(0)}](N)$ would have the constant value of $2\pi/\alphas \approx 53$.

For the gluon coefficient-function, where the \lo contribution is zero, the \aversa and \krk schemes are negative at
small-$N$ whereas the others are positive.
At large-$N$ all converge to zero due to the absence of $a_0,a_1,\Delta$ terms.

\subsubsection{Drell--Yan}
\label{subsubsec:CFsDY}

The coefficient functions for the Drell--Yan process are transformed as described in
\cref{subsubsec:coefffuncs},
\begin{align}
    D_{q\qbar}^{\rFS(0)}
    & {} = 
    \delta(1-z)
    &
    D_{qg}^{\rFS(0)}
    & {} = 
    0
    \\
    D_{q\qbar}^{\rFS(1)}
    & {} = 
    D_{q\qbar}^{\msbar(1)} - 2 \rK_{qq}^{\rFS}
    &
    D_{qg}^{\rFS(1)}
    & {} = 
    D_{qg}^{\msbar(1)} - \rK_{qg}^{\rFS}
\end{align}
and are given in \cref{tab:CoDY} and their Mellin-transforms plotted in \cref{fig:NspaceCDY}.

In this case, the \krk scheme exactly removes the divergent $\mathcal{D}_1$ distribution terms
from the $D_{qq}$ coefficient function,
while this contribution is partially mitigated (halved) in the \aversa/\dis and \phys schemes.

This can be seen in \cref{fig:NspaceCDY}, where
for the $q\qbar$-channel coefficient function,
the large-$N$ behaviour divides the schemes into three main groups:
the \krk-scheme, which is asymptotically constant;
the \phys and \aversa/\dis schemes, which diverge as $2 \cf \log^2 N$;
and the \msbar and \pos-type schemes, which diverge as $4 \cf \log^2 N$.

In this respect the \krk scheme can be seen to have a similar function 
for the Drell--Yan process at large-$N$ as the \dis scheme has for DIS,
removing the logarithmic terms from the coefficient function to leave it asymptotically-constant.
This is due to the $\mathcal{D}_1$ terms being entirely removed by the choice $a_1 = 4$ in 
$\rK_{qq}^{\krk}$.

For the $qg$-channel coefficient function, the \mpos and \krk schemes exhibit positivity in Mellin
space, whereas the others are negative.
All are asymptotically zero due to the absence of $a_0,a_1$ and $\Delta$ terms.

\subsubsection{Higgs}
\label{subsubsec:CompCFs}

The coefficient functions for the gluon-fusion Higgs-production process are again transformed as described in \cref{subsubsec:coefffuncs},
\begin{align}
    C_{gg}^{\rFS(0)}
    & {} = 
    \delta(1-z)
    &
    C_{gq}^{\rFS(0)}
    & {} = 
    0
    \\
    C_{gg}^{\rFS(1)}
    & {} = 
    C_{gg}^{\msbar(1)} - 2 \rK_{gg}^{\rFS}
    &
    C_{gq}^{\rFS(1)}
    & {} = 
    C_{gq}^{\msbar(1)} - \rK_{gq}^{\rFS}.
\end{align}
They are given in \cref{tab:CoHiggs} and their Mellin-transforms are plotted in \cref{fig:NspaceCH}.

As for Drell--Yan, in the flavour-diagonal channel (here $gg$),
the \krk scheme exactly removes the distribution
terms $\mathcal{D}_1$ from the $C_{gg}$ coefficient function,
while the contribution is partially mitigated in the \aversa and \phys schemes,
due to the respective choices $a_1=4$ and $a_1=2$ for $\rK_{gg}^{\rFS}$.
This can be seen in the plots of \cref{fig:NspaceCH},
where the \krk-scheme coefficient function is again rendered asymptotically constant,
in contrast to the others which either diverge as
$  4 \ca \log^2 N $ (\msbar, \dis, \pos-type),
or 
$  2 \ca \log^2 N $ (\aversa, \phys).

For $C_{gq}$ the coefficients are again asymptotically-zero at large-$N$ for all schemes,
save for the \dis scheme due to its conventional local-momentum-conserving definition
in terms of the flavour-diagonal $\rK_{qq}^{\dis}$ in \cref{eq:DIS_def_qg},
which contains $\mathcal{D}_0$ and $\mathcal{D}_1$ contributions.

\section{Phenomenological impact of scheme choice}
\label{sec:pheno}

\enlargethispage{2\baselineskip}

As described in \cref{subsec:CompCFs}, the choice of factorisation scheme
for an NLO calculation is compensated to the same perturbative order
by a modification of the partonic cross-section,
and so induces only formally-NNLO changes to an NLO calculation,
akin to factorisation- and renormalisation-scale variation.%
\footnote{This has previously been studied
for the inclusive jet cross-section
for a parametrised transition between the
\msbar and DIS schemes,
\cite{Anandam:1999aq,Klasen:1996yk}
where the variation at LO was found to be 40\%
and at NLO 8\%.}
However, some observables of phenomenological interest
may only be non-zero in the real-emission kinematics,
leading to an effectively-LO calculation
for which the factorisation-scheme dependence is uncompensated.

The effect of the factorisation-scheme choice on these observables
for an NLO calculation of $pp \to X$
is identical to their effect on an LO calculation of
$pp \to X + j$.
Therefore, in this section we present differential distributions corresponding to
a leading-order calculation of
$p p \to Z + j$ 
and
$p p \to H^0 + j$
as an example.

This illustrates the 
effect 
of varying the factorisation scheme
on the leading-jet distributions of an NLO calculation of the Drell--Yan and gluon-fusion Higgs processes.
These have been chosen because they proceed via the 
$q\qbar$ and $gg$ channels respectively at leading-order.

The general observations made may be expected to generalise
to a wider range of processes, since they relate essentially
to the kinematic region in $(x,Q)$ probed within each
PDF.

For the calculations in this section, 
we use \herwigseven \cite{Bellm:2015jjp,Bewick:2023tfi}
with the \nnpdfmc PDF set \cite{Cruz-Martinez:2024cbz},
transformed into the different factorisation schemes
and stored in \lhapdf6 format \cite{Buckley:2014ana}; 
accordingly, we adopt $\alphas (M_Z) = 0.118$. 
All calculations are done at fixed-order without any parton shower resummation or other corrections.

\subsection{$Z$-plus-jet}
\label{subsec:pheno_Zj}

\enlargethispage*{3\baselineskip}

We use a set-up appropriate for 
LHC Run II, with a centre-of-mass energy of 13 TeV
and fiducial cuts
\begin{subequations}
\begin{align}
\label{eqn:Zjcuts}
    \ptl{1,2} &> 25 \;\GeV \,, 
    & \left\vert\eta^{\ell_{1,2}}\right\vert &< 3.5,  \\
    M_{\ell\ell} & \in \left[ 66, 116 \right] \, \GeV,
\end{align}
\end{subequations}
similar to those used by 
\atlas \cite{ATLAS:2017sag}
and \cms \cite{CMS:2018mdf}.
We identify jets using the anti-$k_\rT$ algorithm
\cite{Cacciari:2008gp}
with jet-radius $R=0.4$ and minimum
transverse momentum of 10 $\GeV$.
We use
the invariant mass of the lepton pair as the
dynamic renormalisation and factorisation scale,
\begin{align}
	\mur = \muf = M_{\ell\ell}.
\end{align}
The shaded uncertainty band illustrates the
(\msbar-scheme) factorisation-scale uncertainty and corresponds to the envelope given by
factorisation-scale variation by a factor of two in each directon,
$\muf \in \{ \frac{1}{2}, 1, 2 \}M_{\ell\ell}$.

\begin{figure*}[p]
\centering
\begin{subfigure}[t]{\textwidth}
\makebox[\textwidth][c]{
    \includegraphics[width=0.47\textwidth]{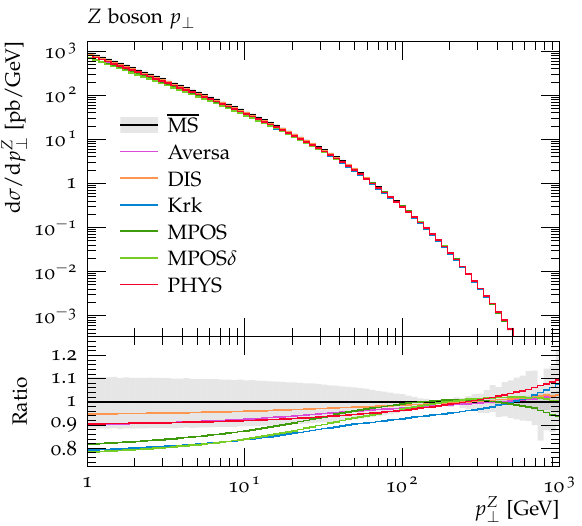} \hfill
    \includegraphics[width=0.47\textwidth]{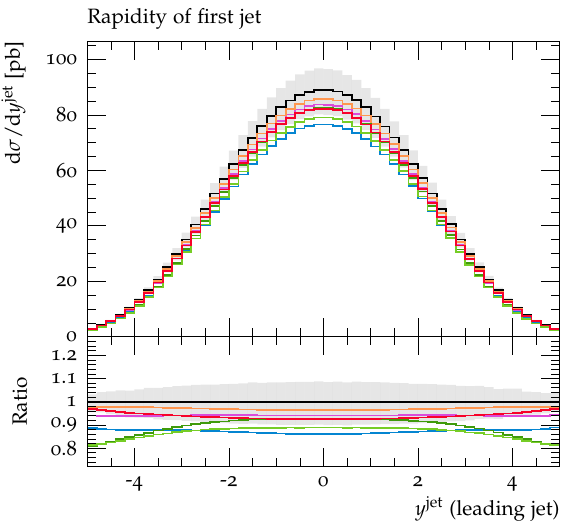}
}
\caption{ $p p \to \ell^+ \ell^- + j$
\label{fig:Zj_pp}}
\end{subfigure}
\\
\begin{subfigure}[t]{\textwidth}
\makebox[\textwidth][c]{
    \includegraphics[width=0.47\textwidth]{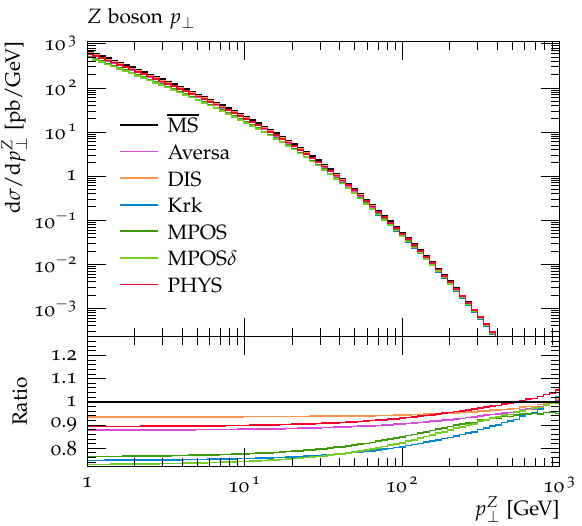} \hfill
    \includegraphics[width=0.47\textwidth]{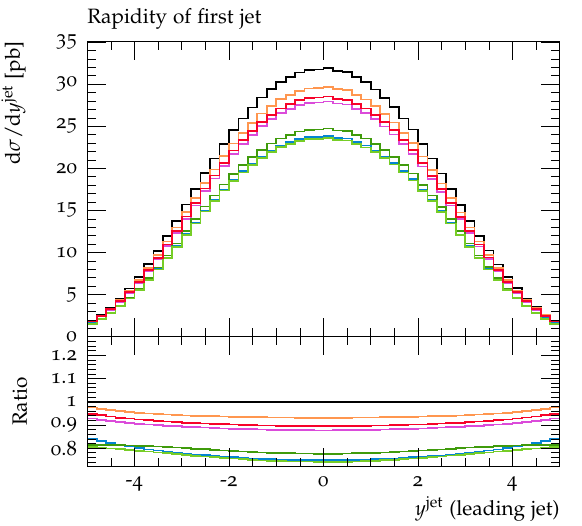}
}
\caption{ $ \left\{ q \qbar , \qbar q \right\} \to \ell^+ \ell^- + j$
\label{fig:Zj_qqb}}
\end{subfigure}
\\
\begin{subfigure}[t]{\textwidth}
\makebox[\textwidth][c]{
    \includegraphics[width=0.47\textwidth]{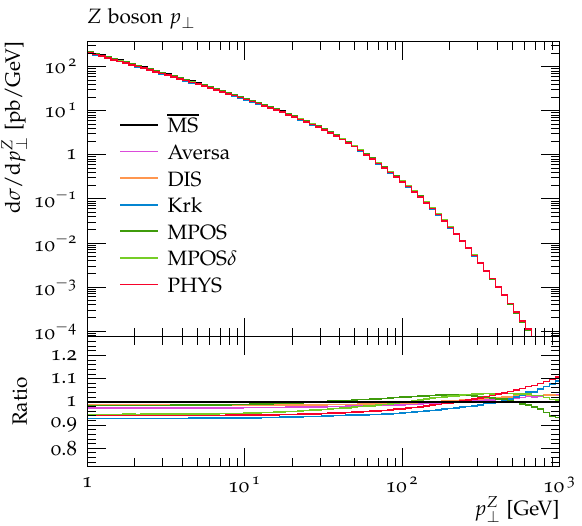} \hfill
    \includegraphics[width=0.47\textwidth]{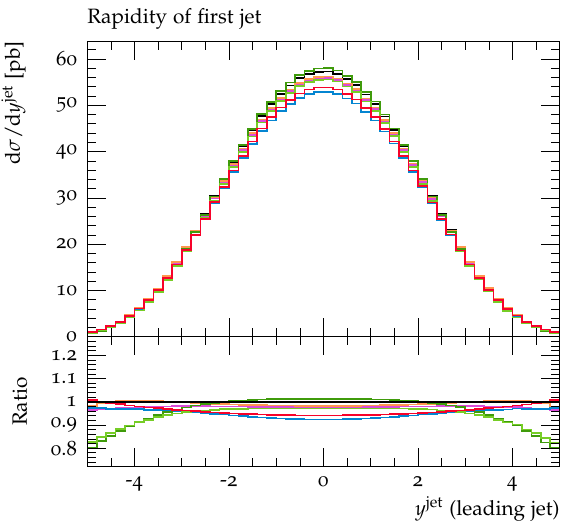}
}
\caption{ $\left\{ qg, gq, \qbar g, g \qbar \right\} \to \ell^+ \ell^-+ j$
\label{fig:Zj_qg}}
\end{subfigure}
\caption{Factorisation-scheme dependence of
differential cross-sections for $Z + j$ production at the LHC.
Leading-order predictions for these observables correspond to
an NLO calculation of the Drell--Yan process.
}
\label{fig:pheno_Z}
\end{figure*}

The results are shown in 
\cref{fig:pheno_Z}
for the transverse momentum of the $Z$-boson
and the rapidity of the first jet (at LO,
this is the only jet, and comprises a single parton).
In addition to the total contribution from proton-proton
scattering (\cref{fig:Zj_pp}),
the distributions have been further subdivided into
the contributions from the quark-antiquark (\cref{fig:Zj_qqb})
and the quark-gluon channels (\cref{fig:Zj_qg}) separately.

The uncertainty on the distributions arising from the
factorisation-scheme variation is largest in the
$q\qbar$-channel, in which it ranges from approximately
30\% at low-$\pt$ to around 5\% at high-$\pt$.
The jet-rapidity distribution shows a modest rapidity-dependence
on the factorisation-scheme uncertainty.
The effect in the $qg$ channels is in the same
direction with the same scale dependence, but at
most 10\% at low-$\pt$;
the scheme-uncertainty of the rapidity distribution grows strongly
with rapidity for the \mpos and \mposd schemes (to $\sim 20$\%) whereas
for other schemes it is modestly suppressed.

Distributions calculated with factorisation schemes other than \msbar are consistently smaller
than the \msbar distributions, 
as might be expected from the plots of the transformed quark PDFs alone.
The hierarchy of schemes matches that of the PDFs for the relevant 
$(x,Q)$ region;
this can ultimately be traced back to the coefficient of $\pqg \log(1-z)$
in the $qg$ transformation kernels,
which separates the schemes into three emergent groups
(i) $\msbar$, 
(ii) \dis, \phys and \aversa, and 
(iii) \krk, \mpos and \mposd.

As can be seen from \cref{fig:Zj_pp},
the factorisation-scheme uncertainty here exceeds the
scale-uncertainty
(here both formally at relative order $\order{\alphas}$), especially at low-$\pt$
and in the extremes of the jet-rapidity distribution.

\subsection{Higgs-plus-jet}
\label{subsec:pheno_Hj}

\begin{figure*}[p]
\centering
\begin{subfigure}[t]{\textwidth}
\makebox[\textwidth][c]{
    \includegraphics[width=0.47\textwidth]{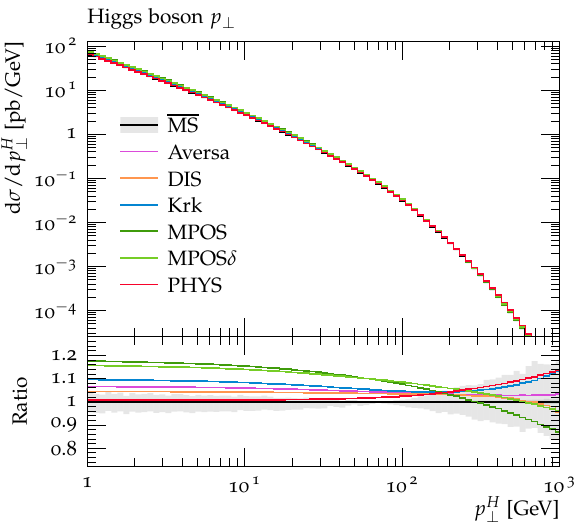} \hfill
    \includegraphics[width=0.47\textwidth]{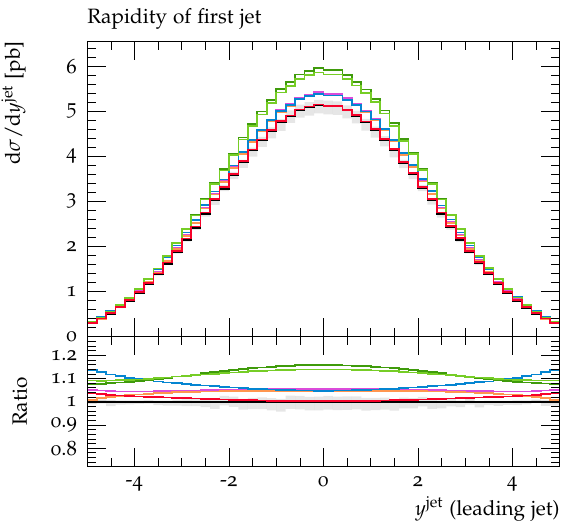}
}
\caption{ $p p \to H^0 + j$
\label{fig:Hj_pp}}
\end{subfigure}
\\
\begin{subfigure}[t]{\textwidth}
\makebox[\textwidth][c]{
    \includegraphics[width=0.47\textwidth]{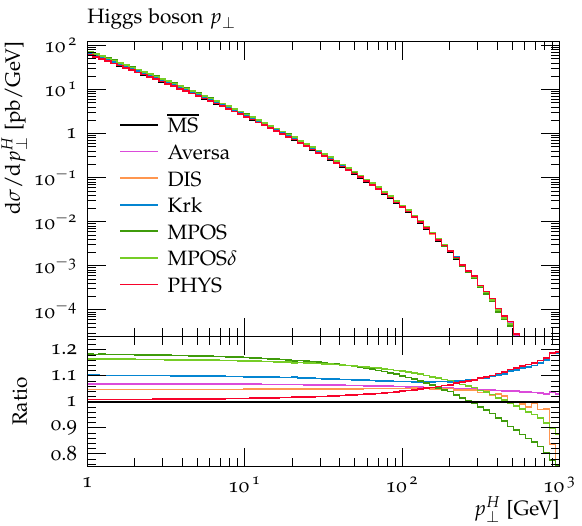} \hfill
    \includegraphics[width=0.47\textwidth]{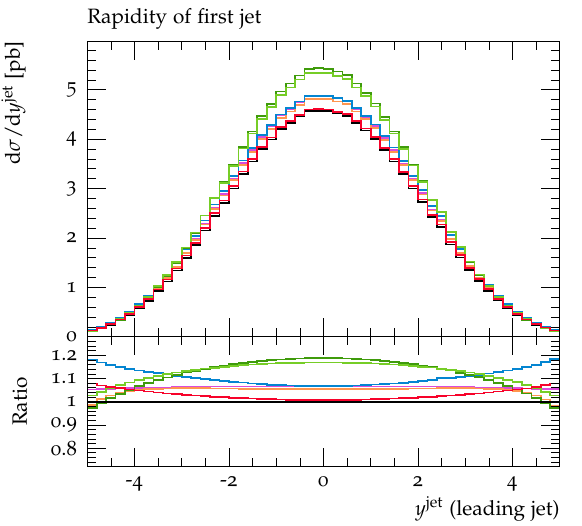}
}
\caption{ $ g g \to H^0 + j$
\label{fig:Hj_gg}}
\end{subfigure}
\\
\begin{subfigure}[t]{\textwidth}
\makebox[\textwidth][c]{
    \includegraphics[width=0.47\textwidth]{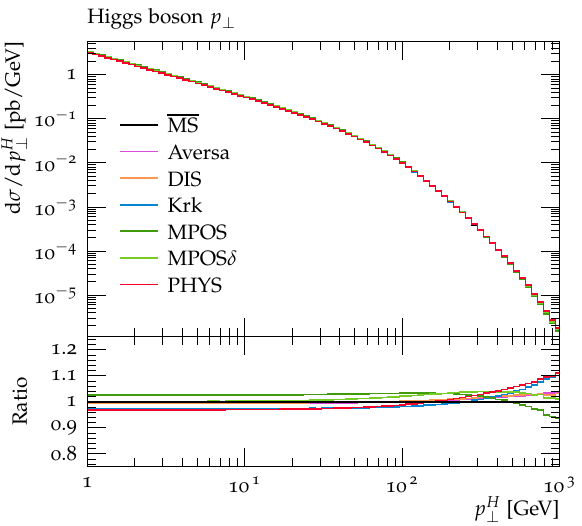} \hfill
    \includegraphics[width=0.47\textwidth]{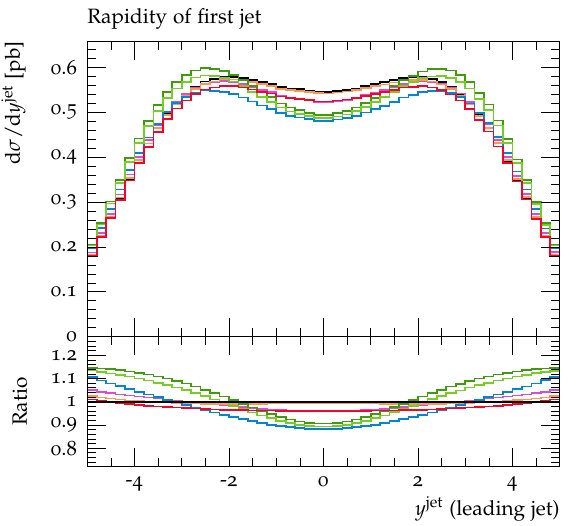}
}
\caption{ $\left\{ qg, gq, \qbar g, g \qbar \right\} \to H^0 + j$
\label{fig:Hj_gq}}
\end{subfigure}
\caption{Factorisation-scheme dependence of
differential cross-sections for $H + j$ production at the LHC.
Leading-order predictions for these observables correspond to
an NLO calculation of the Higgs-production process.
}
\label{fig:pheno_H}
\end{figure*}

Again we use a set-up appropriate for LHC Run II, 
here for $pp \to H^0 \to \tau^+ \tau^-$
with a centre-of-mass energy of 13 TeV
and fiducial cuts
\begin{subequations}
\begin{align}
\label{eqn:Hjcuts}
    \pttau{1,2} &> 25 \;\GeV \,, 
    & \left\vert\eta^{\tau_{1,2}}\right\vert &< 3.5,  \\
    M_{\tau\tau} & \in \left[ 115, 135 \right] \, \GeV
\end{align}
\end{subequations}
and identify jets using the anti-$k_\rT$ algorithm
\cite{Cacciari:2008gp}
with jet-radius $R=0.4$ and minimum
transverse momentum of 10 $\GeV$.
We use the invariant mass of the tauon-pair as the
dynamic renormalisation and factorisation scale,
\begin{align}
	\mur = \muf = M_{\tau\tau}
\end{align}
and work in the Higgs effective theory in the infinite top-quark-mass limit.
The shaded uncertainty band again corresponds to \msbar-scheme
factorisation-scale variation by a factor of two in each directon,
$\muf \in \{ \frac{1}{2}, 1, 2 \} M_{\tau\tau}$.

The results are shown in 
\cref{fig:pheno_H}
for the transverse momentum of the Higgs-boson
and the rapidity of the first jet.
In addition to the total result for proton-proton
scattering (\cref{fig:Hj_pp}),
the distributions have been subdivided into
contributions from the gluon-gluon (\cref{fig:Hj_gg})
and the quark-gluon channels (\cref{fig:Hj_gq}) separately.

In contrast to the $Z+j$ process we observe that all the schemes lead to higher cross-sections than that using the \msbar scheme; this is in accordance
with the general directional effect of the scheme transformations shown
in \cref{fig:PDF_compare}.
The $gg$-channel dominates the total and hence also the scheme-uncertainty.

For \mpos and \mposd we observe deviations from \msbar
of up to 20\%  in the central-rapidity region and at low transverse momenta.
The deviation for \krk is smaller, not exceeding 10\% in most of the kinematic regions.
Again we see grouping of the schemes, however, slightly different from that observed for $Z+j$: (i) \mpos \& \mposd, (ii) \krk, \aversa, \dis, and (iii) \phys \& \msbar.
Another prominent difference compared to the $Z+j$ is related to the shape of the rapidity distribution in the $qg$ channel. For $Z+j$ we observe single peak shape, whereas, for the $H+j$ we see a double-hump structure with a deep in the central rapidity region. This shape is present already for the \msbar scheme but it is further enhanced in the \mpos, \mposd, and \krk schemes (leading to lower minimum in the central rapidity and maxima at $y^{\mathrm{jet}}=\pm2.5$).

As can be seen from \cref{fig:Hj_pp},
the factorisation-scheme uncertainty again exceeds the
scale-uncertainty, especially at low-$\pt$
and in the centre and extremes of the jet-rapidity distribution.
The scale-variation uncertainty is notably smaller in magnitude here
than for $Z + j$, despite the scheme-variation
uncertainty being comparable,
implying that scale-variation uncertainty
is an unreliable guide to scheme-variation uncertainty.

\section{Discussion and conclusions}
\label{sec:summary}

In this work we have investigated for the first time the relationship between many of 
the different factorisation schemes proposed in the literature for NLO QCD calculations.
In doing so we have identified features common to all such schemes.
This hints at the possibility of a consensus scheme achieving multiple
objectives simultaneously.

In general the transformations are dominated by the distribution and logarithmic terms,
with finer details of the polynomial piece $P(z)$ mostly numerically suppressed.
Setting this to zero for convenience, the schemes discussed here may be
considered special cases of the following general expressions:%
\footnote{With the exception of $\rK_{ga}^{\msbar\to\dis}$, which
are unique in being determined by \cref{eq:DIS_localsumrule} rather
than from consideration of the relevant splitting channels,
and $\rK_{gg}^{\msbar\to\aversa}$, which uses only a
single (rational) term from
$\pgg$ in its coefficient of $\log z$.}
\begin{align}
    \rK_{qq}^{\msbar\to\rFS} &=
        \cf \biggl[
            a_{qq} \biggl( 2 \mathcal{D}_1 - (1+z) \log(1-z) \biggr) 
            \\ \notag & \qquad
            - \frac{3}{2} b_{qq} \mathcal{D}_0
            - c_{qq} \pqq \log z
            - \Delta_{qq} \delta(1-z)
        \biggr]
    \\
    \rK_{qg}^{\msbar\to\rFS} &= 
        \tr \biggl[
            a_{qg} \pqg \log(1-z) - c_{qg} \pqg \log z
        \biggr]
    \\
    \rK_{gq}^{\msbar\to\rFS} &=
        \ca \biggl[
            a_{gq} \pgq \log(1-z) - c_{gq} \pgq \log z
        \biggr]
    \\
    \rK_{gg}^{\msbar\to\rFS} &=
        \ca \biggl[
            a_{gg} \biggl( 2 \mathcal{D}_1 + \left(\frac{1}{z} - 2 + z(1-z)\right)
            \\ \notag & \qquad \qquad
             \times \log(1-z) \biggr) 
            - 2 c_{gg} \pgg \log z
            \\ \notag & \qquad \qquad
            - \Delta_{gg} \delta(1-z)
        \biggr]
\end{align}
where $a_{qq},a_{gg} \in \{ 0, 1, 2\}$,
$a_{qg}, a_{gq} \in \{1,2\}$,
$b_{qq} \in \{0,1\}$,
$c_{qq}, c_{qg}, c_{gq}, c_{gg} \in \{0, 1\}$,
and $\Delta_{qq}, \Delta_{gg}$ are fixed by 
the momentum sum-rule as in \cref{eq:momsumrule_deltas}.
Note that the functions appearing in the coefficients
of $\log(1-z)$ in $\rK_{qq}^{\msbar\to\rFS}$
and $\rK_{gg}^{\msbar\to\rFS}$ correspond to the remainder after the subtraction
of divergent (plus-distribution) contribution $\mathcal{D}_0$ from $\pqq$ and $\pgg$
respectively.
Whilst we do not claim that all possible factorisation schemes of interest
fit this pattern, it is notable that the domain 
of interest for factorisation-scheme variation within the literature
is much smaller than might initially be assumed and
appears to be parametrised by a handful of discrete variables.

Following the Mellin-space arguments of \cref{subsec:CompCFs},
for the hadron--hadron Drell--Yan and Higgs
coefficient functions considered there
to be bounded in Mellin-space we require $a_{qq} = a_{gg} = 2$
(with the consequence that the DIS coefficient function $C_q^{\rFS(1)}$ is unbounded).

This inclusion of the threshold logarithms within the PDFs substantially changes
them (in particular the gluon at low-$x$) but could in principle be extended
to higher orders in QCD or adapted to use resummed expressions in the PDF transformations
and their perturbative truncations in the coefficient functions.

The question of the relationship between factorisation-scheme choice and PDF positivity remains
open, but we have observed empirically the extent to which it is violated for transformed
\msbar PDFs, as well as studying the argument of \cite{Candido:2023ujx} applied
to schemes other than \msbar.
We defer a detailed consideration of the provable positivity properties
of factorisation schemes to future work.

Finally, as highlighted in \cref{sec:schemeComparisons}, our study considers only
PDFs transformed from \msbar input PDFs at each scale, while PDFs in alternative
factorisation schemes may also be obtained by performing the evolution,
or the evolution and the fit,
in the scheme of interest.
The conclusions reached here may not apply to PDFs obtained by these methods.
We defer consideration of this to future work.

\section*{Acknowledgements}
The authors wish to thank John Collins and Jonathan Gaunt
for valuable discussions about
the PDF sum-rules
and the significance of their scheme-dependence, and 
Stefano Forte and Felix Hekhorn for productive correspondence about
the argument and results of \cite{Candido:2023ujx}.
We are grateful to the late Staszek Jadach for the initial discussions
that prompted us to pursue this work, and to Wies\l{}aw P\l{}aczek for
comments on the manuscript.

The work of S.D.\ and A.K.\ was supported by National Science Centre
Poland under the Sonata Bis grant No 2019/34/E/ST2/00186. 
The work of A.S.\ and J.W.\ was supported by the National Science
Centre Poland grant No 2019/34/E/ST2/00457. 
A.S.\ is also supported by the Priority Research Area Digiworld
under the program `Excellence Initiative -- Research University'
at the Jagiellonian University in Krakow.
AS thanks the CERN Theoretical Physics Department for
hospitality while part of this research was being carried out.
We gratefully acknowledge the Polish high-performance computing infrastructure PLGrid (HPC Centre: ACK Cyfronet AGH) for providing computing facilities and support within computational grant PLG/2023/016880.

\clearpage
\appendix

\settocdepth{section}

\section{Notation and conventions}
\label{sec:app_notation}

The leading-order colour-factor-stripped DGLAP splitting functions are given in four dimensions by%
\footnote{Note that the flavour indices are transposed here
relative to a recent paper by some of the same authors \cite{Sarmah:2024hdk}.
The convention employed here allows \cref{eq:fstransformations_faFSvec}
to be considered as matrix-multiplication without a transposition.}
\begin{align}
	\pqq &= \frac{1+z^2}{1-z}	  %
	\\
	\pqg &= z^2 + (1-z)^2 
	\\
	\pgq &= \frac{1 + (1-z)^2}{z} 
	\\
	\pgg &= %
	  \frac{1}{1-z} + \frac{1}{z} - 2 + z(1-z)
\end{align}
Perturbative expansions are indexed according
to the convention of \cref{eq:pertexpconv},
\begin{align}
\notag
F(\alphas, \mu)
&=
F^{(0)}(\mu) + \left(\frac{\alphas(\mu)}{2\pi}\right) F^{(1)}(\mu) + \dots 
\\ &= 
\sum_{k} 
\left(\frac{\alphas(\mu)}{2\pi}\right)^k 
F^{(k)}(\mu).
\end{align}
The leading-order expansion of the \msbar DGLAP kernels referred to in
\cref{subsubsec:fsdglap}
is then
\begin{align}
    \rP^{\msbar(1)}_{qq} (z) &= \cf \, \left[ \pqq \right]_+
    \\
    \rP^{\msbar(1)}_{qg} (z) &= \tr \; \pqg
    \\
    \rP^{\msbar(1)}_{gq} (z) &= \cf \; \pgq
    \\ \notag
    \rP^{\msbar(1)}_{gg} (z) &= 2 \ca \; \left(
    \left[\frac{1}{1-z}\right]_+ + \frac{1}{z} - 2 + z(1-z)
    \right)
    \\
    & \qquad - \frac{4 \nf \tR - 11 \ca}{6} \delta(1-z).
\end{align}
In the convention of \cref{subsubsec:fsdglap}, the QCD $\beta$-function 
has the perturbative expansion
\begin{align}
    \beta^{(0)} &= 0 \\
    \beta^{(1)} &= 0 \\
    \beta^{(2)} &= \frac{\pi}{3} \left( 4 \nf \tR - 11 \ca \right).
\end{align}

The decomposition of a kernel in the convention of \cref{eq:tableDecomp},
as in \cref{tab:Kqq,tab:Kqg,tab:Kgq,tab:Kgg,tab:CoDIS,tab:CoDY,tab:CoHiggs},
allows for the direct calculation of its Mellin transform using the following results,
collected here for reference with respect to \cref{subsubsec:coefffuncs}:
\begin{align}
    \mathcal{M}[\mathcal{D}_0(z)](N) &= - H_{N-1}
    \\
    \mathcal{M}[\mathcal{D}_1(z)](N) &= 
    \frac{1}{2}\left( H_{N-1}^2 + H_{N-1}^{(2)} \right)
    \\
    \mathcal{M}[ z^k ](N) &= \frac{1}{N+k}
    \\
    \mathcal{M}[ z^k \log (1-z) ](N) &= - \frac{H_{N+k}}{N+k}
    \\
    \mathcal{M}[ z^k \log z ](N) &= - \frac{1}{(N+k)^2}
    \\
    \mathcal{M}\left[z^k \frac{\log z}{1-z}\right](N) &=
    H_{N+k-1}^{(2)} - \frac{\pi^2}{6}
    \\
    \mathcal{M}[\delta(1-z)](N) &= 1.
\end{align}
Here $H_N^{(r)}$ denotes the $N$\textsuperscript{th} generalised harmonic number of order $r$,
\begin{align}
    H_N^{(r)} = \sum_{k=1}^{N} \frac{1}{k^{r}};
\end{align}
additionally, we omit the order for the harmonic numbers of order 1: $H_N^{(r)}\equiv H_N$.
Note that 
\begin{align}
    H_N = \log N + \gamma_{\mathrm{E}} + \mathcal{O}\left(\frac{1}{N}\right).
\end{align}

\section{Supplementary transformed PDFs}
\label{app:PDF_plots}

\begin{figure*}[tp]
\centering
\begin{subfigure}[t]{\textwidth}
\makebox[\textwidth][c]{
    \includegraphics[width=0.47\textwidth]{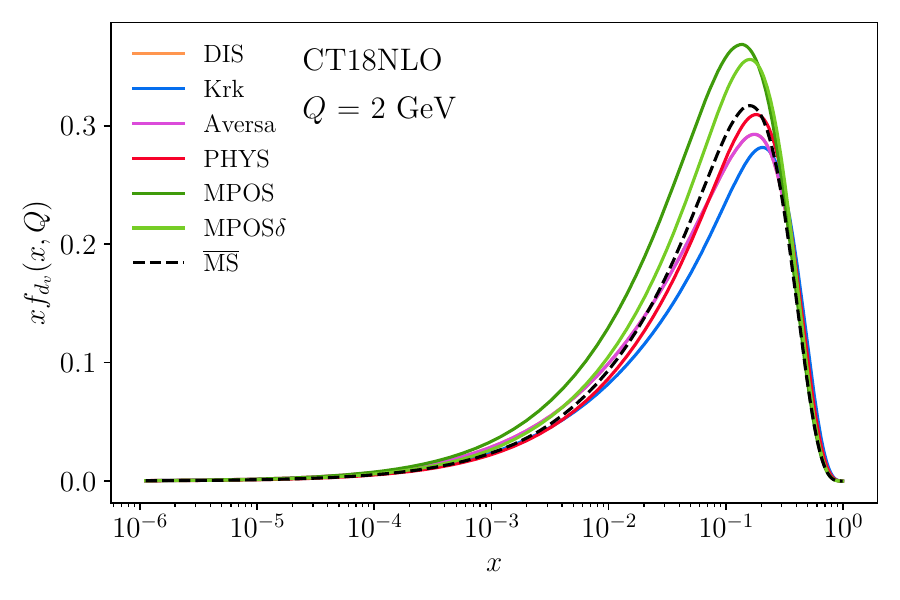} \hfill
    \includegraphics[width=0.47\textwidth]{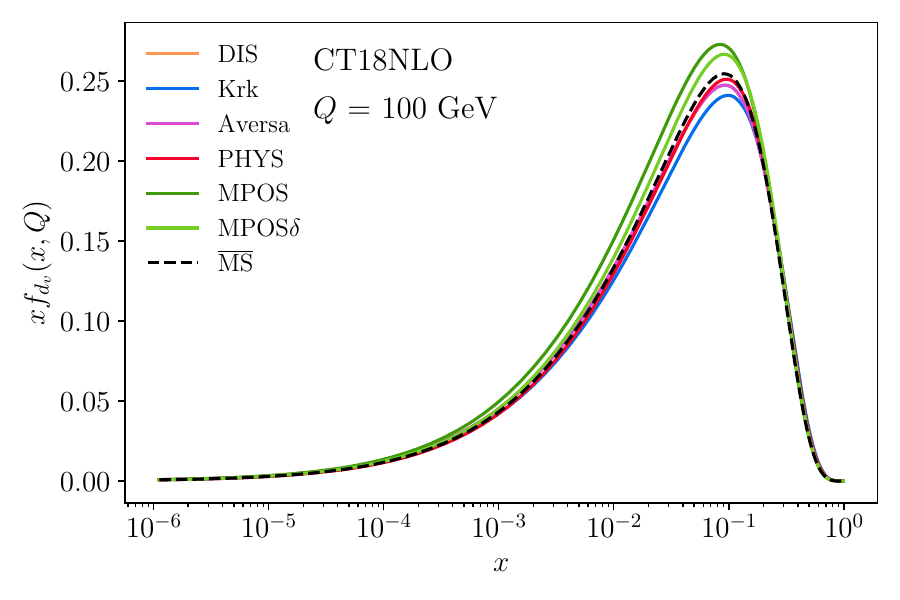}
}
\caption{$d$-valence quark\label{fig:xfdvxQ_CT18_2}}
\end{subfigure}
\\
\begin{subfigure}[t]{\textwidth}
\makebox[\textwidth][c]{
\includegraphics[width=0.47\textwidth]{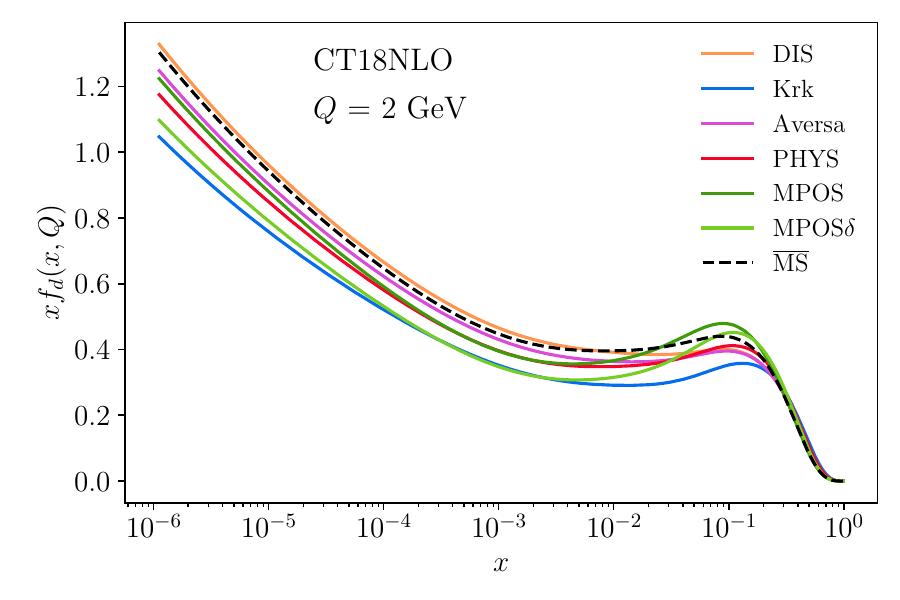} \hfill
\includegraphics[width=0.47\textwidth]{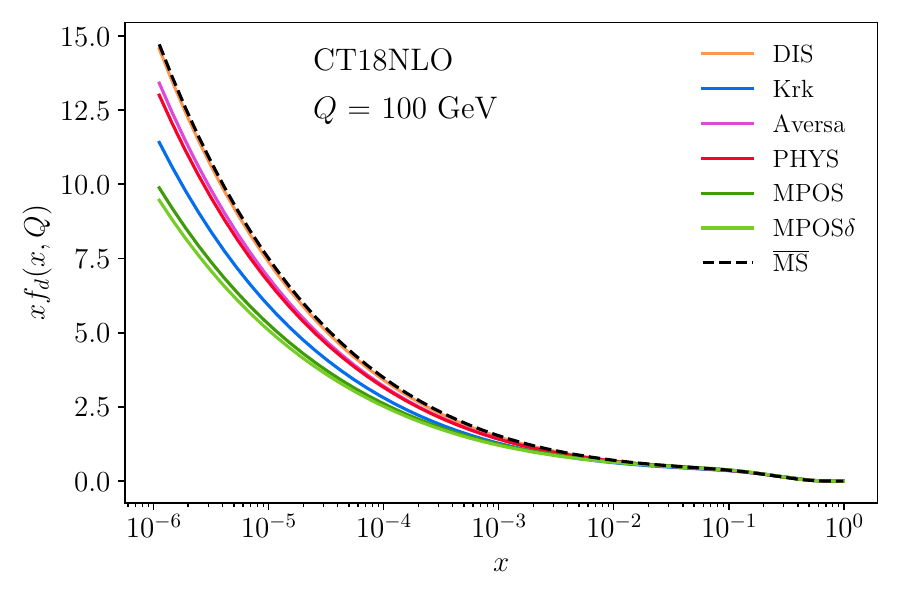}
}
\caption{$d$ quark\label{fig:xfdxQ_CT18_2}}
\end{subfigure}
\\
\begin{subfigure}[t]{\textwidth}
\makebox[\textwidth][c]{
\includegraphics[width=0.47\textwidth]{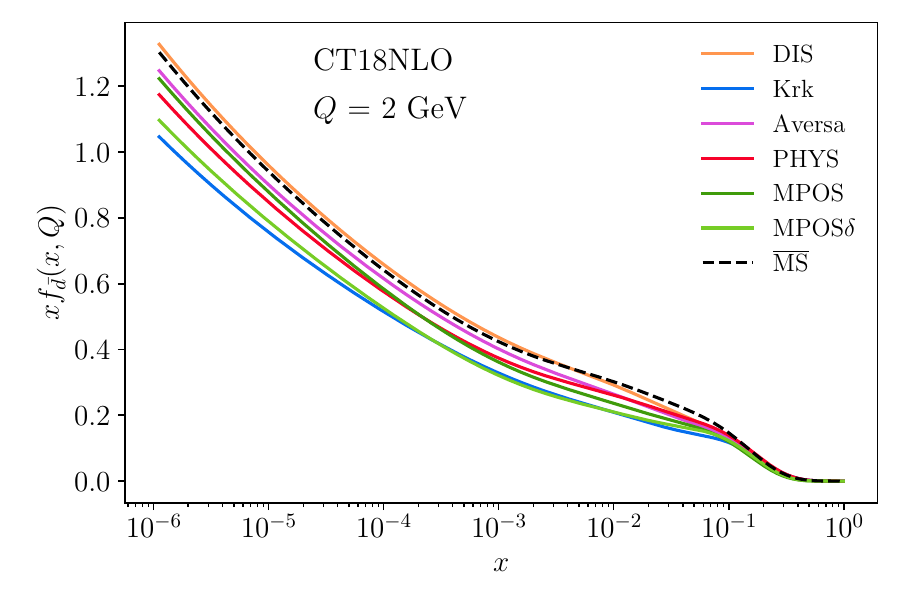} \hfill
\includegraphics[width=0.47\textwidth]{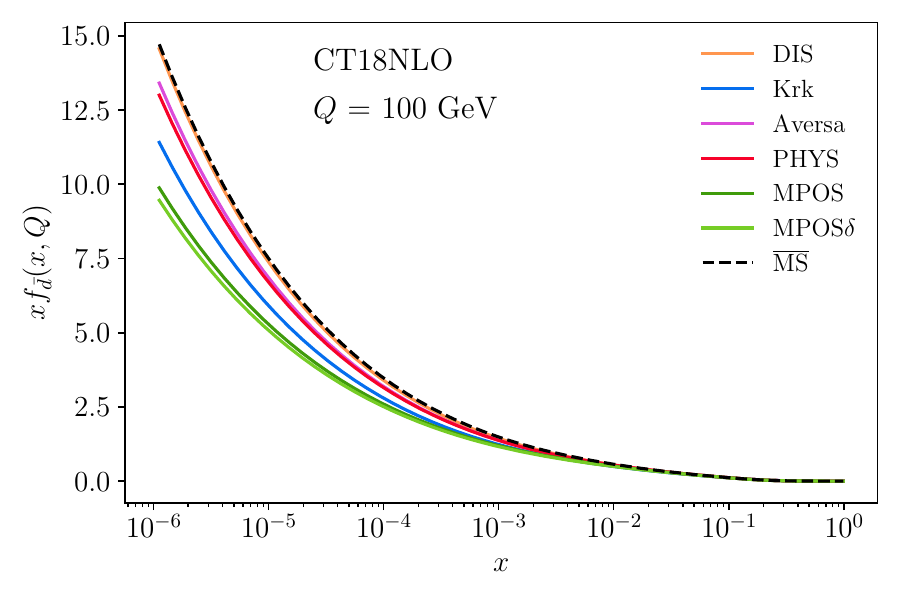}
}
\caption{$\bar{d}$-quark\label{fig:xfdbxQ_CT18_2}}
\end{subfigure}
\\
\begin{subfigure}[t]{\textwidth}
\makebox[\textwidth][c]{
\includegraphics[width=0.47\textwidth]{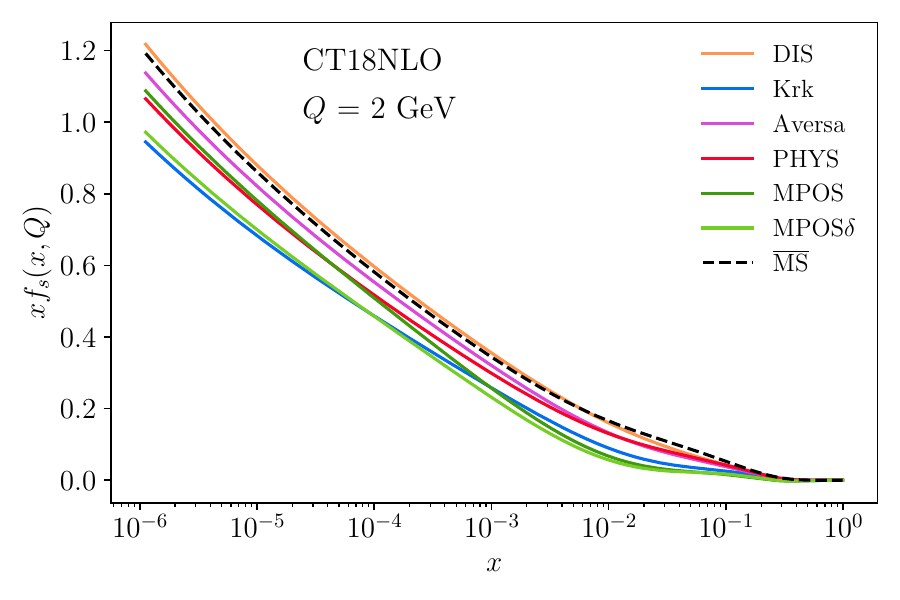} \hfill
\includegraphics[width=0.47\textwidth]{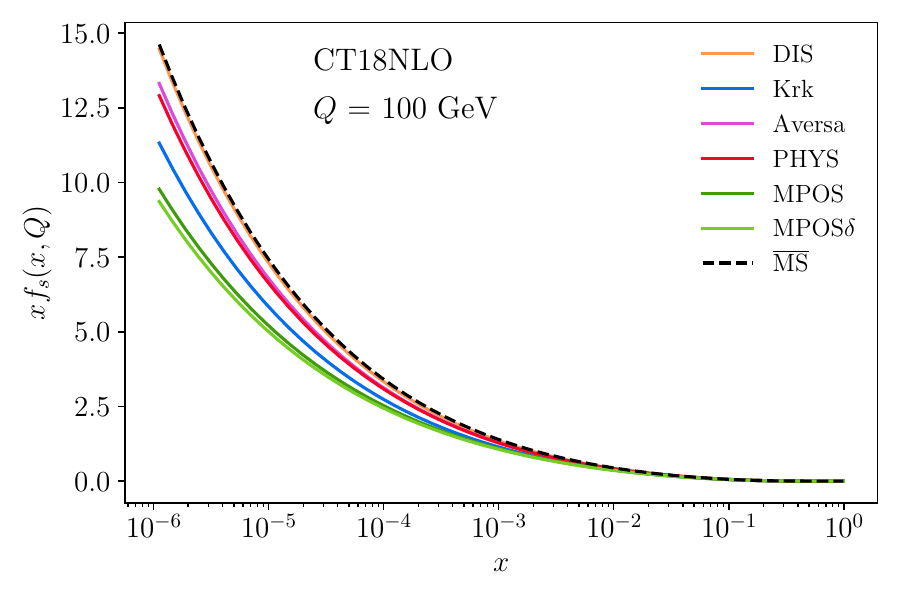}
}
\caption{strange quark\label{fig:xfsxQ_CT18_2}}
\end{subfigure}
\caption{
Comparison of transformed \ctnlo PDFs in different schemes for $d$-valence, $d$, $\bar{d}$, and $s$,
at $Q=2$ GeV (left) and 100 GeV (right).
Companion to \cref{fig:PDF_compare}.}
\label{fig:PDF_compare_CT18_2}
\end{figure*}

\begin{figure*}[tp]
\centering
\begin{subfigure}[t]{\textwidth}
\makebox[\textwidth][c]{
    \includegraphics[width=0.47\textwidth]{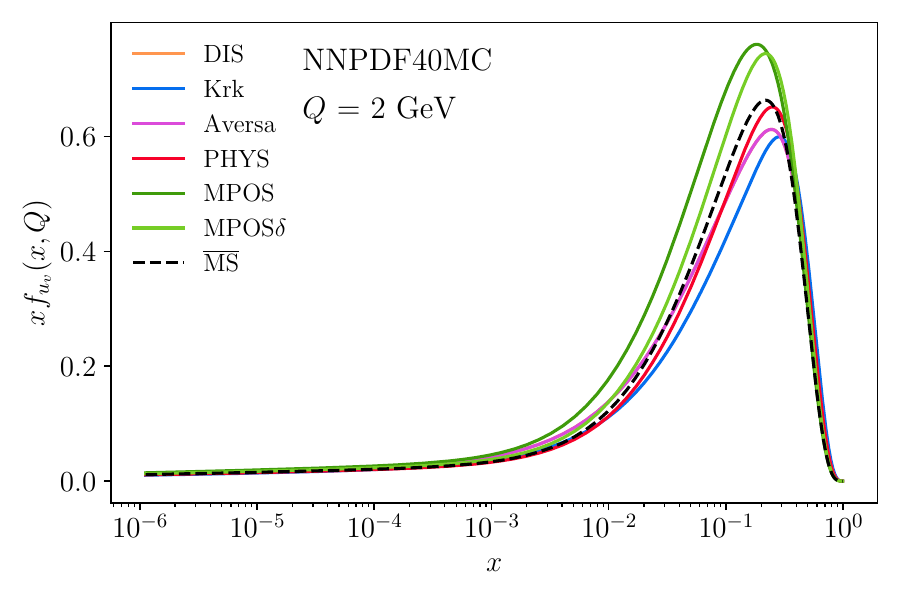} \hfill
    \includegraphics[width=0.47\textwidth]{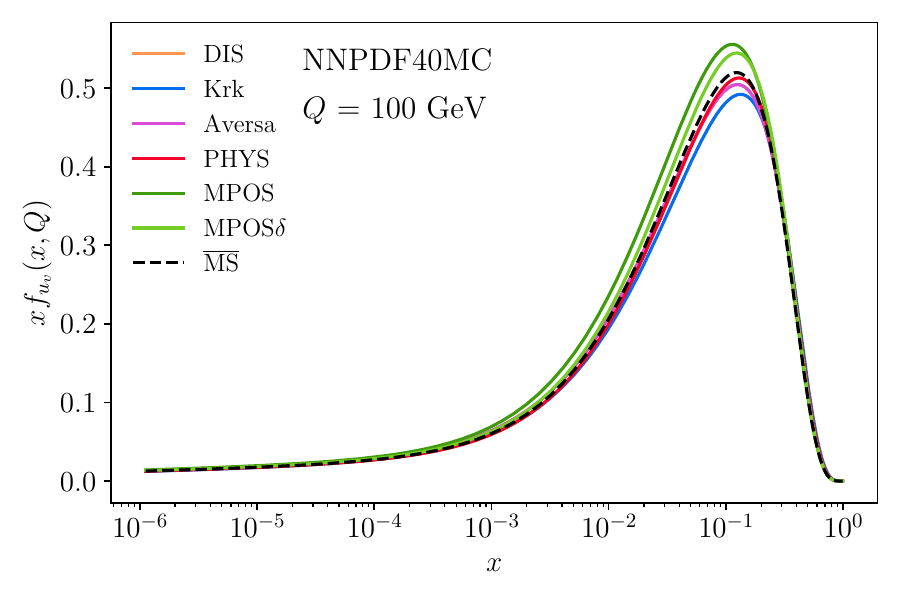}
}
\caption{$u$-valence quark\label{fig:xfuxQ_NNPDF40MC}}
\end{subfigure}
\\
\begin{subfigure}[t]{\textwidth}
\makebox[\textwidth][c]{
\includegraphics[width=0.47\textwidth]{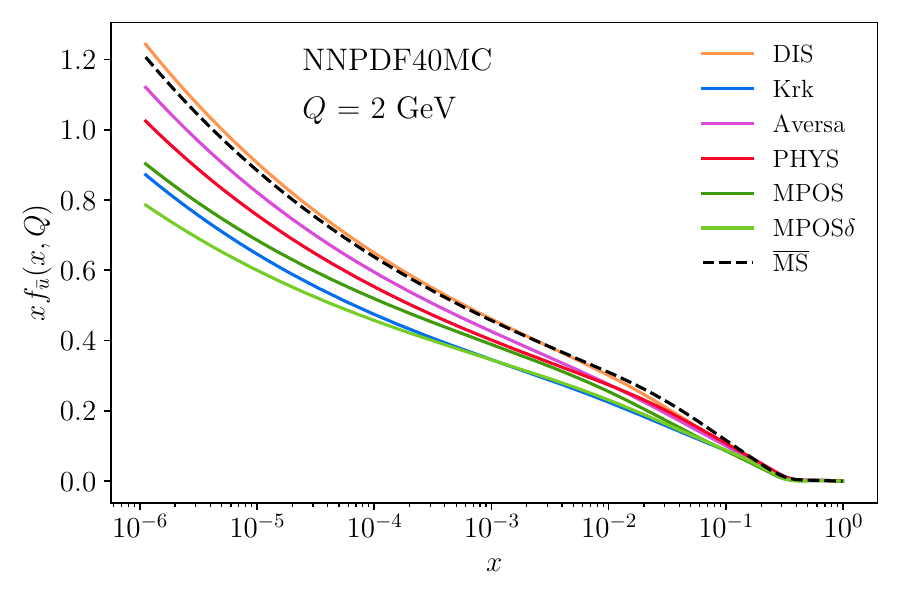} \hfill
\includegraphics[width=0.47\textwidth]{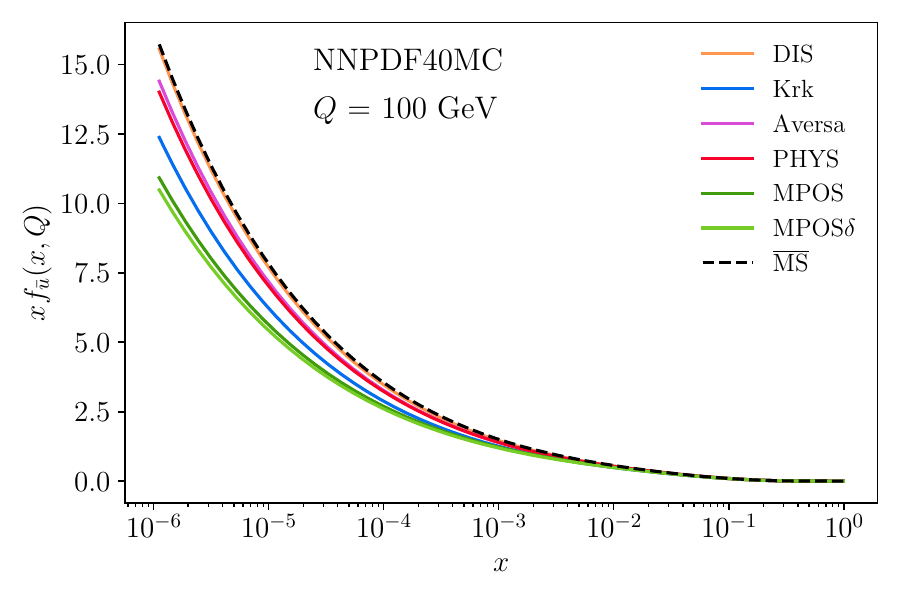}
}
\caption{$\bar{u}$-quark\label{fig:xfubxQ_NNPDF40MC}}
\end{subfigure}
\\
\begin{subfigure}[t]{\textwidth}
\makebox[\textwidth][c]{
\includegraphics[width=0.47\textwidth]{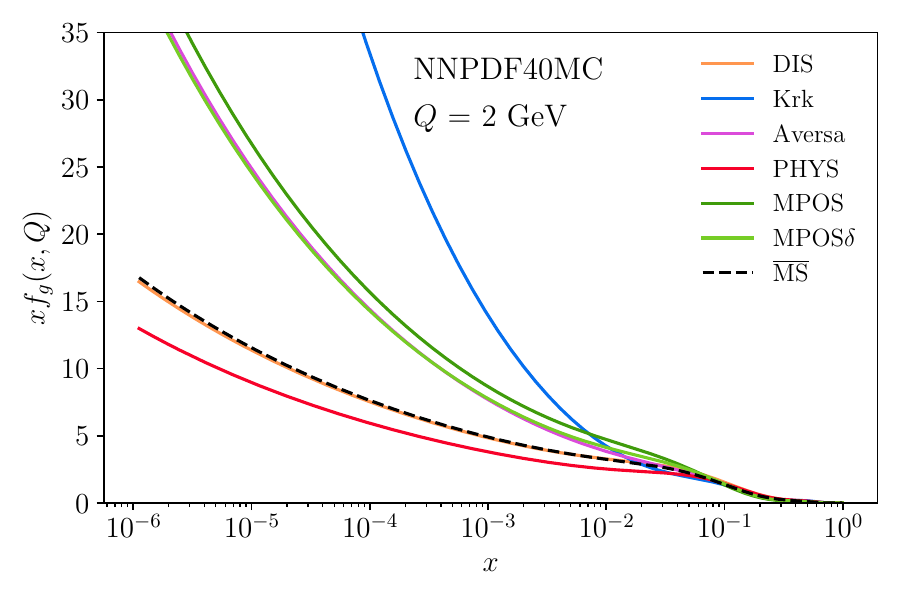} \hfill
\includegraphics[width=0.47\textwidth]{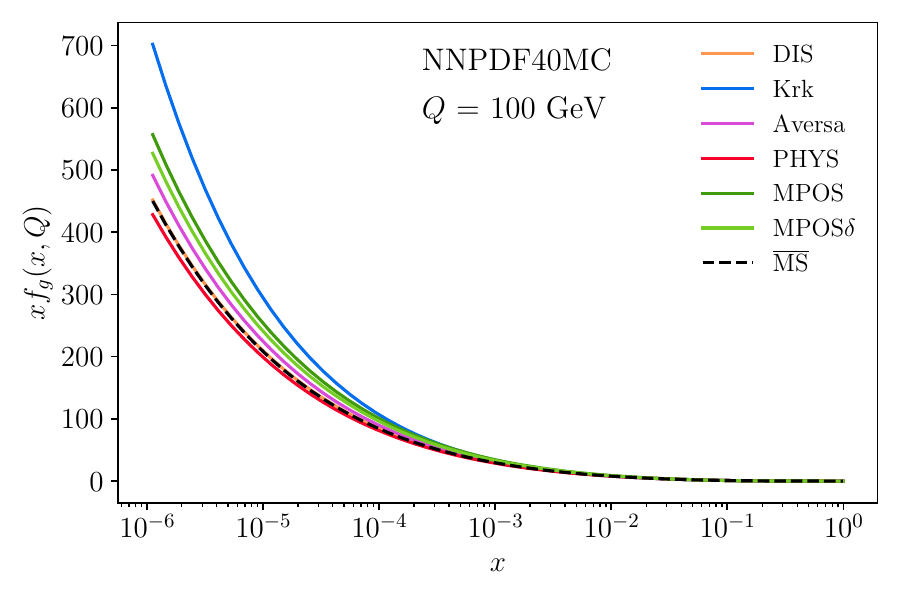}
}
\caption{gluon\label{fig:xfgxQ_NNPDF40MC}}
\end{subfigure}
\\
\begin{subfigure}[t]{\textwidth}
\makebox[\textwidth][c]{
\includegraphics[width=0.47\textwidth]{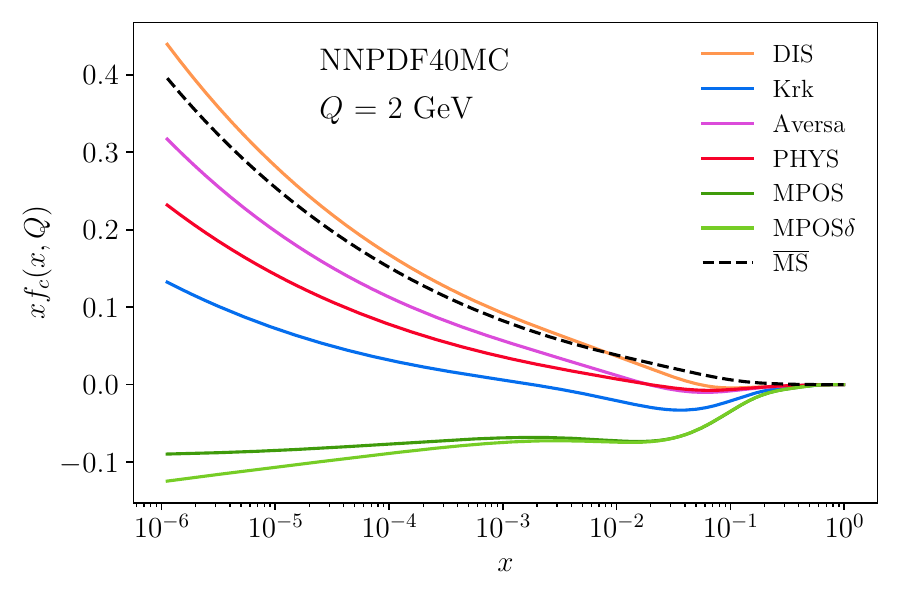} \hfill
\includegraphics[width=0.47\textwidth]{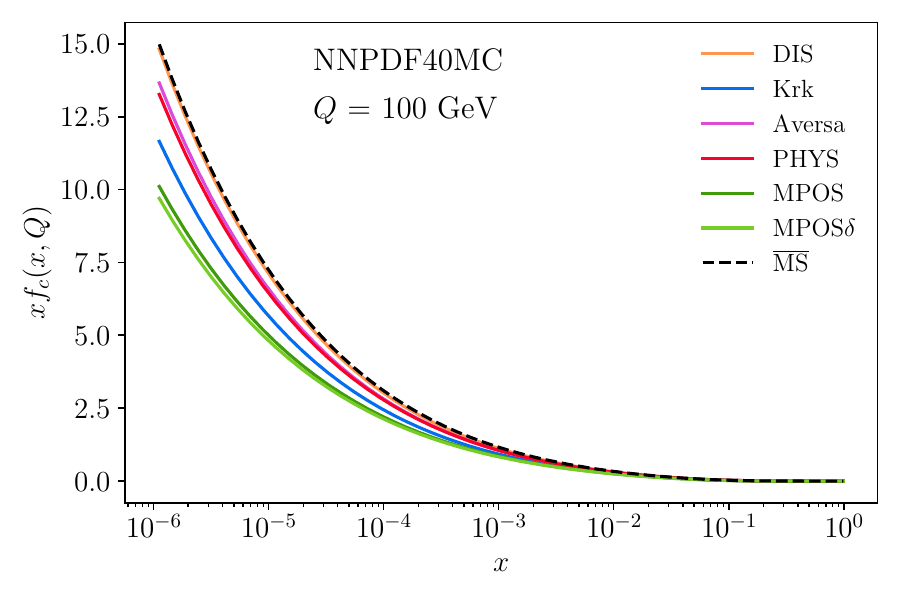}
}
\caption{charm-quark distribution, $c$\label{fig:xfcxQ_NNPDF40MC}}
\end{subfigure}
\caption{
Comparison of transformed \nnpdfmc PDFs in different schemes for $u$-valence, $\bar{u}$, gluon, and charm
at $Q=2$ GeV (left) and 100 GeV (right).
Companion to \cref{fig:PDF_compare}.}
\label{fig:PDF_compare_NNPDF40MC}
\end{figure*}

\begin{figure*}[tp]
\centering
\begin{subfigure}[t]{\textwidth}
\makebox[\textwidth][c]{
    \includegraphics[width=0.47\textwidth]{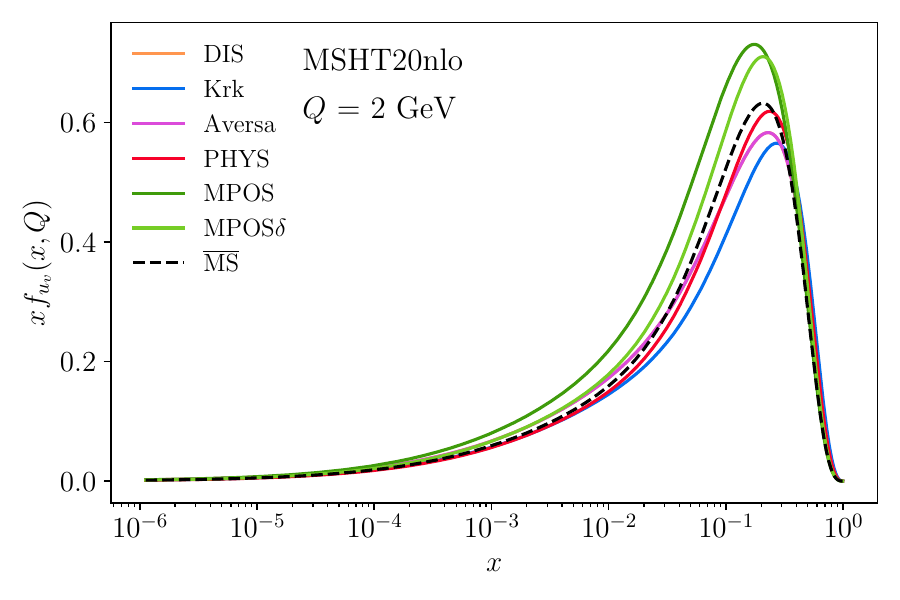} \hfill
    \includegraphics[width=0.47\textwidth]{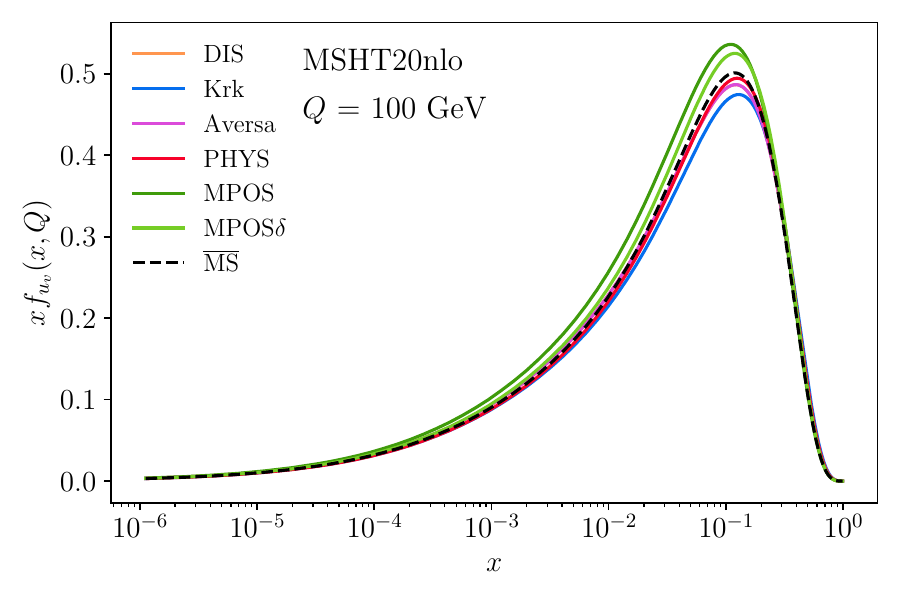}
}
\caption{$u$-valence quark\label{fig:xfuxQ_MSHT20nlo}}
\end{subfigure}
\\
\begin{subfigure}[t]{\textwidth}
\makebox[\textwidth][c]{
\includegraphics[width=0.47\textwidth]{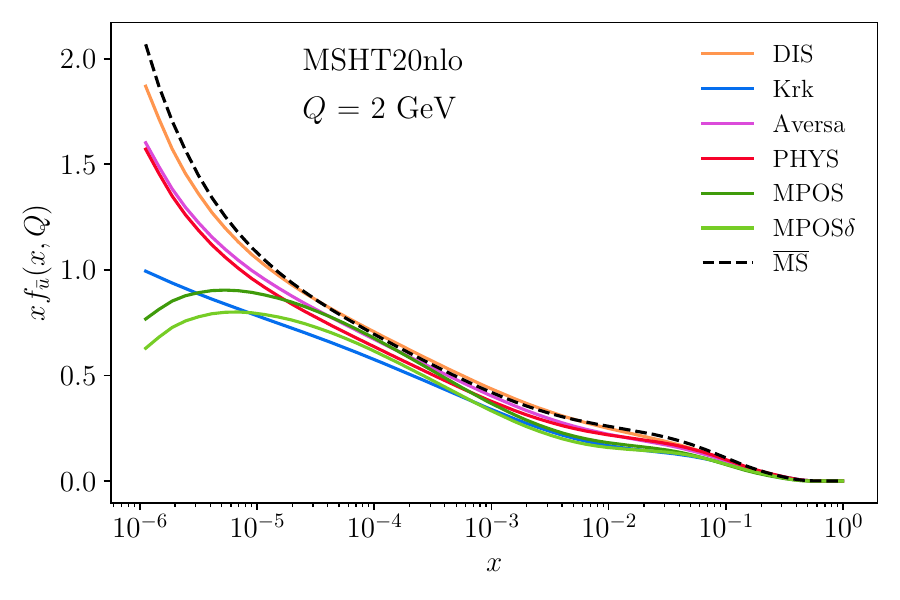} \hfill
\includegraphics[width=0.47\textwidth]{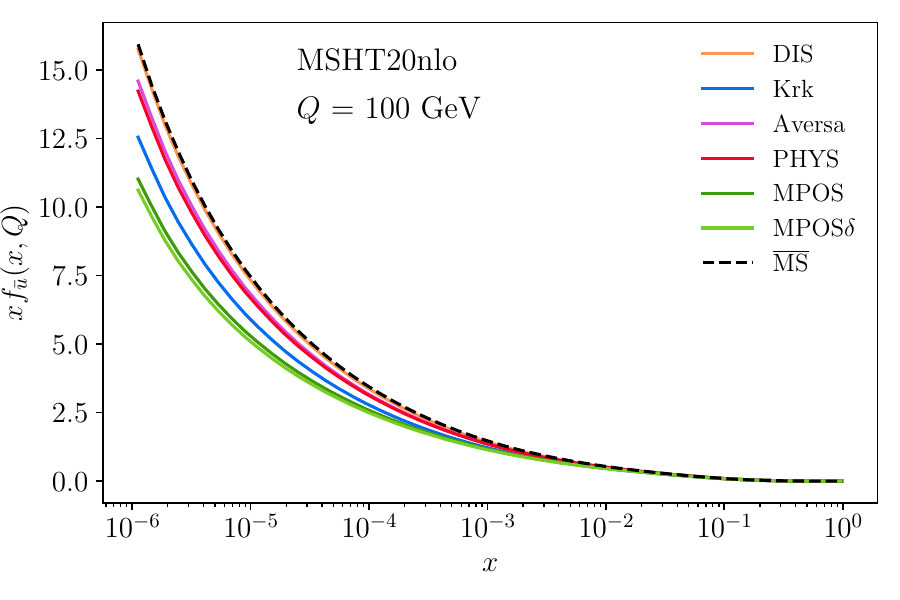}
}
\caption{$\bar{u}$-quark\label{fig:xfubxQ_MSHT20nlo}}
\end{subfigure}
\\
\begin{subfigure}[t]{\textwidth}
\makebox[\textwidth][c]{
\includegraphics[width=0.47\textwidth]{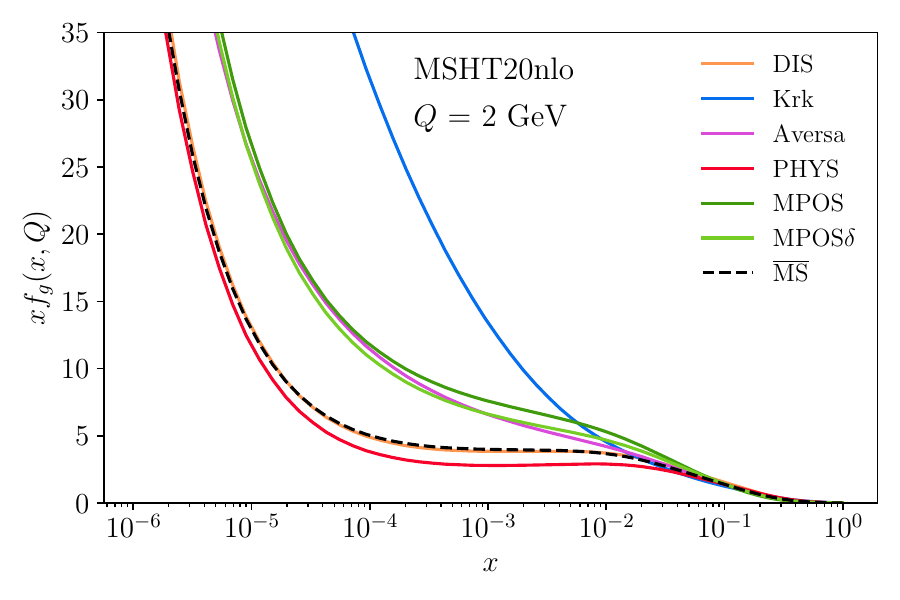} \hfill
\includegraphics[width=0.47\textwidth]{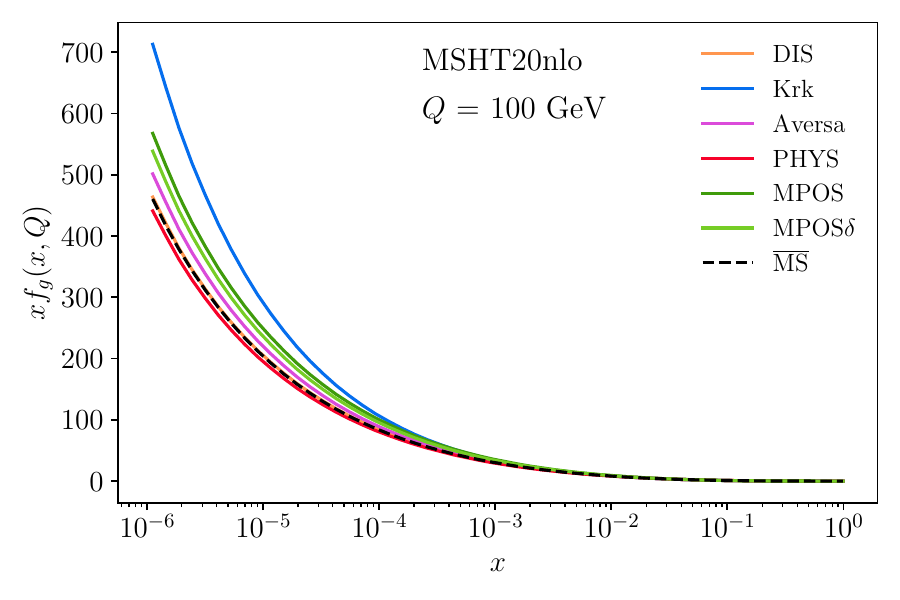}
}
\caption{gluon\label{fig:xfgxQ_MSHT20nlo}}
\end{subfigure}
\\
\begin{subfigure}[t]{\textwidth}
\makebox[\textwidth][c]{
\includegraphics[width=0.47\textwidth]{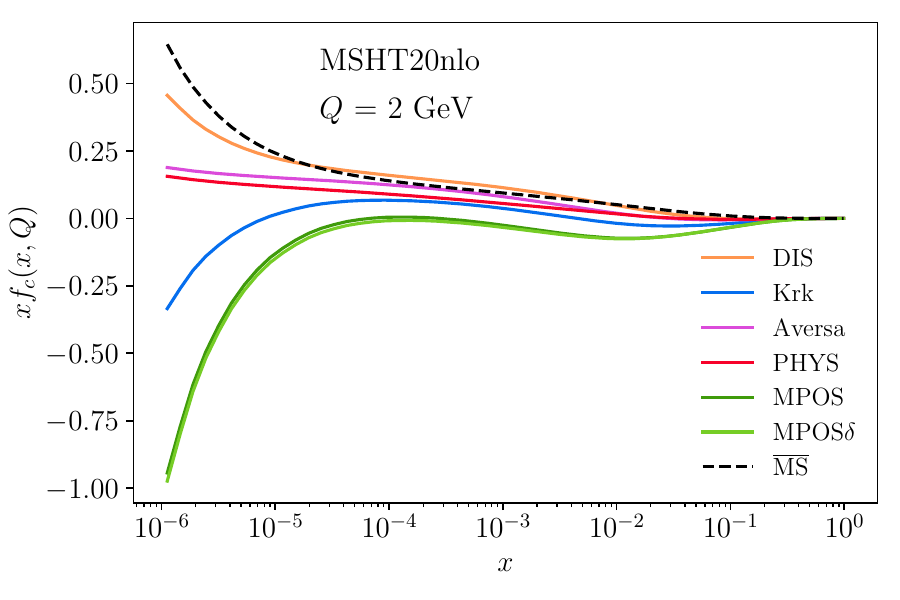} \hfill
\includegraphics[width=0.47\textwidth]{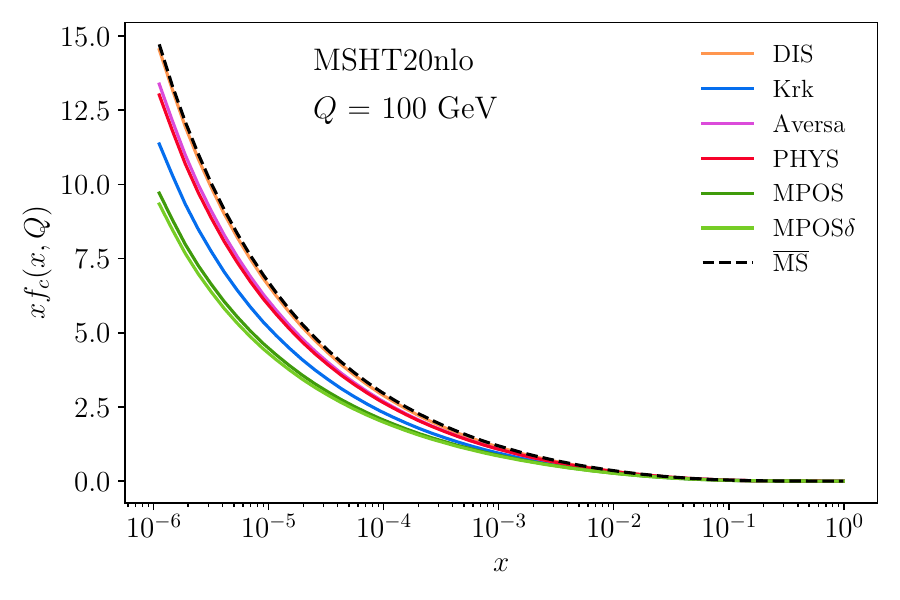}
}
\caption{charm-quark distribution, $c$\label{fig:xfcxQ_MSHT20nlo}}
\end{subfigure}
\caption{
Comparison of transformed \mshtnlo PDFs in different schemes for $u$-valence, $\bar{u}$, gluon, and charm,
at $Q=2$ GeV (left) and 100 GeV (right).
Companion to \cref{fig:PDF_compare}.}
\label{fig:PDF_compare_MSHT20nlo}
\end{figure*}

In this section we present plots showing additional transformed PDFs to complement the selection of plots included in the main text.
In \cref{fig:PDF_compare_CT18_2} we show the remaining partonic flavours omitted from \cref{fig:PDF_compare}, using the \ctnlo \msbar PDF set.
In \cref{fig:PDF_compare_NNPDF40MC,fig:PDF_compare_MSHT20nlo} we show counterparts to the transformed PDFs presented in the
main text using \nnpdfmc\cite{Cruz-Martinez:2024cbz} and \mshtnlo~\cite{Bailey:2020ooq}
instead of \ctnlo.

\section{Supplementary transformation decompositions}
\label{app:extra_decomp_plots}

In this section we present additional decomposition plots
to support the discussion of \cref{subsec:CompPDFsDecomp}.
For ease of reference an overview of all included decomposition plots
is given in \cref{tab:decompPlots}.

\begin{figure*}[p]
\centering
\includegraphics[width=\textwidth]{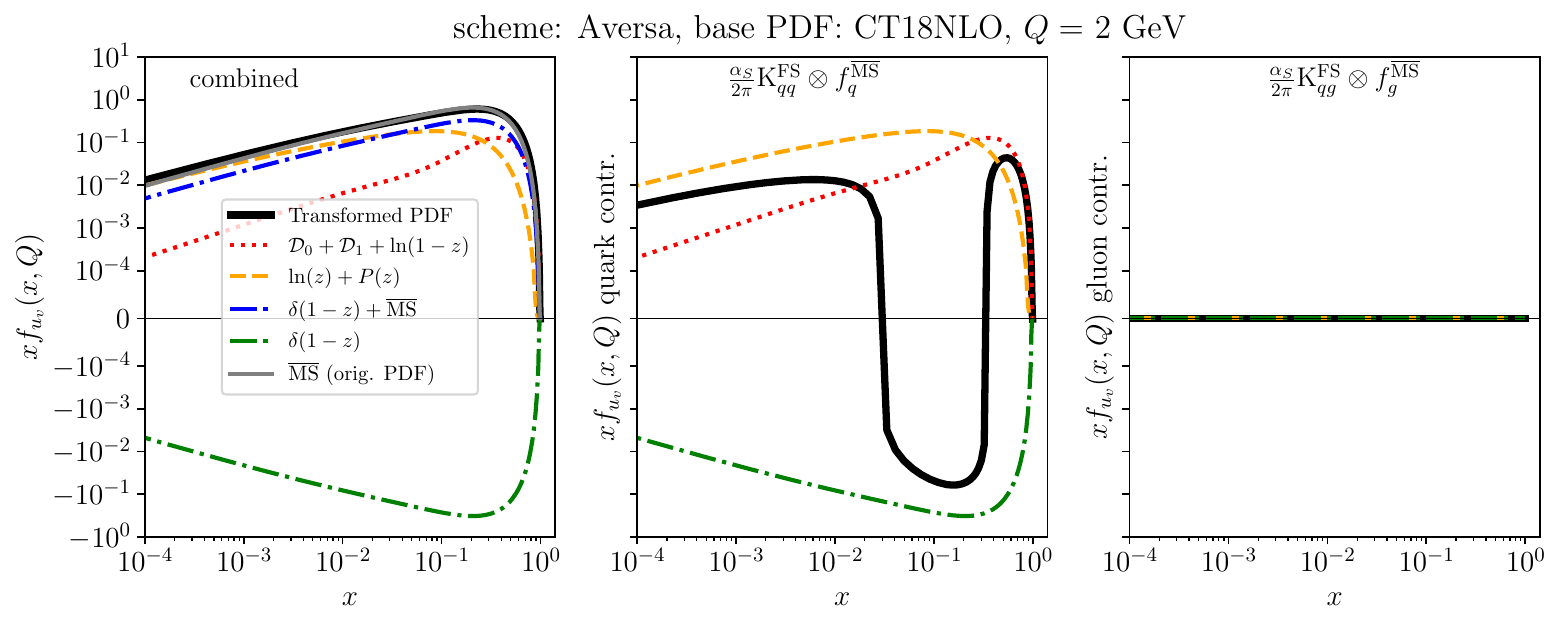}
\\
\includegraphics[width=\textwidth]{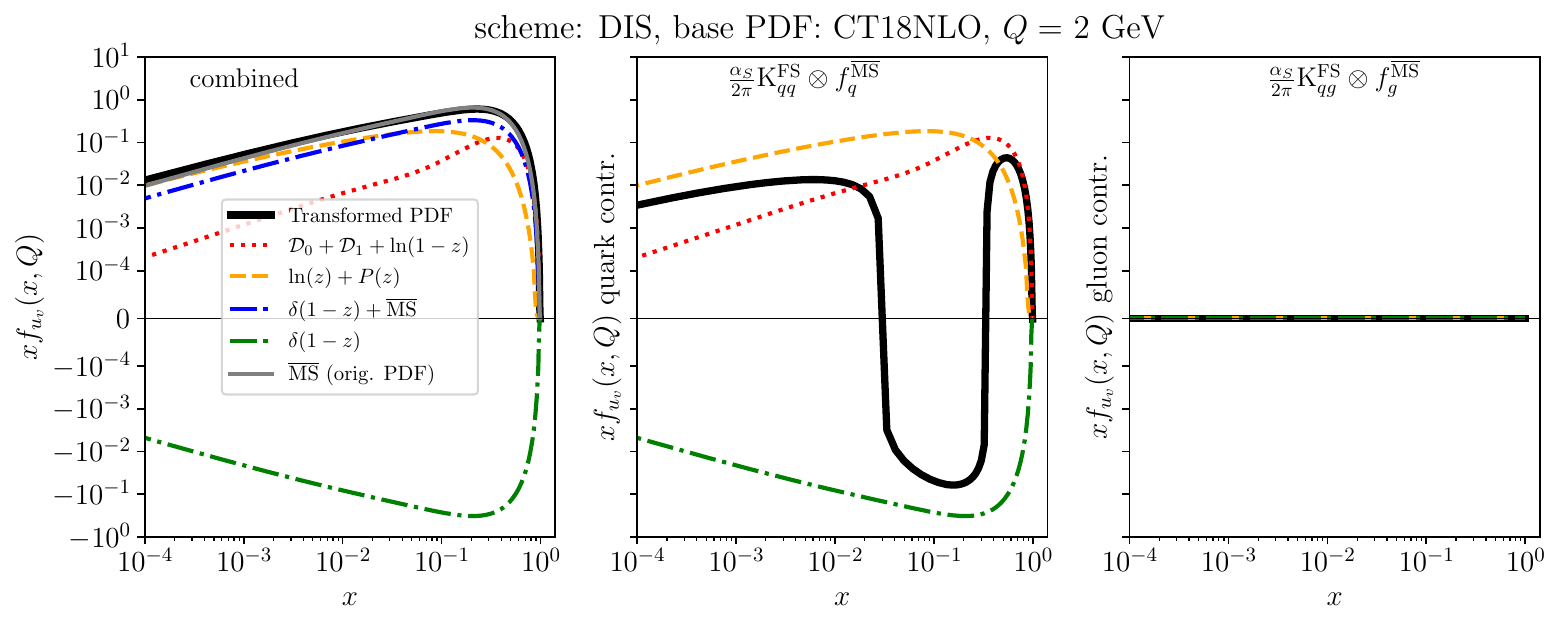}
\\
\includegraphics[width=\textwidth]{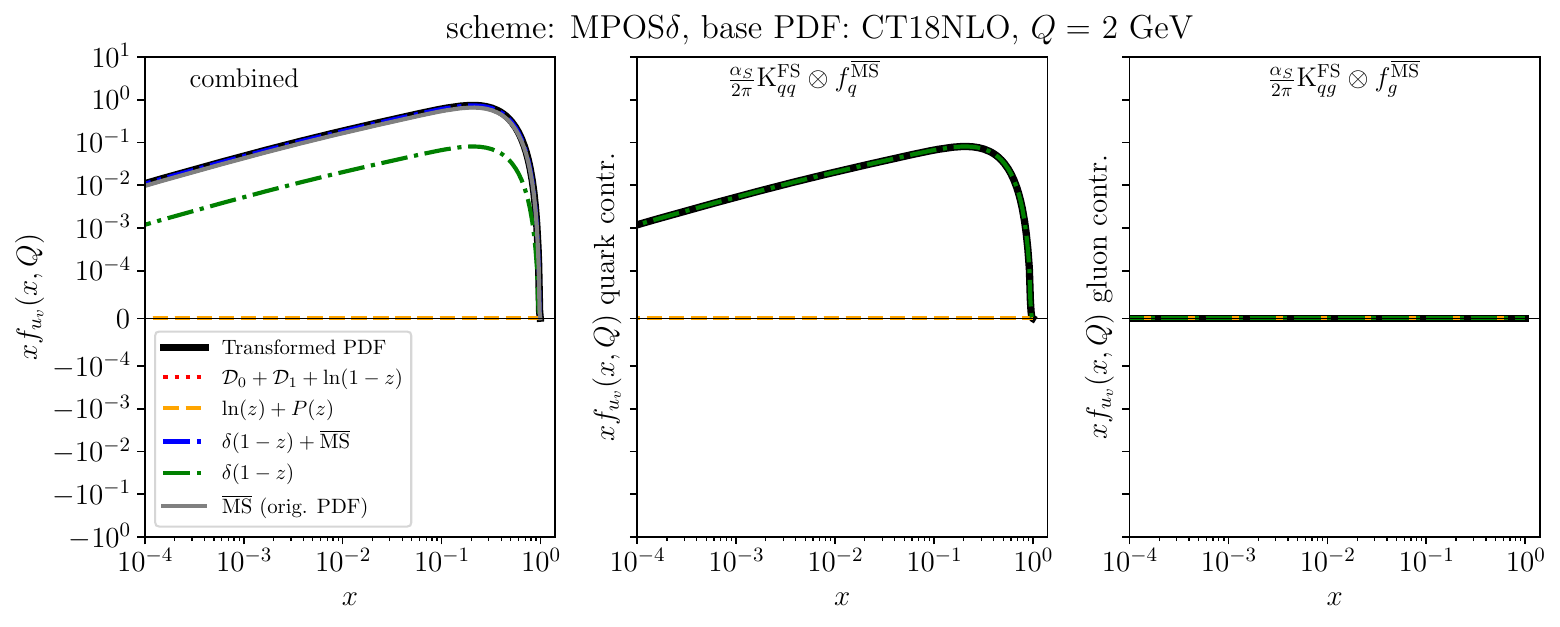}
\caption{Decomposition of transformed $u_v$-quark PDF in the 
\aversa, \dis and \mposd schemes at factorisation scale $Q=2$ GeV,
as described in \cref{subsec:CompPDFsDecomp}.
Companion to \cref{fig:PDF_decomposition_uv}.}
\label{fig:PDF_decomposition_uv_2}
\end{figure*}

\begin{figure*}[p]
\centering
\includegraphics[width=\textwidth]{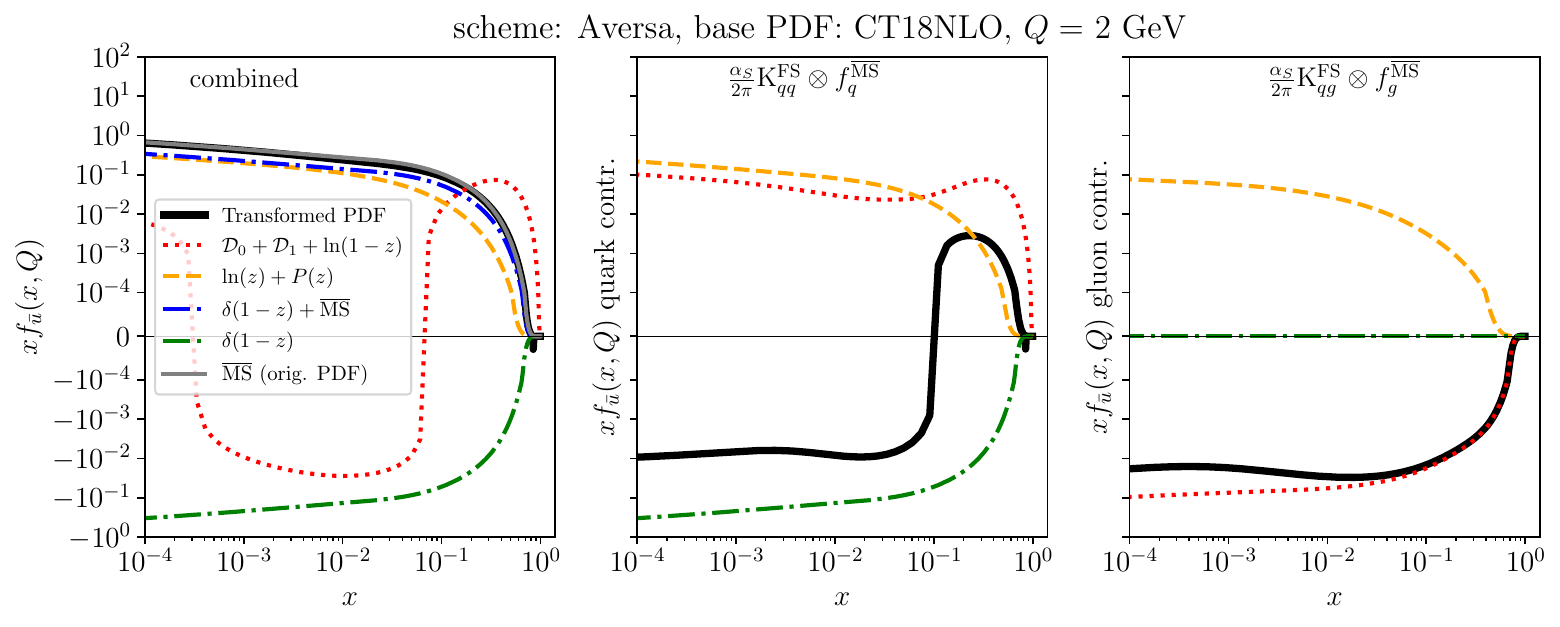}
\\
\includegraphics[width=\textwidth]{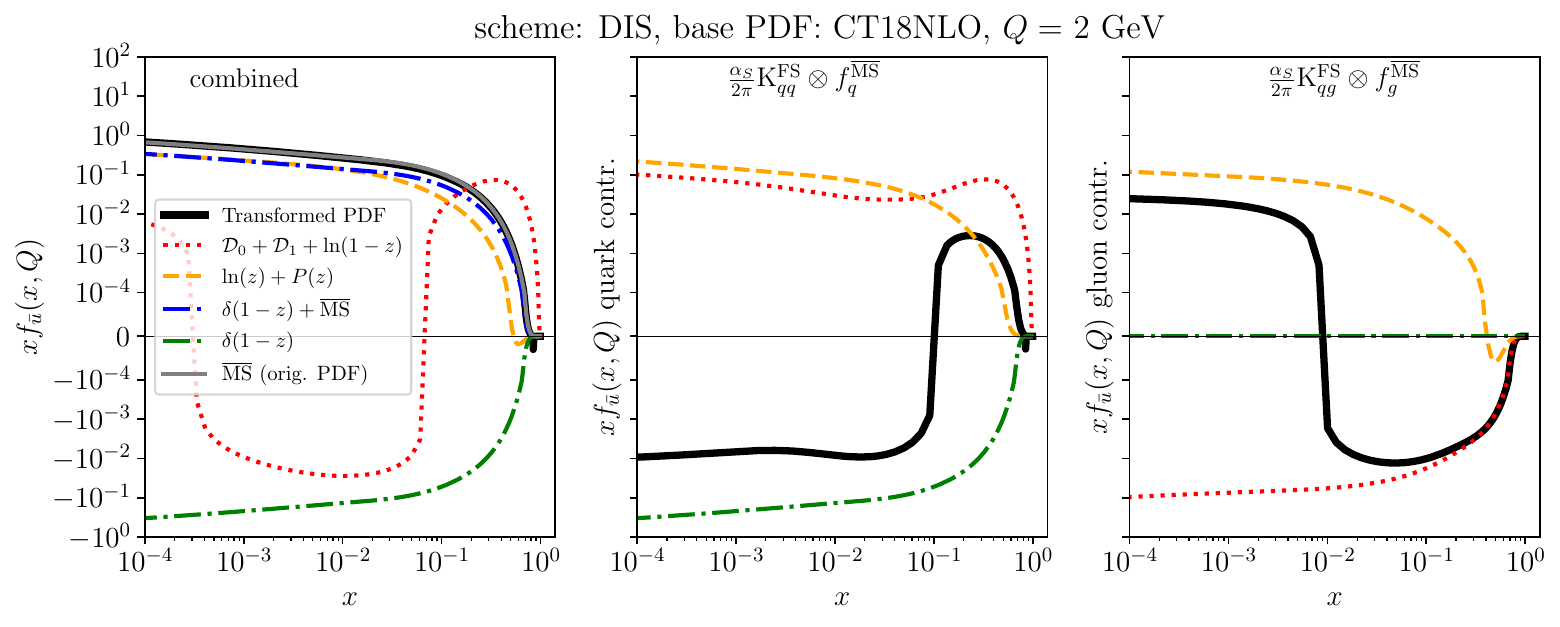}
\\
\includegraphics[width=\textwidth]{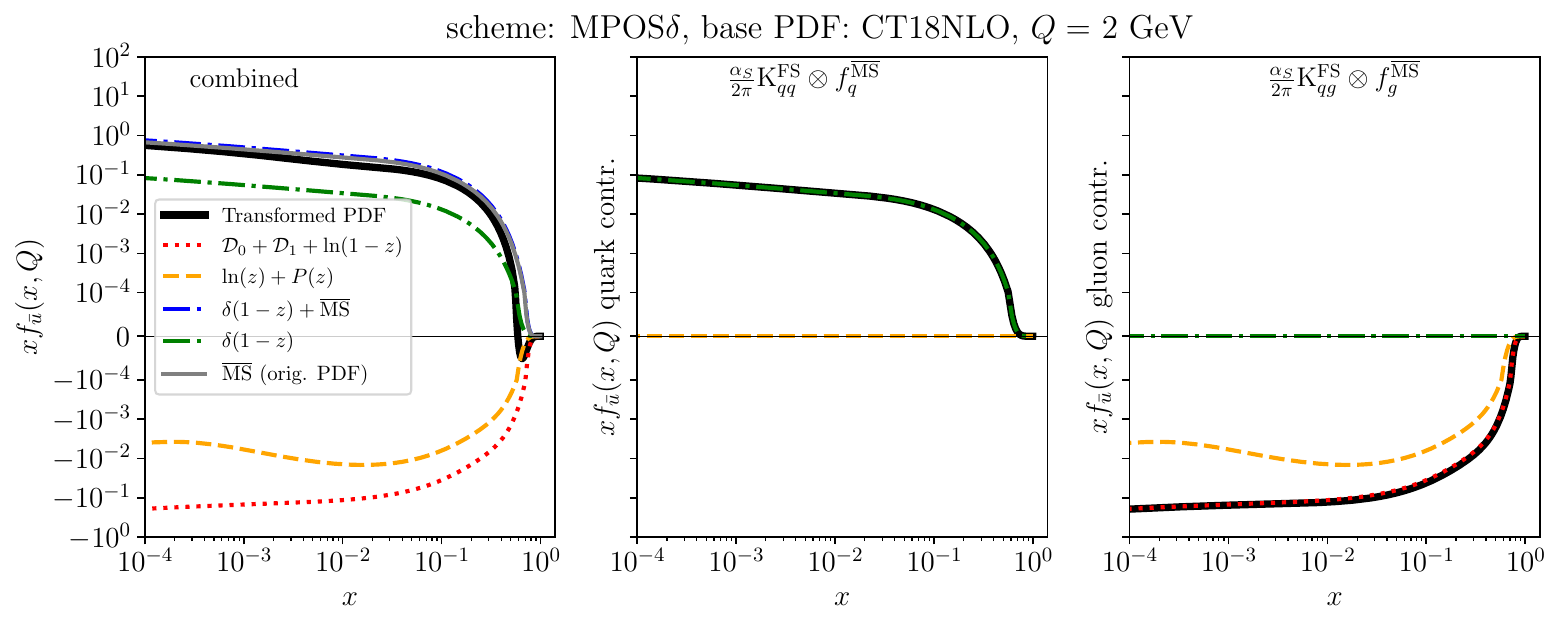}
\caption{Decomposition of transformed $\bar{u}$-quark PDF as a representative of the light sea-quark PDFs,
in the \aversa, \dis and \mposd schemes at factorisation scale $Q=2$ GeV,
as described in \cref{subsec:CompPDFsDecomp}.
Companion to \cref{fig:PDF_decomposition_ub}.}
\label{fig:PDF_decomposition_ub_2}
\end{figure*}

\begin{figure*}[p]
\centering
\includegraphics[width=\textwidth]{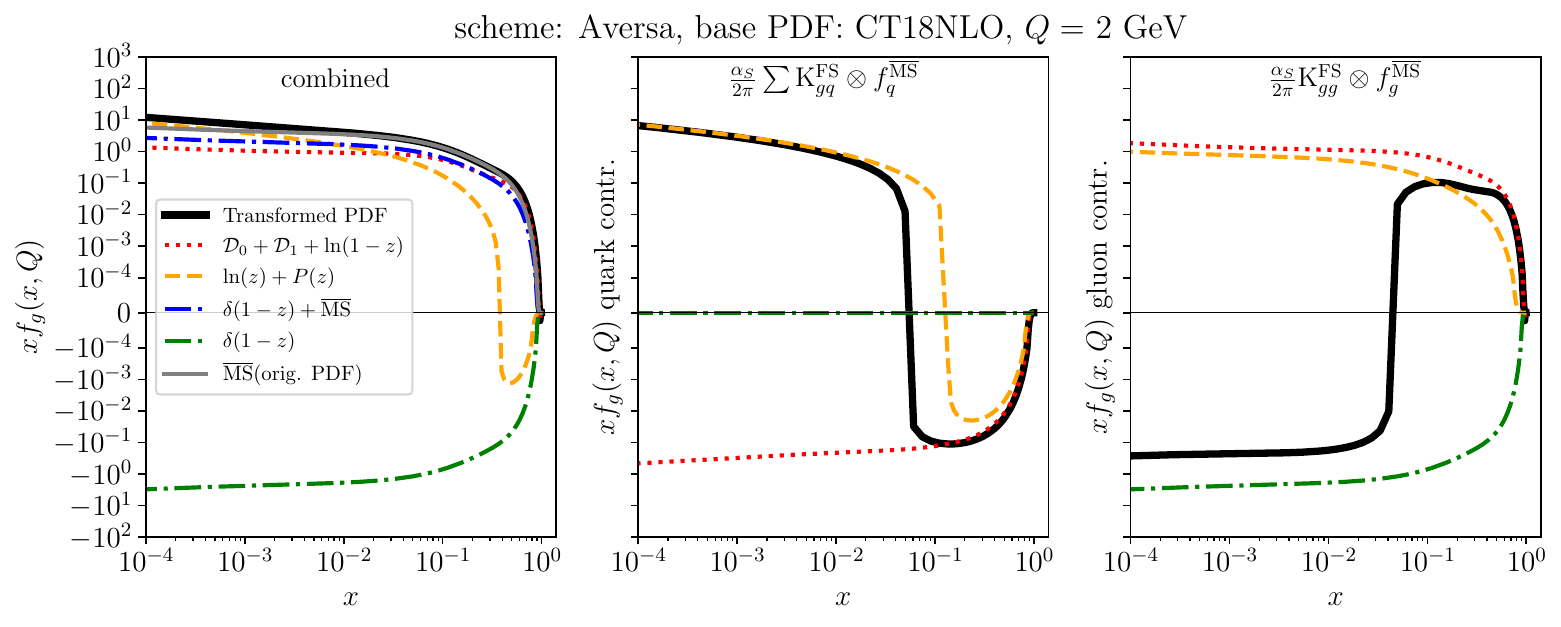}
\\
\includegraphics[width=\textwidth]{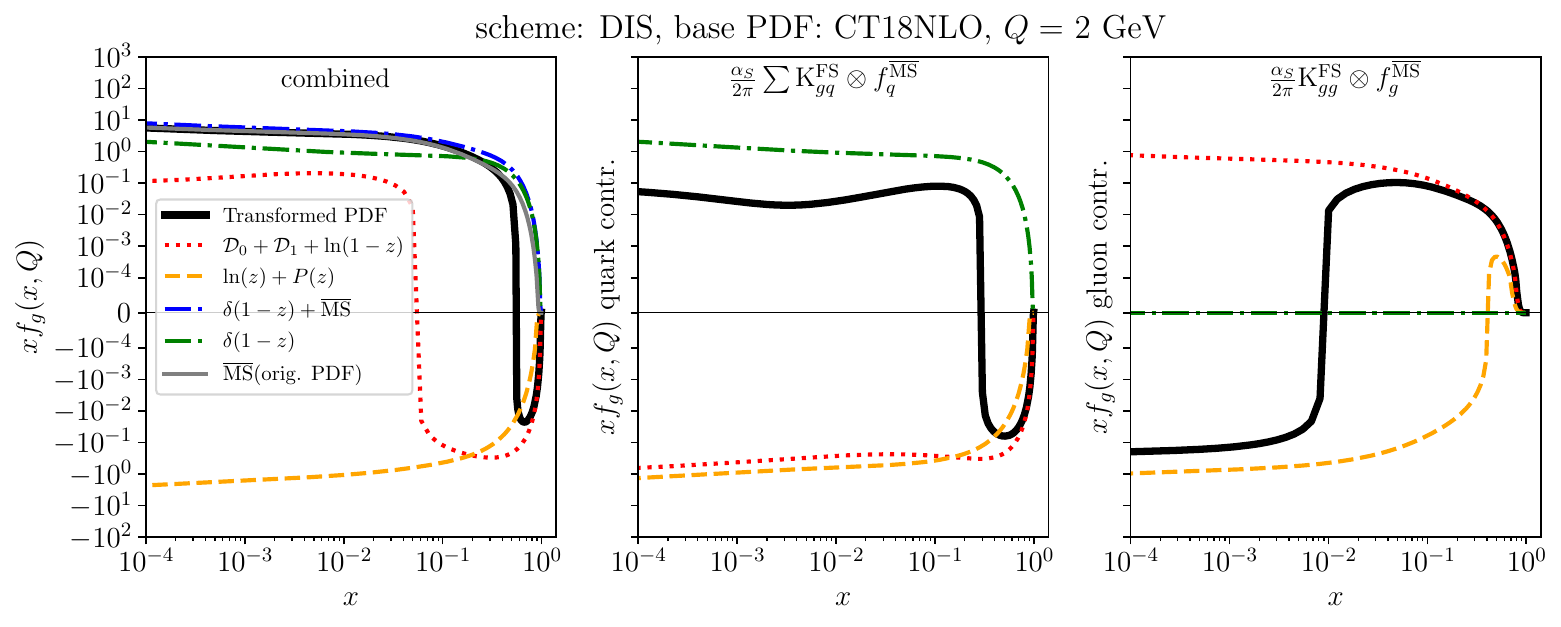}
\\
\includegraphics[width=\textwidth]{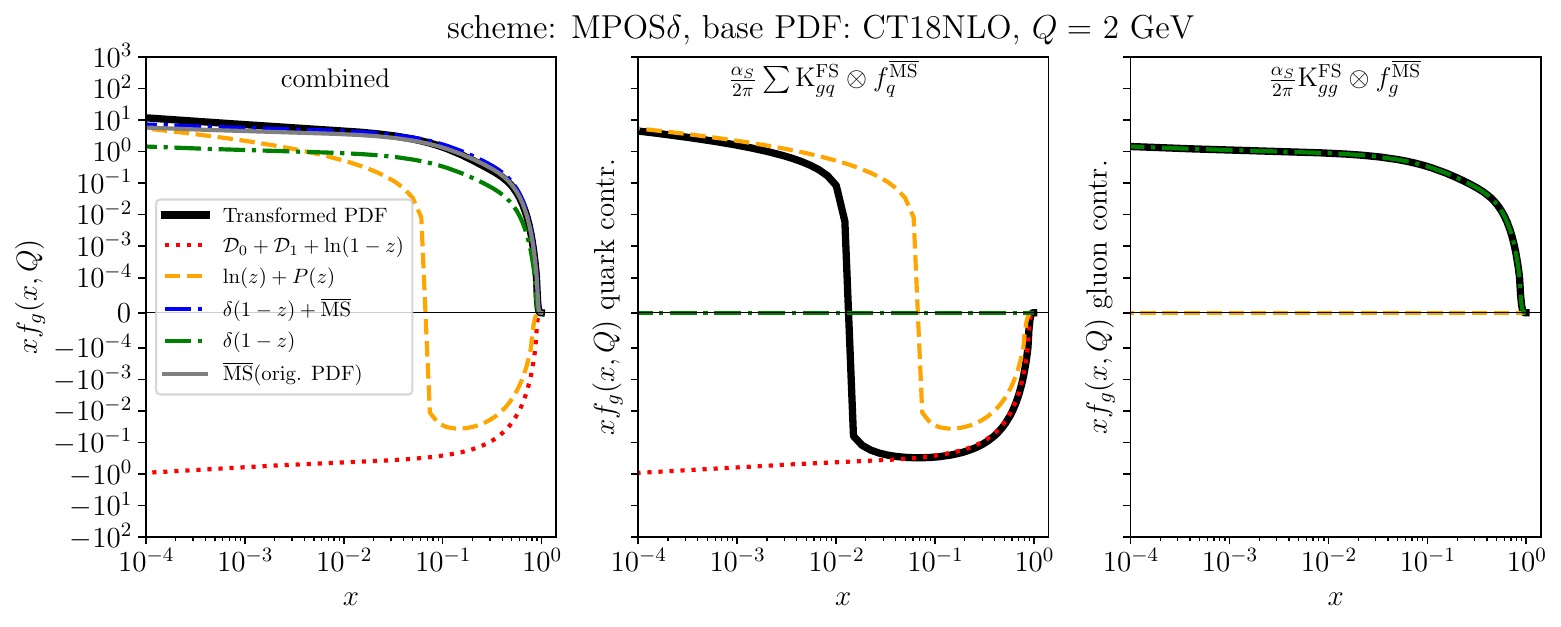}
\caption{Decomposition of transformed gluon PDF
in the \aversa, \dis and \mposd schemes at factorisation scale $Q=2$ GeV,
as described in \cref{subsec:CompPDFsDecomp}.
Companion to \cref{fig:PDF_decomposition_g}.}
\label{fig:PDF_decomposition_g_2}
\end{figure*}

\begin{figure*}[p]
\centering
\includegraphics[width=\textwidth]{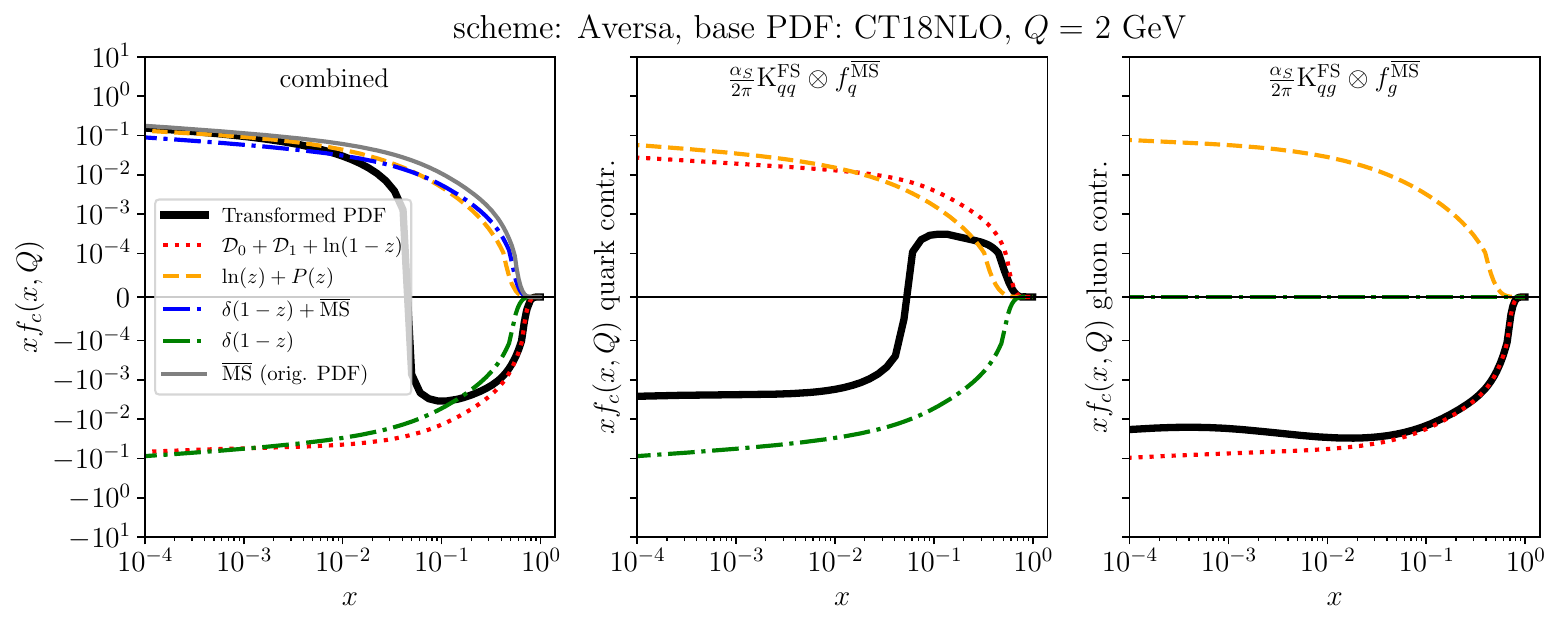}
\\
\includegraphics[width=\textwidth]{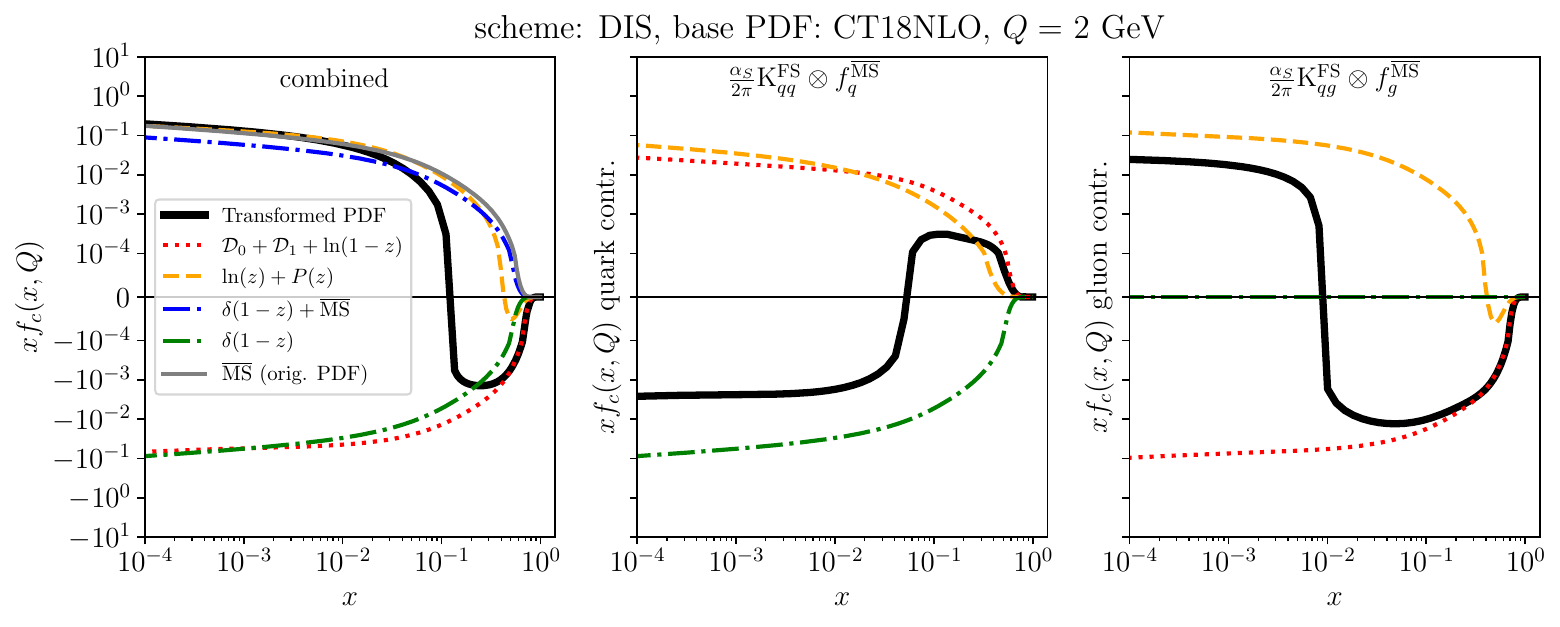}
\\
\includegraphics[width=\textwidth]{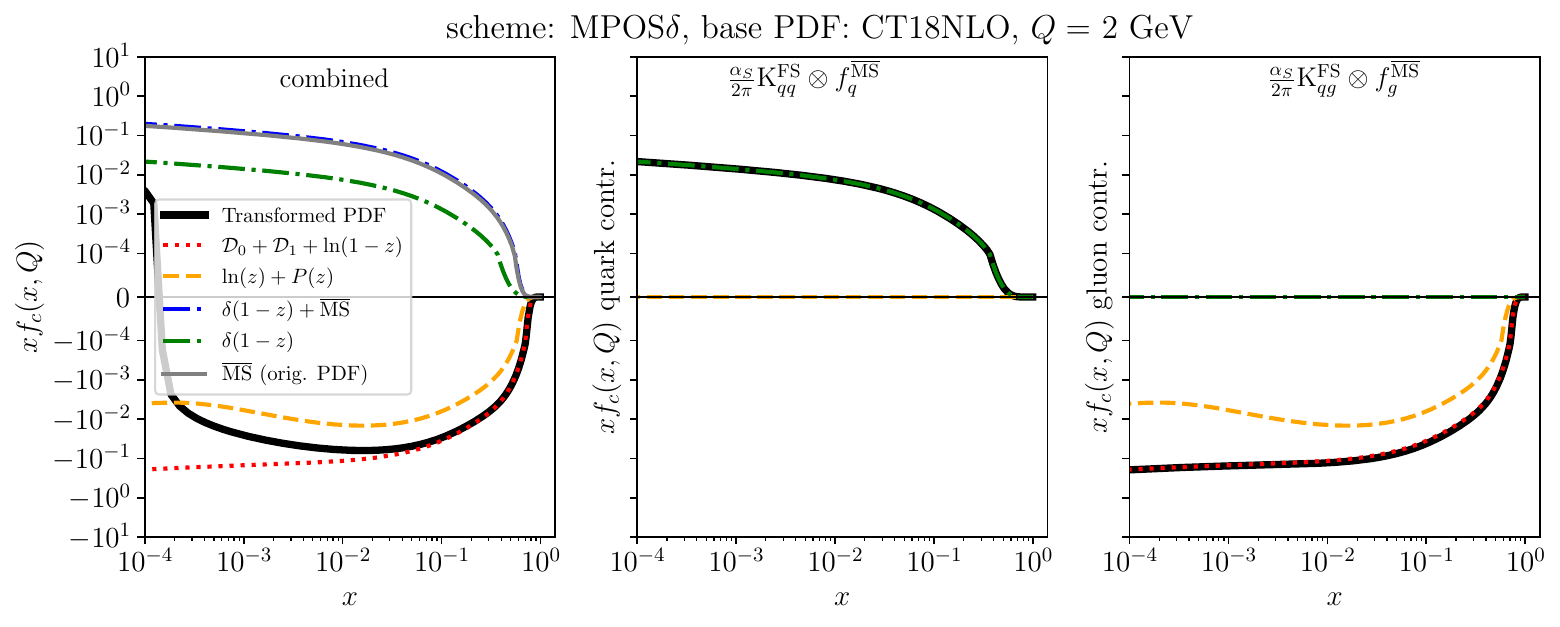}
\caption{Decomposition of transformed $c$-quark PDF as a representative of the heavy flavour PDFs,
in the \aversa, \dis and \mposd schemes at factorisation scale $Q=2$ GeV,
as described in \cref{subsec:CompPDFsDecomp}.
Companion to \cref{fig:PDF_decomposition_c}.}
\label{fig:PDF_decomposition_c_2}
\end{figure*}

In 
\cref{fig:PDF_decomposition_uv_2,fig:PDF_decomposition_ub_2,fig:PDF_decomposition_g_2,fig:PDF_decomposition_c_2}
we show
the remaining three schemes (\aversa, \dis and \mposd)
omitted from the plots of \cref{subsec:CompPDFsDecomp},
again using the \ctnlo PDF set and at $Q=2 \; \GeV$.

\begin{figure*}[p]
\centering
\includegraphics[width=\textwidth]{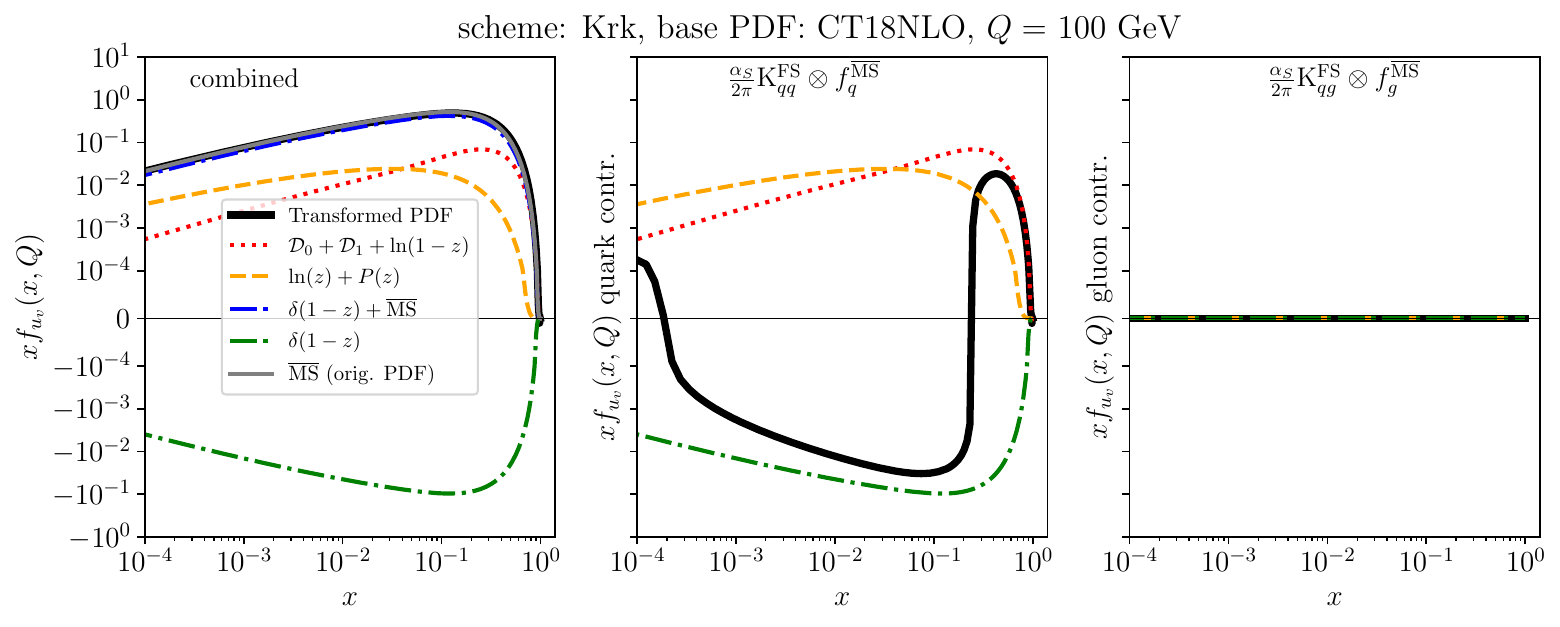}
\\
\includegraphics[width=\textwidth]{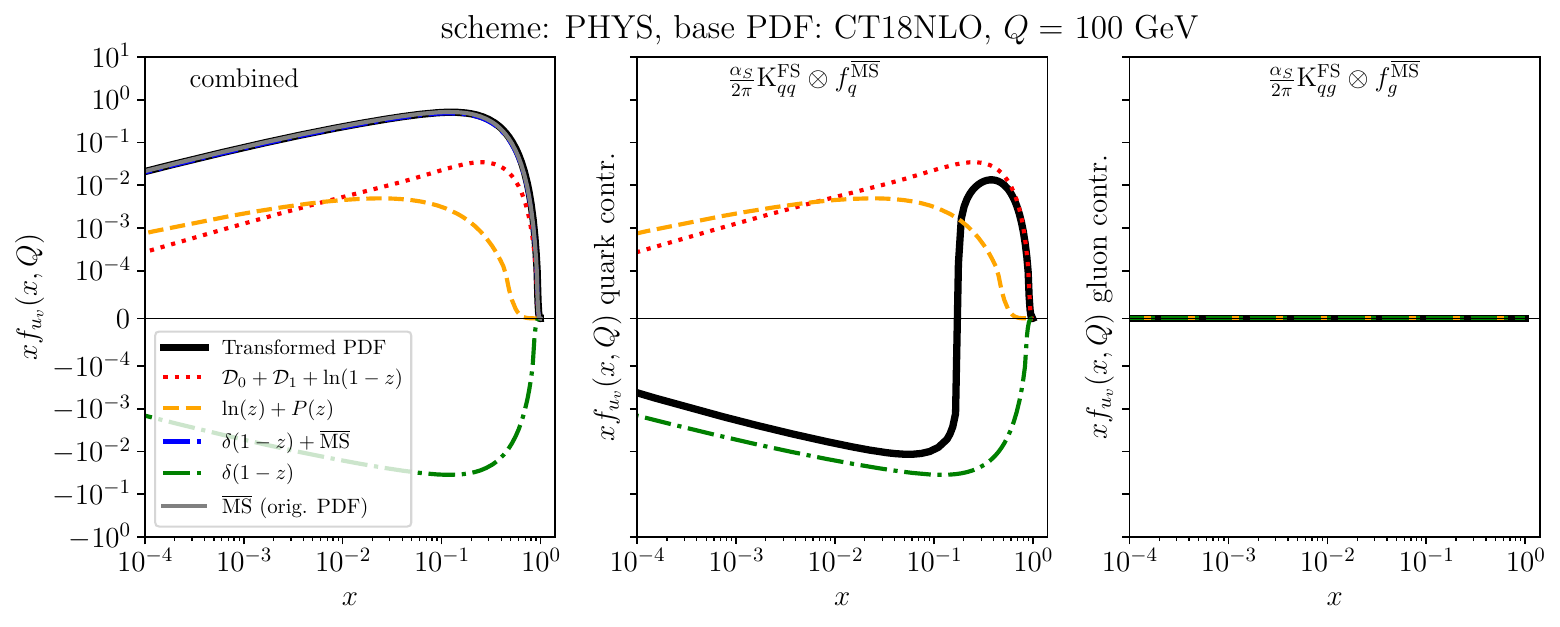}
\\
\includegraphics[width=\textwidth]{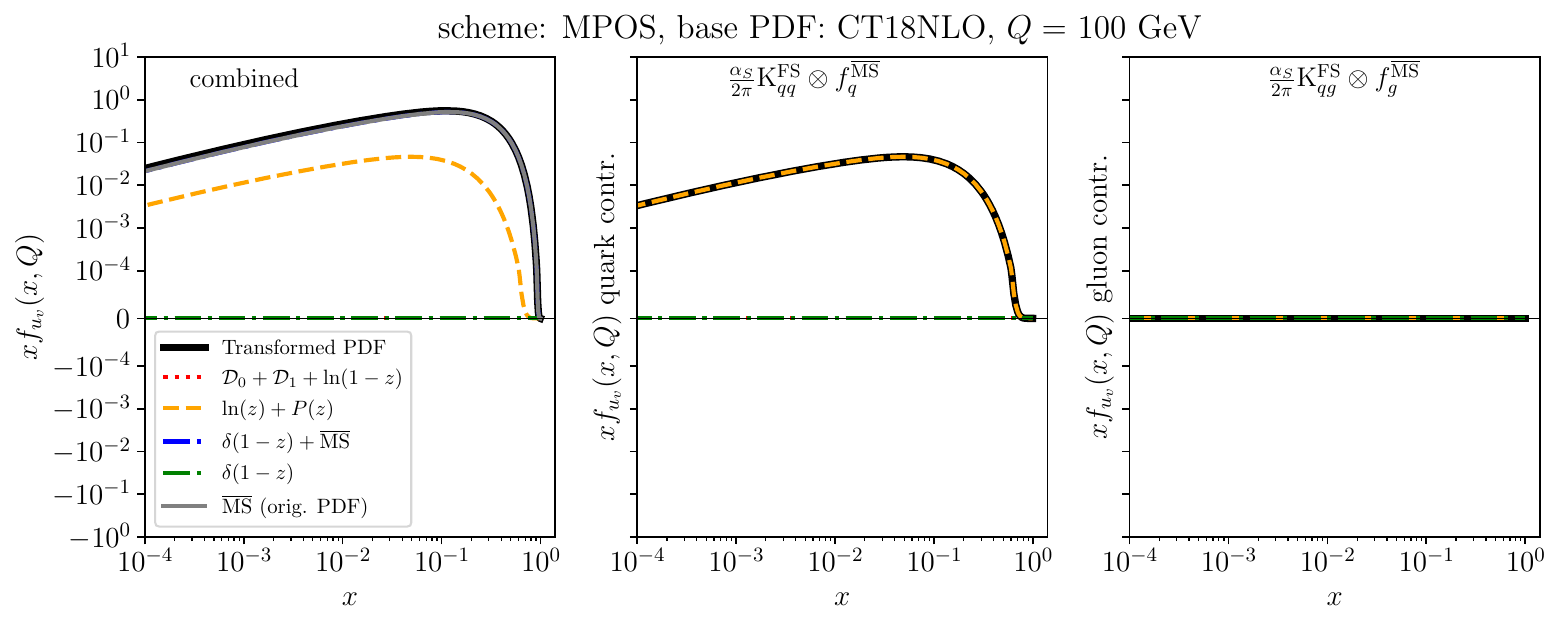}
\caption{Decomposition of transformed $u_v$-quark PDF in the 
\aversa, \dis and \mposd schemes at factorisation scale $Q=100$ GeV,
as described in \cref{subsec:CompPDFsDecomp}.
Companion to \cref{fig:PDF_decomposition_uv}.}
\label{fig:PDF_decomposition_100_uv}
\end{figure*}

\begin{figure*}[p]
\centering
\includegraphics[width=\textwidth]{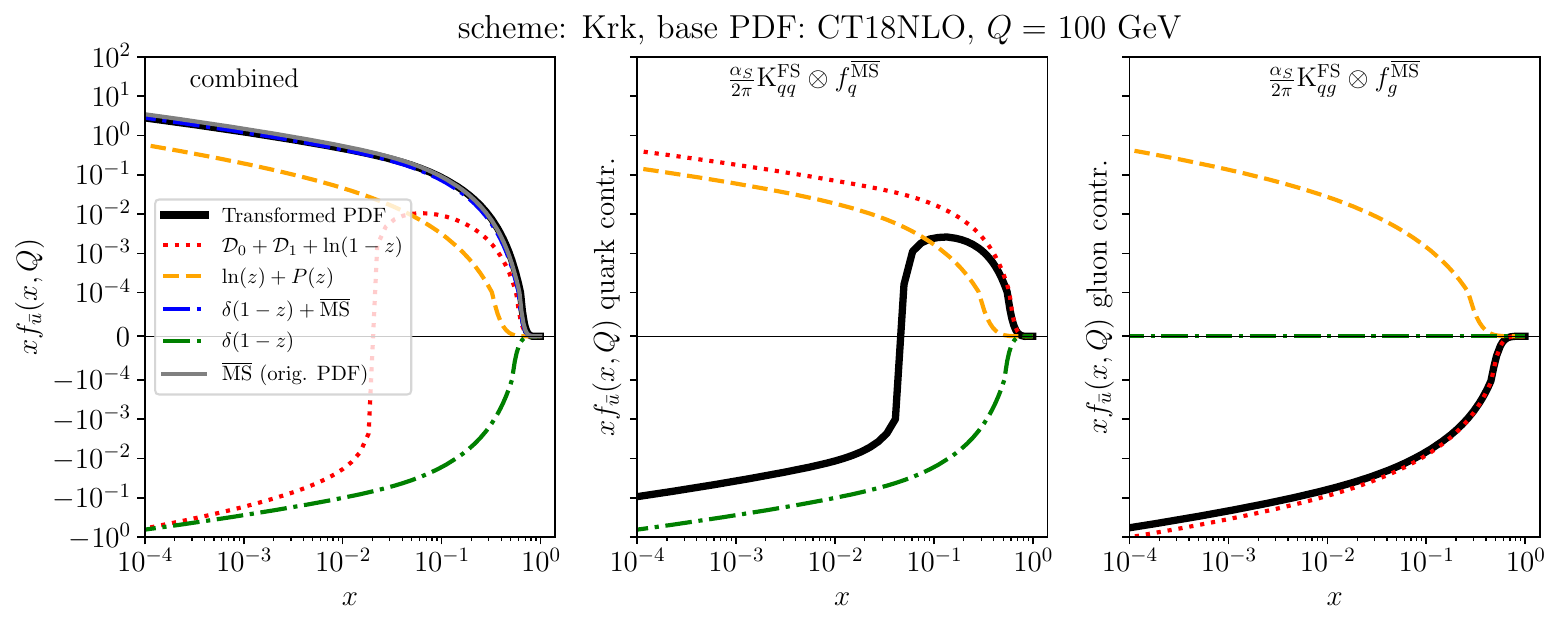}
\\
\includegraphics[width=\textwidth]{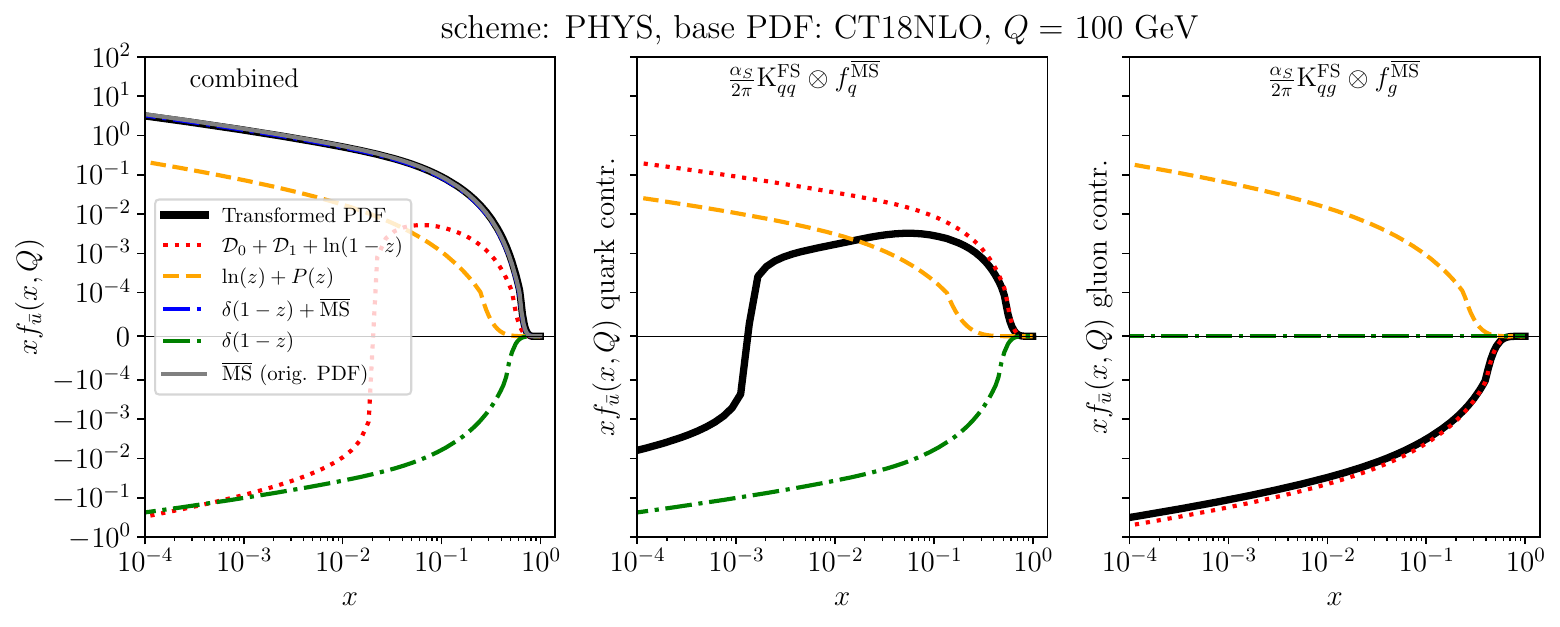}
\\
\includegraphics[width=\textwidth]{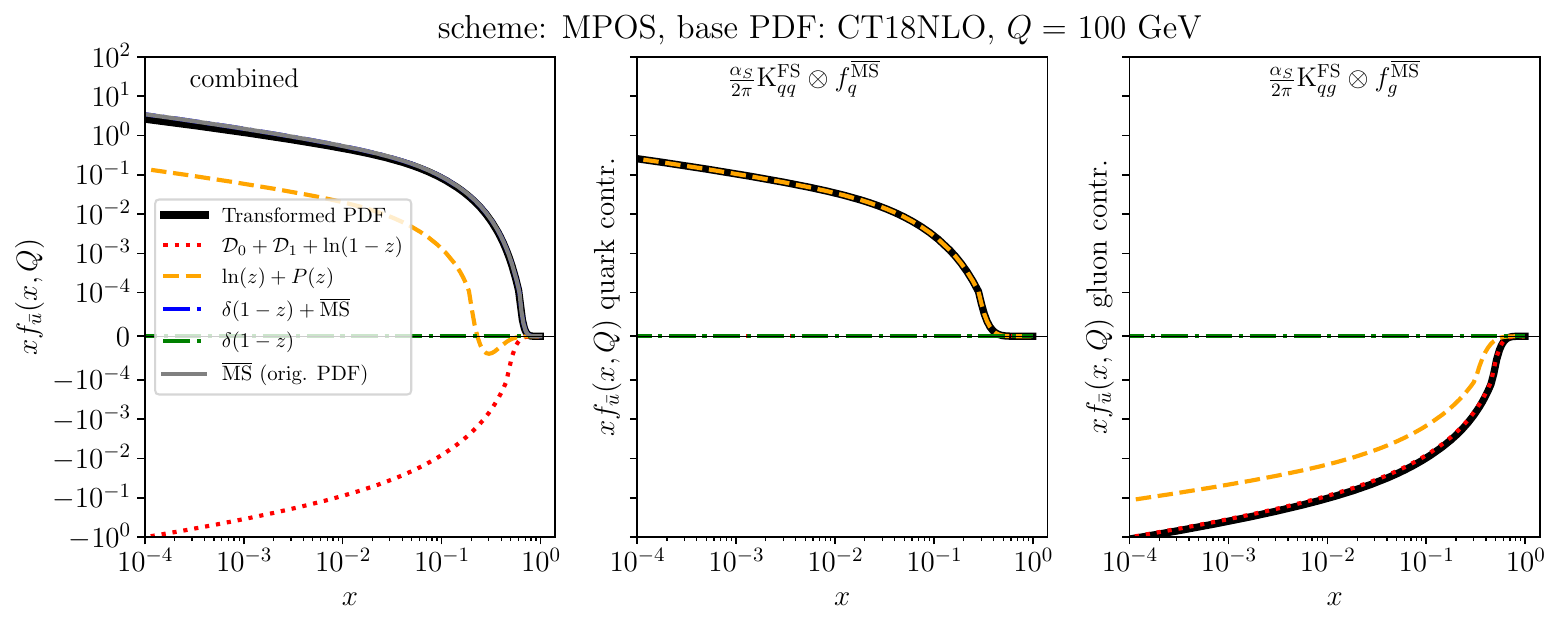}
\caption{Decomposition of transformed $\bar{u}$-quark PDF as a representative of the light sea-quark PDFs,
in the \aversa, \dis and \mposd schemes at factorisation scale $Q=100$ GeV,
as described in \cref{subsec:CompPDFsDecomp}.
Companion to \cref{fig:PDF_decomposition_ub}.}
\label{fig:PDF_decomposition_100_ub}
\end{figure*}

\begin{figure*}[p]
\centering
\includegraphics[width=\textwidth]{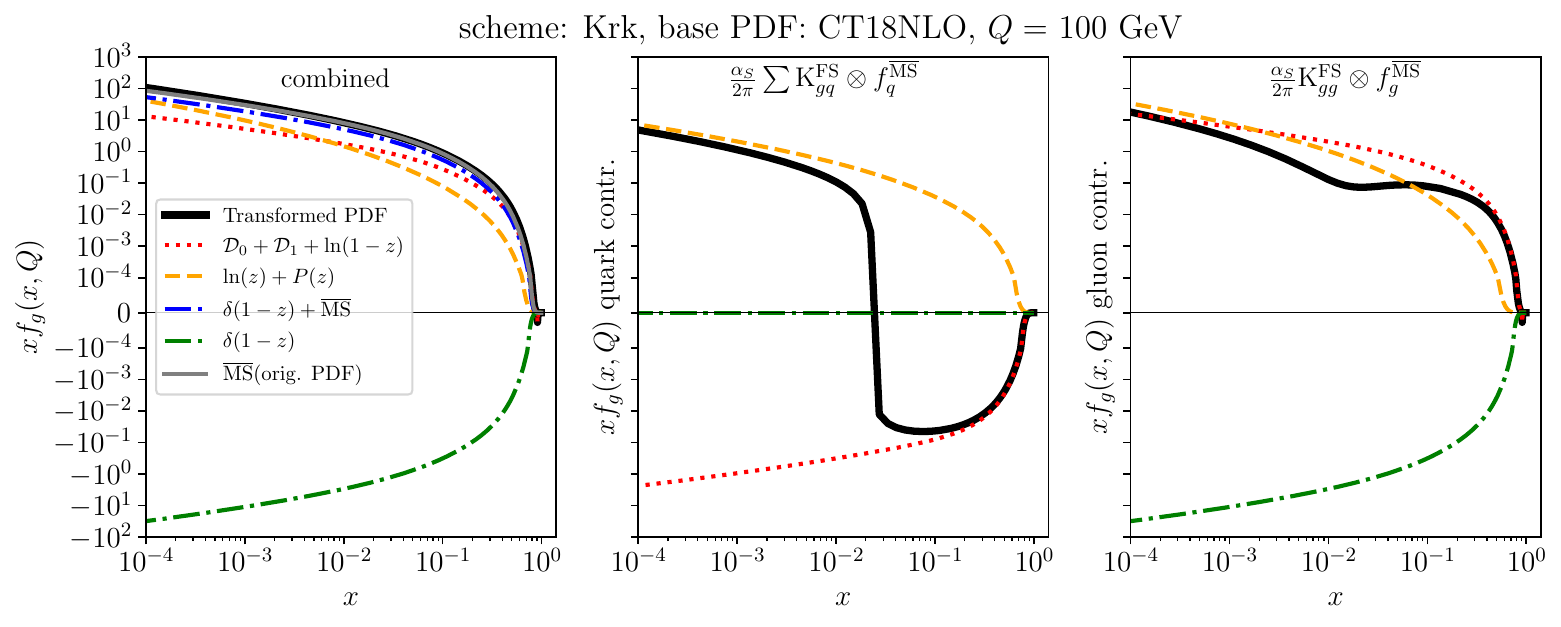}
\\
\includegraphics[width=\textwidth]{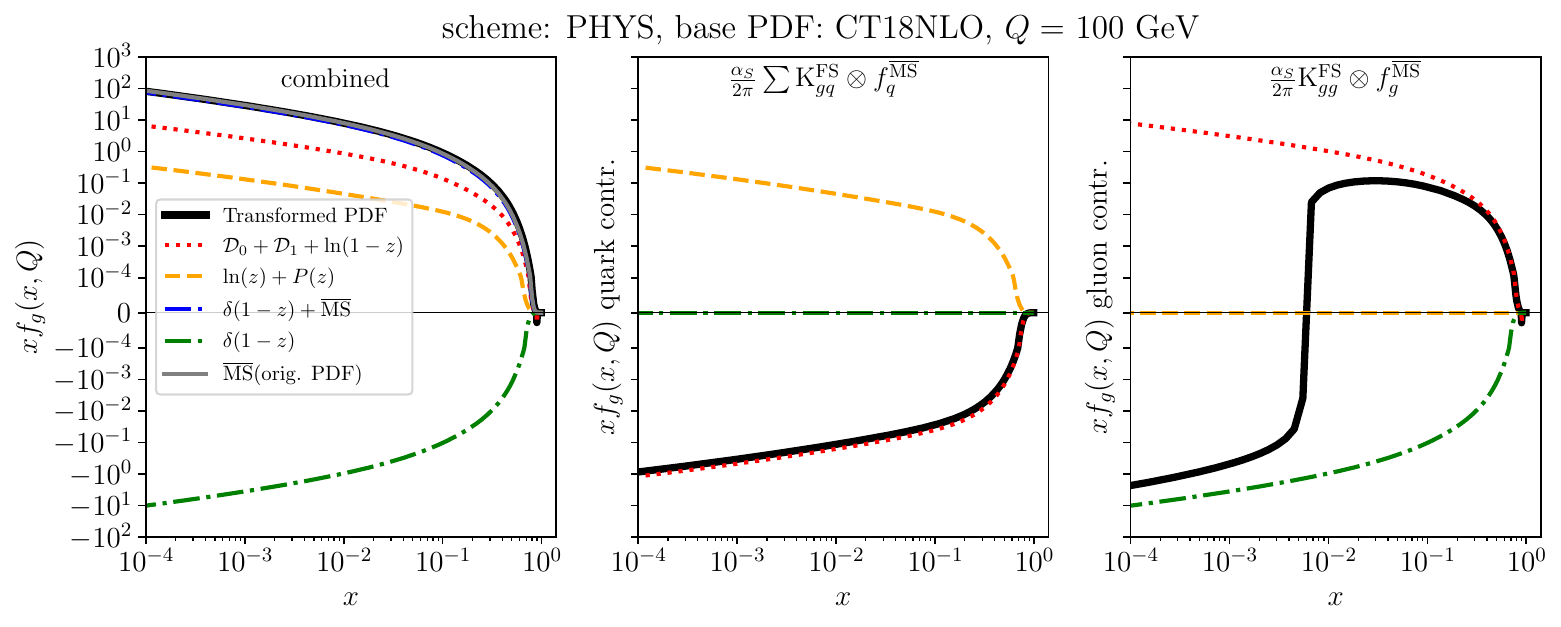}
\\
\includegraphics[width=\textwidth]{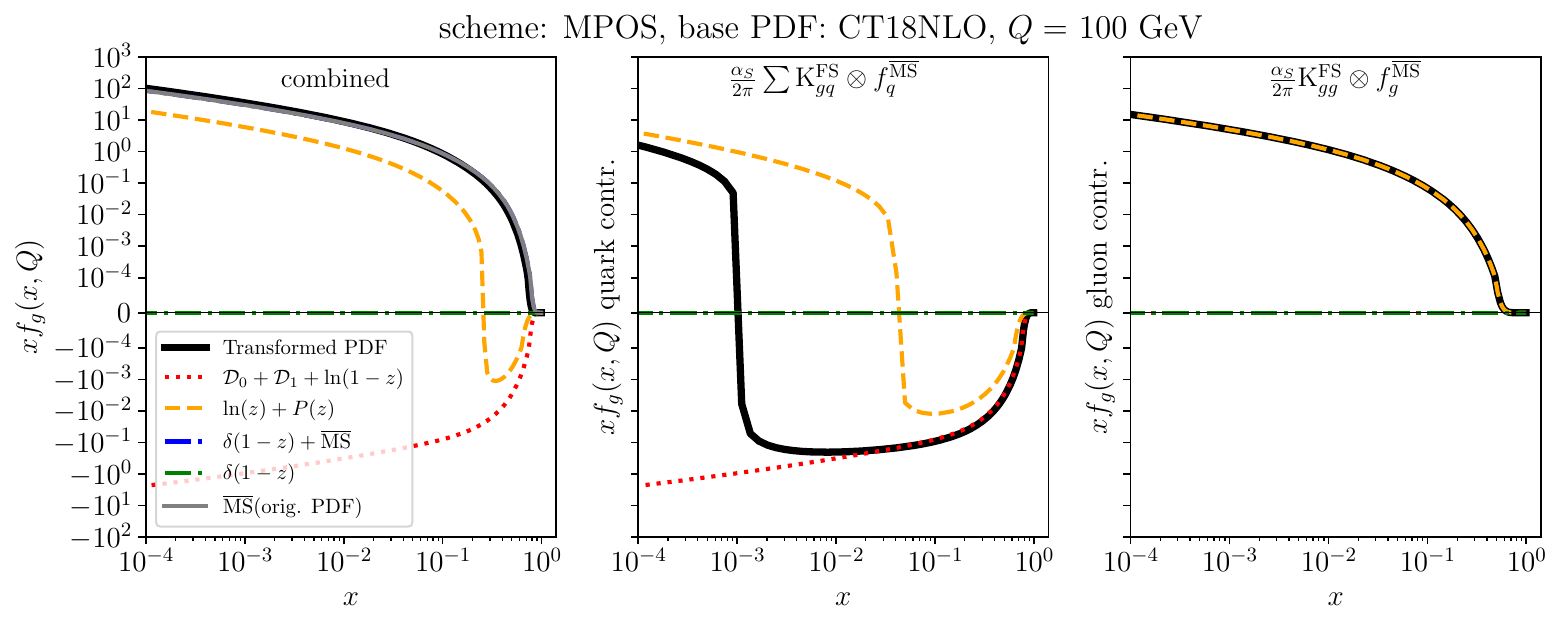}
\caption{Decomposition of transformed gluon PDF
in the \aversa, \dis and \mposd schemes at factorisation scale $Q=100$ GeV,
as described in \cref{subsec:CompPDFsDecomp}.
Companion to \cref{fig:PDF_decomposition_g}.}
\label{fig:PDF_decomposition_100_g}
\end{figure*}

\begin{figure*}[p]
\centering
\includegraphics[width=\textwidth]{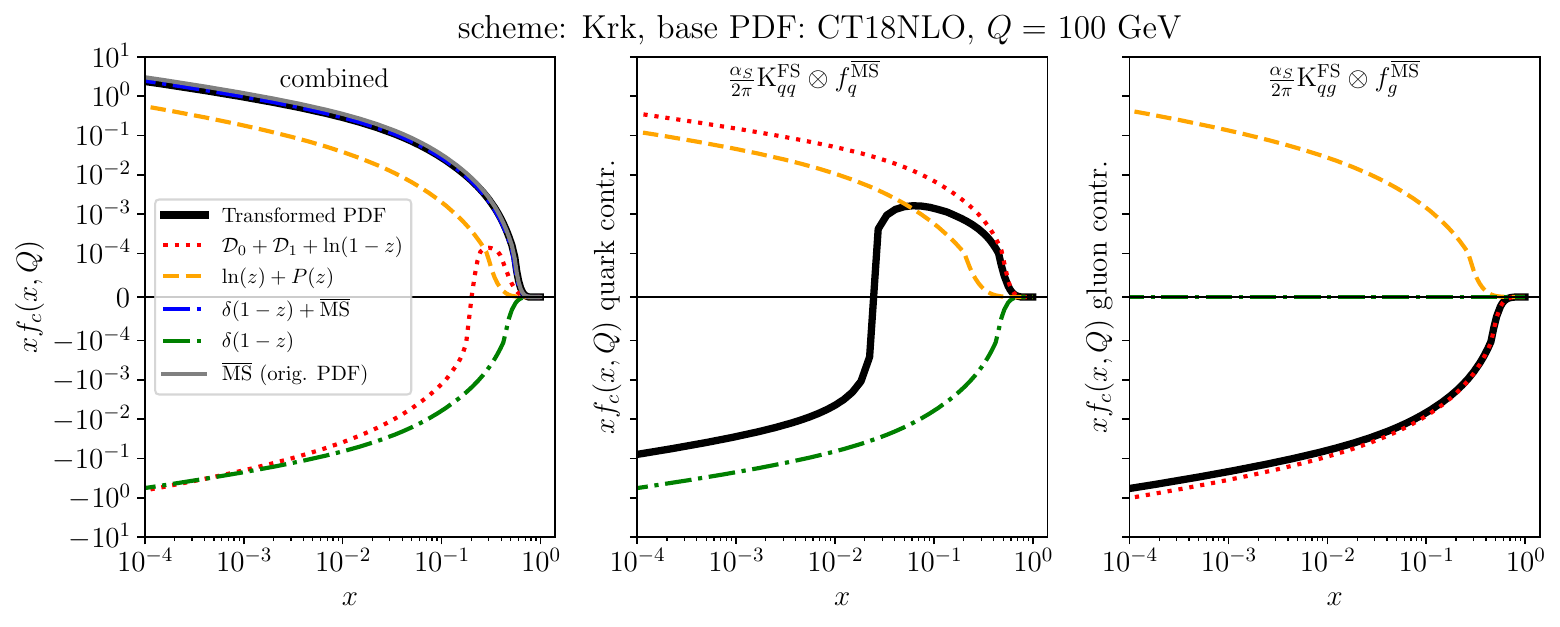}
\\
\includegraphics[width=\textwidth]{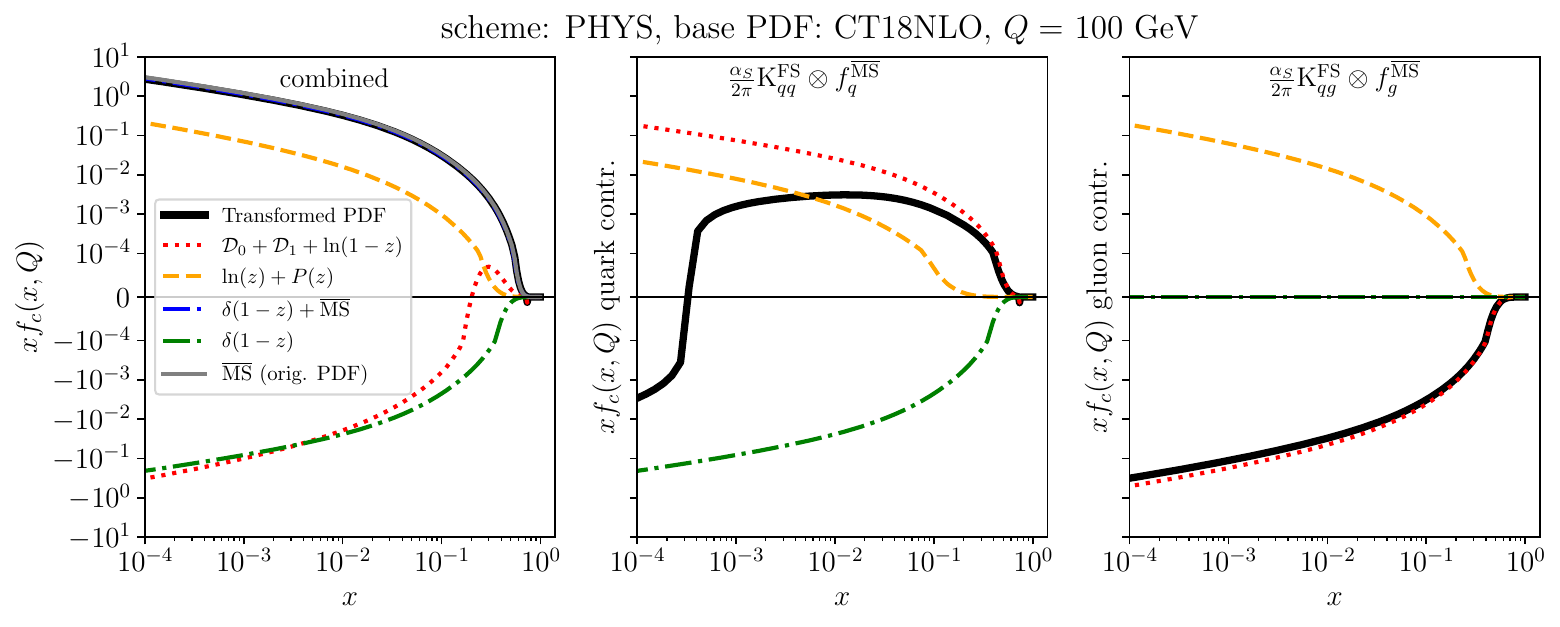}
\\
\includegraphics[width=\textwidth]{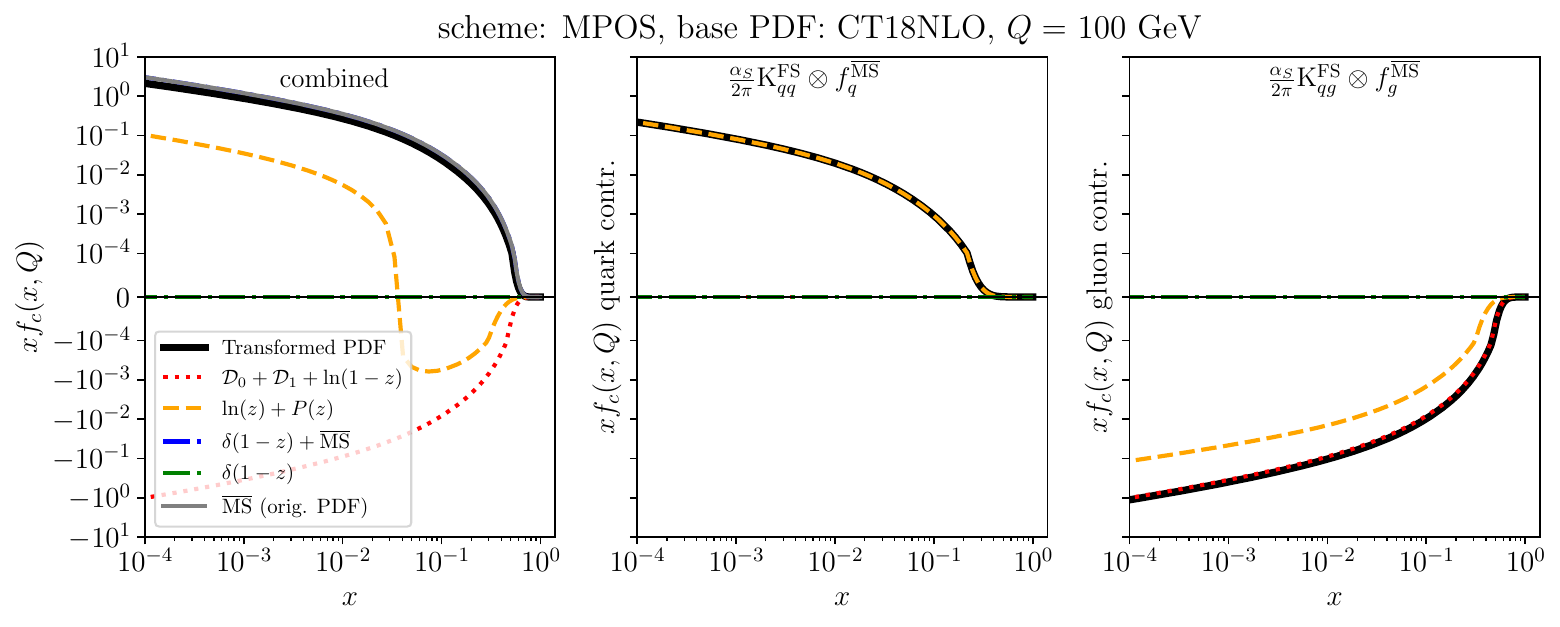}
\caption{Decomposition of transformed $c$-quark PDF as a representative of the heavy flavour PDFs,
in the \aversa, \dis and \mposd schemes at factorisation scale $Q=100$ GeV,
as described in \cref{subsec:CompPDFsDecomp}.
Companion to \cref{fig:PDF_decomposition_c}.}
\label{fig:PDF_decomposition_100_c}
\end{figure*}

In 
\cref{fig:PDF_decomposition_100_uv,fig:PDF_decomposition_100_ub,fig:PDF_decomposition_100_g,fig:PDF_decomposition_100_c}
we present
decomposition plots for the factorisation schemes shown
in the main text (\krk, \phys, \mpos),
using the \ctnlo PDF set
at $Q = 100 \; \GeV$
(in contrast with $Q = 2 \; \GeV$ used in 
\cref{fig:PDF_decomposition_uv,fig:PDF_decomposition_ub,fig:PDF_decomposition_g,fig:PDF_decomposition_c}).

\begin{figure*}[p]
\begin{leftfullpage}
\centering
\includegraphics[width=\textwidth]{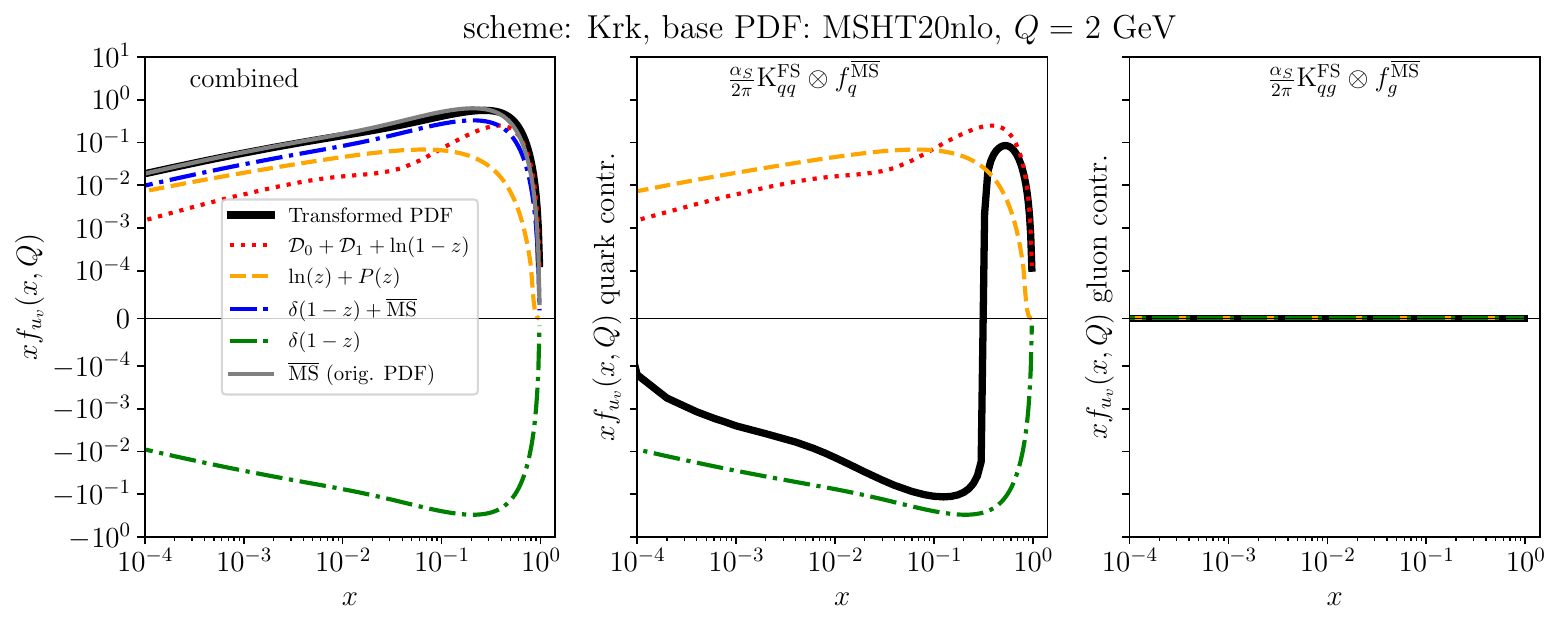}
\\
\includegraphics[width=\textwidth]{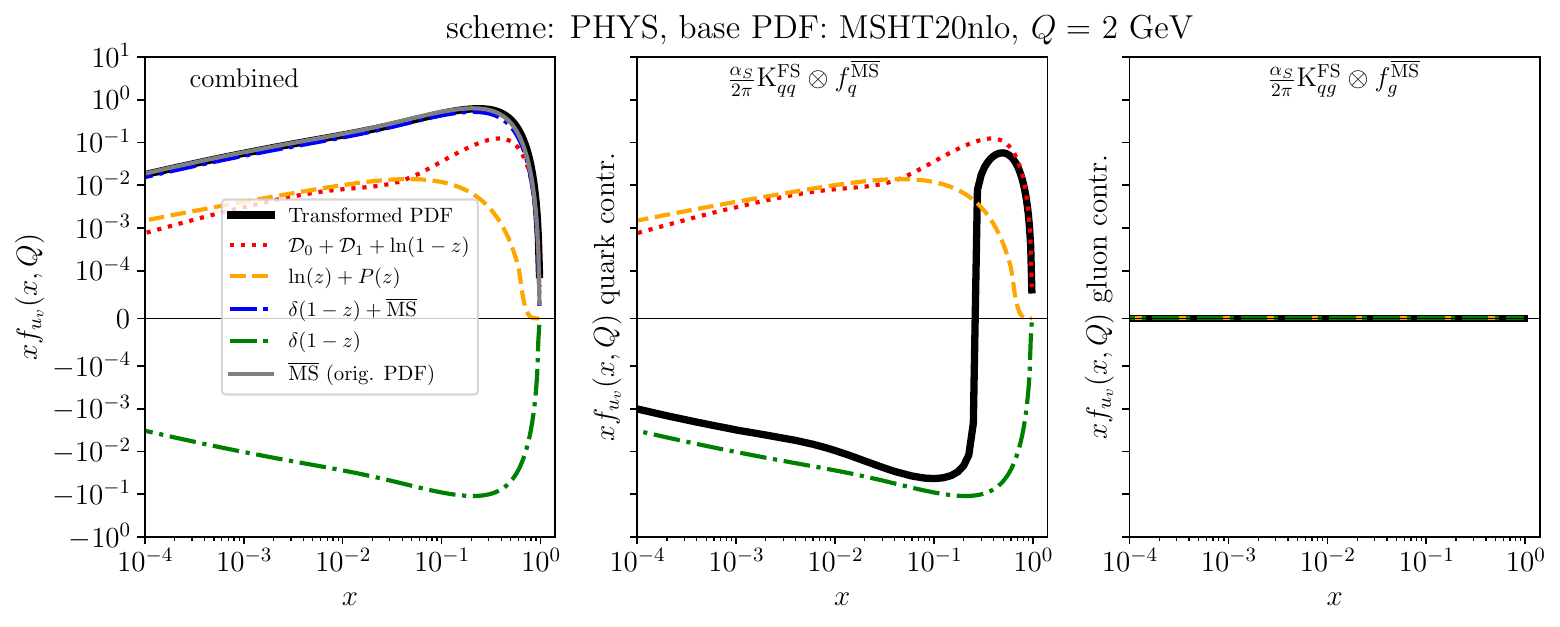}
\\
\includegraphics[width=\textwidth]{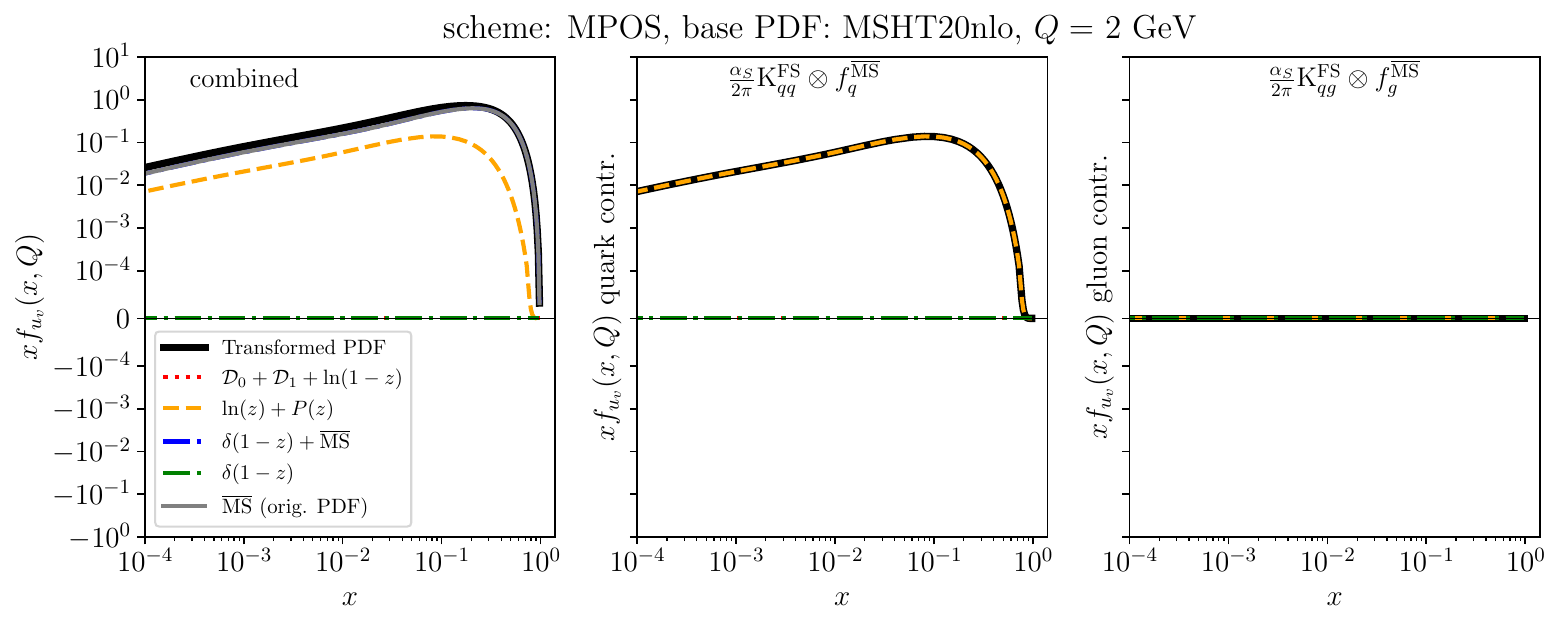}
\caption{Decomposition of transformed $u_v$-quark PDF in the \krk, \mpos, and \phys{} schemes at factorisation scale $Q=2$ GeV,
as described in \cref{subsec:CompPDFsDecomp}, based on \mshtnlo \msbar PDFs.
Companion to \cref{fig:PDF_decomposition_uv}.}
\label{fig:PDF_decomposition_MSHT_uv}
\end{leftfullpage}
\end{figure*}

\begin{figure*}[p]
\begin{leftfullpage}
\centering
\includegraphics[width=\textwidth]{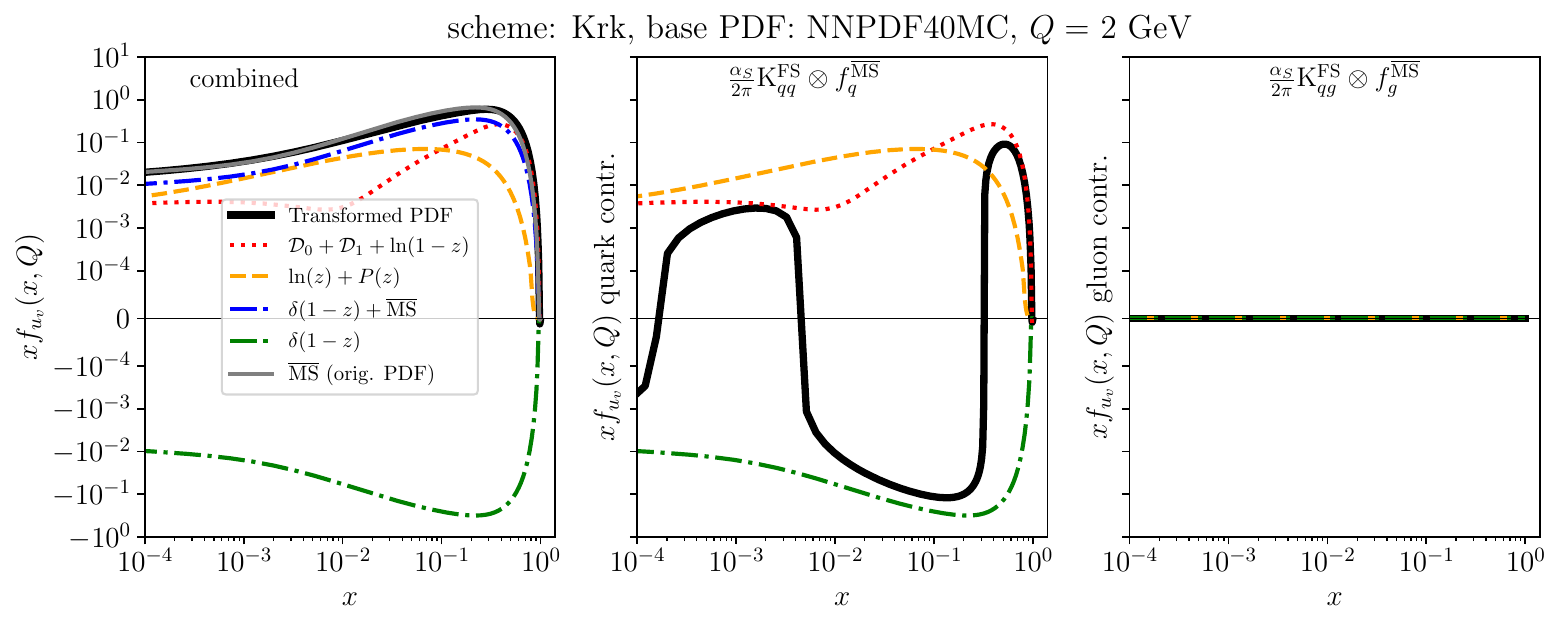}
\\
\includegraphics[width=\textwidth]{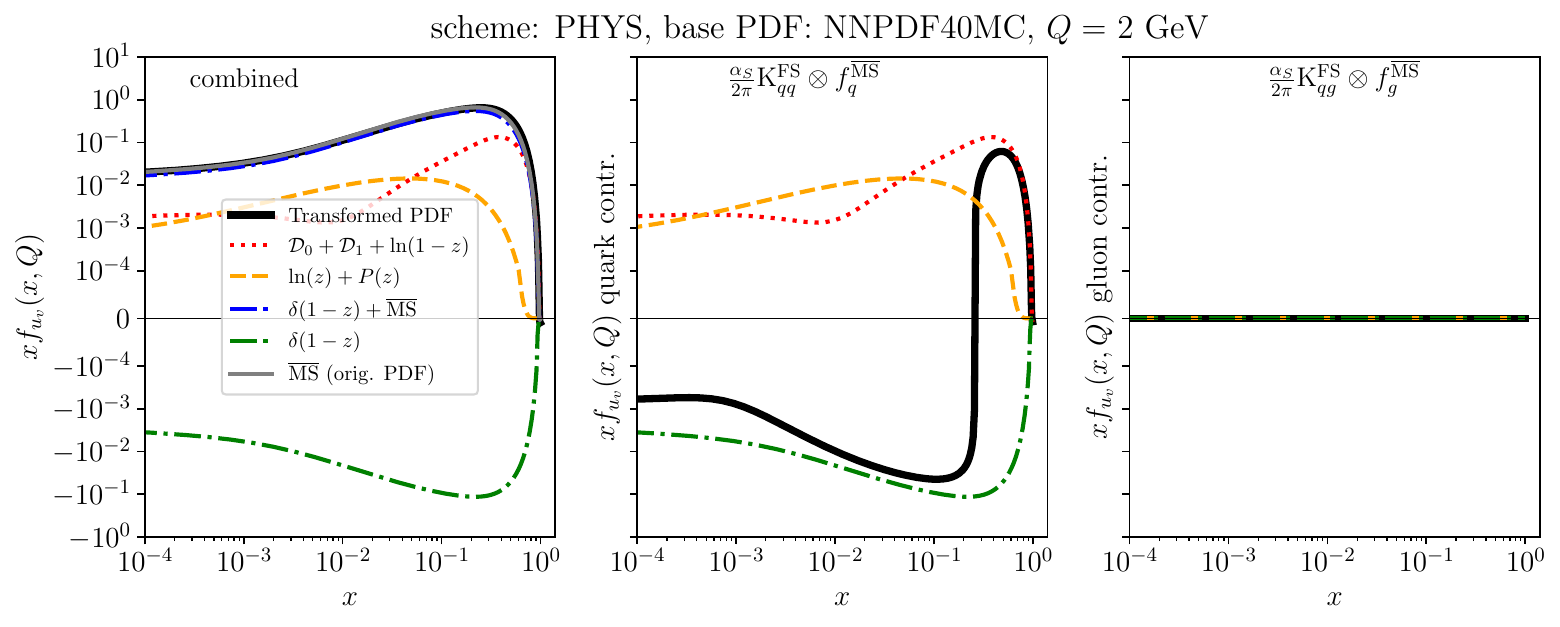}
\\
\includegraphics[width=\textwidth]{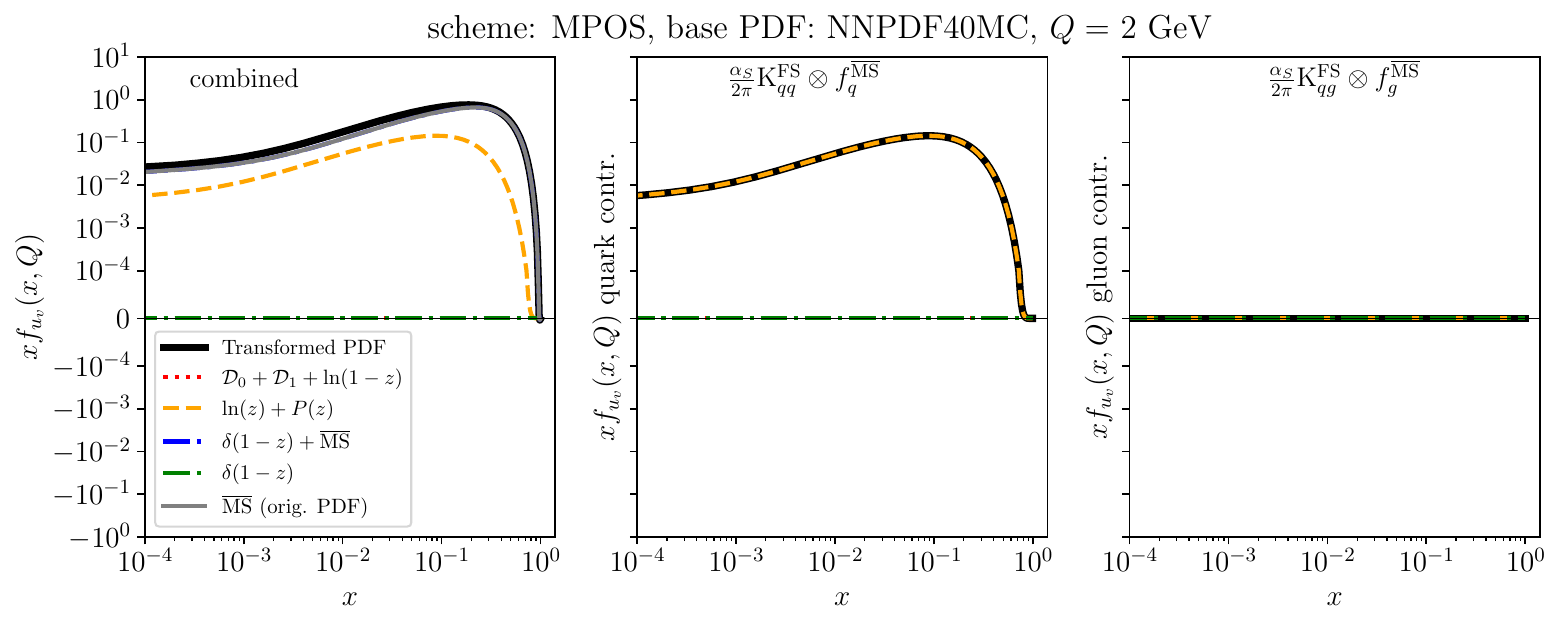}
\caption{Decomposition of transformed $u_v$-quark PDF in the \krk, \mpos, and \phys{} schemes at factorisation scale $Q=2$ GeV,
as described in \cref{subsec:CompPDFsDecomp}, based on \nnpdfmc \msbar PDFs.
Companion to \cref{fig:PDF_decomposition_uv}.
}
\label{fig:PDF_decomposition_NNPDF_uv}
\end{leftfullpage}
\end{figure*}

\begin{figure*}[p]
\centering
\includegraphics[width=\textwidth]{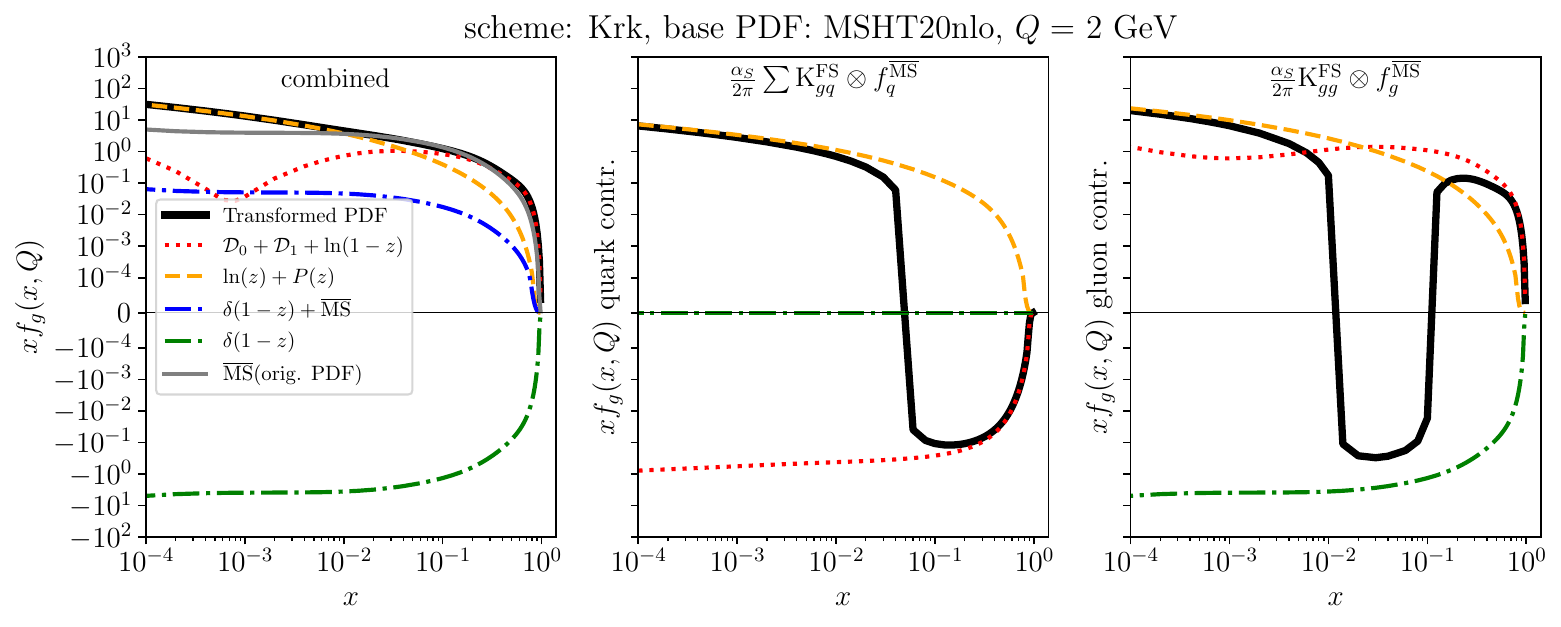}
\\
\includegraphics[width=\textwidth]{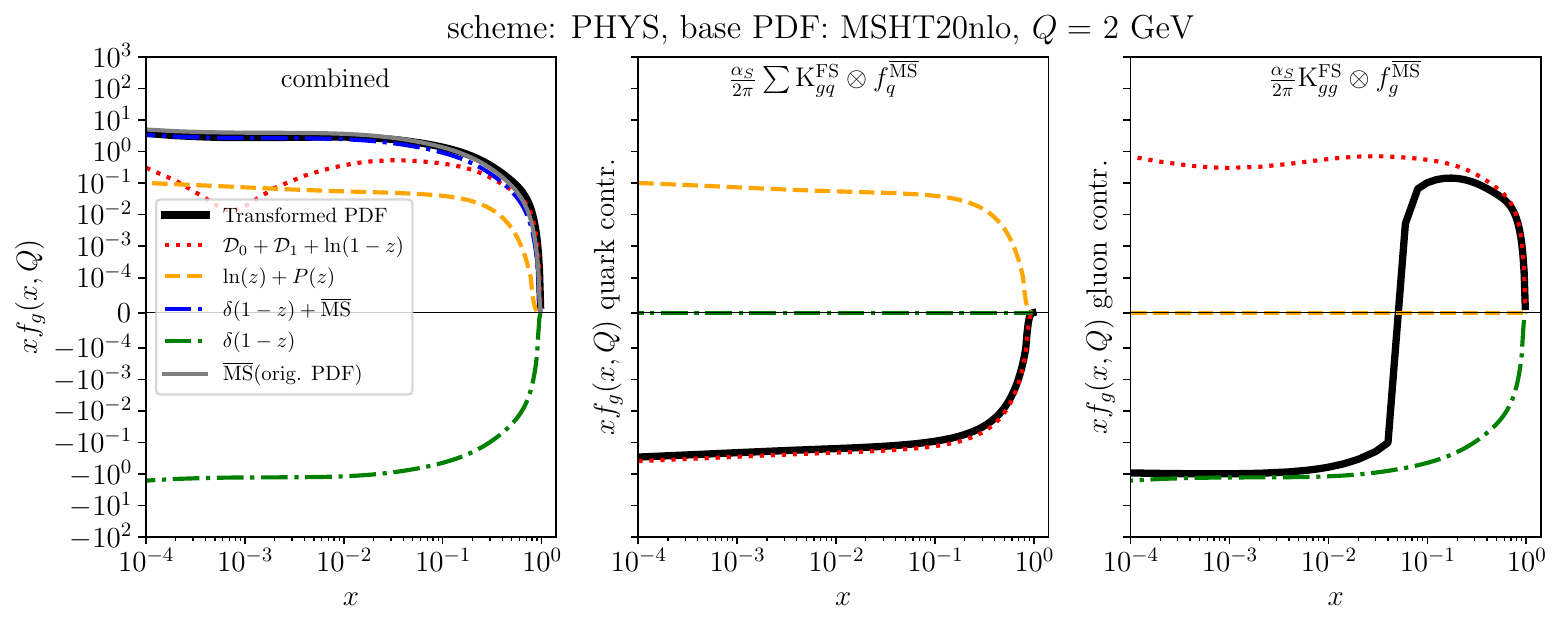}
\\
\includegraphics[width=\textwidth]{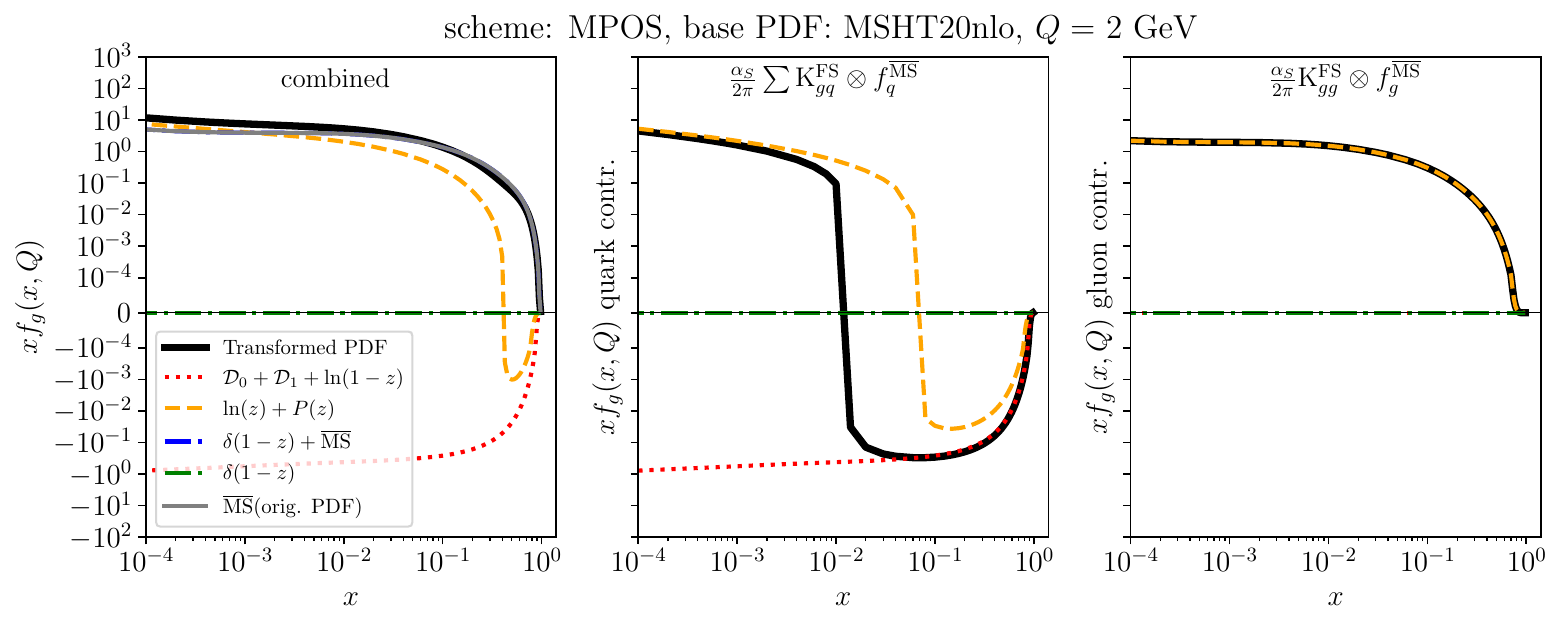}
\caption{Decomposition of transformed gluon PDF in the \krk, \mpos, and \phys schemes at factorisation scale $Q=2$ GeV,
as described in \cref{subsec:CompPDFsDecomp}, based on \mshtnlo \msbar PDFs.
Companion to \cref{fig:PDF_decomposition_g}.
}
\label{fig:PDF_decomposition_MSHT_g}
\end{figure*}

\begin{figure*}[p]
\centering
\includegraphics[width=\textwidth]{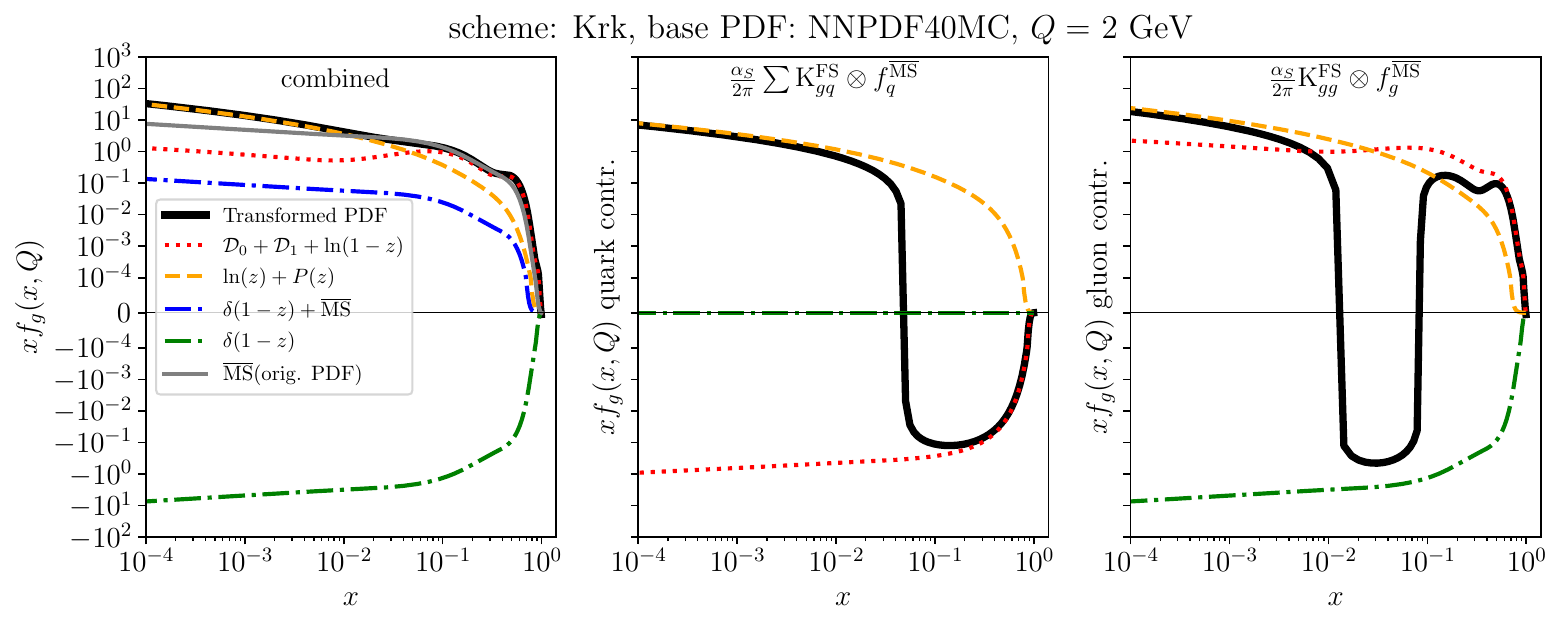}
\\
\includegraphics[width=\textwidth]{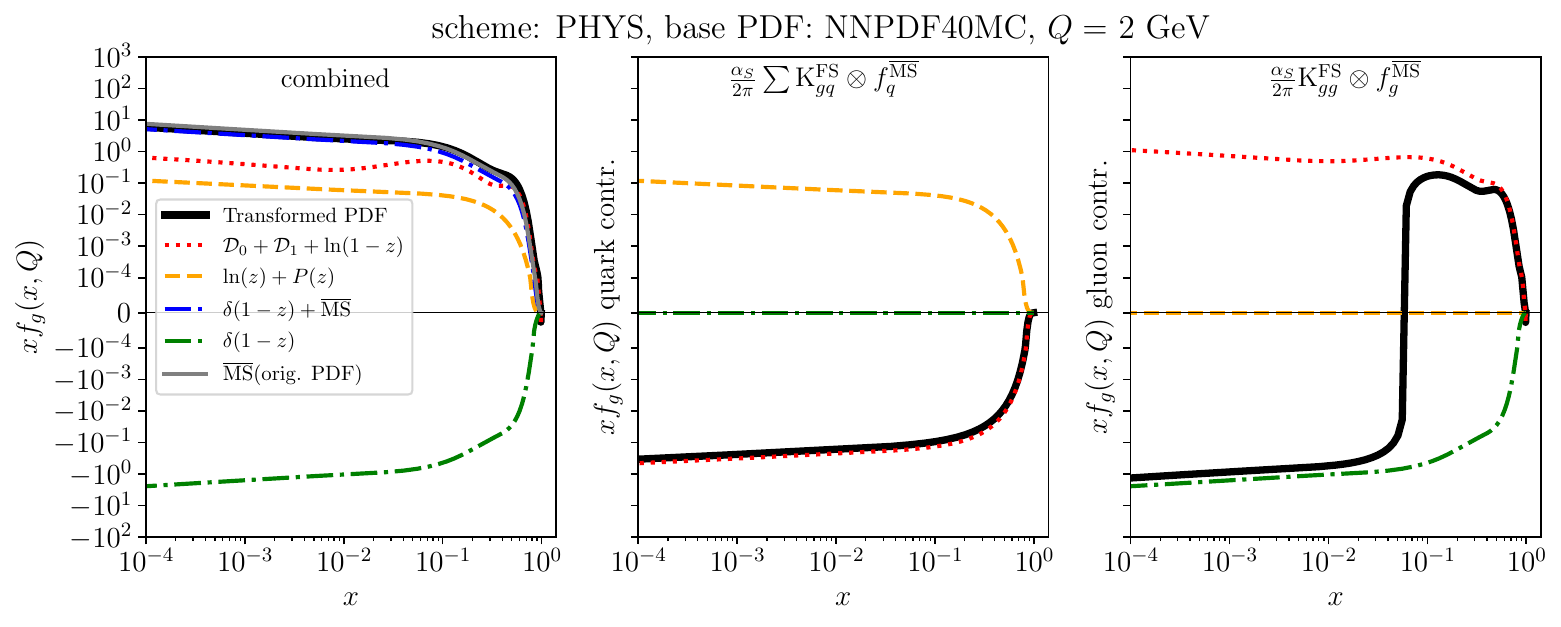}
\\
\includegraphics[width=\textwidth]{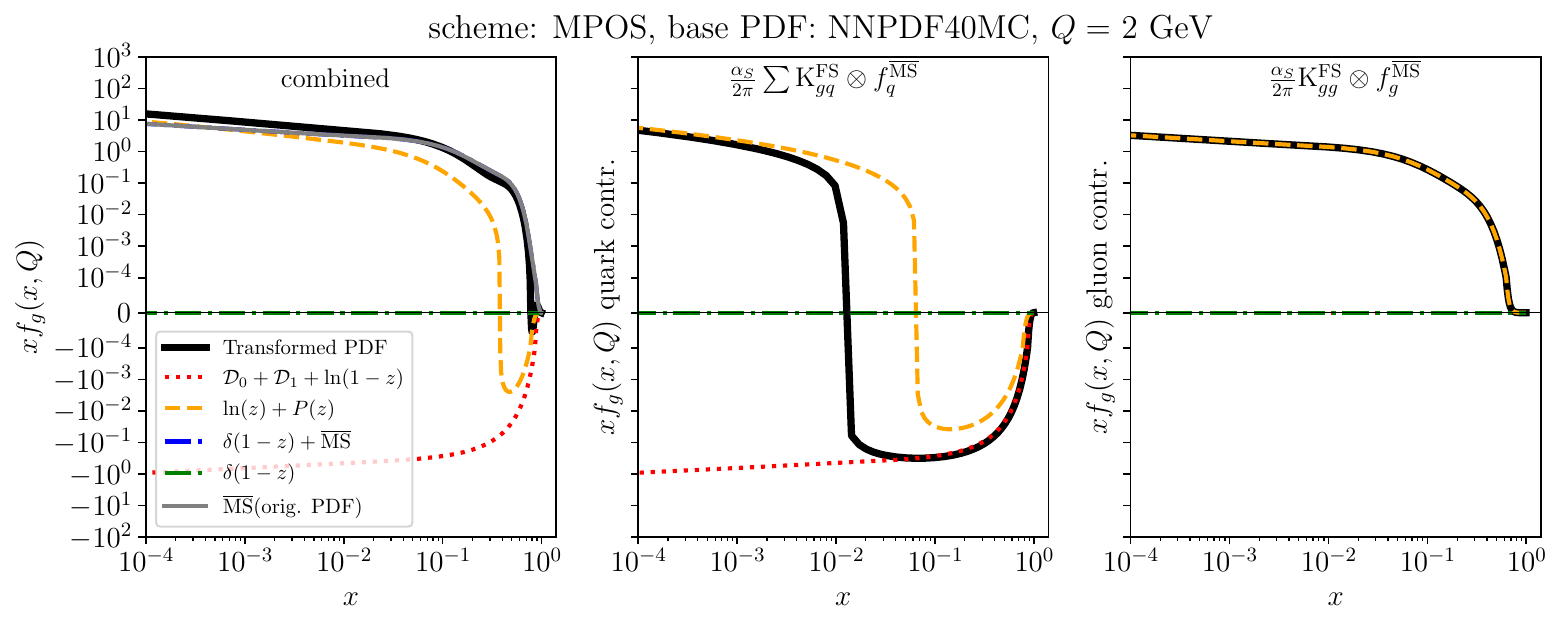}
\caption{Decomposition of transformed gluon PDF in the \krk, \mpos, and \phys schemes at factorisation scale $Q=2$ GeV,
as described in \cref{subsec:CompPDFsDecomp}, based on \nnpdfmc \msbar PDFs.
Companion to \cref{fig:PDF_decomposition_g}.
}
\label{fig:PDF_decomposition_NNPDF_g}
\end{figure*}

In
\cref{fig:PDF_decomposition_MSHT_uv,fig:PDF_decomposition_NNPDF_uv,fig:PDF_decomposition_MSHT_g,fig:PDF_decomposition_NNPDF_g}
we present
decomposition plots analogous to those shown in 
\cref{fig:PDF_decomposition_uv,fig:PDF_decomposition_ub,fig:PDF_decomposition_g,fig:PDF_decomposition_c}
in the main text, using 
\mshtnlo and \nnpdfmc in place of \ctnlo,
at $Q = 2 \; \GeV$.

\begin{table*}[tph]
	\begin{center}
		\begin{tabular}{ c|c|c|c } 
            Flavour
			& Scale $Q$ (GeV)
			& Factorisation scheme(s)
			& PDF set/decomposition plots
			\\ 
			\hline
            \multirowcell{2}{up-valence $u_v$ \\            (\cref{fig:PDF_compare_uval,fig:xfuxQ_MSHT20nlo,fig:xfuxQ_NNPDF40MC}) }
            & 2
            & \krk, \phys, \mpos
            & \texttt{CT18} (\cref{fig:PDF_decomposition_uv}),
              \texttt{MSHT20} (\cref{fig:PDF_decomposition_MSHT_uv}), 
              \nnpdfmc (\cref{fig:PDF_decomposition_NNPDF_uv})
            \\ &
            & \aversa, \dis, \mposd
            & \texttt{CT18} (\cref{fig:PDF_decomposition_uv_2})
            \\
            & 100
            & \krk, \phys, \mpos
            & \texttt{CT18} (\cref{fig:PDF_decomposition_100_uv})
			\\ 
            \multirowcell{2}{up-antiquark $\bar{u}$ (sea) \\            (\cref{fig:PDF_compare_ub,fig:xfubxQ_MSHT20nlo,fig:xfubxQ_NNPDF40MC}) }
            & 2
            & \krk, \phys, \mpos
            & \texttt{CT18} (\cref{fig:PDF_decomposition_ub})
            \\ &
            & \aversa, \dis, \mposd
            & \texttt{CT18} (\cref{fig:PDF_decomposition_ub_2})
            \\
            & 100
            & \krk, \phys, \mpos
            & \texttt{CT18} (\cref{fig:PDF_decomposition_100_ub})
			\\ 
            \multirowcell{2}{gluon \\            (\cref{fig:PDF_compare_g,fig:xfgxQ_MSHT20nlo,fig:xfgxQ_NNPDF40MC})}
            & 2
            & \krk, \phys, \mpos
            & \texttt{CT18} (\cref{fig:PDF_decomposition_g}),
              \texttt{MSHT20} (\cref{fig:PDF_decomposition_MSHT_g}),
              \nnpdfmc (\cref{fig:PDF_decomposition_NNPDF_g})
            \\ &
            & \aversa, \dis, \mposd
            & \texttt{CT18} (\cref{fig:PDF_decomposition_g_2})
            \\
            & 100
            & \krk, \phys, \mpos
            & \texttt{CT18} (\cref{fig:PDF_decomposition_100_g})
			\\ 
            \multirowcell{2}{
            charm (heavy) \\            (\cref{fig:PDF_compare_c,fig:xfcxQ_MSHT20nlo,fig:xfcxQ_NNPDF40MC}) }
            & 2
            & \krk, \phys, \mpos
            & \texttt{CT18} (\cref{fig:PDF_decomposition_c}),
            \nnpdf (\cref{fig:PDF_decomposition_NNPDF40_c})
            \\ &
            & \aversa, \dis, \mposd
            & \texttt{CT18} (\cref{fig:PDF_decomposition_c_2})
            \\
            & 100
            & \krk, \phys, \mpos
            & \texttt{CT18} (\cref{fig:PDF_decomposition_100_c})
			\\
        \end{tabular}
    \end{center}
    \caption{Overview of all factorisation-scheme decomposition plots.}
	\label{tab:decompPlots}
\end{table*}

\section{Test of perturbative inversion}
\label{app:pert_inversion}

In this appendix we test numerically the perturbative inversion of each
transformation kernel, as summarised in \cref{subsubsec:fstransformations}.
For each scheme, we apply the perturbative inverse transformation
to the PDF obtained in that scheme by transformation from \msbar.
An exact inverse would return the input \msbar PDF;
since we calculate the inverse perturbatively
we expect to be discarding terms of order $\order{\alphas^2}$.
However, the logarithmic terms in the transformation kernels
are large and so may in principle not be amenable to perturbative inversion
\cite{Candido:2023ujx}. 

\begin{figure*}[p]
\centering
\begin{subfigure}[t]{0.49\textwidth}
    \centering
    \includegraphics[width=\textwidth]{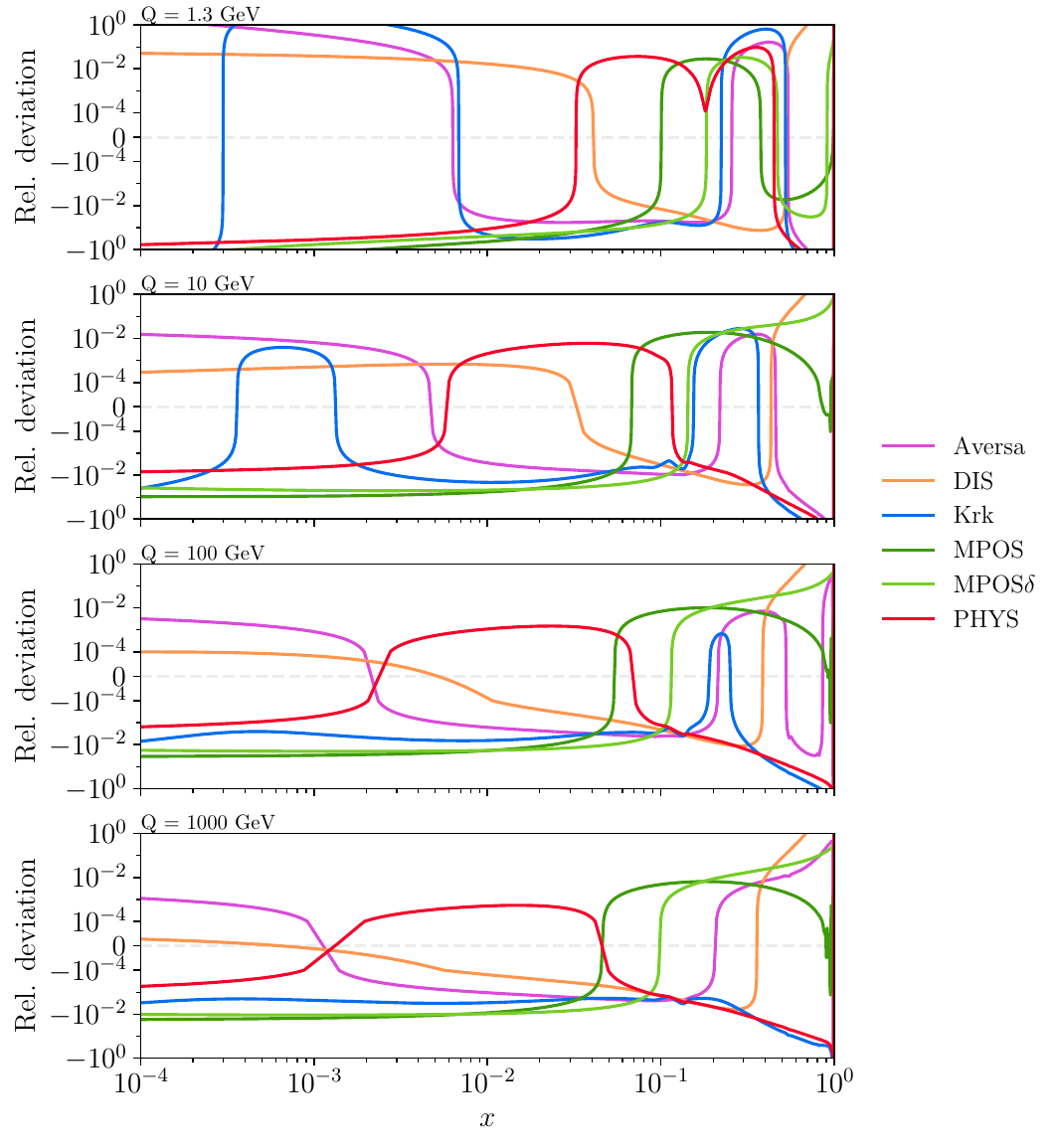}
    \caption{gluon distribution}
\end{subfigure}
\begin{subfigure}[t]{0.49\textwidth}
    \centering
    \includegraphics[width=\textwidth]{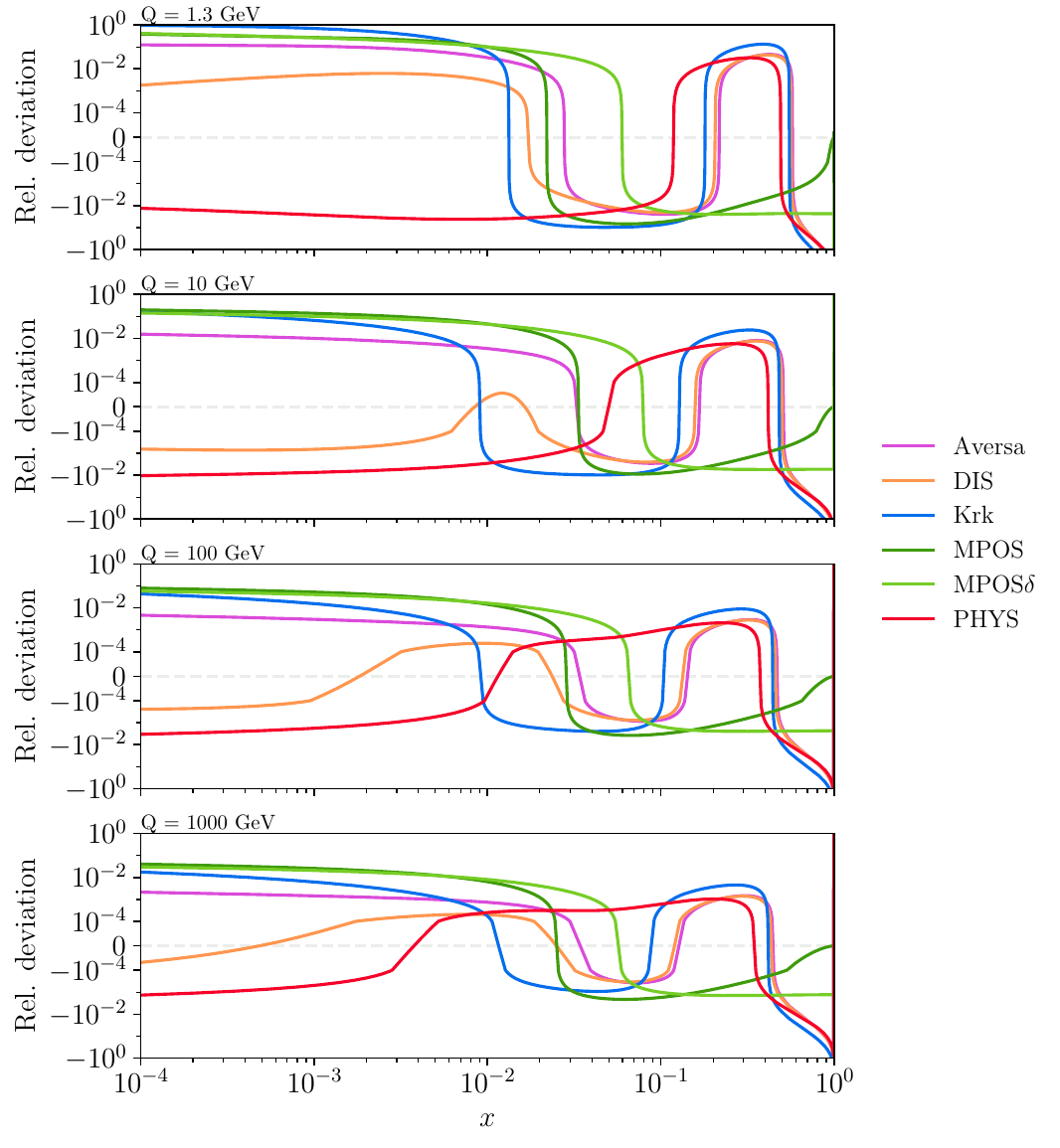}
    \caption{up-quark distribution}
\end{subfigure}
\begin{subfigure}[t]{0.49\textwidth}
    \centering
    \includegraphics[width=\textwidth]{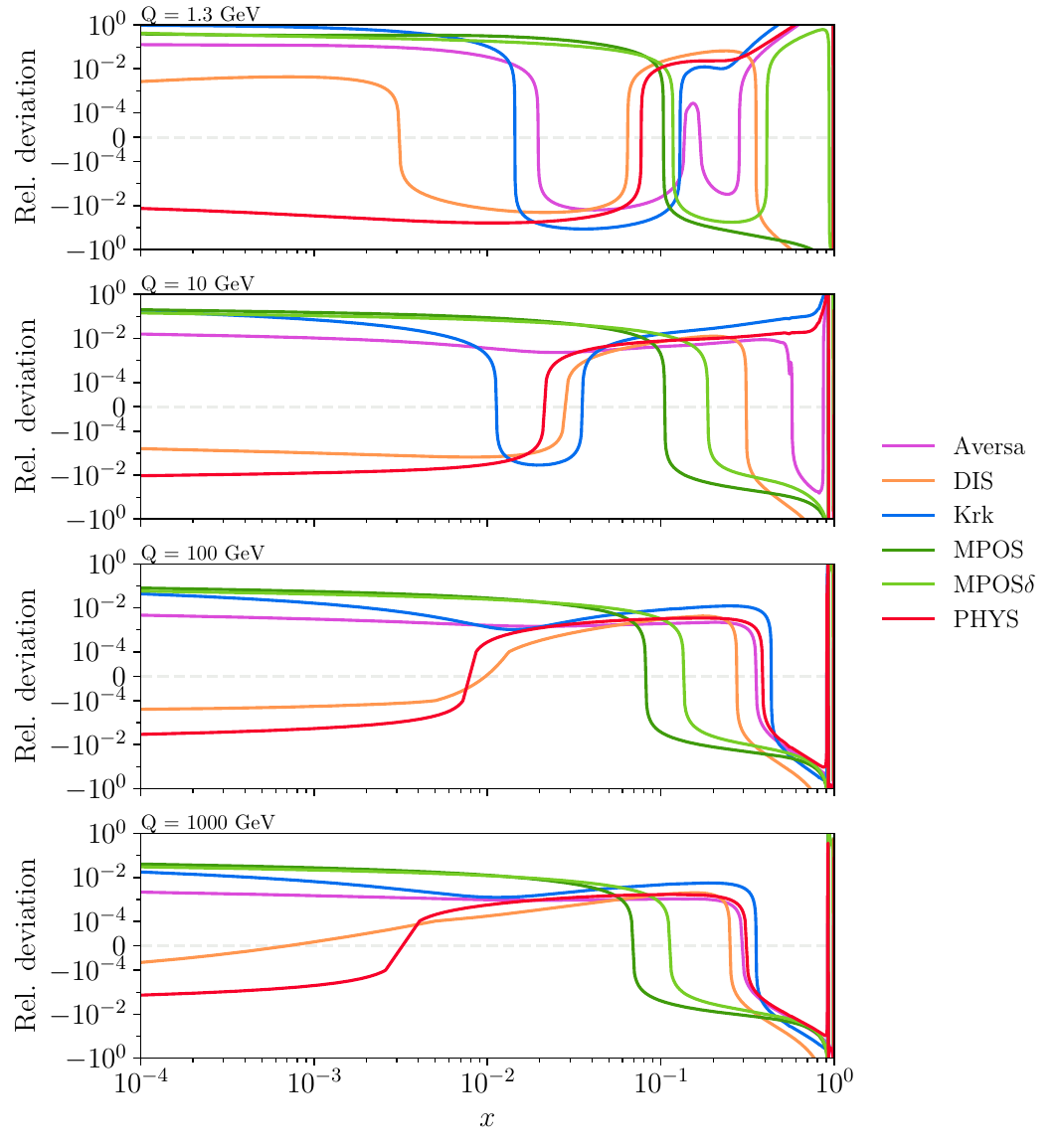}
    \caption{up anti-quark distribution}
\end{subfigure}
\begin{subfigure}[t]{0.49\textwidth}
    \centering
    \includegraphics[width=\textwidth]{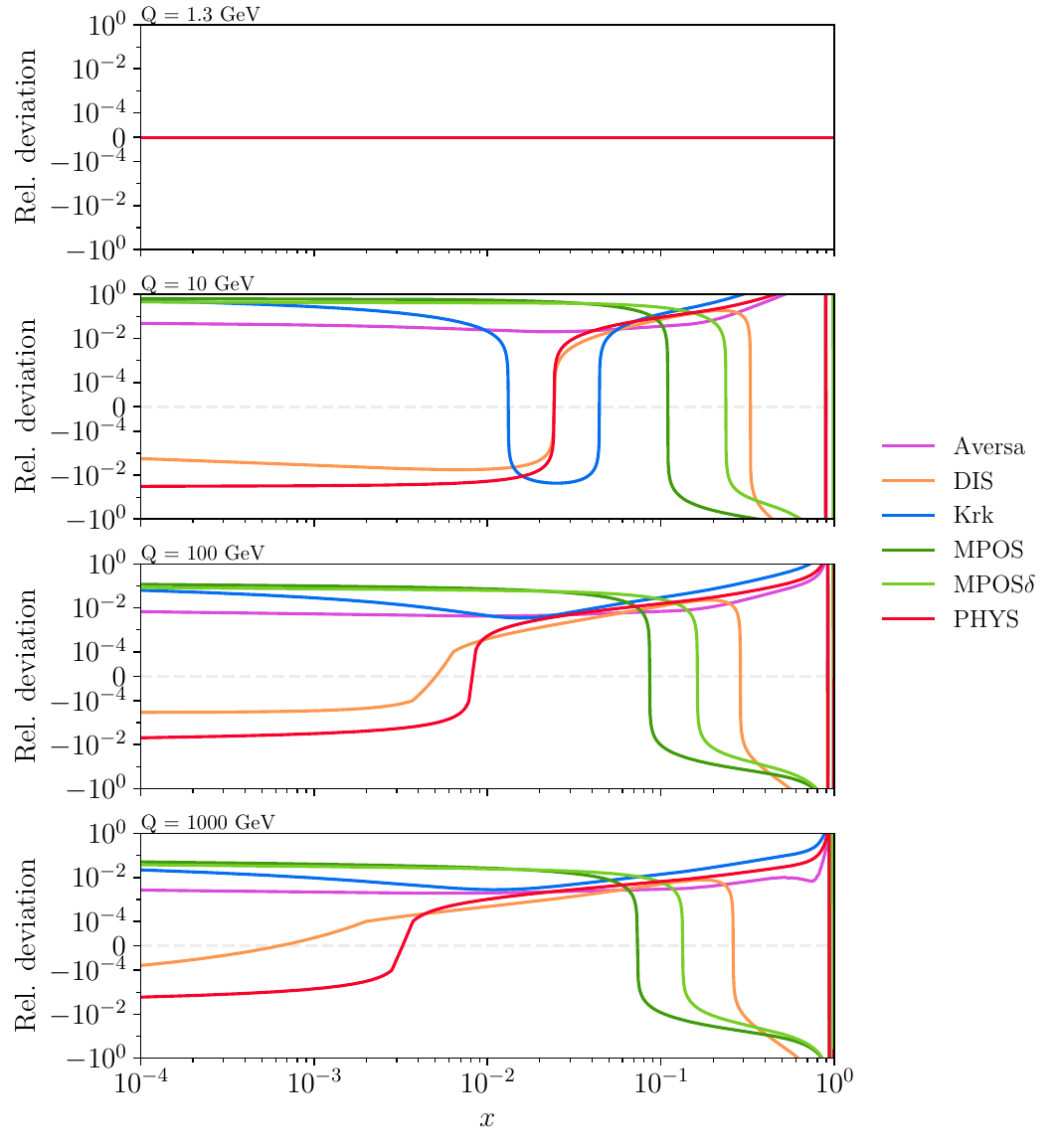}
    \caption{bottom-quark distribution}
\end{subfigure}
\caption{Test of the perturbative invertibility of the transformation kernels
as outlined in \cref{app:pert_inversion}.
For each scheme we plot the relative deviation from unity
of the PDF obtained by
performing the transformation into the displayed scheme, followed by
its perturbative inverse, relative to the original \msbar PDF.
An exact inversion would correspond to a uniform ratio of 1, so a 
relative deviation of 0.
This plot uses the \ctnlo PDF set.
}
\label{fig:pert_inv_test}
\end{figure*}

We show the relative deviation of the resulting PDFs for each
of the schemes considered in \cref{tab:Kqq,tab:Kqg,tab:Kgq,tab:Kgg}
in \cref{fig:pert_inv_test}.
As expected, at low scales the perturbativity of the transformation
breaks down and the error of the perturbative inversion is large.
At higher scales the error is within approximately 1\% for
all schemes.

\section{Validation}
\label{sec:app_validation}

In this section we summarise the technical details of the numerical codes used to produce the results presented throughout this paper.
The convolution codes used to produce the PDF grids for the various factorisation schemes were written in \texttt{python}. In order to implement a robust validation strategy,
two independent codes were developed, using different libraries, the first based on
\texttt{numpy} \cite{harris2020array} and \texttt{scipy} \cite{2020SciPy-NMeth}
for fast computation with floating-point precision,
and the second using 
\texttt{sympy} \cite{10.7717/peerj-cs.103}
for exact symbolic algebra
and \texttt{mpmath} \cite{mpmath}
for numerical calculation with arbitrary-precision arithmetic.

The implementation using \texttt{scipy} and \texttt{numpy} computes the 
convolutions using the \texttt{quad} routine, with a target maximal absolute
error of $10^{-4}$. The integrals that appear in 
\cref{eq:fstransformations_faFS} are not computed up to $1$,
but rather to $1 - 10^{-8}$ to avoid numerical errors due to endpoint-singularities,
which was found to give accurate results. Other values of cut-off (both lower and greater)
were also tested but they lead to less accurate results.

The second implementation uses symbolic representations of the kernels
within \texttt{sympy}, which may be converted automatically
to \texttt{numpy} or \texttt{mpmath} functions for numerical calculations.
The symbolic manipulation of the kernels has been used to
check the coefficient function results given in \cref{tab:CoDIS,tab:CoDY,tab:CoHiggs},
and to produce the plots of their Mellin transforms shown in \cref{fig:NspaceC}.
For numerical calculations shown here \texttt{mpmath}
was used with 200 digits of targeted decimal precision,
with the aim of eliminating
any numerical imprecision due to `loss-of-significance' cancellation errors
or the imperfect convergence of numerical integration routines applied
to pathological integrands.
These were written into \lhapdf output files,
using the interpolation knots of the input PDF,%
\footnote{Where the input PDF used a single grid we have
split the grid at the heavy-flavour thresholds to accommodate
the anticipated discontinuities in the transformed PDFs, as outlined
in \cref{subsubsec:flavthresh}.}
which were used for subsequent plotting .
Where input PDFs are taken from \lhapdf the interpolation of the input
PDFs is performed using floating-point precision, so the result
may be considered to correspond to the exact convolution
to the numerical accuracy of the input (interpolated) PDF
with no loss of precision introduced by the transformation.

In both implementations, the \krk-scheme transformations were checked against the original \texttt{C++} code used for
\cite{Jadach:2015mza,Jadach:2016acv,Jadach:2016qti}
which highlighted the possible sensitivity of the convolutions to the 
numerical integration routine.
This comparison helped to benchmark the convolution codes,
as well as tune the parameters in the \texttt{scipy} implementation.
The convolution codes were checked against each other by comparing the outputs 
for the parton distribution functions, kernel convolutions and momentum sum rules. 
The discrepancy between the results obtained from the two different implementations
were found to be of the order of $10^{-5}$.

The \phys-scheme transformations of the \texttt{MSTW2008}
\cite{Martin:2009iq}
NLO PDF set displayed in \cite{Oliveira:2013aug}
have been repeated using our code and the results
compared graphically against the results displayed there.

\bibliographystyle{utphys_spires}
\bibliography{refs.bib}

\end{document}